\documentclass[oneside]{book}
\usepackage[utf8]{inputenc}
\usepackage{authblk, tocloft, hyperref, tikz, float, url, fancyhdr, marvosym, enumerate, braket, tikzsymbols}

\usepackage{xcolor}
\hypersetup{
    colorlinks,
    linkcolor={red!50!black},
    citecolor={blue!50!black},
    urlcolor={blue!80!black}
}

\usepackage{amssymb, amsmath, dsfont, multicol, appendix}

\usepackage[framemethod = TikZ]{mdframed}

\usepackage{subfig}

\usepackage{graphicx, pdfpages}

\usepackage[sectionbib]{chapterbib}

\usepackage[paperwidth = 8in, paperheight = 10in]{geometry}

\newcommand{\txte}{\textrm{e}}
\newcommand{\grad}{\nabla}
\newcommand{\tderiv}[2][s]{\frac{\textrm{d} #2}{\textrm{d} #1}}
\newcommand{\pderiv}[2][s]{\frac{\partial #2}{\partial #1}}

\newcommand{\figref}[1]{Fig. \ref{#1}}
\newcommand{\equaref}[1]{Eq. \ref{#1}}
\newcommand{\tblref}[1]{Table \ref{#1}}

\title{\huge\textbf{Tunneling Through the Math Barrier} \\ \vspace{0.3in}
\large The Fledgling Physics Student's Field Guide to Essential Mathematics}
\author{\large\vspace{-0.15in}W. J. Meese}
\date{\large\vspace{-0.15in}May 2018}

\usepackage[variablett]{lmodern}
\newcommand*{\Button}[1]{%
  \Acrobatmenu{#1}{\fbox{\texttt{#1}}}%
}
\newcommand*{\Navigation}{
    \Button{GoBack}
}

\definecolor{Lavender}{RGB}{193,120,205}

\newlistof{example}{ex}{\listexamplename}

\newcommand{\listexamplename}{List of Examples}

\renewcommand{\theexample}{\arabic{chapter}.\arabic{example}}

\newenvironment{example}[2][]{%
    \refstepcounter{example}
    \addcontentsline{ex}{example}{\protect{$\;$\theexample$\;\;$ #1}}
 
    \ifstrempty{#1}%
{\mdfsetup{%
    frametitle={%
        \tikz[baseline=(current bounding box.east),outer sep=0pt]
        \node[anchor=east,rectangle,fill=Lavender]
        {\strut Example~\theexample};}
    }%
}{\mdfsetup{%
    frametitle={%
        \tikz[baseline=(current bounding box.east),outer sep=0pt]
        \node[anchor=east,rectangle,fill=Lavender]
        {\strut Example~\theexample:~#1};}%
    }%
}%
\mdfsetup{%
    innertopmargin=10pt,linecolor=Lavender,%
    linewidth=2pt,topline=true,%
    frametitleaboveskip=\dimexpr-\ht\strutbox\relax%
}

\begin{mdframed}[]\relax%
\label{#2}}{\end{mdframed}}

\newcommand{\exref}[1]{Example \ref{#1}}

\definecolor{McGill_Soft_Red}{RGB}{232,137,130}

\newlistof{problem}{prob}{\listproblemname}

\newcommand{\listproblemname}{List of Problems}

\renewcommand{\theproblem}{\arabic{chapter}.\arabic{problem}}

\newenvironment{problem}[2][]{%
    \refstepcounter{problem}
    \addcontentsline{prob}{problem}{\protect{$\;$\theproblem$\;\;$ #1}}
 
    \ifstrempty{#1}%
{\mdfsetup{%
    frametitle={%
        \tikz[baseline=(current bounding box.east),outer sep=0pt]
        \node[anchor=east,rectangle,fill=McGill_Soft_Red]
        {\strut Problem ~\theproblem};}
    }%
}{\mdfsetup{%
    frametitle={%
        \tikz[baseline=(current bounding box.east),outer sep=0pt]
        \node[anchor=east,rectangle,fill=McGill_Soft_Red]
        {\strut Problem ~\theproblem:~#1};}%
    }%
}%
\mdfsetup{%
    innertopmargin=10pt,linecolor=McGill_Soft_Red,%
    linewidth=2pt,topline=true,%
    frametitleaboveskip=\dimexpr-\ht\strutbox\relax%
}

\begin{mdframed}[]\relax%
\label{#2}}{\end{mdframed}}

\newcommand{\probref}[1]{Problem \ref{#1}}

\begin{document}

\renewcommand{\figurename}{FIG.}
\renewcommand{\tablename}{TABLE}

\maketitle

\clearpage
\thispagestyle{plain}
\par\vspace*{.35\textheight}{\centering \emph{For Kyo, who got the ball rolling (hopefully without too much slipping on my part).}\par}

\cleardoublepage
\phantomsection
\chapter*{Acknowledgements}
\addcontentsline{toc}{chapter}{Acknowledgements}
\thispagestyle{plain}
There are a few people whom I definitely need to acknowledge and thank as their help and support made this project possible. First and foremost, I need to thank the Internet and the spirit of Open Source projects.  I definitely would not have been able to learn nearly as much over the years without information being available to me in the giant database that it currently is in.  Being able to come up with questions and research and then think over possible solutions helped make this project possible, and I would be remiss to ignore this fact. Thus, I would also like to thank everyone who continues to make the Internet the a great tool for learning that it currently is.

Secondly, I would like to thank Professor Peter Persans (Rensselaer Polytechnic Institute) for taking the time to edit this work and provide helpful suggestions to improve its utility. I don't know if he quite knew what he was agreeing to by signing on to be my Master's Project adviser, and so I really do appreciate all the extra time he had to invent to thoroughly read my work.  Additionally, I would like to thank him for distributing snippets of this book as it was being written to his Quantum Physics II and Honors Physics II classes so I might get helpful feedback from those I aim to help. Luckily for me, Linus Koepfer from the Honors Physics II $\beta$-tested the chapter on Complex Algebra and brought a bunch of my typos to my attention, and so I would also like to thank him as well for finding the time during the semester to read my work. 

I would also like to thank my sister, Morgan Meese, D.P.T., for her help with brainstorming ideas for my title.  She also brought up a pretty important point to me that for a lot of people, this book will be a shock because, as she says, it is \textit{Math Without Numbers}. Without this reminder, I definitely would have forgotten to make mention of how infrequently I use numbers, so that you, my reader, is not too taken aback by their absence. 

Additionally, I'd like to thank my parents, Colleen and Bill Meese, and my sister, Katee Meese, for their support throughout the semester and for their recommendations of who my target audience could be. Their thoughts and helpful discussions, particularly early on, definitely helped me collect all of my incoherent thoughts out of the ether so that they could be written down into book form.

Nicholas Smieszek and Erika Nelson also deserve a shout-out for being incredibly patient friends and roommates while I was almost always on my computer typing away. I know that I would often forget conversations we would have because my mind was on this project and my brain only has a finite RAM, and so I really do appreciate how supportive of me and this project they were throughout the semester.

Last but not least, I would like to thank Jennifer Freedberg for everything that she contributed to this project. First off, she was the one to originally encourage me to propose such a project. As I began to write this book, she would kindly remind me to cut down on my rambly-ness and get to the point. She also brainstormed many useful examples or would check my work when I thought I had a cool way of deriving something. She also had to listen to me blabber about this book for four months which definitely could not have always been easy.

\newpage

\chapter*{README.tex}
\addcontentsline{toc}{chapter}{README.tex}

When I started my undergrad in 2014, I expected that there would be some potential barrier for me to overcome between high school and college as far as physics goes, but I didn't think that there would be one for math.  I thought, if anything, I'd be ahead of the game since I came into college already placing out of Calc II. To be clear, coming into college I might have known a lot of math through the Calc II level, but I did not think about it in the way a physicist does. The way in which physicists think about things is weird, like I mean \textit{really} out there.  Or at least that was the way it seemed to me as a freshman physics student.  The first striking thing was that physicists, at least those I had just met, do \textit{not} remember a lot of formulas for things. In fact, they rarely remember any unless they are immediately relevant to the problem at hand.  As I was just coming from the world of AP classes, this realization was heretical and confusing.  But the more shocking thing, even without remembering all of those formulas, physicists could still \textit{smoothly} move from one mathematical idea to the next \textit{effortlessly}.  It was as if I were watching a physicist \textit{speak} mathematics fluently. Sure they may forget a word or two here or there, but it was no different than when I had talked to someone in the past and inserted an \textquotedblleft umm...\textquotedblright$\,$ or \textquotedblleft uhh...\textquotedblright$\,$ here or there.  So it was pretty clear to me that the way I had learned to think about math was wrong.  The physicists I saw did not work within the confines of the rules I had so carefully followed to get decent AP scores in high school calculus.  They used what formulas they knew, but instead of working within some predetermined, College-Board-approved template, they translated their thoughts into equations.

My goal with this book is to provide some kind of bridge for mathematics between the high-school-level and college-level for physics students. From my perspective, our job as physicists is to observe and understand the universe around us. Unfortunately our universe happens to be pretty complicated, at least from a mathematical point-of-view.  However, a lot of the underlying physics | the underlying set of rules surrounding how things move and behave | is usually not too complicated.  Sure, when we couple things together, everything turns disgusting, and we need to turn to really powerful computers or simulators to get a lot accomplished. But the basic rules are not that bad. My hope is to provide enough of a conceptual framework for you to learn to \textit{read} math as I saw a lot of other physicists do.  I plan to cover some of the topics that tend to trip up a lot of physics students, although I could not cover these exhaustively. When I found that I should have covered more, but could not due to finite time constraints, I left a reference to other work that I think you may find helpful. However, unlike a lot of works, my focus is to help modify your thinking of how math is used, rather than just pummel you with algorithms for you to memorize without giving you the proper context for such algorithms.  Whenever you find my explanations long-winded, feel free to proceed through faster. If you get stuck, you are always free to go back and see what I might have referenced before.

I am not going to lie to you | I expect a lot out of you as a reader.  It's not that I expect you be able to do everything out on your own | I \textbf{hate} when textbooks do that. I will show you the mechanics of a lot of basic calculations, or argue to you how I might interpret an equation based on its logical underpinnings. I might also ask you to imagine certain scenarios, or really push your inherent ability to reason beyond plugging in numbers whenever they appear\footnote{I actually loathe numbers (I am \textit{really} bad at them), and so I work almost exclusively with symbols. Luckily, almost everything in physics (and math, too) is symbolic because numerical values tend to hide general trends.}. That, in some sense, may be harder than just having you do problems out without any kind of help.  I recognize that learning is very hard (I am still a student after all), and so I wrote a lot of this book as though I was sitting next to you and explaining it in person.  I hope that I can serve as a guide, or a significant perturbation, so your transition from a high school understanding of math to a physics-level understanding of math is not as tough as it otherwise could be.

I really do think that many people with some hard work are more than capable of understanding most of the intricacies in physics; unfortunately mathematics can have a tendency to clutter everything, especially for people who may struggle to read it as any other human language. But once you can use it as such, it becomes any other human tool designed for us to make sense of everything else in our lives. This book is long, but I wrote in in mind that it can serve as a reference guide for you as you move throughout your undergraduate career | although it honestly will probably be most helpful when you are a first- and second-year student. Later on, this book may serve as a memory bank for some interesting references or a place where you can find a formula or two that you will forget (yes, as a physics major, you will forget a lot of math; it's part of the job). If you happen to have a PDF version, then there are hyperlinks everywhere for you to use. If you have a hardcopy, then feel free to still try and click on the equation numbers to activate the hyperlinks. Worth a shot, right? 

So, without further ado, let's get started.

\vspace{0.15in}\noindent
\emph{William \textquotedblleft Joe\textquotedblright}$\,$\emph{Meese} $\sim$ \emph{May 2018}\newline\emph{M.Sc. Physics, Rensselaer Polytechnic Institute, 2018}\newline\emph{B.S. Physics \& Mathematics, Rensselaer Polytechnic Institute, 2017}


\newpage
\tableofcontents
\newpage

\pagestyle{fancy}
\lhead{}
\chead{}
\rhead{}
\lfoot{\Navigation}
\cfoot{\thepage}
\rfoot{}
\renewcommand{\headrulewidth}{0pt}
\renewcommand{\footrulewidth}{0pt}

\phantomsection
\addcontentsline{toc}{chapter}{\listfigurename}
\listoffigures\newpage

\phantomsection
\addcontentsline{toc}{chapter}{\listtablename}
\listoftables\newpage

\phantomsection
\addcontentsline{toc}{chapter}{\listexamplename}
\listofexample\newpage

\phantomsection
\addcontentsline{toc}{chapter}{\listproblemname}
\listofproblem\newpage

\setcounter{example}{0}
\setcounter{problem}{0}

\chapter{Vectors}

Vectors are everywhere in physics. Their utility comes in many different varieties, from helping us measure position in classical mechanics, to allowing us to describe the distribution of available momentum states for electrons in semiconductors, to aiding us as we describe the interactions between particles and fields.  Since they are everywhere, we must tackle them first.  They will probably seem pretty difficult/annoying/infuriating to you, particularly if you have never dealt with vectors before.  But truth be told, vectors allow us to manipulate and transport a bunch of otherwise cumbersome information in a nice compact way.  For example, in Maxwell's original work on electromagnetism, he needed \emph{twenty} coupled equations to fully capture all of electricity, magnetism, and light because vectors were not yet a mathematical tool \cite{charap_2011}. Now we can write his famous equations\footnote{To have a full picture of the dynamics between the fields and any nearby particles, we need a couple more equations, but this is besides the point. } down rather quickly as
\begin{align*}
    \grad\cdot \vec{D} &= \rho_{free}
    \\
    \grad\cdot\vec{B} &= 0
    \\
    \grad\times \vec{E} &= -\pderiv[t]{\vec{B}}
    \\
    \grad\times \vec{H} &= \vec{j}_{free} + \pderiv[t]{\vec{D}}
\end{align*}
The specific details of what each of these equations means are not within the scope of this book, so I will leave their explanation for your electromagnetic theory coursework (or you can go to for a fairly quick overview at \cite{hyperphysics_maxwells_equations}).

The purpose of this chapter is twofold. Firstly, and most importantly, I want to help you learn how to use vectors for implementation in your study of the natural world.  Another goal I have in mind is to help you think about mathematical objects, like numbers, sets, or functions, in ways that you may not have thought of before. In many ways, mathematics is a game.  It has players (you and me), it has pieces (mathematical objects), and it has rules for how the players and pieces can interact.  Hopefully, by the end of this chapter you will begin to see this for yourself, too.

\section{Mathematical Objects}
\begin{quote}
    \textbf{Disclaimer:} this section will be pretty rudimentary as far as mathematical concepts go | we will will only talk about the basis of mathematical reasoning, properties of real numbers, and functions.  However, the level of \textit{abstractness} is pretty high. In other words, this level of abstractness is not commonly taught in most high school math classes, nor is it usually taught in introductory calculus courses (although there are exceptions!). I understand that the formalism that I use below may be a little difficult to get through for a lot of you the first time you see it, so do not worry when it looks like we start at a much higher level than you are expecting.  This section is not absolutely necessary for one to understand vectors; meanwhile I think it is worthwhile to peruse so that you can gain some familiarity with the idea of the \textquotedblleft math game,\textquotedblright$\,$ and how we establish rules and then use those rules to prove mathematical ideas.  I chose to start with something as elementary as the real numbers partly because they are used to build essentially the rest of mathematics, and vectors by extension, but I mostly chose them because most of you will be pretty familiar with the real numbers.  Thus, you will already have some intuition about them that you can use as a reference for when I build them from a set of rules.  If this section is too mathy for you, feel free to proceed to Section \ref{subsec: Vector Operations}. Later on in the book, if I ever use ideas proven in this section, just refer back here and brush up on what you need.  Otherwise, especially if you are very mathematically-inclined\footnote{Like more mathematically inclined than a physicist.  Or even an applied mathematician...  So proceed to read this section basically if you really like pure logic.}, sit back and enjoy constructing mathematical reasoning and the real numbers!
\end{quote}

Before we get going any further, we need to lay down some ground rules.  These rules will help us understand what we are doing, and at worst, will be something to fall back on if we get lost. The first two rules are the most fundamental, but they are necessary for us to move forward.
\begin{enumerate}
    \item We are allowed to define new mathematical objects that have a specified set of properties.  
    \item Mathematical objects can only be used consistently with their given properties. If we need a new property for our object, we must redefine the object to incorporate it.
\end{enumerate}
However, this is not a pure mathematics book, so I will not build mathematics from set theory.  I will lean on a lot of your prior knowledge of math (maybe up through algebra, geometry, and trigonometry), but whenever I will be talking about math in this much more elementary way, I will warn you.  

To illustrate how we might use these rules, let's say we wanted to invent Euclidean geometry for the first time.  I might say something like, \textquotedblleft there exists an object called a point\textquotedblright. After that statement, the only thing we know about points | i.e. their only predefined property | is that they are extant.  We conclude then that points, by themselves, are \textbf{boring}.  We decide to define a means for points to interact with each other.  We do so by saying \textquotedblleft any two points belong to a common mathematical object called a line that runs through both\textquotedblright.  Now we have a means of talking about two points with respect to one another, since they are connected by a line. We may move to say that there are then infinitely many different points on the line since they are all extant mathematical objects.  This would not be quite right, though, as we have not given our points a property that they can be distinguished.  To try to do this, we would have to redefine our point object and say something along the lines of \textquotedblleft there exists only one object called a point for every location\textquotedblright.  We do start to get into trouble here because we would need to be clear about what a \textquotedblleft location\textquotedblright$\,$ object is in order for this new definition to make sense.  To proceed with the infinite number of points idea, we would eventually have to come up with some kind of length and throw in integers somehow to argue that a line has at least as many points as integers, therefore is comprised of infinitely many points.  The point is that we could build up Euclidean geometry in our own way, our only limitation is the way in which we define our gamepieces. But after we have our pieces in order, we continue to play the game to see if we can beat our previous score.

\section{Numbers, Lists, and Functions}
We will build up the idea of vectors in a much less geometric way than many other people do, mostly because I think that the geometry can sometimes muddle the concision that vectors bring to the table.  Once we have a few rules in place for vectors, then I will bring up some geometric interpretations for what we have found.

\subsection{Real Numbers \label{subsec: Vectors - Real Numbers}}
I am going to assume that you know the difference between types of numbers, such as integers and irrational numbers, or at least I will assume that you could type them into a calculator if someone asked you to.  What I will quickly review though, is a few defining properties of how real numbers interact with each other. The set of all real numbers is denoted by $\mathds{R}$, and whenever we say \textquotedblleft $x$ is in the set of real numbers\textquotedblright, we say write it mathematically as $x\in\mathds{R}$, where the symbol $\in$ means \textquotedblleft in\textquotedblright. 

Let's choose three real numbers, $a$, $b$, and $c$. It does not matter what these numbers are, as long as they are real. And since they are all real numbers, we would write $a,b,c\in\mathds{R}$.  If you need to at this point, whip out your calculator and pick your three favorite real numbers and test the following properties to show yourself that they hold (recall that subtraction is denoted by $a-b$ or $a+(-b)$ and multiplication is denoted by either $ab$ or $a\cdot b$).

\begin{enumerate}
\setlength{\columnsep}{30pt}
\begin{multicols}{2}
    \item Addition is commutative: \\ $a + b = b + a \in \mathds{R}$
    \item Addition is associative: \\ $(a+b)+c = a+(b+c)\in\mathds{R}$ 
    \item There exists an identity operator  for addition, namely $0\in\mathds{R}$ such that \\ $a+0 = 0+a = a$.
    \item There exists an additive inverse for every $a\in\mathds{R}$, denoted by $-a$ such that \\ $a+(-a) = (-a)+a = 0$.
    
    \item Multiplication is commutative: \\ $ab = ba\in \mathds{R}$
    \item Multiplication is associative: \\ $(ab)c = a(bc)\in\mathds{R}$
    \item There exists an identity operator  for multiplication, namely $1\in\mathds{R}$ such that \\ $a\cdot 1 = 1\cdot a = a$.
    \item There exists an multiplicative inverse for every $a\neq 0\in\mathds{R}$, denoted by $\frac{1}{a}$ such that \\ $a\cdot\frac{1}{a} = \frac{1}{a}\cdot a = 1$.
    \end{multicols}
    \item There exists a distributive property between addition and multiplication, namely $$a(b+c) = ab + ac\in\mathds{R}$$
\end{enumerate}

These properties, although written in perhaps a rather abstract way, are our initial rules for the real numbers. They outline everything that we are allowed to do with real numbers as far as calculations are concerned.  These rules tell us how to add, subtract, multiply, and divide real numbers.  They also tell us that \textbf{we will always get a real number back by adding, subtracting, multiplying, or dividing other real numbers}.  This is important because it is sometimes more important just for us to know what type of object we are dealing with, rather than knowing the specific value for the object.  Thus, we will always know that when we are combining real numbers according to the 9 rules above, our final result will always be just another real number.

Real numbers also have the following \emph{Ordering Property}: if $a,b\in\mathds{R}$, then only one of the following statements is true. Either $a - b < 0$, $-a + b < 0$, or $a = b$. These could equivalently be written as $a<b$, $b < a$, or $a = b$.  This ordering property essentially says that some real numbers have lower values than other real numbers ($a < b$ or $b < a$), unless the two numbers we are talking about are the same ($a = b$). Furthermore, there are two rules governing how addition and multiplication work with ordering.  For example,
\begin{align*}
    1.& \textrm{ If } b < c \textrm{, then } a+b<a+c.
    \\
    2.& \textrm{ If } 0 < a,b \textrm{, then } 0<ab.
\end{align*}
Using these rules, it can be \emph{proven} that $0<1$.  This may sound silly at first, but remember, we only defined 1 and 0 as the multiplicative and additive identities.  We did not define them as \emph{something} and \emph{nothing} as is usually done. To show that this is true, we must actually show a few other things first (these things that facilitate the execution of proof are called \textbf{lemmas}).

\begin{enumerate}
    \item If $0 < a$, then $-a < 0$ and if $a < 0$ then $0 < -a$.\\
     To prove the first part, we note that $0 = -a + a$.  But $0 < a$, so by the first rule $-a + 0 < -a + a = 0$.  Thus, $ -a + 0 = -a < 0$.  I recommend you use the same line of reasoning to prove the second part.
    
    \item $-ab = -(ab)$. \\
     Consider $-ab + ab = (-a + a)b = 0b = 0$.  We conclude then that since $-ab + ab = 0$, then it must be true that $-ab$ is the the additive inverse of $ab$, otherwise denoted by $-(ab)$.  Hence, $-ab = -(ab)$.  (This shows that our rules for real numbers allow us to permute the negative sign around products.)
     
     \item $-(-a) = a$. \\
     This proof may be a little confusing because of the negative signs (they confuse me, at least).  Since $(-a)\in\mathds{R}$ whenever $a\in\mathds{R}$, then $(-a) + -(-a) = 0$.  But by definition, $(-a) + a =0$. Thus, $$-(-a) = -(-a) + 0 = -(-a) + (-a) + a = 0 + a = a$$. We conclude that $-(-a) = a$.
     
     \item $1^2 = 1$.\\
     Given that $1\in\mathds{R}$, then we apply the multiplicative identity to it to get $1 = 1\cdot 1 = 1^2$.
     
     \item $(-a)(-b) = ab$. \\
     We use the second lemma above to move the negative on $a$ to the outside: $(-a)(-b) = -[a(-b)]$.  Then we apply it again to get $-[a(-b)] = -[(-b)a] = (-)(-)ba = -[-(ab)]$.  By the third lemma, $-[-(ab)] = ab$.
\end{enumerate}

Finally, we have enough lemma-backup to proceed.  We will do so by showing that if $a\in\mathds{R}$, then $a^2 > 0$ if $a\neq 0$.  We start with any real number $a\neq 0$, so there are two cases ($0<a$ or $-a<0$).  If $0 < a$, then by the multiplication ordering rule, $0 < aa = a^2$.  Next, if $a < 0$, then by the first lemma, $0 < -a$.  Thus, $0 < (-a)(-a)$.  By the fifth lemma, $(-a)(-a) = aa = a^2$, so $0 < (-a)(-a) = a^2$, proving that when $a\neq 0$, then $0 < a^2 $ for all other real numbers. Now we choose $a = 1$, since by the fourth lemma, $1^2 = 1$.  Therefore, $0 < 1^2 = 1$, completing our proof.

\begin{figure}
    \centering
    \includegraphics[width = 5.5in, keepaspectratio]{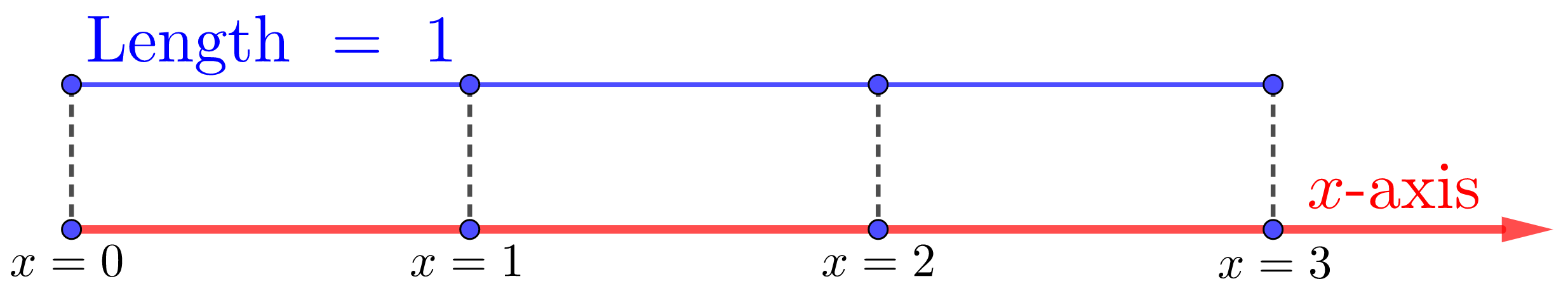}
    \caption{We assign real numbers to a number line by determining what length should be 1 unit and then project those lengths onto a line (or vector). }
    \label{fig:Ruler_x_axis}
\end{figure}

Again, this little exercise may have seemed silly because we have always been told that $0 < 1$, or equivalently $ 1 > 0$. But all of that was based on numbers being representations of physical things.  Now we have showed that this holds for our new \textit{abstract} set of real numbers, given our rules for them. Hence, our numbers no longer have to represent sets of things for them to have any meaning to us.  They now can stand on their own. Furthermore this allows us to interpret our abstract real numbers geometrically so that we can regain the physical intuition we all had about numbers before reading this book.  First off, we can define something of unit length, as is shown in Fig. \ref{fig:Ruler_x_axis}.  Then we call some point $0$ and measure out to a point $1$ using the unit length we have. But since we know that $1>0$, then the position of $1$ relative to $0$ shows us which direction represents an increase in our abstract real numbers.  From here, we can assign these numbers to represent whatever we want, whether it be money, distance, energy, complex numbers, complex quantum states, \textit{et cetera}.  The power we have now obtained from this set of rules for real numbers is that we control how and when we use the real numbers to help us in explaining physical phenomena, rather than having to wait until we understand how a number line can somehow be morphed into something that may not be strictly geometrical.


The final important property that I want to address is that the real numbers are considered mathematically \emph{complete}.  This essentially means that for any $a,b\in\mathds{R}$, there exists a real number $x\in\mathds{R}$ such that $a\leq x\leq b$.  The implications here are actually quite surprising.  This means that the real numbers form a continuum, meaning we can zoom in between any two points on a number line indefinitely.  There will always be another real number there no matter how far in we zoom.  This property is essential for continuum mathematics and physics overall. If the reals were not complete, then every once in a while we might measure a length and fail because our ruler could suddenly run into a gap in the real numbers! 

\subsection{Collections of Real Numbers: Lists and Functions}
Sometimes it is important for us to connect numbers to each other.  For example, let's say we have money and want to buy things.  To determine how many things we could buy with our money, we might divide our money by how much it costs per thing. For example, to use a case that is probably fairly relatable, if you only have \$100 and a textbook for Physics 1 costs \$200, then you can buy exactly \$100/\$200 per textbook which is exactly half of a Physics 1 textbook. What we have done though is we have implicitly created a list of real numbers, $(\textrm{money}, \textrm{textbooks})$, and we created a function to join the numbers together in the list.  Using this example, we come up with a couple more gamepieces to use and we define them below.

\begin{itemize}
    \item The \textbf{Cartesian Product} of $n$ sets of real numbers, formally written as $\mathds{R}^n = \mathds{R}\times\mathds{R}\times \dots\times\mathds{R} $, is the set of all $n$-tuples, where any element $a\in\mathds{R}^n$ is written as $a = (a_1,a_2,\dots,a_n)$.
    
    \item Consider any two sets of objects, $A$ and $B$. A \textbf{function} (or \textit{mapping}) $f$ from $A$ into $B$, written $f:A\rightarrow B$, relates the two sets $A$ and $B$ such that every object $a\in A$ is assigned a unique object $b\in B$, denoted by $f(a) = b$. The set $A$ is called the \textit{domain} of $f$ while the set of all $f(a)\in B$ is called the \textit{range} of $f$\footnote{Note that the range of $f$ is not necessarily all of $B$. If $\textrm{range}(f) = B$, then we say that $f$ is an \textit{onto} function, meaning it is a mapping from $A$ \textit{onto} $B$ instead of just $A$ into $B$. }.
\end{itemize}

\begin{figure}
    \centering
    \includegraphics[width = 5.5in, keepaspectratio]{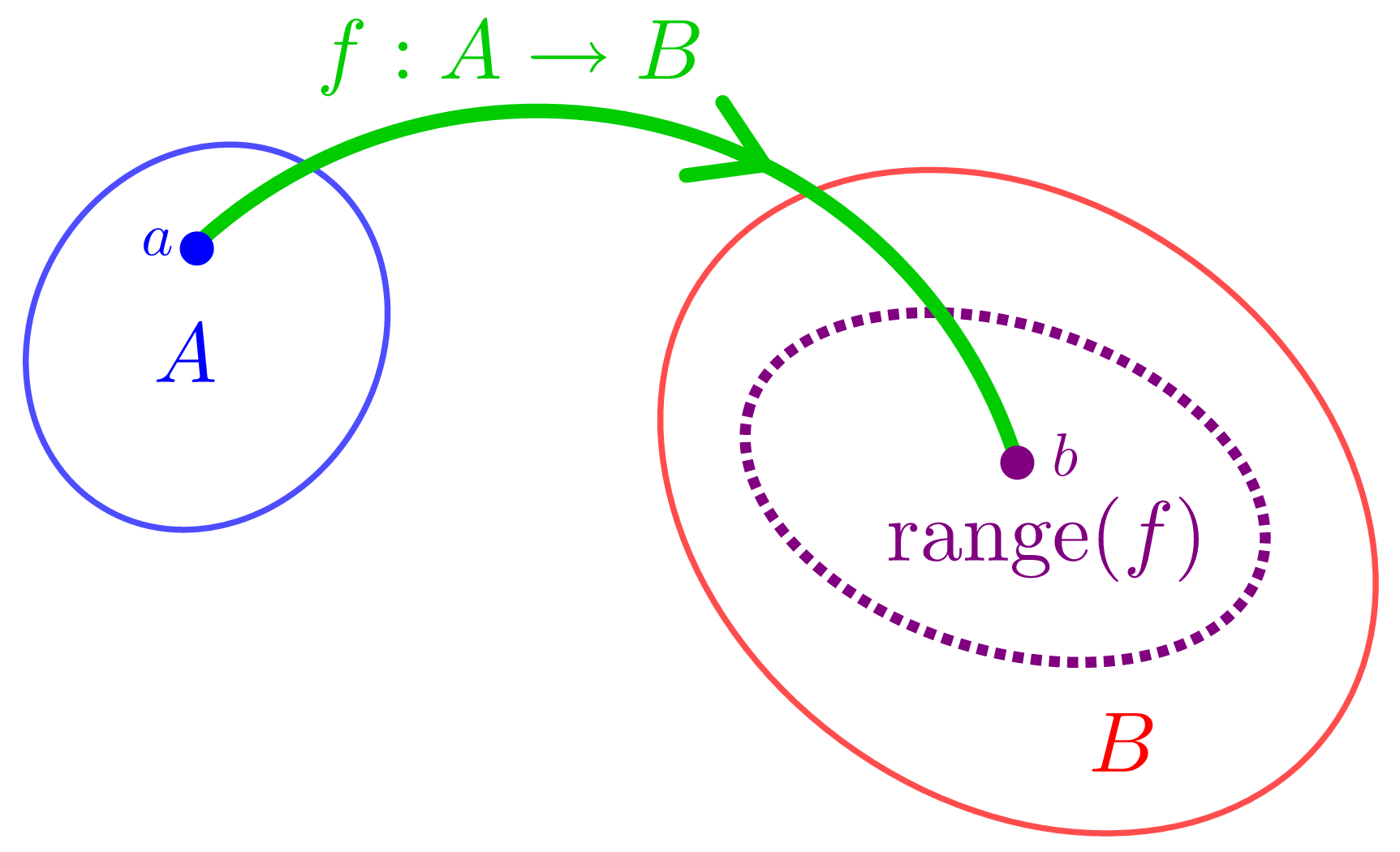}
    \caption{A function $f: A\rightarrow B$ takes every single point $a\in A$ and maps it into exactly one point $f(a) = b\in B$. The set of all $f(a)\in B$ is the $\textrm{range}(f)$. }
    \label{fig:Function_AtoB}
\end{figure}

The Cartesian Product allows us to group collections of real numbers together, like we did before with money and textbooks with $\mathds{R}^2 = \mathds{R}\times\mathds{R}$, and the function allows us to find how many textbooks $b=f(a)\in B$ we can buy with our money $a\in A$.  A schematic of what a function looks like is given in \figref{fig:Function_AtoB}. Now there are MANY special properties of functions, and we will only cover a few throughout the course of this book.  Whenever we must use them, I will define/derive them if they are relatively easy, but an in-depth study about the nature and structure of functions will ultimately distract us from physics, and so anyone interested in learning about these topics should either add a pure mathematics major or reference one of the following \cite{wolfram_function,rudin_1976,wikipedia_functions}. For now, the key take away from functions is that \textbf{every function has an input and it will give us exactly one output}.


\section{Vector Operations \label{subsec: Vector Operations}}
\begin{figure}
    \centering\qquad
    \subfloat[A vector $\vec{v}$ defined solely as a direction.]{
        \centering
        \includegraphics[width = 1.05in, keepaspectratio]{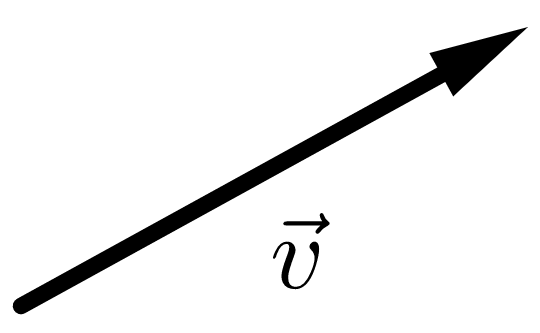}
        
        \label{subfig: Vector}
        }
    \qquad
    \subfloat[When two vectors, or directions, $\vec{v}$ and $\vec{u}$ are added together, they form a third vector $\vec{v}+\vec{u}$.]{
        \centering
        \includegraphics[width = 3.5in, keepaspectratio]{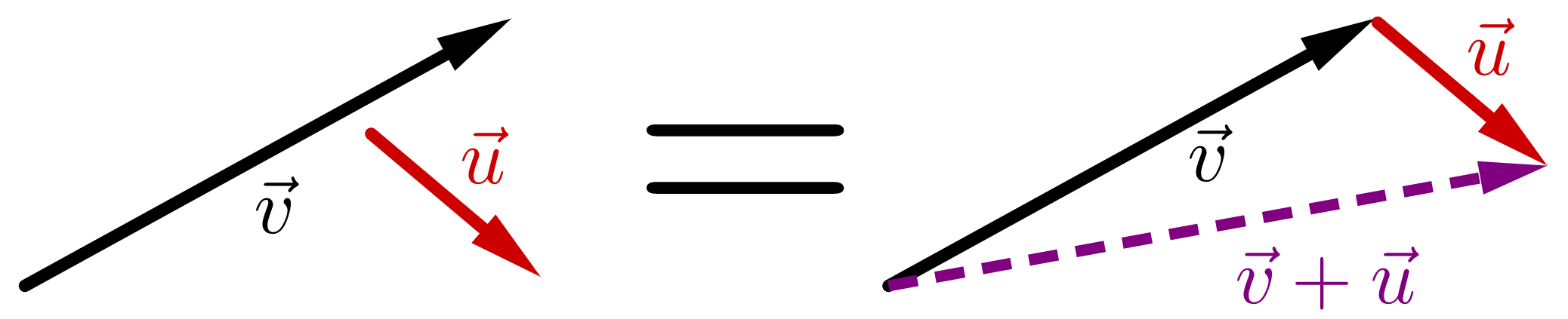}
        
        \label{subfig: Vector Addition}
        }
    \qquad
    \subfloat[When two vectors $\vec{u}$ and $\vec{v}$ are added together, they form a third vector $\vec{u}+\vec{v}$.]{
        \centering
        \includegraphics[width = 3.5in, keepaspectratio]{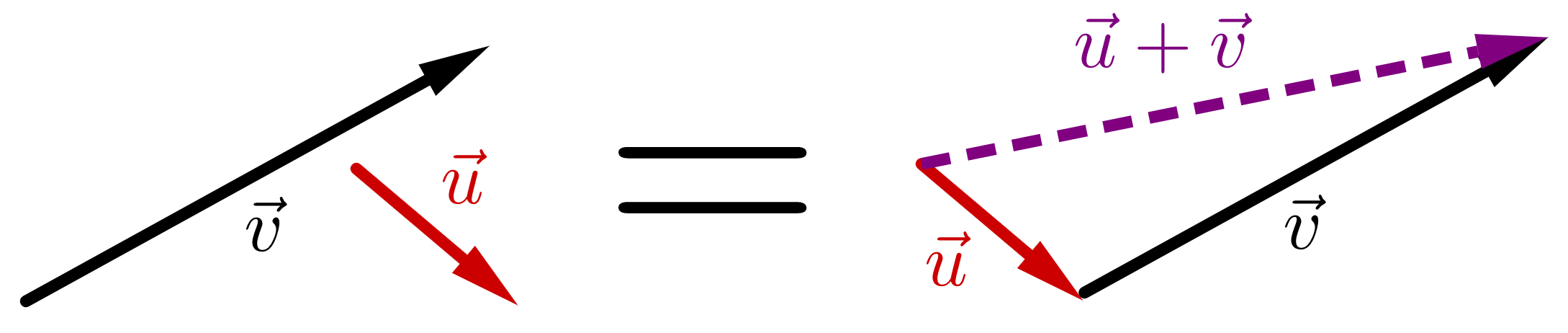}
        
        \label{subfig: Vector Addition 2}
        }
    \qquad
    \subfloat[The superposition of either addition method.]{
        \centering
        \includegraphics[width = 1.05in, keepaspectratio]{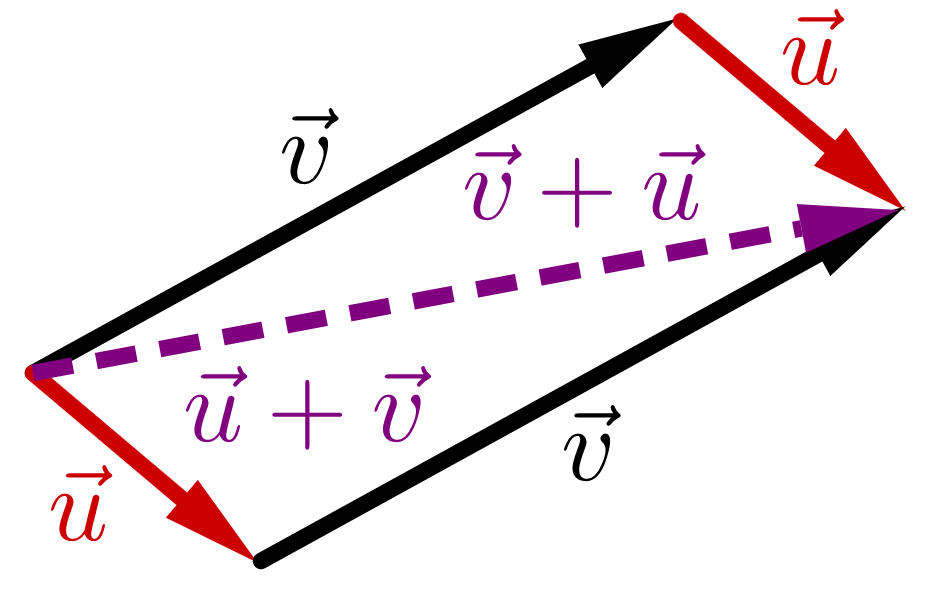}
        
        \label{subfig: Vector Parallelogram}
        }
    \caption{Vectors are simply just mathematical objects that describe directions.  This definition leads to an interesting way to add them.}
\end{figure}

Now that we have developed a little bit of prerequisite mathematical analysis, we will proceed with our discussion of vectors.  For right now, we will define a vector simply as a length in a particular \textquotedblleft direction,\textquotedblright$\,$ like the thing drawn in \figref{subfig: Vector}, and we denote a vector with a little arrow on top, like $\vec{v}$.  Many other texts write vectors as $\boldsymbol{v}$, but since bold-faced type is hard for most people to write by hand, I will stick to arrows.


\subsection{The Geometry Behind Vector Addition}
For right now, the important property of vectors is that they have directions.  This means that they represent, for example, the arrow that connects your eyes and these words. Since vectors only have directions, they do not belong to any particular points in whatever space they live in\footnote{The term \textquotedblleft direction\textquotedblright$\,$ is not clearly defined yet, but for right now, just consider it a ray-segment instead of a line-segment.}.  Now let's say that two vectors exist, namely $\vec{v}$ and $\vec{u}$.  These vectors (directions) are shown in the left-hand side of the equality in \figref{subfig: Vector Addition}.  Since these vectors have defined lengths and directions, but are allowed to be translated in space, we are free to align $\vec{v}$ and $\vec{u}$, as shown in the right-hand side of \figref{subfig: Vector Addition} such that the tip (arrowhead) of $\vec{v}$ touches the tail (not arrowhead) of $\vec{u}$. This is equivalent to looking at these words, and then following the arrow between them and something else, until you end up looking at something else.  The end results is that you ended up at the something else even though you started at those words.  Since we started somewhere and ended up elsewhere, we will call the final result $\vec{v}+\vec{u}$ which is the \textit{resultant} vector drawn from the tail of $\vec{v}$ to the tip of $\vec{u}$ when $\vec{v}$ and $\vec{u}$ are aligned \textbf{tip-to-tail}.

But if this is a summation, it is natural for us to ask whether vector addition is commutative like regular old real number addition. For example, does something like $3+4=4+3$ hold for vectors? To check this, we look at $\vec{u}+\vec{v}$ by translating $\vec{v}$ such that it is aligned with $\vec{u}$ tip-to-tail, as is shown in \figref{subfig: Vector Addition 2}. If we compare the direction of the vector $\vec{v}+\vec{u}$ from \figref{subfig: Vector Addition} with that of the vector $\vec{u}+\vec{v}$ from \figref{subfig: Vector Addition 2}, we can start to see that they look very similar, if not identical.  To see that the directions are the same, and further that the vector sum is indeed commutative with respect to this definition, then we superimpose the right-hand sides of \figref{subfig: Vector Addition} with \figref{subfig: Vector Addition 2} to get the parallelogram shown in \figref{subfig: Vector Parallelogram}.  By looking at this parallelogram, it is clear that adding $\vec{u}$ to $\vec{v}$ ($\vec{v}+\vec{u}$) tip-to-tail and adding $\vec{v}$ to $\vec{u}$ ($\vec{u}+\vec{v})$ leave us with a vector that has the same length and direction.  This discovery allows us to conclude two \textbf{very important} things about vectors:

\begin{enumerate}
    \item Vector addition is \textbf{commutative} when we \textbf{add vectors tip-to-tail}.
    
    \item Two vectors $\vec{A}$ and $\vec{B}$ are equivalent if they have the same length and direction. 
\end{enumerate}

The first point is important because in nature we observe all kinds of phenomena, such as something's position and displacement, whose directions are commutative. That is, the resultant directions due to a sum are independent of the order in which we add them. For example, if you were to draw a square on some paper and then follow the sides from one corner to the opposite corner (a lot like what's shown in \figref{subfig: Vector Parallelogram}), you'd see that you would end up in that corner regardless of which direction you first moved along the sides.  In other words, the addition of either sets of sides is commutative.  But now we know definitively that \textbf{adding vectors tip-to-tail} geometrically gives us the result that we know from our everyday experience.  The second point is important because it gives us a way to specify uniqueness in our vectors.  In the physical world, if we are describing something's velocity at any particular point, we intuitively would only need one measurement of it to be complete.  We should not have to name a bunch of different velocities if an object only has one.

\subsection{The Geometry Behind Vector Subtraction}
Before we continue, I want to make note of something important.  To define vector addition, I said that we would \textbf{add vectors tip-to-tail}. But you may ask, what happens if instead we add them tail-to-tail? Why can't this also be vector addition?

\begin{figure}
    \centering
    \includegraphics[width = 4.5in, keepaspectratio]{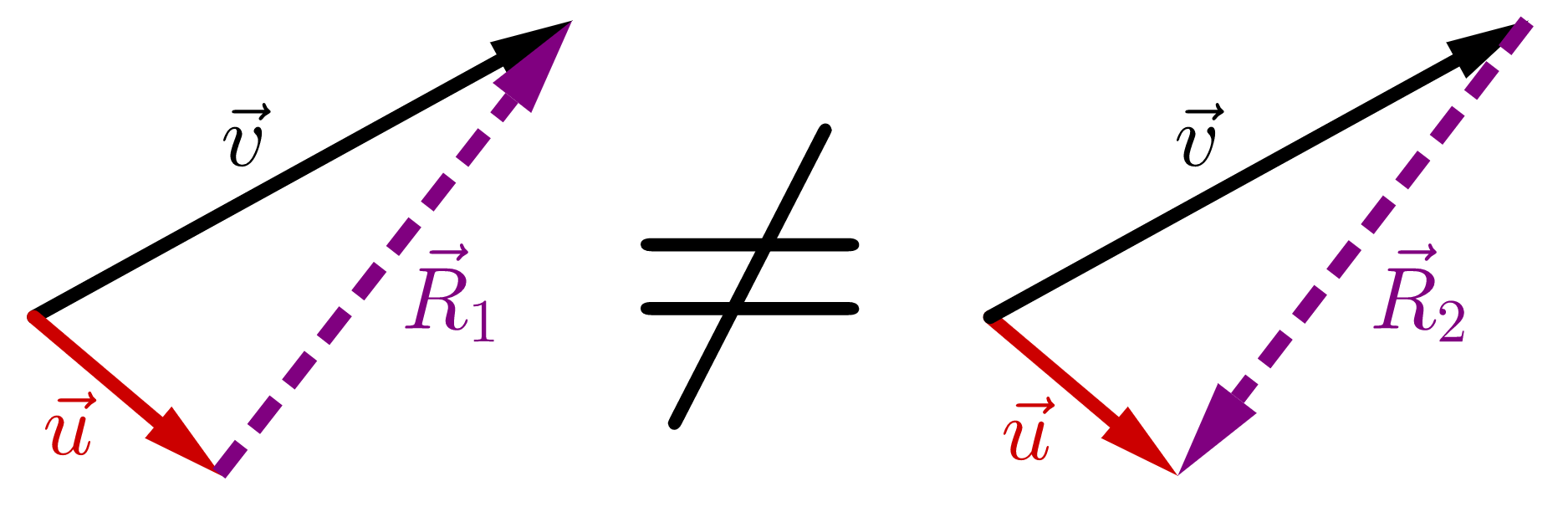}
    \caption{The the \emph{function} of aligning the vectors $\vec{v}$ and $\vec{u}$ tail-to-tail leaves the resultant (output) vector $\vec{R}$ ambiguous. Drawing the resultant vector from tip-to-tip could then yield two different answers.}
    \label{fig: Vector Subtraction}
\end{figure}

This is a very good question, and it actually involves the definition of \emph{functions} we used above.  However, now we will construct a function that takes in \textit{two} vectors arguments and outputs \textit{one}.  By the definition of a function, there can only be one output for every input in the domain, so we need our one vector to be unique; thus, the \textit{resultant} vector of this function must have a well-defined length and direction.  Consider the same two vectors $\vec{v}$ and $\vec{u}$ from \figref{subfig: Vector Addition}, \figref{subfig: Vector Addition 2}, and \figref{subfig: Vector Parallelogram}.  Now we align them tail-to-tail as is shown in \figref{fig: Vector Subtraction} and draw the resultant vector tip-to-tip.  But which way do we go?  Do we draw from $\vec{u}$ to $\vec{v}$ (as is done in the left-hand side of the figure), or draw from $\vec{v}$ to $\vec{u}$ (as is done in right-hand side)?


Let the function that aligns vectors tail-to-tail be named $\vec{R}$. Since it aligns vectors tail-to-tail, the function takes in two vectors as inputs.  Now let us say that it has the following possible output vectors
\begin{align*}
    \vec{R} = \begin{cases}
    \vec{R}_1 = \vec{R}(\vec{u},\vec{v}), &\textrm{ if drawn from }\vec{u}\textrm{ to }\vec{v}
    \\
    \vec{R}_2 = \vec{R}(\vec{v},\vec{u}),&\textrm{ if drawn from }\vec{v}\textrm{ to }\vec{u}
    \end{cases}
\end{align*}
It is clear from the geometry in \figref{fig: Vector Subtraction} that the lengths of the two outputs are the same, however the directions are reversed.  Thus $\vec{R}(\vec{u},\vec{v}) \neq \vec{R}(\vec{v},\vec{u})$.  Hence, this particular function is NOT commutative, unlike aligning vectors tip-to-tail.  therefore, if we chose this particular function to be vector addition, our answers would not be commutative, and we would have a hard time describing a lot of physics.

This non-commutative function is reminiscent of subtraction from arithmetic though.  For example, $4-3\neq 3-4$.  However, from the properties of the real numbers above,
\begin{align*}
    3-4 =  -(-3) + (-4) = - (-3 + 4) = -(4-3),
\end{align*}
or more generally, for any two real numbers $a$ and $b$, $a-b = -(b-a)$ (prove it).  If we consider the \textit{length} or \textit{magnitude} of a number as its distance from $0$ of a number line, then both $a-b$ and $b-a$ have the same length even though their signs are opposite.  By analogy, $\vec{R}_1$ and $\vec{R}_2$ in \figref{fig: Vector Subtraction} both have the same length, however their directions are opposite.  Thus, we use this analogy to define vector subtraction by aligning vectors tail-to-tail. But just like $4-3 + 3 = 4$, or $a-b+b = a$, we need $\vec{v} - \vec{u} + \vec{u} = \vec{v}$ and  $\vec{u} - \vec{v} + \vec{v} = \vec{u}$.  In other words, the tip of the resultant vectors $\vec{R}$ should fall on the vector being subtracted from and the tail should fall on the vector being subtracted.  This means that
\begin{align*}
    \vec{R}_1 &= \vec{R}(\vec{u},\vec{v}) = \vec{v} - \vec{u},
    \\
    \vec{R}_2 &= \vec{R}(\vec{v},\vec{u}) = \vec{u} - \vec{v}.
    \\
\end{align*}
Again, by analogy, $a-b = -(b-a)$ for real numbers.  By looking at $\vec{R}_1$ and $\vec{R}_2$ we see that 
\begin{align*}
    \vec{R}_2 = \vec{u} - \vec{v} = - (\vec{v} - \vec{u}) = -\vec{R}_1.
\end{align*}
if we allow ourselves to factor out negative signs in an identical way as we did with real numbers.  Since $\vec{R}_2$ is in the opposite direction as $\vec{R}_1$, and if we factor out the negative sign like we did above, then we conclude that multiplying a vector by a negative sign reverses its direction!  This then means that we can eliminate a vector completely by subtracting it from itself, or by aligning it with a copy of itself tail-to-tail and then drawing the resultant vector tip-to-tip. Since the two tips will occupy the same point, the vector will have zero length, and we call this the \textbf{null} or \textbf{zero vector}, defined by $$\vec{u} - \vec{u} = \vec{u} + (-\vec{u}) = \vec{0}.$$
To summarize this point, we can subtract a vector $\vec{u}$ from a $\vec{v}$ by aligning the two tail-to-tail and drawing the resultant vector from $\vec{u}$ to $\vec{v}$.  Meanwhile, since multiplying by a negative reverses a vector's direction, we can equivalently add $-\vec{u}$ to $\vec{v}$ to get $\vec{v}-\vec{u}$.  Hence, we can reverse the direction of $\vec{u}$ and then \textbf{add it tip-to-tail} to $\vec{v}$. The ability to subtract vectors is really helpful in physics, particularly whenever we want to describe the change in a system.  When something changes, for example, position, it moves from one point to another.  The vector that connects the initial and final points would then represent a \textit{displacement}, that is, a change in position.  But we will talk more about this point later. Now that we established a visual (geometrical) representation of how vectors can combine, we move on to study more complicated systems of vectors.

\subsection{Vector Addition and Coordinate Systems \label{subsec: vector addition and coordinate systems}}
In physics, the real power of vectors comes from their simplification of otherwise annoying or intangible systems of coordinates. Sometimes it is really useful to talk about a problem in terms of three perpendicular directions (hint hint vectors) | electrons that swim through a solid chunk of metal, for instance, are described very well with a system of three perpendicular axes.  Other times, however, we need to use a set of coordinates that have one direction that emanates radially outward from some center and then two others that are mutually perpendicular with each other and the radial direction | these spherical coordinates are used in a lot of places, from electromagnetic or gravitational radiation problems to black holes to proton-proton collisions (and more). It actually turns out that some incredibly interesting physics comes out of the study of coordinate systems; for example, time and space can dilate and contract because of a difference in two observer's coordinate systems (thank you, Einstein \cite{einstein_menendez}). So a thorough understanding of coordinate systems is imperative for a thorough understanding of the universe, and vectors facilitate such an understanding. 


\begin{figure}
    \centering
    \includegraphics[width = 4.5in, keepaspectratio]{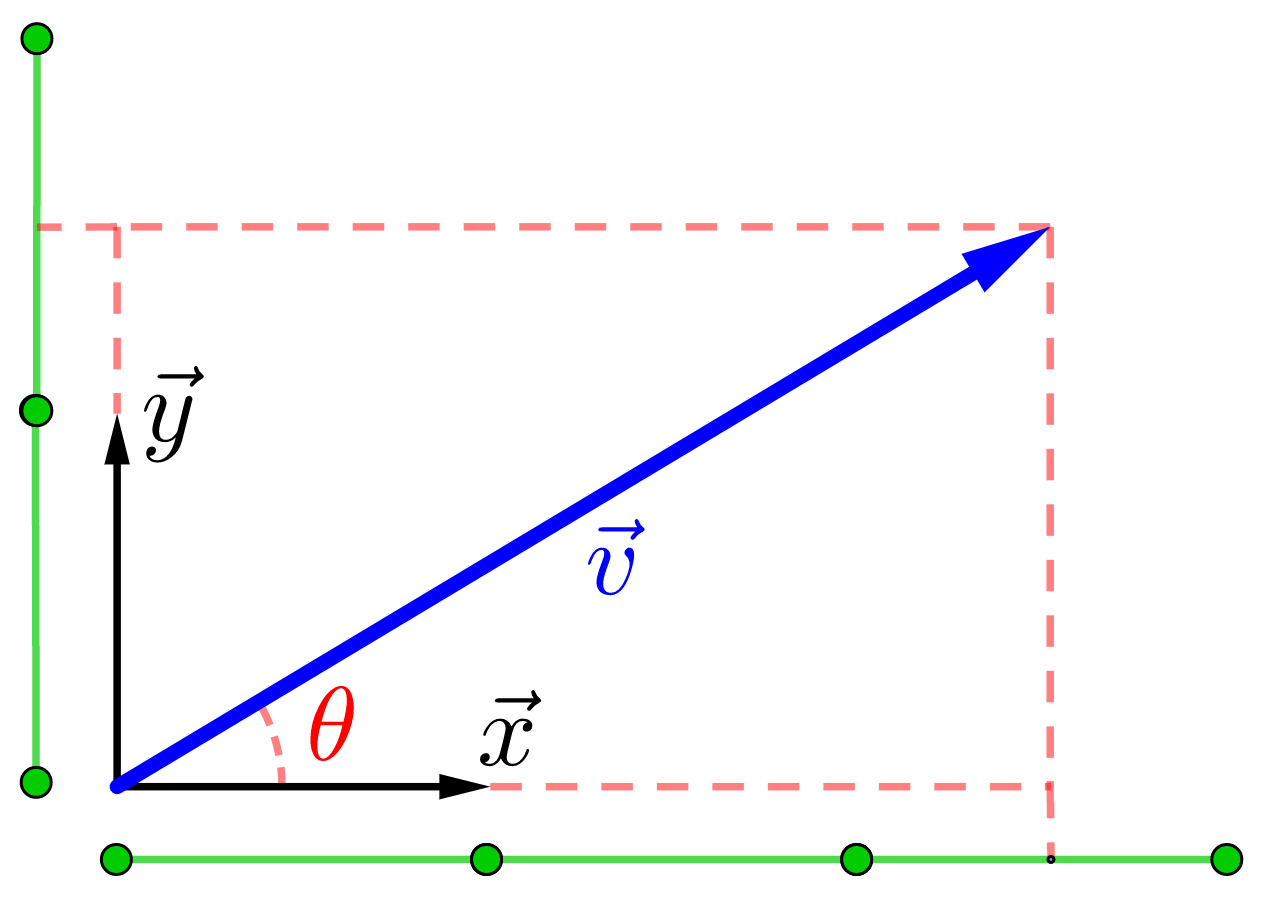}
    \caption{If we choose to use the vectors $\vec{x}$ and $\vec{y}$ of known lengths and directions, then we can rewrite vector $\vec{v}$ as the sum of $\vec{x}$ and $\vec{y}$. In this case, $\vec{v} = 2.5\vec{x} + 1.5\vec{y}$.}
    \label{fig: Coordinates}
\end{figure}


Since we have established what vector addition is (remember \textbf{tip-to-tail}\footnote{In case you haven't caught on, this should be a geometrical mantra of yours.}), we can apply it to define a coordinate system for ourselves.  This is because that any vector in a system of coordinates is represented by a list of lengths in some predetermined set of directions. But no particular system is universal over the others | any one that is used to describe a vector works\footnote{To get a better idea of this artifact in the mathematics, draw an arrow on a piece of paper. Now draw a little $xy$-axis on the side of the arrow. Then rotate the page and draw another little $xy$-axis. Here, neither axis is necessarily better than the other one; whichever we \textit{choose} to use is purely that | a \textit{choice}.}.  Thus, we start by choosing a set of directions that have some particular kind of meaning to us.  As can be seen in \figref{fig: Coordinates}, we choose $\vec{x}$ to represent the rightward horizontal direction and $\vec{y}$ to represent the upward vertical direction (when it comes to constructing our coordinate system, we align them tail-to-tail). We will call these coordinate-defining vectors \textbf{basis vectors}. For reasons that will become clearer later and in the Linear Algebra chapter, we intentionally choose $\vec{x}$ and $\vec{y}$ to be perpendicular.  Suppose that we know the lengths of both $\vec{x}$ and $\vec{y}$ so these vectors are clearly defined.  Now whenever we have another vector $\vec{v}$ living in the same space as $\vec{x}$ and $\vec{y}$, we can use vector addition to write $\vec{v}$ in terms of $\vec{x}$ and $\vec{y}$.

To do this, we start by extending $\vec{x}$ and $\vec{y}$, as is shown with the dashed lines in \figref{fig: Coordinates}.  Then we draw lines from the tip (arrowhead) of $\vec{v}$ onto the extensions of $\vec{x}$ and $\vec{y}$.  One of the reasons why it helps to have perpendicular basis vectors is that these lines that we drew define two right triangles with $\vec{v}$ as the hypotenuse.  All we have to do now is count how many $\vec{x}$ vectors fit along the horizontal extension and how many $\vec{y}$ vectors fit along the vertical extension. In \figref{fig: Coordinates}, the horizontal extension is $2.5$ $\vec{x}$s long, while the vertical extension is $1.5$ $\vec{y}$s long.  Therefore, we would write\footnote{If I were being 100\% truthfull, I would technically need to precisely define what \textit{scalar multiplication} is before we multiply vectors by numbers, so for right now consider it like we have 2.5 US dollars and 1.5 pecan pies. The idea is that we can assign some numerical amount to some fixed quantity, such as money and pies. However, the money and pies are not necessarily the same (we would need some \textit{function} to connect the two). }
$$ \vec{v} = 2.5\vec{x} + 1.5\vec{y} $$
Furthermore, by exploiting the following trigonometric relationships,
\begin{align*}
    \sin (\textrm{angle}) &= \frac{\textrm{opposite side}}{\textrm{hypotenuse}},
    \\
    \cos (\textrm{angle}) &= \frac{\textrm{adjacent side}}{\textrm{hypotenuse}},
    \\
    \tan (\textrm{angle}) &= \frac{\sin(\textrm{angle})}{\cos(\textrm{angle})} =\frac{\textrm{opposite side}}{\textrm{adjacent side}},
\end{align*}
we have
\begin{align*}
    \sin \theta &= \frac{1.5y}{v},
    \\
    \cos \theta &= \frac{2.5 x}{v},
    \\
    \tan \theta &= \frac{1.5y}{2.5 x},
\end{align*}
where $x$, $y$, and $v$ are the lengths of the vectors $\vec{x}$, $\vec{y}$, and $\vec{v}$, respectively.  The tangent relationship is pretty useful, because if we now think of the direction of the vector $\vec{v}$ as the angle $\theta$ between it and $\vec{x}$ when aligned tail-to-tail as is drawn in \figref{fig: Coordinates}, then we find that the \textbf{direction is independent of the length} $v$. This is reassuring because we can intuitively reason that we can stretch or shrink a vector's length arbitrarily without changing its direction.  For example if you pick a point somewhere near you at eye-level and stare at it, and then you take a few steps forwards or backwards while still looking at it, you will see you never really have to turn your head or readjust to keep fixed on the point\footnote{This is assuming you take normal-sized steps of course. If you wiggle your head too much then this experiment will not help at all.}.  Hence, we keep this new definition for the \textquotedblleft direction\textquotedblright$\,$ of a vector.

We can actually learn a couple more things from the trigonometry.  First, if you remember the Pythagorean identity
$$ \sin^2(\textrm{angle}) + \cos^2(\textrm{angle}) = 1, $$
Then,
\begin{align*}
    1 &= \sin^2\theta + \cos^2\theta
    \\
    &= \left(\frac{1.5y}{v} \right)^2 + \left(\frac{2.5 x}{v} \right)^2
    \\
    &= \frac{1}{v^2}\left[ (1.5y)^2 + (2.5x)^2 \right].
\end{align*}
By multiplying both sides by $v^2$ we find that 
\begin{align*}
    v^2 = (2.5x)^2 + (1.5y)^2 \Rightarrow v = \sqrt{(2.5x)^2 + (1.5y)^2}.
\end{align*}
which is an expression showing that the \textbf{length of a vector is independent of its direction} $\theta$ (also notice that is this the geometry version of the Pythagorean identity, a.k.a the Pythagorean Theorem).  Ergo, since a vector is only defined as an object with a length and a direction, the establishment of our coordinate system allows us to talk about them in three ways:
\begin{align}
    \vec{v} &= 2.5\vec{x} + 1.5\vec{y}, \nonumber
    \\
    \tan \theta &= \frac{1.5y}{2.5 x}, \label{eq: coordinates fig vector direction}
    \\
    v &= \sqrt{(2.5x)^2 + (1.5y)^2}. \label{eq: coordinates fig vector length}
\end{align}
All that we had to do is agree upon a set of perpendicular basis vectors of known lengths.

\subsection{Unit Vectors}
The discussion of coordinates systems relies pretty heavily on the choice of basis vectors $\vec{x}$ and $\vec{y}$. In order to actually take out a calculator and get a number for lengths or directions, it is necessary for us to specify the lengths $x$ and $y$. By looking at \equaref{eq: coordinates fig vector direction}, we may be motivated to try and simplify things by making the vectors $\vec{x}$ and $\vec{y}$ have the same length, i.e. we assume that $y = x$.  This would certainly make our lives easier because \equaref{eq: coordinates fig vector direction} would simplify to
\begin{align*}
    \tan \theta = \frac{1.5x}{2.5 x} = \frac{1.5}{2.5} = \frac{3}{5}.
\end{align*}
This is great if we wanted to find a numerical value for $\theta$ since we can isolate $\theta$ as follows,
\begin{align*}
    \theta = \arctan \left( \frac{3}{5} \right) \approx 0.540 \textrm{ rad} \approx 31.0^{\textrm{o}}.
\end{align*}
I want to emphasize though that this numerical value for $\theta$ only applies to the vector in \figref{fig: Coordinates}.  We can now reduce \equaref{eq: coordinates fig vector length} as well
\begin{align*}
    v = \sqrt{(2.5x)^2 + (1.5x)^2} = \sqrt{x^2\left((2.5)^2 + (1.5)^2\right)} = x\,\sqrt{(2.5)^2 + (1.5)^2}.
\end{align*}
It is at this last step though that we realize that to go any further (numerically) we need to specify what $x$ is. 

If we look at the equation above, we have an arbitrary length $x$ and then a numerical factor.  But that numerical factor contains the values 2.5 and 1.5 | I did not square either intentionally.  These numbers are important because they again represent the number of times the vectors $\vec{x}$ and $\vec{y}$ need to be extended to intersect the lines drawn from the tip of the arrow in \figref{fig: Coordinates}. It would be incredibly helpful to be able to describe a vector exclusively by reading off the set of analogous numbers from any system of basis vectors.  In other words, it would be the most helpful to NOT have to carry around either factors of $x$ or $y$.  However, we are not free to simply ignore $x$ or $y$ to remain consistent with our definition of vectors since we make them have lengths.  Further, since we use basis vectors to build our coordinate systems, in order for our coordinate systems to make any sense we need lengths associated with our basis.  Otherwise, again, our work would be inconsistent.  So all that is left is for us to choose $x = y = 1$ because $1\cdot a = a\cdot 1 = a$ for all $a\in\mathds{R}$.  When we do this, we no longer have to lug the $x$s and $y$s around our equations because they will always multiply other numbers without changing their value.  And finally would we be able to write the length of $\vec{v}$ from \figref{fig: Coordinates} as
\begin{align*}
    v = x\,\sqrt{(2.5)^2 + (1.5)^2} = 1\cdot \sqrt{(2.5)^2 + (1.5)^2} = \sqrt{(2.5)^2 + (1.5)^2} \approx 2.92.
\end{align*}

Vectors that have $\textrm{length} = 1$ are defined as \textbf{unit vectors}.  As we have seen, their definition and implementation comes about rather naturally when we use vectors to build a coordinate system.  They have many other useful vector properties that we will discuss more a bit later.  For right now, it is very important to note that unit vectors are so special that physicists and mathematicians gave them their own \textquotedblleft hats\textquotedblright,
\begin{align*}
    \vec{x} \rightarrow \Hat{x} \textrm{ if and only if } x = 1.
\end{align*}
and so from now on, whenever I am talking about unit vectors, they will have the little triangular hat to distinguish them from regular vectors with the arrows.  By extension then, we could write that same vector $\vec{v}$ from \figref{fig: Coordinates} as 
\begin{align*}
    \vec{v} = 2.5\Hat{x} + 1.5\Hat{y} = v_x\Hat{x} + v_y\Hat{y}
\end{align*}
where the symbols $v_x$ and $v_y$ stand for the \textbf{$x$ and $y$ components of the vector}, repectively.  This notation generalizes quite nicely, too. If our vector lives in three dimensions, for example, we would write it as
\begin{align}
    \vec{v} = v_x\Hat{x} + v_y\Hat{y} + v_z\Hat{z}, \label{eq: 3d Cartesian vector}
\end{align}
or if our vector lives in $n$-dimensions and we call the $j^{\textrm{th}}$ mutually perpendicular unit vector $\Hat{x}_j$, then we would write $\vec{v}$ as
\begin{align}
    \vec{v} = v_1\Hat{x}_1 + v_2\Hat{x}_2 + v_3\Hat{x}_3 + \dots + v_{j-1}\Hat{x}_{j-1} + v_j\Hat{x}_j +  \dots + v_{n-1}\Hat{x}_{n-1} + v_n\Hat{x}_n, \label{eq: nd Cartesian vector}
\end{align}
where $v_j$ is the $j^{\textrm{th}}$ the components of the vector $\vec{v}$.  Using the components of vectors, we can then rewrite the direction and length formulas given by \equaref{eq: coordinates fig vector direction} and \equaref{eq: coordinates fig vector length} as follows
\begin{align}
    \tan \theta &= \frac{v_y}{v_x} \label{eq: theta xy plane}
    \\
    v &= \sqrt{v_x^2 + v_y^2} \label{eq: 2d length}
\end{align}
in two dimensions for any general set of coordinates $\{v_x,v_y\}$.  In $n$ dimensions these formulas generalize to
\begin{align}
    \tan \theta_{jk} &= \frac{v_k}{v_j} \label{eq: theta jk}
    \\
    v &= \sqrt{v_1^2 + v_2^2 + v_3^2 + \dots + v_{j-1}^2 + v_j^2 + \dots + v_{n-1}^2 + v_n^2} \label{eq: nd length}
\end{align}
where the notation $\theta_{jk}$ means represents the angle measured with respect to the $j^{\textrm{th}}$ unit vector in the $jk$-plane (notice that if you substitute in $x$ for $j$ and $y$ for $k$, you get \equaref{eq: theta xy plane} back). I know this formula may look a little clunky, but it is a way to generalize the notion of an angle | an object that only exists between two axes $j$ and $k$ | to any dimensional space. For example, in three dimensions, we can have angles measured above the $x$-axis in either the $xy$-plane or the $xz$-plane.

Knowing when to switch back and forth between talking about a vector's components or its length and direction can get confusing sometimes.  In practice, it is important to remember that although vectors do have components, those \textbf{components depend on the coordinate system built from one's choice of unit vectors}.  Even though these components can have different values in different coordinate systems, \textbf{the vector should always remain the same}. For example, if we are talking about a vector drawn between the positions of two points, it does not matter if we are looking directly at it that vector, or we rotate ourselves, or me move ourselves further from that vector. Its location is invariant. However, the numerical values we would use to describe it in each of our coordinate systems would vary wildly.  When coordinate systems get weird, it is useful to stick to thinking of the vector as its own object that has many different sets of components (or representations). Problems like these occur in relativity and quantum mechanics.  However, when it suffices to only deal with one coordinate system per problem, like a lot of introductory physics, it is usually safe to think of the vector as only a set of components in that one coordinate system.

\subsection{Component-wise Addition and Subtraction of Vectors}
Since we have discovered a way to represent vectors in terms of a set of components in a particular coordinate system, it is useful to go through and define addition and subtraction in terms of these components.  We start with addition.

\begin{figure}
    \centering
    \includegraphics[width = 4.25in, keepaspectratio]{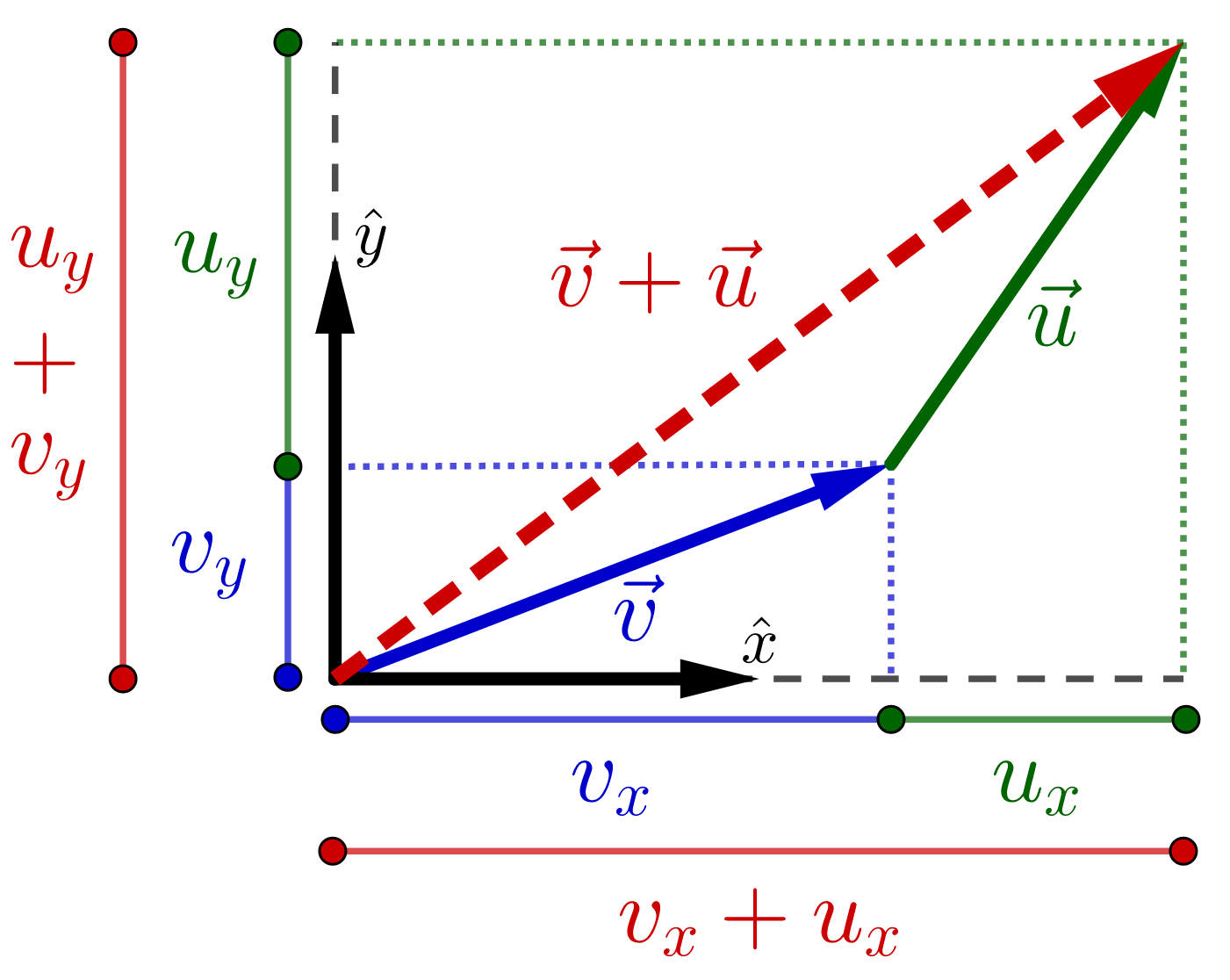}
    \caption{Here we combine adding vectors \textbf{tip-to-tail} with the components we can measure in our $\Hat{x},\Hat{y}$ coordinate system.}
    \label{fig: Coordinate Addition}
\end{figure}

Recall that vector addition is a geometric function that aligns two vectors $\vec{v}$ and $\vec{u}$ \textbf{tip-to-tail} and outputs the vector drawn from the tail of $\vec{v}$ to the tip of $\vec{u}$.  Meanwhile, we can write $\vec{v}$ and $\vec{u}$ in terms of their $x$ and $y$ components\footnote{We are starting in two dimensions for simplicity.} as
\begin{align}
    \vec{v} &= v_x\Hat{x}+v_y\Hat{y}
    \\
    \vec{u}&= u_x\Hat{x}+u_y\Hat{y}
\end{align}
once we establish a set of perpendicular basis vectors $\Hat{x}$ and $\Hat{y}$.  By looking at \figref{fig: Coordinate Addition}, we can see that there is a rather convenient way to express vector addition using the components.  From the figure we can conclude that the $x$-component of the sum $\vec{v}+\vec{u}$ is $(v+x)_x = v_x + u_x$, while the $y$-component of the sum is $(v+u)_y = v_y + u_y$. Therefore,
\begin{align}
    \vec{v} + \vec{u} = (v_x+u_x)\Hat{x} + (v_y+u_y)\Hat{y}\label{eq: 2d vector additon}
\end{align}
Note that by using \equaref{eq: 2d vector additon}, we can show that component-wise vector addition is commutative, just like the geometrical version of vector addition.
\begin{align*}
    \vec{v} + \vec{u} = (v_x+u_x)\Hat{x} + (v_y+u_y)\Hat{y} = (u_x+v_x)\Hat{x} + (u_y+v_y)\Hat{y} = \vec{u} + \vec{v}
\end{align*}
Here these equalities hold because $u_x,u_y,v_x,v_y\in\mathds{R}$.


Using the same idea with vector subtraction, the $x$-component of $\vec{v}-\vec{u}$ is $(v-u)_x = v_x-u_x$ and the $y$-component is $(v-u)_y = v_y-u_y$.  Therefore, in component notation, vector subtraction becomes
\begin{align}
        \vec{v} - \vec{u} = (v_x-u_x)\Hat{x} + (v_y-u_y)\Hat{y}\label{eq: 2d vector subtraction}
\end{align}
I leave it to you to show that component-wise subtraction is \textit{anti-commutative} in agreement with the geometrical version of vector subtraction in Problem \ref{prob: vector subtraction anticommutative}.

\vspace{0.15in}
\begin{problem}[Vector Subtraction is Anti-Commutative]{prob: vector subtraction anticommutative}
Show that $\vec{v} - \vec{u} =  (\vec{v} - \vec{u})$ using the vector components in Eq. \ref{eq: 2d vector subtraction}. (Hint: factor out a negative sign from each term.)
\end{problem}
\vspace{0.15in}

We can generalize these two dimensional formulas pretty easily.  If we have $n$ mutually perpendicular unit vectors, $\{\Hat{x}_j\}_{j=1}^n$, then to get the individual components, we draw the relevant $\Hat{x}_j$,$\Hat{x}_k$ coordinate systems like we did with $j=x$ and $k=y$ in \figref{fig: Coordinate Addition}.  Then we add \textbf{tip-to-tail} component-wise.  Thus, if 
\begin{align*}
    \vec{v} &= v_1\Hat{x}_1 + v_2\Hat{x}_2 + \dots + v_n\Hat{x}_n = \sum_{j=1}^n v_j\Hat{x}_j,
    \\
    \vec{u} &= u_1\Hat{x}_1 + u_2\Hat{x}_2 + \dots + u_n\Hat{x}_n = \sum_{j=1}^n u_j\Hat{x}_j,
\end{align*}
then the sum and difference between these vectors will be
\begin{align}
    \vec{v} + \vec{u} &= (v_1 + u_1)\Hat{x}_1 + (v_2 + u_2)\Hat{x}_2 + \dots + (v_n + u_n)\Hat{x}_n = \sum_{j=1}^n (v_j + u_j)\Hat{x}_j, \label{eq: nd vector addition}
    \\
    \vec{v} - \vec{u} &= (v_1 - u_1)\Hat{x}_1 + (v_2 - u_2)\Hat{x}_2 + \dots + (v_n - u_n)\Hat{x}_n = \sum_{j=1}^n (v_j - u_j)\Hat{x}_j. \label{eq: nd vector subtraction}
\end{align}
where we have made use of summation notation only to write a potentially gigantic sum rather concisely\footnote{In summation notation, the dummy index $j$ takes on values of all of the integers between $1$ and $n$, inclusive, inside the \textit{summand}, where each time we change a $j$-value, we then add it to all the terms we had before. For example, $$\sum_{j=1}^5 j = 1 + 2 + 3 + 4+5 = 15.$$ In the sum above, $n=5$ and the summand is $j$.}. As we can see, using components to add and subtract vectors is a rather powerful tool.  Using these vector operations as inspiration, we will search for useful geometrical functions with vector components for the remainder of this chapter.

\subsection{Scaling Vectors}
In this section, we will study what happens if we set $\vec{u} = \vec{v}$ in the addition formulas given by \equaref{eq: 2d vector additon} and \equaref{eq: nd vector addition}. Making a direct substitution, we have
\begin{align*}
    \vec{v} + \vec{v} &= (v_x+v_x)\Hat{x} + (v_y+v_y)\Hat{y} = (1+1)v_x\Hat{x} + (1+1)v_y\Hat{y} = 2v_x\Hat{x} + 2v_y\Hat{y},
    \\
    \vec{v} + \vec{v} &=  \sum_{j=1}^n (v_j + v_j)\Hat{x}_j =  \sum_{j=1}^n (1+1)v_j\Hat{x}_j = \sum_{j=1}^n 2v_j\Hat{x}_j.
\end{align*}
Hence, adding $\vec{v}$ to itself \textit{scales} each of the components of $\vec{v}$ by a factor of 2. In regular old real number arithmetic, whenever we add a number to itself, that is the equivalent of \textit{scaling} the number by a factor of 2.   By analogy, we could factor the 2 out of the equations above to get $\vec{v}+\vec{v} = 2\vec{v}$. Then, if we add $\vec{v}$ to $2\vec{v}$, again, by analogy, we would get $\vec{v}+2\vec{v} = 3\vec{v}$, and so on for all the integers (prove this to yourself using the same argument for positive integers, and then use the subtraction formulas for the negative integers.). But the question remains of whether we can scale a vector by any number even if that number is not an integer.

One could try to answer this question by adding some series of vectors together to get non-integer scaling, however this would ultimately require us to be able to scale vectors by non-integer numbers anyway, and so the logic would kind of be circular.  To fix this, we instead make use of vector components to help us understand what it means to scale vectors outside of adding or subtracting the same vector over and over again.  We just need two pieces of information from $\vec{v}+\vec{v} = 2\vec{v}$ to generalize. First, we notice that integer scaling keeps the vector \textit{colinear}, meaning it leaves \equaref{eq: theta xy plane} and \equaref{eq: theta jk} invariant.
\begin{align*}
    \tan\theta_{2\vec{v}} &= \frac{2v_y}{2v_x} = \frac{v_y}{v_x} = \tan\theta_{\vec{v}}
    \\
    \tan(\theta_{2\vec{v}})_{jk} &= \frac{2v_k}{2v_j} = \frac{v_k}{v_j} =  \tan(\theta_{\vec{v}})_{jk}
\end{align*}
where the vector subscripts on $\theta$ differentiate between the directions for either vectors. Further, scaling $\vec{v}$ by $2$ scales the length of the vector by $2$, as can be seen from \equaref{eq: 2d length} and \equaref{eq: nd length}:
\begin{align*}
    v_{2\vec{v}} &= \sqrt{(2v_x)^2 + (2v_y)^2} = \sqrt{2^2\left[ (v_x)^2 + (v_y)^2 \right]} = 2\left(\sqrt{v_x^2 + v_y^2}\right) = 2v_{\vec{v}}
    \\
    v_{2\vec{v}} &= \sqrt{\sum_{j=1}^n (2v_j)^2} = \sqrt{2^2\sum_{j=1}^n v_j^2} = 2\left(\sqrt{\sum_{j=1}^n v_j^2}\right) = 2v_{\vec{v}}
\end{align*}
Therefore, when we end up scaling vectors in a more general sense, we want to make sure that the scaled vector is also colinear and the length is scaled by the length (magnitude) of same factor.


\begin{figure}
    \centering
    \includegraphics[width = 4.25in, keepaspectratio]{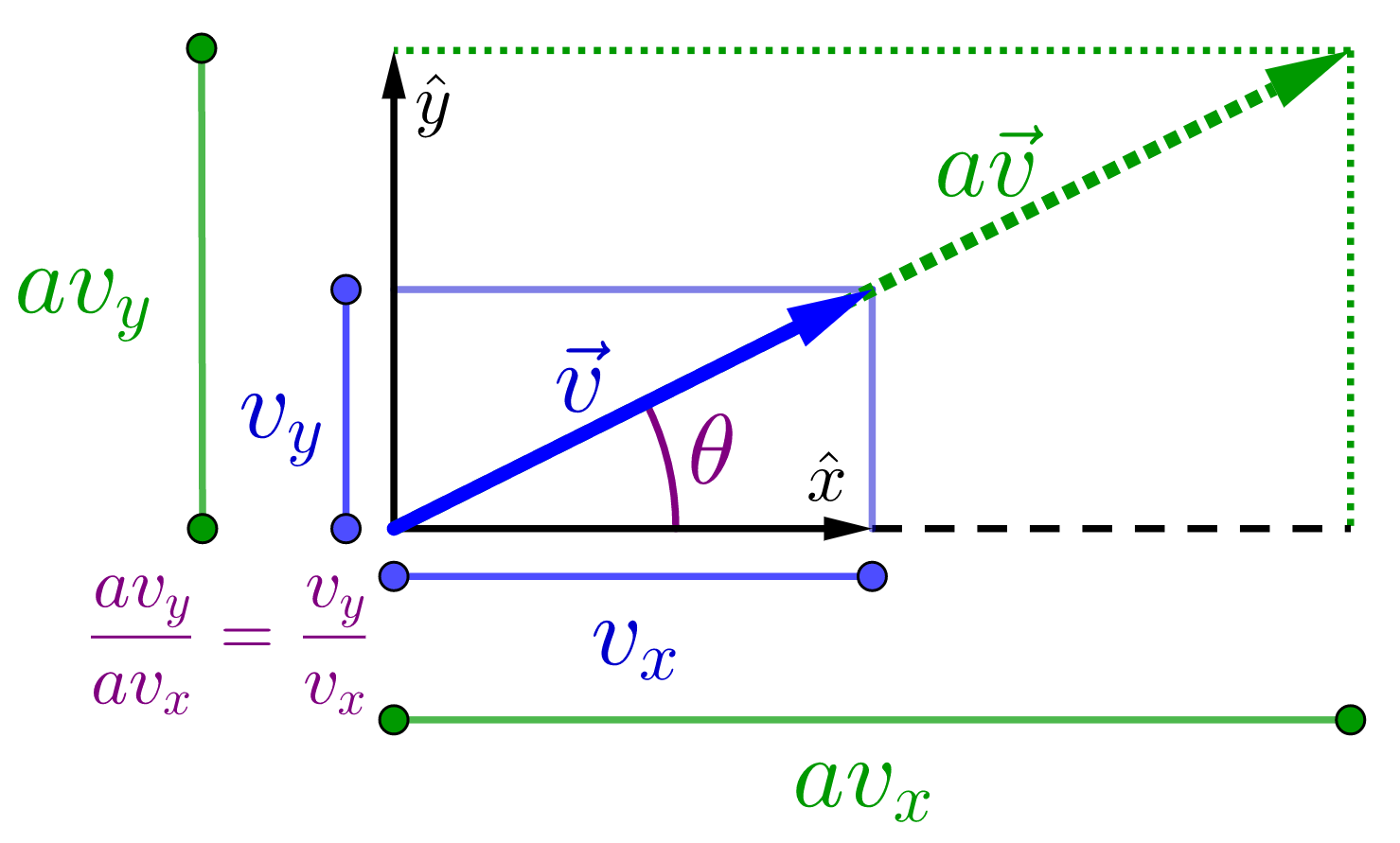}
    \caption{Whenever we have a function that scales all of the components of a vector $\vec{v}$ by the same real number $a$, then we interpret this as $a\vec{v}$. Notice that this leaves \equaref{eq: theta xy plane} unchanged.}
    \label{fig: scaling vector}
\end{figure}


It is important to realize that the reason why the direction and length of a vector behave as they do under scaling by a factor of $2$ is because each individual component is scaled by that same factor.  Before, this was because we added $\vec{v}$ to itself tip-to-tail, and the scaling of the components was a consequence of our use of coordinates.  Now, however, we will use an alternative approach, and seek a function that scales all the components by the same amount, as is shown in \figref{fig: scaling vector}.  In this figure, we consider a function $f_a$ that takes in vector $\vec{v}$ with a set of component $\{v_x,v_y\}$ and multiplies each component by a real number $a$.  Therefore, under this definition,
\begin{align*}
    f_a: \{v_x,v_y\} \rightarrow \{av_x,av_y\}.
\end{align*}
Notice we can do this for all $a\in\mathds{R}$ because $v_x,v_y\in\mathds{R}$, and any two real numbers can be multiplied together to get another real number, meaning this function is defined component-wise.  This function also keeps the vector $\vec{v}$ colinear\footnote{The only limitation here is that $a\neq 0$ because there is no multiplicative inverse of 0.  In other words, we are only guaranteed to have a $1/a\in\mathds{R}$ if $a\neq 0$.} while scaling the vector's length.
\begin{align*}
    \tan\theta_{a\vec{v}} &= \frac{av_y}{av_x} = \frac{v_y}{v_x} = \tan\theta_{\vec{v}}
    \\
    v_{a\vec{v}} &= \sqrt{(av_x)^2 + (av_y)^2} = \sqrt{a^2\left[ (v_x)^2 + (v_y)^2 \right]} = \sqrt{a^2}\left(\sqrt{v_x^2 + v_y^2}\right) = \vert a \vert v_{\vec{v}}.
\end{align*}
We must take care to remember that negative signs reverse the direction of a vector.  This is equivalent to $\tan\theta$ switching between the first and third quadrants of a Cartesian plane, when both the numerator and denominator switch sign. However, since the reversed vector falls on the same line, the resultant vector of this scaling is still colinear.  Additionally, since we can insert $a = 2$ into the function above and obtain the same results as we had before, we conclude that in two dimensions, multiplying a vector's components by any real number is equivalent to multiplying the whole vector by the same factor. In other words, our generalization holds!  I leave it to you to use a similar argument to show that in $n$-dimensions,
\begin{align}
    a\vec{v} = (av_1)\Hat{x}_1 + (av_2)\Hat{x}_2 + \dots + (av_n)\Hat{x}_n = \sum_{j=1}^n (av_j)\Hat{x}_j, \label{eq: nd vector scaling}
\end{align}
is the correct scaling function $f_a$ because it also keeps the vector colinear (all the $\tan\theta$ are invariant) while scaling the length of $\vec{v}$ by $\sqrt{a^2}$.

This picture of vector scaling is actually what gives normal numbers their name of \textbf{scalar}.  Many other texts describe scalars as object with only magnitude or length.  In this context, our scalar is $a$ and the magnitude of $a$ is $\sqrt{a^2} = \vert a\vert$.  I chose to introduce scalars as a function that act on vectors because that is all we can do with scalars and vectors.  We can never add a scalar to a vector because our definition of vector addition involves aligning vectors \textbf{tip-to-tail}.  We can never subtract a scalar from a vector (or vice versa) because vector subtraction involves aligning vectors tail-to-tail.  What we are allowed to do with scalars is multiply them by vectors.  This then either stretches or shrinks the vector, but it always keeps the vector colinear.  It may reverse the vector's direction, but scalar multiplication can never change one component without changing all of the others by the same proportion.

A useful application of vector scaling is in finding a formula to transform any vector into a unit vector that points in the same direction as the original.  To do this, we recall that a vector is only a unit vector if its length is one. Thus, if we have a vector $\vec{v}$, we seek a scalar $a$ such that the length of $a\vec{v}$ is given by
\begin{align*}
    \vert a\vec{v}\vert = \sqrt{\sum_{j=1}^n (av_j)^2} = 1.
\end{align*}
The use of absolute value bars is used to represent length in analogy with the absolute value of real numbers used to represent the length or magnitude of that number.  But since the scalar multiplies every term in the sum then we can factor out the $a^2$ inside the radicand.
\begin{align*}
    \vert a\vec{v}\vert = \sqrt{a^2\sum_{j=1}^n v_j^2} = \vert a\vert \sqrt{\sum_{j=1}^n v_j^2} = 1.
\end{align*}
But if we choose our set of coordinates, then we should know all of the components of that vector $\{v_j\}$, and those are just numbers, therefore if we knew their values, we could in principle square them and then add them and then take their square-root.  But ultimately, that is just a number that we know\footnote{Also, we assume it to be nonzero.}. Remember that we want $\vert a\vec{v}\vert = 1$ and we want the vectors to point in the same direction, hence $a>0$.  Then we conclude
\begin{align*}
    a = \frac{1}{\sqrt{\sum\limits_{j=1}^n v_j^2}} = \frac{1}{v},
\end{align*}
where the last equality holds from the length of the $n$-dimensional vector $\vec{v}$, written as $v = \vert v\vert$, and given by \equaref{eq: nd length}.  Hence, for any nonzero vector $\vec{v}$, we have
\begin{align}
    \hat{v} = \left(\frac{1}{v}\right) \vec{v} = \frac{\vec{v}}{v}, \label{eq: General Unit Vector}
\end{align}
which is the general way to write any unit vector from a known vector.

\subsection{Dot Product}
\begin{figure}
    \centering
    \includegraphics[width = 4.25in, keepaspectratio]{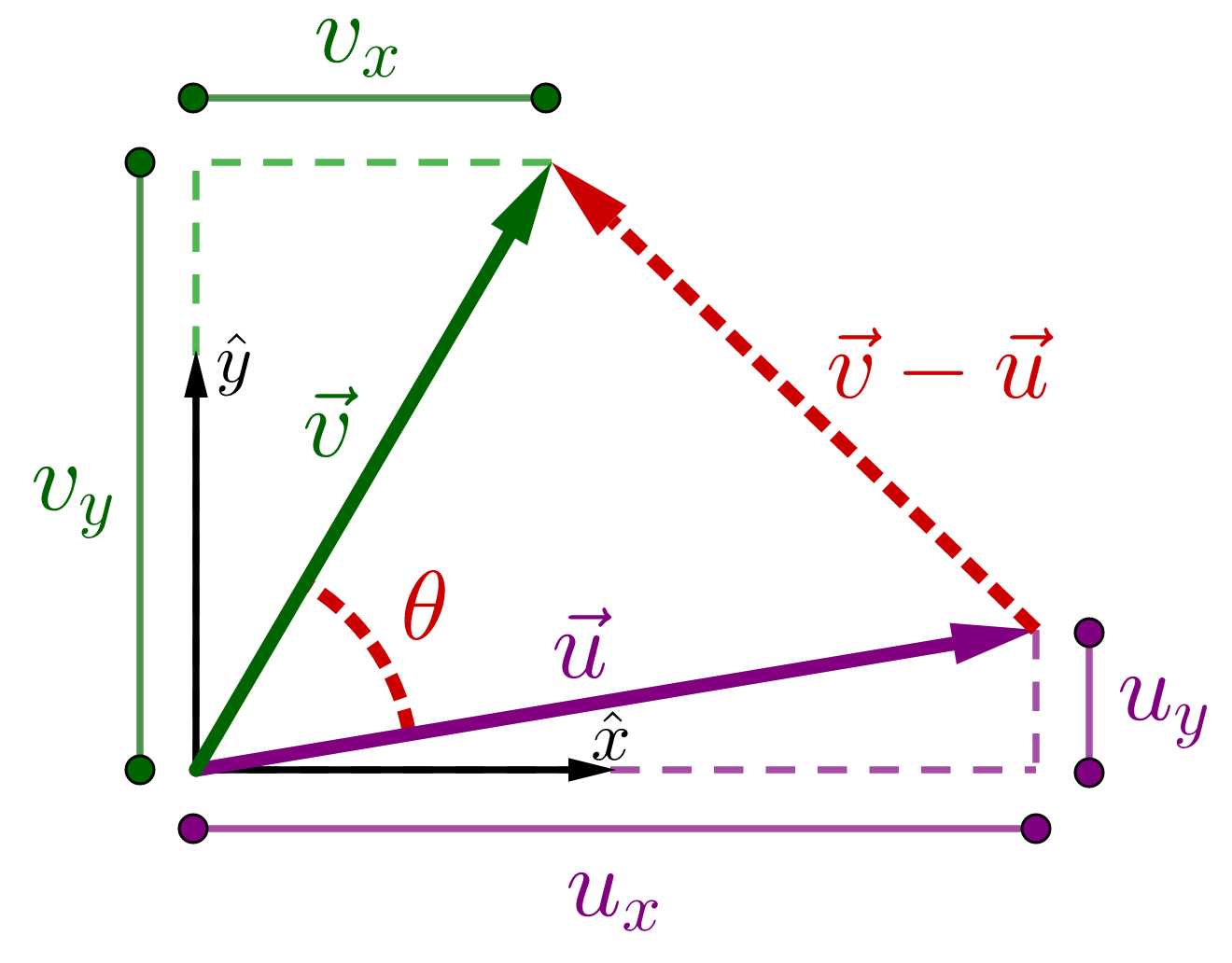}
    \caption{By the Law of Cosines, we can relate the lengths of three legs of a triangle to their angles.  Likewise, we can relate the lengths of these legs to their components.  Therefore, we should be able talk about angles in terms of the components of the legs.}
    \label{fig: Law of Cosines}
\end{figure}

We return now to a consequence of combining geometry and coordinate systems to talk about vectors.  Specifically, we will look at vector subtraction of $\vec{u}$ from $\vec{v}$, as is shown in \figref{fig: Law of Cosines}. Using \equaref{eq: 2d vector subtraction}, we can find $\vec{v}-\vec{u}$ as 
\begin{align*}
    \vec{v}-\vec{u} = (v_x-u_x)\Hat{x} + (v_y-u_y)\Hat{y}.
\end{align*}


\noindent Therefore the length of the difference, denoted by absolute value bars just like as is done for the length of real numbers, is 
\begin{align*}
    \vert \vec{v}-\vec{u}\vert = \sqrt{\left(v_x - u_x \right)^2 + \left(v_y - u_y \right)^2},
\end{align*}
given \equaref{eq: 2d length}.  Now we employ the Law of Cosines\footnote{For a refresher on the Law of Cosines, please check out \cite{hyperphys_law_of_cosines, wikipedia_law_of_cosines}.} to write 
\begin{align*}
    \vert \vec{v}-\vec{u}\vert = \sqrt{v^2 + u^2 - 2uv\cos\theta}
\end{align*}
where $\theta$ is the angle between $\vec{v}$ and $\vec{u}$, as is shown in \figref{fig: Law of Cosines}.  But these two expressions for $ \vert \vec{v}-\vec{u}\vert $ must be equivalent.  To proceed, we square both expressions\footnote{We square the square-roots because square-roots are algebraic nightmares (hard to handle); so it is, in general, a good idea to get rid of them if you can in any derivation.} and equate them
\begin{align*}
    \left(v_x - u_x \right)^2 + \left(v_y - u_y \right)^2 = v^2 + u^2 - 2uv\cos\theta
\end{align*}
By rearranging terms, we can write the angular part in terms of the squares as
\begin{align*}
    2uv\cos\theta &= v^2 + u^2 - \left(v_x - u_x \right)^2 - \left(v_y - u_y \right)^2
    \\
    &= \left(\sqrt{v_x^2 + v_y^2} \right)^2 + \left(\sqrt{u_x^2 + u_y^2}\right)^2 - \left(v_x - u_x \right)^2 - \left(v_y - u_y \right)^2 \\
    &= v_x^2 + v_y^2 + u_x^2 + u_y^2 - \left(v_x - u_x \right)^2 - \left(v_y - u_y \right)^2.
\end{align*}
If we recall that $(a-b)^2 = a^2 + b^2 - 2ab$ when $a,b\in\mathds{R}$, then we have
\begin{align*}
    2uv\cos\theta &= v_x^2 + v_y^2 + u_x^2 + u_y^2 - \left( v_x^2 + u_x^2 - 2u_xv_x\right) - \left( v_y^2 + u_y^2 - 2u_yv_y \right)
    \\
    &= v_x^2 + v_y^2 + u_x^2 + u_y^2 - v_x^2 - u_x^2 + 2u_xv_x - v_y^2 - u_y^2 + 2u_yv_y
    \\
    &= (v_x^2 - v_x^2) + (v_y^2 - v_y^2) + (u_x^2 - u_x^2) + (u_y^2 - u_y^2) + 2u_xv_x + 2u_yv_y
    \\
    &= 0 + 0 + 0 + 2(u_xv_x + u_yv_y)
    \\
    &= 2(u_xv_x + u_yv_y).
\end{align*}
After dividing both sides by 2 then we can conclude
\begin{align}
    uv\cos\theta = u_xv_x + u_yv_y.
\end{align}
This formula is very powerful because it tells us how the angle between any two, two dimensional vectors is related to the components of those vectors.  Namely,
\begin{align}
    \cos\theta = \frac{u_xv_x + u_yv_y}{\sqrt{u_x^2 + u_y^2}\,\sqrt{v_x^2 + v_y^2}}, \label{eq: 2d theta dot product}
\end{align}
when we divide both sides by the lengths of $\vec{u}$ and $\vec{v}$. The quantity in the numerator is pretty important.  For example, if it happens to be the case that $u_xv_x = - u_yv_y$, then 
\begin{align}
    \cos\theta = \frac{u_xv_x + u_yv_y}{\sqrt{u_x^2 + u_y^2}\,\sqrt{v_x^2 + v_y^2}} = \frac{-u_yv_y + u_yv_y}{\sqrt{u_x^2 + u_y^2}\,\sqrt{v_x^2 + v_y^2}} = \frac{0}{\sqrt{u_x^2 + u_y^2}\,\sqrt{v_x^2 + v_y^2}} = 0.
\end{align}
Then we would conclude that $\theta = \arccos(0) = 90^{\textrm{o}}$.  This actually turns out to be a necessary and sufficient condition to tell if two vectors are perpendicular, or in fancy linear algebra talk, this condition tells us if two vectors are \textbf{orthogonal}. Because of this seeming utility, we give that quantity the name of the vector \textbf{dot product}\footnote{Other courses or textbooks call this quantity a \textit{scalar product} because it itself is a function that returns a scalar.  However, I think this will probably lead to confusion for anyone first learning there is a difference between a scalar and a vector.}.  The two dimensional dot product is defined below
\begin{align}
    \vec{u}\cdot\vec{v} = u_xv_x + u_yv_y = uv\cos\theta.\label{eq: 2d dot product}
\end{align}
where the little $\cdot$ symbol is explicitly written; hence the name \textquotedblleft dot\textquotedblleft product.  In $n$-dimensions, the dot product formula becomes
\begin{align}
    \vec{u}\cdot\vec{v} = u_1v_1 + u_2v_2 + \dots + u_nv_n = \sum_{j=1}^n u_jv_j =uv\cos\theta. \label{eq: nd dot product}
\end{align}
Actually it is pretty straightforward to derive the $n$-dimensional case in the same way that the two dimensional case was. It is worth it to try it out in Problem \ref{prob: nd dot product}.

\vspace{0.15in}
\begin{problem}[$n$-Dimensional Dot Product]{prob: nd dot product}
In this problem, I will walk you through how to derive the $n$-dimensional dot product cosine formula.  I would recommend adhering to the following steps, but by all means, if you have a better way to derive it, definitely use it instead.
\begin{enumerate}[(a)]
    \item Define two $n$-dimensional vectors $\vec{u}$ and $\vec{v}$ according to \equaref{eq: nd Cartesian vector}.
    \item Write out $\vert \vec{u} - \vec{v}\vert = \sqrt{u^2 + v^2 - 2uv\cos\theta}$ according to the Law of Cosines.
    \item Square both sides.
    \item Expand the left-hand side as $\vert \vec{u} - \vec{v}\vert^2 = (\vec{u} - \vec{v})\cdot (\vec{u} - \vec{v})$ using the $n$-dimensional dot product (only the components though).
    \item Expand the right hand side's $u^2 + v^2$ in terms of their components.
    \item Compare both side and divide by any of the 2's that may appear at the end.
\end{enumerate}
\end{problem}
\vspace{0.15in}

Now again, \textbf{if the dot product between two nonzero vectors is zero, then they are mutually orthogonal (perpendicular)}.

Given that we know $\vec{u}\cdot\vec{v}$, is it possible to determine what $\vec{v}\cdot\vec{u}$ is?  Based off of our geometrical understanding of the dot product as being a way to measure the angle in between two vectors, it intuitively makes sense that this angle should be the same in both $\vec{u}\cdot\vec{v}$ and $\vec{v}\cdot\vec{u}$.  This intuition turns out to be completely correct as the following line of reasoning shows
\begin{align*}
    \vec{u}\cdot\vec{v} = u_1v_1 + u_2v_2 + \dots + u_nv_n = v_1u_1 + v_2u_2 + \dots + v_nu_n = \vec{v}\cdot\vec{u}.
\end{align*}
In other words, the \textbf{dot product is commutative}.

Now that I have made the claim that any two vectors with a vanishing dot product are orthogonal, we should test it with two vectors that we know to be perpendicular: $\Hat{x}$ and $\Hat{y}$.  If we write out these unit vectors in terms of their components, then we have
\begin{align*}
    \Hat{x} &= 1\Hat{x} + 0\Hat{y},
    \\
    \Hat{y} &= 0\Hat{x} + 1\Hat{y}.
\end{align*}
Since \equaref{eq: nd dot product} says that the dot product is the sum of the products of all the individual components of a vector, then
\begin{align*}
    \Hat{x}\cdot\Hat{y} = 1(0) + 0(1) = 0.
\end{align*}
We then conclude that the angle between $\Hat{x}$ and $\Hat{y}$ is indeed $90^{\textrm{o}}$, which shows that this new vector function is also consistent with the framework we have built up so far.  


\begin{figure}
    \centering
    \includegraphics[width = 4.5in, keepaspectratio]{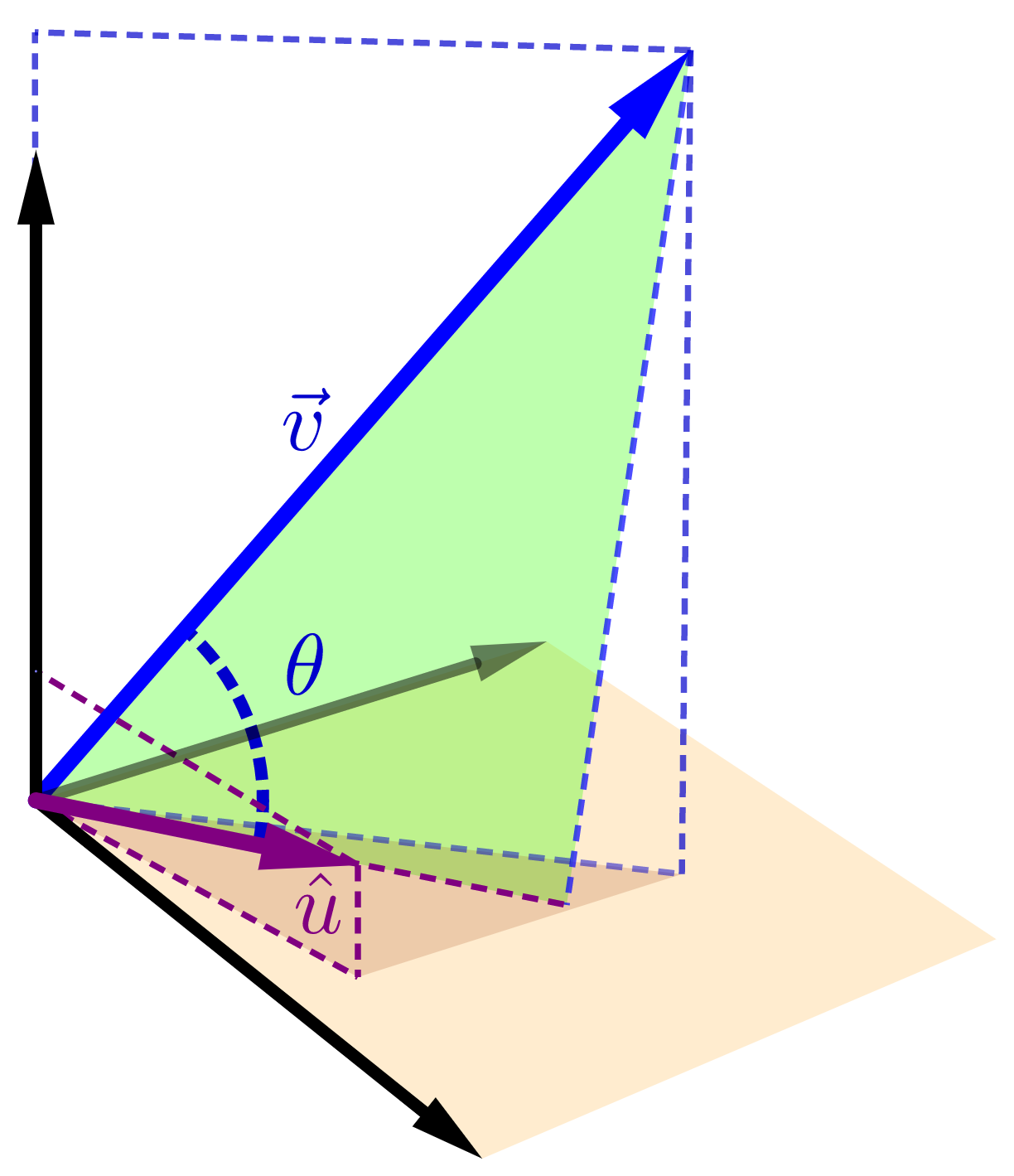}
    \caption{When we take the dot product of any vector $\vec{v}$ with any unit vector $\Hat{u}$, then we obtain $v\cos\theta$ by \equaref{eq: nd dot product}.  This situation corresponds geometrically to the diagram above, where the value $v\cos\theta$ is the length of $\vec{v}$ along the $\hat{u}$ direction in the (green) plane spanned by $\vec{v}$ and $\hat{u}$. If $v\cos\theta > 1$ ($v\cos\theta < 1$), this length will be longer (shorter) than $\hat{u}$.}
    \label{fig: Projection}
\end{figure}


We move on now to something slightly different. Since we have an expression for the dot product in terms of the lengths of two vectors being \textit{dotted}, let's consider the case where one of the vectors, $\vec{u}$, is actually a unit vector.  Thus $\vec{u}\rightarrow \hat{u}$ and $u \rightarrow 1$ in \equaref{eq: nd dot product} as shown below
\begin{align*}
    \hat{u}\cdot\vec{v} = v\cos\theta. 
\end{align*}
(We cannot use the component notation here because we have not specified any basis vectors for our coordinates, yet.)  This problem is drawn in \figref{fig: Projection}.  In the figure, the angle between $\vec{v}$ and $\hat{u}$ is $\theta$.  Notice that this angle is in the plane spanned by the two vectors.  It is not necessarily in the (orange-ish) horizontal plane, nor is it necessarily in either vertical planes.  This would physically correspond to the length of the shadow of $\vec{v}$ cast on $\Hat{u}$ if a light were shined behind it at the unit vector. I want to emphasize here that we have not established a set of coordinates yet.  In fact, I really don't even need the black coordinate vectors to talk about the cosine side of the dot product | I put them in because it helps me draw a three dimensional picture in only two dimensions. It turns out that we don't need components to talk about the dot product representing directions as it is here because the cosine side of \equaref{eq: nd dot product} is derived entirely from coordinate-free geometry (Law of Cosines). Hence, this length, or \textbf{projection}, of $\vec{u}$ in the $\hat{u}$ direction allows us to talk about the directionality of $\vec{v}$ without necessarily specifying our coordinates!  Formally, the projection function\footnote{Some authors say that the projection is a vector, whereas the way I have written it in \equaref{eq: projection} makes it scalar.  I have chosen not to include the vector form in this chapter to try and mitigate as much potential confusion as possible.} is given by 
\begin{align}
    \textrm{proj}_{\hat{u}} (\vec{v}) = \hat{u}\cdot\vec{v} = v\cos\theta, \label{eq: projection}
\end{align}
where the function is read as \textquotedblleft the projection of $\vec{v}$ onto the unit vector $\hat{u}$\textquotedblleft.  In this sense, we are holding the unit vector fixed while we see how much different $\vec{v}$ vectors point in its direction.  If we recall from \equaref{eq: General Unit Vector} that any $n$-dimensional unit vector is 
\begin{align*}
    \hat{u} = \frac{\vec{u}}{u} = \frac{u_1\hat{x}_1 + u_2\hat{x}_2 + \dots + u_n\hat{x}_n}{\sqrt{u_1^2 + u_2^2 + \dots + u_n^2}},
\end{align*}
then
\begin{align}
    \textrm{proj}_{\hat{u}} (\vec{v}) = \left(\frac{\vec{u}}{u} \right)\cdot \vec{v} = \frac{\vec{u}\cdot \vec{v}}{u}, \label{eq: proj v onto u}
\end{align}
which then implies the dot product can be written in terms of a projection as
\begin{align}
    \vec{u}\cdot\vec{v} = u\,\textrm{proj}_{\hat{u}} (\vec{v}).\label{eq: dot product projection}
\end{align}
Since the dot product is really just a number $u$ multiplied by $\textrm{proj}_{\hat{u}}(\vec{v})$, then many people interchangeably refer to taking a dot product as finding a projection of one vector in the direction of another.

This comes up all the time in classical mechanics, electromagnetism, relativity, but probably most interestingly, using the idea of a dot product to tell us something about a vector's direction is everywhere in quantum mechanics. When systems behave quantum mechanically, they are only allowed to occupy certain states in a many, or infinitely, dimensional \textit{Hilbert} space, and we associate each possible state with its own unit vector, or direction in that space.  For example, we can run things like \textit{Stern-Gerlach experiments}\footnote{These are really interesting experiments that I highly recommend checking out the following sources to try and understand them \cite{gerlach1922experimental,stern_gerlach_exp_revisited}. A user-friendly applet to experiment with is here \cite{phet_dubson_mckagan_wieman_2011}.} to find out that electrons can only ever be spin-up or spin-down when we measure them, but can never be both at once.  We then would use spin-up as something similar to an $\hat{x}$ and spin-down as something similar to a $\hat{y}$, since we have seen that $\hat{x}\cdot\hat{y} = 0$, which implies that these vectors have no projection in the direction of the other. Typically, we \textit{normalize} state vectors in quantum mechanics, meaning we make every state vector unit-length (so they are unit vectors). By doing this, we make a state's \textbf{projection} along a particular direction related to the probability of that state of the system being equivalent to the physical meaning behind that direction.  For example, if we have a state vector $\vec{\psi}$ given by
\begin{align*}
    \vec{\psi} = \vec{\mathrm{up}},
\end{align*}
then it only is projected in the spin-up direction, meaning the state of the electron is spin-up. We could also ask what is the probability of this state being spin-down using the square of the projection operation, \equaref{eq: dot product projection}.
\begin{align*}
    \mathrm{Probability} = \left[\mathrm{proj}_{\mathrm{down}}(\vec{\psi}) \right]^2= \left(\vec{\mathrm{down}}\cdot \vec{\mathrm{up}}\right)^2 = 0^2 = 0.
\end{align*}
We are also free to study electrons whose spins may be less clearly defined for us, but more on that when you get there in a quantum mechanics course. If you are very eager to read ahead right now,  some very thorough introductory resources on this matter are Townsend's and Sakurai's quantum mechanics books \cite{townsend_2012, sakurai_napolitano_2011}.

A more immediate application of the dot product's ability to select the projection of one vector onto another vector is in the physical description of \textbf{work}.  In physics, we define work as the amount of force exerted along a particular distance.  The term \textquotedblleft along\textquotedblright, though, implies that direction is involved.  If we think about this intuitively, if we push an object so that it moves in a particular direction, we can only possible associate the object's motion with our push.  But if we push an object one way, and it totally moves in an orthogonal direction, then the distance the object moved along our push is zero. Therefore, we cannot say we were responsible for the object's motion at all.  But this is is the exact behavior that the dot product describes.  As is shown in \figref{fig: Projection}, \textbf{the projection selects the amount one vector points in the direction of another}.  Then \textbf{the dot product scales the projection by the length of the other vector}. 

An application of the dot-product-as-projection idea comes into play when we calculate the $\textrm{proj}_{\Hat{v}}(\vec{v})$.  This function should intuitively yield $v = \vert \vec{v}\vert$ because the projection measures how much a vector points in the direction of the unit vector.  But here the unit vector is $\Hat{v}$, which by \equaref{eq: General Unit Vector} points in the same direction as $\vec{v}$.  So we are really just measuring the length of $\vec{v}$ in the direction of $\vec{v}$, which is exactly the length of $\vec{v}$.  This, however, is just our intuition based on our (wordy) English interpretations of our math.  We need to verify that our work does indeed return the same result with math, not just with words.  By \equaref{eq: projection}, we have
\begin{align*}
    \textrm{proj}_{\Hat{v}}(\vec{v}) = \Hat{v}\cdot\vec{v} = v\cos\theta = v.
\end{align*}
The last equality holds because the angle between $\hat{v}$ and $\vec{v}$ is $\theta = 0^{\textrm{o}}$. Hence, our projection function does indeed return $v$, which means that our intuition from before is consistent.  Now, however, if we combine \equaref{eq: dot product projection} with $\textrm{proj}_{\Hat{v}}(\vec{v})$, then we find
\begin{align*}
    \vec{v}\cdot\vec{v} = v\,\textrm{proj}_{\Hat{v}}(\vec{v}) = v\cdot v = v^2,
\end{align*}
which is again, the projection of $\vec{v}$ onto $\Hat{v}$ scaled by the length of $\vec{v}$.  More importantly, this shows that dot products are deeply related to the lengths of vector, especially because $\vec{v}\cdot\vec{v}$ was derived without specifying any components.  It holds that in general
\begin{align}
    v = \vert v\vert = \sqrt{\vec{v}\cdot \vec{v}}. \label{eq: dot product magnitude}
\end{align}
For the sake of completeness/honesty, this is not the only way to establish \equaref{eq: dot product magnitude}.  We could have done it using the component form of the dot product, but I wanted to show this using the projection formula without components to provide more intuition as to why this relationship exists. The generality that exists from the dot product, or the more general \textit{inner products}, is physicists use to talk about lengths.

\subsection{Cross Product}
To finish up this chapter, we just need to talk about one more incredibly useful (geometrical) vector operation.  So far, we have added and subtracted vectors, scaled vectors, and found a way to multiply vectors so that we can talk about how much they point in the directions of others.  But whatever happened to multiplying two lengths to get an area?

This also turns out to be a physically interesting quantity, but it must be distinct from the dot product, since the dot product only scales the length of one vector by the length of the other in the direction of the first.  It is not explicitly considered as the area of the parallelogram created by two vectors; hence we cannot conclude that the dot product is in general that particular area.  We seek now to find this area.


\begin{figure}
    \centering
    \includegraphics[width = 5in, keepaspectratio]{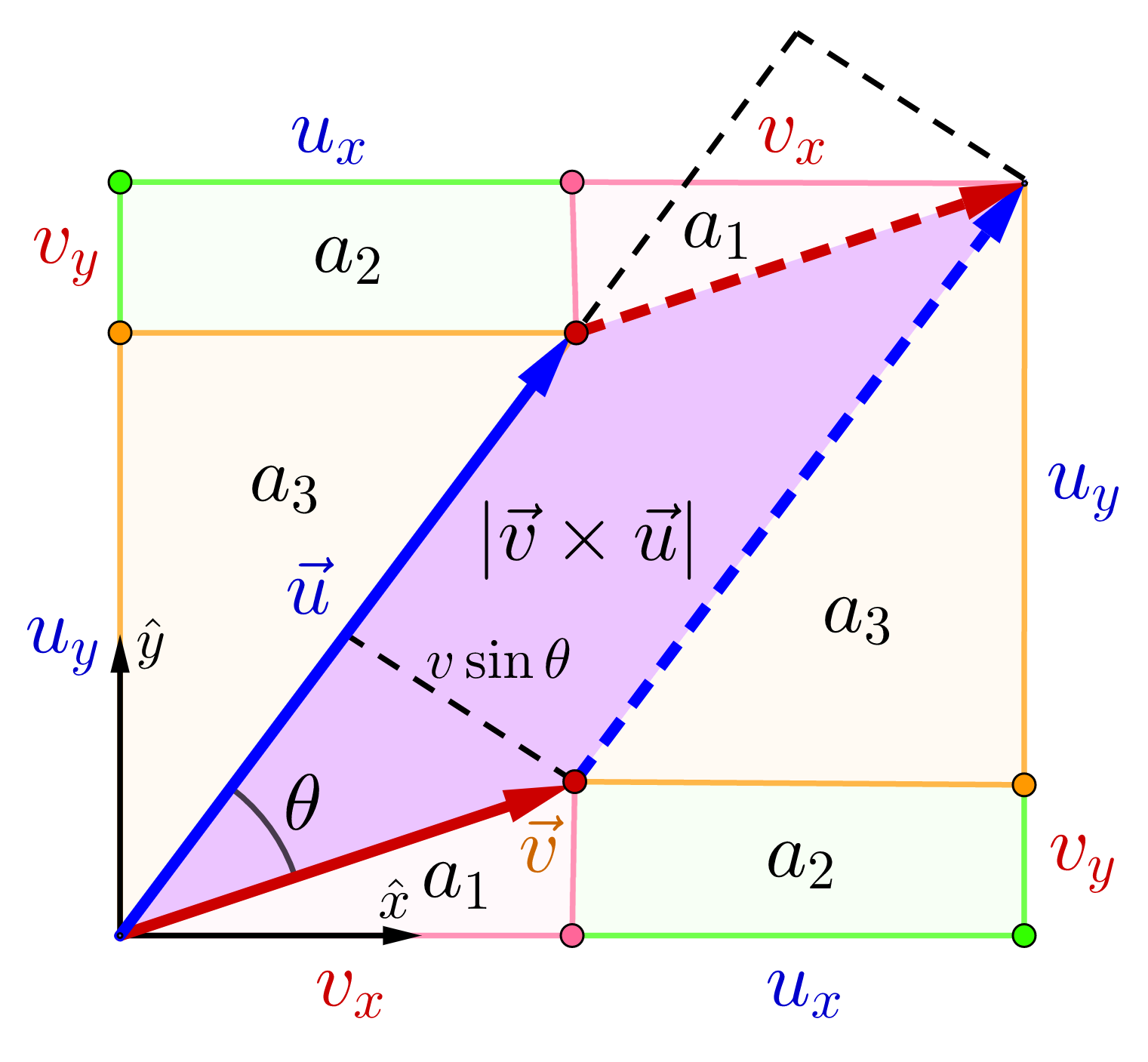}
    \caption{A geometric calculation of the area of the central parallelogram created by $\vec{v}$ and $\vec{u}$, denoted by $\vert \vec{v}\times\vec{u}\vert $. The combination of geometry with the idea of vector components leads to a peculiar property of the area: without careful considerations, the resultant area may become negative! }
    \label{fig: Cross Product}
\end{figure}


Consider the two vectors $\vec{u}$ and $\vec{v}$ drawn in \figref{fig: Cross Product}.  If we line them up tail-to-tail, then we see that they span a (purple-ish) parallelogram. From simple geometry, we remember that the area of this parallelogram, labeled $\vert \vec{v}\times\vec{u}\vert$, is
\begin{align}
    \vert \vec{v}\times\vec{u}\vert = uv\sin\theta,
\end{align}
where the quantity $v\sin\theta$ is the \textquotedblleft height\textquotedblright$\,$ in the formula $\textrm{area } = \textrm{ base }\times\textrm{ height}$. We can also find the area of the parallelogram though by using the components of $\vec{u}$ and $\vec{v}$ in the $\hat{x}\Hat{y}$ coordinate system.  Specifically,
\begin{align*}
    \vert \vec{v}\times\vec{u}\vert &= (v_x + u_x)(u_y + v_y) - 2a_1 - 2a_2 - 2a_3,
    \\
    &=  v_xu_y + v_xv_y + u_xu_y + u_xv_y - 2a_1 - 2a_2 - 2a_3.
\end{align*}
where the quantities $a_1$, $a_2$, and $a_3$ correspond to the areas of each of the non-parallelogram shaded regions in \figref{fig: Cross Product}.  Using the components in the figure, we determine those areas to be 
\begin{align*}
    a_1 &= \frac{1}{2}v_xv_y.
    \\
    a_2 &= u_xv_y,
    \\
    a_3 &= \frac{1}{2}u_xu_y
\end{align*}
where the area of the triangles are $(\textrm{base }\times\textrm{ height})/2$. Substituting these areas into the previous equation yields
\begin{align*}
     \vert \vec{v}\times\vec{u}\vert &= v_xu_y + v_xv_y + u_xu_y + u_xv_y - v_xv_y - 2u_xv_y - u_xu_,
     \\
     &= v_xu_y + (u_xv_y - 2u_xv_y) + (v_xv_y - v_xv_y) + (u_xu_y - u_xu_y,
     \\
     &= v_xu_y - u_xv_.
\end{align*}
Hence we have 
\begin{align}
    \vert \vec{v}\times\vec{u}\vert = uv\sin\theta = v_xu_y - u_xv_y, \label{eq: area of parallelogram}
\end{align}
which is the way to relate the area of the parallelogram to the lengths of the vector that span it to the components of those vectors.  So naturally, we need to check if this area is commutative, just like we did for the dot product.
\begin{align*}
    \vert \vec{u}\times\vec{v}\vert = u_xv_y - v_xu_y = - (v_xu_y - u_xv_y) = - \vert \vec{v}\times\vec{u}\vert.
\end{align*}
I hope you have issues with the equation above. In it, I claim that a nonnegative area $\vert \vec{u}\times\vec{v}\vert$ is equal to a negative area $-\vert \vec{v}\times\vec{u}\vert$. The only way this would hold for arbitrary areas is if the area of parallelograms is always zero! What happened? Technically, there isn't anything actually wrong here besides my sloppy notation.  I \textit{should have} included absolute value bars on everything to guarantee nonnegativity. However, leaving it out highlights a problem here with this area.  There is a nontrivial sign change that accompanies the component form of evaluating the parallelogram's area.  Furthermore this sign change arose when we switched the order of the vectors being multiplied | a property called \textbf{anti-commutativity} | and so there is a significant mathematical difference between multiplying the vectors one way versus the other.

If we think of this sign problem along the lines of vector scaling, then we would remember that we can completely reverse a vector's direction by multiplying it by $-1$.  This leads us to think that maybe we could encapsulate this odd sign behavior if we treated the area of the parallelogram in \figref{fig: Cross Product} as a vector instead of a scalar. But if this area is the length of some vector, what is the vector's direction?  By some dimensional analysis, the area must have different units than either constituent vector if these vectors are going to have any physical utility for us. For example, if $\vec{u}$ represents position and $\vec{v}$ represents velocity, then the area would have dimensions of $\textrm{length } \times \textrm{ speed}$ | a quantity that is clearly different than the dimensions of either. Additionally, by looking at \figref{fig: Cross Product}, the area of the parallelogram is invariant when we \textit{rotate} the whole picture by some angle clockwise or counterclockwise. Even though the constituent vectors change direction under rotations, the magnitude \textit{and} sign of the cross product do not as long as the order of the vectors is the same. These two facts seem to indicate that the cross product behaves differently than either constituent vector\footnote{This is true in general because the cross product is an \textit{axial vector} rather than a \textit{polar vector} like all of those we have been talking about up until this point. For more information, see \cite{axial_vector}.} it makes sense that this new vector should be in a direction that is totally distinct from either constituent vector\footnote{Actually, in a unit-less space (one where the axes do not have units ascribed to them), the rotation argument is sufficient to establish the orthogonality of the cross product.}.  By \equaref{eq: projection} and the discussion that follows it, we know that two vectors are totally distinct if and only if they are \textit{orthogonal} | that is, the angle between the vectors is $90^{\textrm{o}}$. But that means our new \textquotedblleft area vector\textquotedblright $\,$ must be perpendicular to \textit{both} of its constituent vectors.  This idea holds when we consider that there is a perpendicular line running through \figref{fig: Cross Product}, and since scaling the vector by $-1$ keeps a vector \textit{colinear}, the anti-commutativity of this \textquotedblleft area vector\textquotedblright $\,$ preserves the orthogonality. Thus, we \textit{define} the \textbf{vector cross product}.  Explicitly, \textbf{the cross product of the vectors } $\vec{v}$ \textbf{and} $\vec{u}$ \textbf{ is a vector whose magnitude is the area spanned by both vectors, and whose direction is orthogonal to both vectors}.

Hence, we could introduce a new basis vector, let's call it $\hat{z}$, and use it to write down the cross product of the vectors $\vec{v}$ and $\vec{u}$ in \figref{fig: Cross Product}
\begin{align}
    \vec{v}\times\vec{u} = \left(v_x\hat{x} + v_y\hat{y} \right) \times \left(u_x\hat{x} + u_y\hat{y} \right) = 0\hat{x} + 0\hat{y} + (v_xu_y - v_yu_x)\hat{z}, \label{eq: xy cross product}
\end{align}
where the $\times$ symbol is explicitly written. Now, as long as $\hat{x}\cdot\hat{z} = \hat{y}\cdot\hat{z} = 0$, then this definition of the cross product will be orthogonal to both $\vec{v}$ and $\vec{u}$ and this vector will have the length of the parallelogram spanned by the two | I leave it to you to prove this using \equaref{eq: nd dot product} for the orthogonality and \equaref{eq: nd length} for the length.  Furthermore, the \textit{anti-commutativity} of the area calculation is satisfied for this definition as well.  Written explicitly, 
\begin{align}
    \vec{u}\times\vec{v} &=  0\hat{x} + 0\hat{y} + (u_xv_y - u_yv_x)\hat{z} \nonumber
    \\
    &= 0\hat{x} + 0\hat{y} - (u_yv_x - u_xv_y)\hat{z}\nonumber
    \\
    &= -\left[ 0\hat{x} + 0\hat{y} + (v_xu_y - v_yu_x)\hat{z} \right] \nonumber
    \\
    &= -\left(\vec{v}\times\vec{u} \right), \label{eq: anti-commutativity of cross product}
\end{align}

By looking at \equaref{eq: xy cross product}, we can deduce a couple of properties.  First, we consider the cross product of a vector with itself, namely $\vec{v}\times \vec{v}$.  By \equaref{eq: xy cross product}, we have
\begin{align}
    \vec{v}\times\vec{v} = 0\hat{x} + 0\hat{y} + (v_xv_y - v_yv_x)\hat{z} =  0\hat{x} + 0\hat{y} + (v_xv_y - v_xv_y)\hat{z} =  0\hat{x} + 0\hat{y} + 0\hat{z} = \vec{0}. \label{eq: xy v cross v = 0}
\end{align}
Now this normally may seem weird\footnote{It is actually a more general property of anti-commutative operators.}, but in terms of our geometrical picture given by \figref{fig: Cross Product}, this interpretation makes complete sense: the area of the parallelogram spanned by $\vec{v}$ and $\vec{v}$ is zero, because the area of a line is zero! In other words, if the angle between the two vectors $\theta$ aligned tail-to-tail approaches zero (or 180$^{\textrm{o}}$), the vectors being crossed span a thinner and thinner parallelogram.  Another property that is important to us to consider is what happens when we scale one vector by a scalar $a$ and then cross them.  In other words, what is $(a\vec{v})\times\vec{u}$? By \equaref{eq: xy cross product} and \equaref{eq: nd vector scaling}, we have
\begin{align}
    (a\vec{v})\times\vec{u} &= \left(av_x\hat{x} + av_y\hat{y} \right) \times \left(u_x\hat{x} + u_y\hat{y} \right) \nonumber
    \\
    &= 0\hat{x} + 0\hat{y} + (av_xu_y - av_yu_x)\hat{z} \nonumber
    \\
    &= \hat{x} + 0\hat{y} + a(v_xu_y - v_yu_x)\hat{z} \nonumber
    \\
    &= a \left[ \hat{x} + 0\hat{y} + (v_xu_y - v_yu_x)\hat{z} \right] \nonumber
    \\
    &= a\left(\vec{v}\times\vec{u} \right). \label{eq: av cross u }
\end{align}
What we conclude from \equaref{eq: av cross u } is that scaling one of the vectors being crossed effectively scales the entire cross product, itself, by that same exact factor. But this again makes sense in terms of parallelograms because if we were to stretch or shrink one of the sides of the parallelogram by a factor of $a$, then we would expect the area to increase or decrease by $\vert a \vert$.  I recommend that you show that
\begin{align}
    (a\vec{v})\times (b\vec{u}) = ab\left(\vec{v}\times\vec{u} \right), \label{eq: av cross bu}
\end{align}
using a very similar argument (except now both vectors are scaled).

\subsection{Unit Vectors are Ambidextrous (but we like them to be Right-Handed)}
Let's dive a little deeper into what \equaref{eq: xy cross product} implies if our coordinate system be consistent with itself.  To do this, we expand the first equality in the equation just like we would for binomial multiplication:
\begin{align*}
    \left(v_x\hat{x} + v_y\hat{y} \right) \times \left(u_x\hat{x} + u_y\hat{y} \right) = v_x\hat{x}\times u_x\hat{x} + v_x\hat{x}\times u_y\hat{y} + v_y\hat{y}\times u_x\hat{x} + v_y\hat{y}\times u_y\hat{y}
\end{align*}
By \equaref{eq: av cross bu}, we can simplify the equation above by factoring out the scalar values of the components to get
\begin{align*}
    \left(v_x\hat{x} + v_y\hat{y} \right) \times \left(u_x\hat{x} + u_y\hat{y} \right) = v_x u_x \left(\hat{x}\times \hat{x}\right) + v_xu_y\left(\hat{x}\times \hat{y}\right) + v_yu_x\left(\hat{y}\times \hat{x}\right) + v_yu_y\left(\hat{y}\times \hat{y}\right)
\end{align*}
By \equaref{eq: xy v cross v = 0}, both terms with $\hat{x}\times\hat{x}$ and $\hat{y}\times\hat{y}$ are identically zero (again, because the parallelogram spanned by only a single vector is really just a line which has vanishing area).  Then, by \equaref{eq: anti-commutativity of cross product}, $\hat{y}\times\hat{x} = -\left( \hat{x}\times\hat{y}\right)$. Thus,
\begin{align*}
    \left(v_x\hat{x} + v_y\hat{y} \right) \times \left(u_x\hat{x} + u_y\hat{y} \right) = v_x u_x \left(\vec{0}\right) + v_xu_y\left(\hat{x}\times \hat{y}\right) - v_yu_x\left(\hat{x}\times \hat{y}\right) + v_yu_y\left(\vec{0}\right).
\end{align*}
Given that $a\vec{0}=\vec{0}$, for every scalar $a\in\mathds{R}$ (prove this for yourself | remember that the zero vector $\vec{0}$ is really just a set of components that are all $0$), and $\vec{0} + \vec{v} = \vec{v}$, we conclude
\begin{align}
    \left(v_x\hat{x} + v_y\hat{y} \right) \times \left(u_x\hat{x} + u_y\hat{y} \right) &=  \vec{0} + \left(v_xu_y - v_yu_x\right)\left(\hat{x}\times \hat{y}\right) + \vec{0} \nonumber
    \\
    &= \left(v_xu_y - v_yu_x\right)\left(\hat{x}\times \hat{y}\right) \nonumber
\end{align}
Furthermore, since $0\vec{v} = \vec{0}$, it must be true that
\begin{align*}
    \vec{v}\times\vec{u} = \left(v_x\hat{x} + v_y\hat{y} \right) \times \left(u_x\hat{x} + u_y\hat{y} \right) = 0\hat{x} + 0\hat{y} + \left(v_xu_y - v_yu_x\right)\left(\hat{x}\times \hat{y}\right).
\end{align*}
Upon comparing this result with \equaref{eq: xy cross product}, we are forced to conclude that
\begin{align}
    \hat{x}\times\hat{y} = \hat{z},\label{eq: xhat cross yhat zhat}
\end{align}
in order for the cross product to be consistent with the way we write vectors using components and the basis vectors $\hat{x}$ and $\hat{y}$.

But even though \equaref{eq: xhat cross yhat zhat} must be true for consistency, it actually does not tell us anything about the \textit{geometrical} direction of $\hat{z}$.  Remember that up until now, the cross product is not really geometrically unique | all that is required up until this point is that the cross product be orthogonal to the two vectors being crossed.  Since there is always a perpendicular line running through the parallelogram spanned by two vectors\footnote{This is only true is three and seven dimensions. In other dimensions, a more general object called the \textit{wedge} or \textit{exterior} product must be used in lieu of the cross product.}, then the direction of our cross product up until now is still ambiguous.  It could be either one direction along the line or the other and everything cross-product-wise would still check out. In \figref{fig: Cross Product}, the two directions would be either \textit{out-of-the-page} or \textit{into-the-page}. This ambiguity leads us to the idea of \textbf{handedness} in coordinate systems.


\begin{figure}
    \centering
    \includegraphics[width = 5in, keepaspectratio]{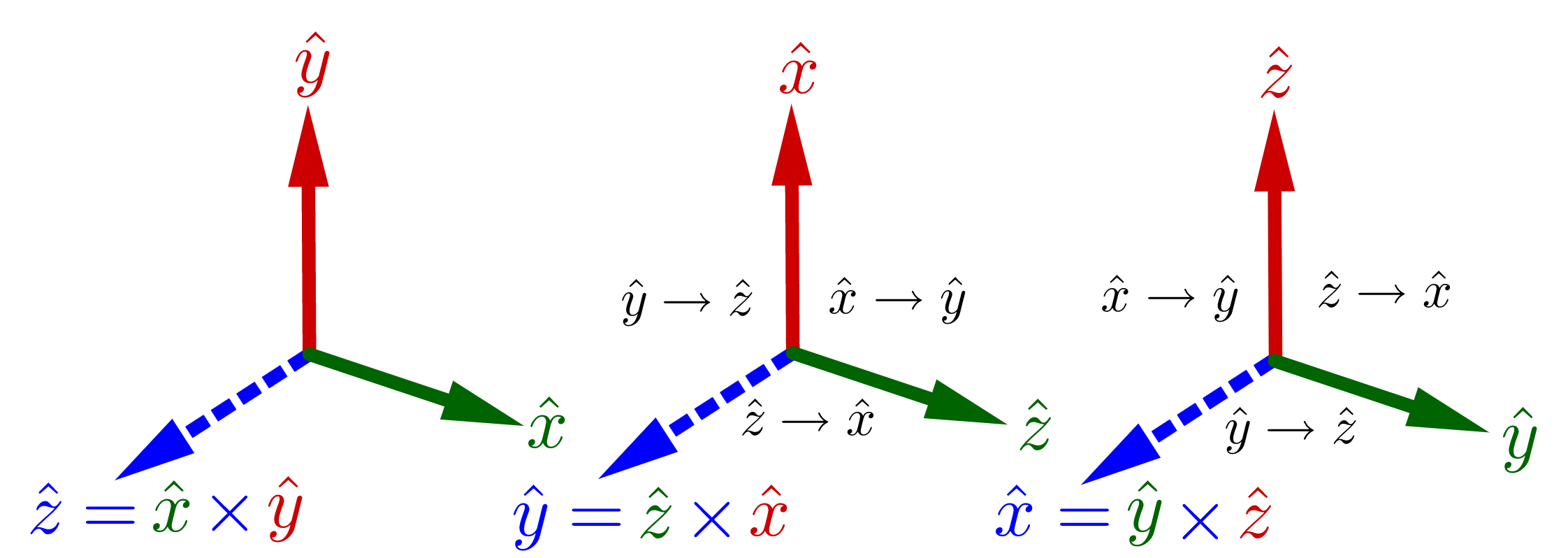}
    \caption{A diagram of how we define \textit{right-handedness} among the Cartesian unit vectors $\{\hat{x},\hat{y},\hat{z}\}$.  Throughout this diagram, we choose to label each direction differently, but the direction of each vector remains the same in each set of axes.  Starting from the leftmost set of axes, we have $\hat{x}\times\hat{y} = \hat{z}$, where the dashed (blue) vector defines the otherwise ambiguous cross product $\hat{z}$.  By rotating the labels \textbf{counter-clockwise} and then taking (green vector) $\times$ (red vector) we can define $\hat{z}\times\hat{x} = \hat{y}$ and $\hat{y}\times\hat{z} = \hat{x}$.}
    \label{fig: Right-Handedness}
\end{figure}


Since there are two options for our $\hat{z}$ in \figref{fig: Cross Product} to have, either \textit{out-of-the-page} or \textit{into-the-page}, we should have two options for handedness in our coordinate systems. In practice, we must choose our coordinates to be either \textbf{right-handed} or \textbf{left-handed}. In a right-handed coordinate system, $+\hat{z}$ \textit{defined} to be \textit{out-of-the-page}.  In a left-handed coordinate system, $+\hat{z}$ is \textit{defined} to be \textit{into-the-page}\footnote{This may not always be true. Sometimes, $+\hat{z}$ is still out-of-the-page, but in these cases, $\hat{x}\times\hat{y} = -\hat{z}$, which would point into-the-page. In those cases, the coordinate system is still left-handed, but the axis labeling scheme is different than the one I adopt. It heavily depends on the author because left-handed coordinates are so infrequently used that there really isn't a convention established as well as there is for right-handed coordinates.}. The reason why coordinate systems are given handedness is because three orthogonal unit vectors behave like our thumb, pointer finger, and middle finger when we orient them to be all mutually orthogonal.

To illustrate this point, consider the set of perpendicular basis vectors $\{\hat{x},\hat{y},\hat{z}\}$ drawn in the leftmost set of axes in \figref{fig: Right-Handedness}.  Since these are mutually orthogonal and fixed in space, these vectors are called \textbf{Cartesian basis vectors}\footnote{Not all basis vectors have to be Cartesian.  For example, there exist coordinate systems that have cylindrical or spherical symmetry as opposed to rectangular-prism symmetry.  However, these \textit{curvilinear} coordinate systems have basis vectors that are functions of the Cartesian basis vectors.}.  By \equaref{eq: xhat cross yhat zhat}, we know that $\hat{z} = \hat{x}\times\hat{y}$.  To remove the ambiguity in the direction of $\hat{z}$, we choose our system of coordinates to be \textbf{right-handed} so that $+\hat{z}$ would point \textit{out-of-the-page}.  Now take your pointer finger \textit{on your right hand} and align it with the $\hat{x}$-axis. Next take your right middle finger and align it with the $\hat{y}$-axis. Finally, as you should see, if you stick your right thumb straight out, then it will be pointing along the $\hat{z}$ direction. This method for finding the direction of a cross product of two vectors is called the \textbf{right-hand rule}, and therefore all systems that abide by it are \textit{right-handed}.  Notice that if you did the exact same procedure with your \textit{left} pointer and middle fingers, then your left thumb would point in the exact opposite direction as your right thumb, and therefore the leftmost set of coordinates in \figref{fig: Right-Handedness} cannot be left-handed.

I want to emphasize something now: the choice of Roman letters, $x$, $y$, or $z$, that are used to describe each direction established by our basis vectors is nothing special.  Letters are just letters.  We could have easily enough chosen to use \Sun, \Smiley, and \S $\,$ instead. The important things are the directions that these \textit{labels} represent, namely the green solid vector, the red solid vector, and the blue dashed vector in \figref{fig: Right-Handedness}, respectively.  So since the directions are the things that matter, we could really say something like
\begin{align*}
    (\textrm{green solid direction}) \times (\textrm{red solid direction}) = (\textrm{blue dashed direction}),
\end{align*}
as long as we understand that the cross product relationship above is between the three mutually orthogonal directions sketched in \figref{fig: Right-Handedness}. This then leads us to another property of right-handed coordinate systems: we must keeps the right-hand rule symmetry in $\hat{z} = \hat{x}\times\hat{y}$ whenever we change the axis labels in our coordinate system.  This is required so that our geometric and component interpretations of vectors are consistent.  Thus, if the green solid direction is \textit{labeled} as $\hat{z}$, then if we were to align our right thumb with the green solid vector in the middle set of axes in the figure and our right pointer finger with the red solid vector, we would conclude from the right-hand rule that the red solid direction (our right pointer finger) would have to be \textit{labeled} $\hat{x}$ and the blue dashed direction would have to be \textit{labeled} $\hat{y}$. Notice that even with our new labeling-scheme, the directions the vectors point in are identical to the ones they pointed in before.  Then by crossing the colors in the second set of axes, we conclude that
\begin{align}
    \hat{z}\times\hat{x} = \hat{y}. \label{eq: zhat cross xhat yhat}
\end{align}
By looking to the third set of axes, relabeling the red solid direction as $\hat{z}$ and aligning our right thumbs with it, and so on, we would see then that 
\begin{align}
    \hat{y}\times\hat{z} = \hat{x}.\label{eq: yhat cross zhat equals xhat}
\end{align}
Hence we have found the following cross product relationship for the right-handed Cartesian basis vectors:
\begin{align}
    \begin{cases}
        \hat{x} \times\hat{y} &= \hat{z}
        \\
        \hat{z}\times\hat{x} &= \hat{y}
        \\
        \hat{y}\times\hat{z} &= \hat{x}
    \end{cases}
    \label{eq: Right-Handed Cartesian Basis}
\end{align}
Notice that these relationship have a \textit{cyclical} symmetry; we can permute each of the basis vectors rightward (the rightmost basis vector then is moved to the leftmost) and proceed from one relationship to the next.  This cyclical property is used extensively in physics and mathematics, particularly when dealing with observable quantities in Quantum Mechanics.   

I want to take the time to note that a similar cross product relationship holds for left-handed Cartesian basis vectors.  However, it is conventional that coordinate systems be right-handed in physics, and so I will neither diagram it nor write the relationships because I risk confusing myself (and probably you, too).  You can totally decide to change your conventions though if you wanted | although your graders will definitely not like it! The reason you can is that the \textit{into-the-page} direction that I \textit{arbitrarily decided to NOT be the }$+\hat{z}$\textit{ direction} is still a physical thing.  The physical direction of does not change whether we decide to use right-handed coordinates or left-handed ones.  The physics is usually the same either way. However, again, \textbf{the convention in physics is that coordinate systems are right-handed}.

Oddly enough, there are physical processes that are exclusively left-handed.  For example, the weak force governing the particle physics behind nuclear decays will only act on particles that have a left-handed \textit{chirality} which is a measure of the relationship between a particle's momentum and spin angular momentum\footnote{If it just so happens that if there were particles that had right-handed chirality, then they would be undetectable via weak force interactions. The short version of the reason why this is true is weak force carriers can only \textquotedblleft see\textquotedblright$\,$ the left-handed particles. The much longer version is given in \cite{griffiths_2014}. Some of these particles, the \textit{sterile neutrinos}, are hypothesized to exist as candidates for dark matter | a bunch of matter in the universe that is only detectable via gravitational interactions, but outnumbers regular matter 5-to-1 \cite{nasa_science_dark_matter}!} One could argue now that since there is at least one fundamental physical process that prefers left-handedness, we physicists should all learn how to use left-handed coordinate systems so that our physical models have a closer connection to nature. While that may be a fair argument for some people, the truth of the matter is that any natural preference for either handedness is so rare in most physical systems it isn't apparent at all.  This is true for essentially all of undergraduate and a lot of graduate physics. Therefore, since most of our written Laws and Theories of Physics, including Maxwell's Equations written at the very beginning of this chapter, are written within the framework of a right-handed coordinate system, the physics community has stuck with the right-handed coordinate convention.

\begin{figure}
    \centering
    \includegraphics[width = 3.0in, keepaspectratio]{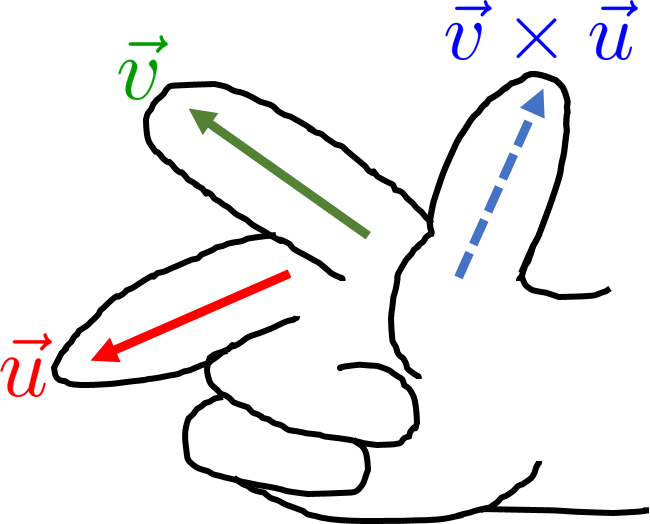}
    \caption{An illustration of the right-hand rule.}
    \label{fig: right hand rule}
\end{figure}

Before finishing up this section, I want to take the time to talk a little bit more about the \textbf{right-hand rule} because of how useful it is in finding directions of cross products.  An algorithm that you can use that will never fail to give you the correct direction of $\vec{v}\times\vec{u}$ for any two three-dimensional vectors $\vec{v}$ and $\vec{u}$ is as follows:
\begin{enumerate}
    \item Align your right pointer finger with the \textit{first} vector in the cross product (remember the order matters) $\vec{v}$.
    \item Point your right middle finger in the direction of the \textit{second} vector $\vec{u}$.
    \item Stick your right thumb out straight so it is perpendicular to both $\vec{v}$ \textit{and} $\vec{u}$.  This is the direction of $\vec{v}\times\vec{u}$.
\end{enumerate}
This algorithm is summarized in \figref{fig: right hand rule}.
It is imperative to remember that the order of the cross product matters.  This is why I italicized the words \textquotedblleft first\textquotedblright$\,$ and \textquotedblleft second\textquotedblright$\,$ in the algorithm above. I recommend proving to yourself that $\vec{u}\times\vec{v}$ has the exact opposite direction of $\vec{v}\times\vec{u}$ using the right-hand rule, just to make sure I'm telling you the truth about the right-hand rule retaining the cross product's antic-commutativity. I also want to note that this is not the only algorithm that you can use. Just like how we changed which of our fingers pointed in the direction of the blue dashed vector in \figref{fig: Right-Handedness} to calculate all of the cross product relations in \equaref{eq: Right-Handed Cartesian Basis}, we could similarly do the same thing with general three-dimensional vectors.  The key is to move from the first vector to the second to the cross product, and align your first right finger to the second right finger to your third right finger in a \textbf{counter-clockwise} rotation.  However, for a lot of people who are learning the right-hand rule for the first time, the more general cyclic properties of cross products can be overwhelming.  So when in doubt, use the algorithm above; since it is a special case of the more general cyclic properties of the cross product, it will never fail you. Just please | and this may sound silly at first but you'd be surprised how many newbies\footnote{This has literally cost me full letter grades on Physics exams before. On the exam I would write with my right hand and use my left to find cross products.  DO NOT DO THIS. IT IS WRONG.} mess this up when they are first learning the right-hand rule | remember to use your \textbf{right} (not your left) hand!

\subsection{Three-Dimensional Cross Products}
So now that we have a relationship between the cross products of our right-handed Cartesian basis vectors, we will use them to calculate a general three-dimensional formula for cross products. Interestingly enough, this formula does NOT generalize as a cross product in higher dimensions, \textit{unlike} all of the other formulas we have derived so far.  The behavior of the cross product as a orthogonal vector whose magnitude is that of a parallelogram only holds in three dimensions and seven dimensions.  There are generalizations called wedge products, but they are outside the scope of this chapter.  Without further ado, we will find the cross product between two three-dimensional vectors in a similar way to the how we found \equaref{eq: xhat cross yhat zhat}.

Consider the two vectors $\vec{v}$ and $\vec{u}$ and their right-handed Cartesian components given below
\begin{align*}
    \vec{v} &= v_x\hat{x} + v_y\hat{y} + v_z\hat{z},
    \\
    \vec{u} &= u_x\hat{x} + u_y\hat{y} + u_z\hat{z}.
\end{align*}
Then the cross product $\vec{v}\times\vec{u}$ is given by
\begin{align*}
    \vec{v}\times\vec{u} &= \left( v_x\hat{x} + v_y\hat{y} + v_z\hat{z} \right)\times \left( u_x\hat{x} + u_y\hat{y} + u_z\hat{z} \right),
    \\
    &= \;\; v_xu_x\left(\hat{x}\times\hat{x}\right) + v_xu_y\left(\hat{x}\times\hat{y}\right) + v_xu_z\left(\hat{x}\times\hat{z}\right)
    \\
    & \;\;\; + v_yu_x\left(\hat{y}\times\hat{x}\right) + v_yu_y\left(\hat{y}\times\hat{y}\right) + v_yu_z\left(\hat{y}\times\hat{z}\right)
    \\
    & \;\;\; + v_zu_x\left(\hat{z}\times\hat{x}\right) + v_zu_y\left(\hat{z}\times\hat{y}\right) + v_zu_z\left(\hat{z}\times\hat{z}\right).
\end{align*}
But since $\hat{x}\times\hat{x} = \hat{y}\times\hat{y} = \hat{z}\times\hat{z} = \vec{0}$, then we can simplify the nine terms above to only six.
\begin{align*}
    \vec{v}\times\vec{u} &= \;\; v_xu_x\left(\hat{x}\times\hat{x}\right) + v_xu_y\left(\hat{x}\times\hat{y}\right) + v_xu_z\left(\hat{x}\times\hat{z}\right)
    \\
    & \;\;\; + v_yu_x\left(\hat{y}\times\hat{x}\right) + v_yu_y\left(\hat{y}\times\hat{y}\right) + v_yu_z\left(\hat{y}\times\hat{z}\right)
    \\
    & \;\;\; + v_zu_x\left(\hat{z}\times\hat{x}\right) + v_zu_y\left(\hat{z}\times\hat{y}\right) + v_zu_z\left(\hat{z}\times\hat{z}\right),
    \\
    &= \;\; v_xu_x\left(\vec{0}\right) + v_xu_y\left(\hat{x}\times\hat{y}\right) + v_xu_z\left(\hat{x}\times\hat{z}\right)
    \\
    & \;\;\; + v_yu_x\left(\hat{y}\times\hat{x}\right) + v_yu_y\left(\vec{0}\right) + v_yu_z\left(\hat{y}\times\hat{z}\right)
    \\
    & \;\;\; + v_zu_x\left(\hat{z}\times\hat{x}\right) + v_zu_y\left(\hat{z}\times\hat{y}\right) + v_zu_z\left(\vec{0}\right),
    \\
    &= \;\;\; v_xu_y\left(\hat{x}\times\hat{y}\right) + v_xu_z\left(\hat{x}\times\hat{z}\right)
    \\
    & \;\;\; + v_yu_x\left(\hat{y}\times\hat{x}\right) + v_yu_z\left(\hat{y}\times\hat{z}\right)
    \\
    & \;\;\; + v_zu_x\left(\hat{z}\times\hat{x}\right) + v_zu_y\left(\hat{z}\times\hat{y}\right).
\end{align*}
We now make use of the anti-commutativity of the cross product to combine terms like $\hat{x}\times\hat{y}$ and $\hat{y}\times\hat{x}$.
\begin{align*}
    \vec{v}\times\vec{u} &= \;\;\; v_xu_y\left(\hat{x}\times\hat{y}\right) - v_xu_z\left(\hat{z}\times\hat{x}\right)
    \\
    & \;\;\; - v_yu_x\left(\hat{x}\times\hat{y}\right) + v_yu_z\left(\hat{y}\times\hat{z}\right)
    \\
    & \;\;\; + v_zu_x\left(\hat{z}\times\hat{x}\right) - v_zu_y\left(\hat{y}\times\hat{z}\right).
\end{align*}
Notice that some of the $+$ signs became $-$ signs. And now we use the cross product relations \equaref{eq: Right-Handed Cartesian Basis}:
\begin{align*}
    \vec{v}\times\vec{u} &= \;\;\; v_xu_y\left(\hat{x}\times\hat{y}\right) - v_xu_z\left(\hat{z}\times\hat{x}\right)
    \\
    & \;\;\; - v_yu_x\left(\hat{x}\times\hat{y}\right) + v_yu_z\left(\hat{y}\times\hat{z}\right)
    \\
    & \;\;\; + v_zu_x\left(\hat{z}\times\hat{x}\right) - v_zu_y\left(\hat{y}\times\hat{z}\right),
    \\
    &= \;\;\; v_xu_y\left(\hat{z}\right) - v_xu_z\left(\hat{y}\right)
    \\
    & \;\;\; - v_yu_x\left(\hat{z}\right) + v_yu_z\left(\hat{x}\right)
    \\
    & \;\;\; + v_zu_x\left(\hat{y}\right) - v_zu_y\left(\hat{x}\right).
\end{align*}
From here, we combine like-terms to obtain the general cross product formula for three-dimensional vectors:
\begin{align}
    \vec{v}\times\vec{u} = \left(v_yu_z -v_zu_y \right)\hat{x} + \left(v_zu_x - v_xu_z \right)\hat{y} + \left( v_xu_y - v_yu_x \right)\hat{z}. \label{eq: 3d cross product}
\end{align}
I leave it to you to show that this cross product is still orthogonal to both constituent vectors, $\vec{v}$ and $\vec{u}$ (Problem \ref{prob: cross product orthogonality}).  For completeness, I want to emphasize that even in three-dimensions,
\begin{align}
    \left\vert \vec{v}\times\vec{u} \right\vert = vu\sin\theta, \label{eq: cross product sin theta}
\end{align}
where $v$ and $u$ are the lengths of the vectors $\vec{v}$ and $\vec{u}$, respectively, and $\theta$ is the angle between the two lined up tail-to-tail, and so the full three-dimensional cross product still has a magnitude that is equal to the parallelogram spanned between $\vec{v}$ and $\vec{u}$. 

\vspace{0.15in}
\begin{problem}[Orthogonality of a Cross Product and Its Constituents]{prob: cross product orthogonality}
Use the component form of the vectors $\vec{v}$ and $\vec{u}$ to show both of the following equations
\begin{align*}
    \vec{v}\cdot\left(\vec{v}\times\vec{u} \right) &= 0,
    \\
    \vec{u}\cdot\left(\vec{v}\times\vec{u} \right) &= 0.
\end{align*}
Since these dot products are zero for both constituent vectors, then we must conclude that the cross product is indeed orthogonal to both of its constituent vectors.
\end{problem}

\subsection{A Note on Linearity}
There is one very useful property that comes up all over the place in physics, and I actually surreptitiously exploited it in a couple different places in this chapter.  This exceedingly useful property is called \textbf{linearity}, and it is so useful because it can greatly simplify calculations with vectors, or derivatives, or electromagnetic waves, or quantum mechanical wavefunctions, \textit{et cetera}.  Remarkably, this property is also incredibly simple to write down.  It essentially goes something like this: consider any operation $\mathcal{O}$ that acts on two vectors $a\vec{u}$ and $b\vec{v}$, where $a$ and $b$ are just numbers.  Then we define $\mathcal{O}$ to be linear if the following statement is true:
\begin{align}
    \mathcal{O}(a\vec{u} + b\vec{v}) = a\mathcal{O}(\vec{u}) + b\mathcal{O}(\vec{v}),\;\;\textrm{for all } a,b\in\mathds{R}. \label{eq: def linearity}
\end{align}
In other words, an operation is \textit{linear} if 
\begin{enumerate}
    \item if it is distributative over a sum,
    \item if it ignores numerical coefficients.
\end{enumerate}

To understand this property, let's consider two operations defined below
\begin{align*}
    \mathcal{O}_1(\vec{c}) &= \vec{c} + \vec{r},\;\; \vec{r}\neq\vec{0}
    \\
    \mathcal{O}_2(\vec{c}) &= \gamma \vec{c}.
\end{align*}
These two operations are ones we have already discussed.  The first is vector addition, where the operation takes in a vector $\vec{c}$ and adds $\vec{r}\neq\vec{0}$ to it.  The second is scalar multiplication, where the operation takes in a vector $\vec{c}$ and stretches or shrinks it by a factor of $\gamma$.  We will now check to see if they fit the definition of linearity. To do this, we essentially act on $a\vec{u} + b\vec{v}$ with each operator, and see if we can massage the resulting equation to look like the right-hand side of \equaref{eq: def linearity}.  We start with $\mathcal{O}_1$.
\begin{align*}
    \mathcal{O}_1(a\vec{u} + b\vec{v}) = (a\vec{u} + b\vec{v}) + \vec{r} \neq a(\vec{u} + \vec{r}) + b(\vec{v} + \vec{r}) = a\mathcal{O}_1(\vec{u}) + b\mathcal{O}_1(\vec{v})
\end{align*}
Since the equality does not hold, we conclude vector addition is \textbf{not} linear.  To see why it does not hold, remember based on the definition of linearity, \equaref{eq: def linearity} must hold \textit{for all} values of $a$ and $b$.  Therefore, if we can even find just one value where the equality does not hold, the operation is not linear (based on our definition).  To do this, let's take $a = b = 1$.  Then we would have
\begin{align*}
    \mathcal{O}_1(a\vec{u} + b\vec{v}) = \mathcal{O}_1(\vec{u} + \vec{v}) = (\vec{u} + \vec{v}) + \vec{r} = \vec{u} + \vec{v} + \vec{r}.
\end{align*}
Meanwhile,
\begin{align*}
    a\mathcal{O}_1(\vec{u}) + b\mathcal{O}_1(\vec{v}) = \mathcal{O}_1(\vec{u}) + \mathcal{O}_1(\vec{v}) = (\vec{u} + \vec{r}) + (\vec{v} + \vec{r}) = \vec{u} + \vec{v} + 2\vec{r}. 
\end{align*}
If these two equations were equal, then by comparing them, we would find $\vec{r} = 2\vec{r}$, which can only be true if $\vec{r} = \vec{0}$ (prove this to yourself). But since $\vec{r}\neq\vec{0}$, then this is a contradiction. Hence we conclude that vector addition cannot be linear.

Now we test scalar multiplication, $\mathcal{O}_2$.
\begin{align*}
    \mathcal{O}_2(a\vec{u} + b\vec{v}) &= \gamma(a\vec{u} + b\vec{v})
    \\
    &= \gamma \left[a(u_1\hat{x}_1 + u_2\hat{x}_2 + \dots) + b(v_1\hat{x_1} + v_2\hat{x}_2 + \dots)\right]
    \\
    &= \gamma \left[(au_1 + bv_1)\hat{x_1} + (au_2 + bv_2)\hat{x}_2 + \dots \right]
    \\
    &= \gamma (au_1 + bv_1)\hat{x_1} + \gamma (au_2 + bv_2)\hat{x}_2 + \dots
    \\
    &= (\gamma au_1 + \gamma bv_1)\hat{x_1} + (\gamma au_2 + \gamma bv_2)\hat{x}_2 + \dots
    \\
    &= \left[a(\gamma u_1\hat{x}_1 + \gamma u_2\hat{x}_2 + \dots) + b(v_1\gamma \hat{x_1} + v_2\gamma \hat{x}_2 + \dots)\right]
    \\
    &= a(\gamma\vec{u}) + b(\gamma\vec{v})
    \\
    &= a\mathcal{O}_2(\vec{u}) + b\mathcal{O}_2(\vec{v}).
\end{align*}
Thus, scalar multiplication is linear! I want to emphasize that it was necessary to expand out both $\vec{u}$ and $\vec{v}$ in terms of the basis vectors $\{\hat{x}_i\}$ because we have only defined scalar multiplication in terms of multiplying all the components equally by the same factor.  Then to finish the proof, we had to rearrange the multiplication of the scalars $a$, $b$, and $\gamma$ (which is possible because they are all numbers that have commutative multiplication rules), and then recombine the components.  So the proof is a little less trivial than it may have initially seemed, but it shows something pretty important: stretching and shrinking is distributive over a sum of vectors and totally ignores the numerical coefficients in front.  Remember, this was absolutely not the case for vector addition!

Using the component-based definitions of the dot product and the cross product, we can see that both of these operations are linear, as well.  I will show the proof of the dot product formula, and I encourage you to show the cross product case.  To begin, we start with the $n$-dimensional dot product formula given by \equaref{eq: nd dot product} and set $\vec{v} = a\vec{s} + b\vec{r}$, for any two vectors $\vec{s}$ and $\vec{r}$.  Then we would have
\begin{align*}
    \vec{u}\cdot(a\vec{s}+b\vec{r}) &= \sum_{j = 1}^n u_j(as_j + br_j)
    \\
    &= u_1(as_1 + br_1) + u_2(as_2 + br_2) + \dots
    \\
    &= u_1as_1 + u_1br_1 + u_2as_2 + u_2br_2 + \dots
    \\
    &= au_1s_1 + bu_1r_1 + au_2s_2 + bu_2r_2 + \dots
    \\
    &= a(u_1s_1 + u_2s_2 + \dots) + b(u_1r_1 + u_2r_2 + \dots)
    \\
    &= a\sum_{j=1}^n u_js_j + b\sum_{j=1}^n u_jr_j
    \\
    &= a(\vec{u}\cdot\vec{s}) + b(\vec{u}\cdot\vec{r}).
\end{align*}
And therefore the dot product is indeed linear because it is distributive and totally ignores numerical coefficients.  As I said before, it is also true that the cross product is linear, therefore
\begin{align*}
    \vec{u}\times(a\vec{s}+b\vec{r}) = a(\vec{u}\times\vec{s}) + b(\vec{u}\times\vec{r}).
\end{align*}
To prove this one, make use of \equaref{eq: 3d cross product}.  It is actually possible to prove both of these operations are linear using only geometry (i.e. without the use of components), but since we have the component machinery in place (and it is much more straightforward than the geometric proof), I figured it was best to use it.

The reason why linearity matters in physics is because it allows us to computer \textit{superpositions} of quantities. You may have heard of superposition in terms of wavefunctions and Schr\"odinger's famous cat, but superposition is not strictly a quantum phenomena.  It occurs in (linear) wave phenomena, finding the net force on a particle, summing over the torques on a system, \textit{et cetera}.  The utility as far as vectors go, comes in from the following example.  Suppose $\vec{u} = u\hat{x}$ and $\vec{v} = v_x\hat{x} + v_y\hat{y}$.  Thus $\vec{u}$ ONLY points in the $\hat{x}$-direction, while $\vec{v}$ points in both the $\vec{x}$- and $\hat{y}$-directions.  We now take the dot product and cross product of these two vectors, starting with the dot product.
\begin{align*}
    \vec{u}\cdot\vec{v} = \vec{u}\cdot(v_x\hat{x} + v_y\hat{y}) = v_x\vec{u}\cdot\hat{x} + v_y\vec{u}\cdot\hat{y} = uv_x(\hat{x}\cdot\hat{x}) + uv_y(\hat{x}\cdot\hat{y}).
\end{align*}
But since $\hat{x}\cdot\hat{x} = \vert\hat{x}\vert^2 = 1$ by \equaref{eq: dot product magnitude} and the definition of a unit vector, and $\vec{x}\cdot\hat{y}= 0 $ since they are perpendicular then we have
\begin{align*}
    \vec{u}\cdot\vec{v} = uv_x.
\end{align*}
But since $\vec{u}$ exclusively points in the $\hat{x}$-direction, the linearity of the dot product \textit{selected} the parallel component of $\vec{v}$ and totally \textit{ignored} the perpendicular components it!  Further, the linearity in the second term shows that it DOES NOT MATTER how big the perpendicular component of $\vec{v}$ is when dotting it with $\vec{u}$ | it is totally zero by the unit-vector construction.  This idea generalizes well in $n$-dimensions, too, actually.  Whenever vectors are dotted together, only the parallel components will survive while all of the perpendicular components will always go to zero.

The cross product actually works in the exact opposite way (except only in three and seven dimensions).  If we take these two vectors again and calculate their cross product, then we have
\begin{align*}
    \vec{u}\times\vec{v} = \vec{u}\times(v_x\hat{x} + v_y\hat{y}) = v_x\vec{u}\times\hat{x} + v_y\vec{u}\times\hat{y} = uv_x(\hat{x}\times\hat{x}) + uv_y(\hat{x}\times\hat{y}).
\end{align*}
But now $\hat{x}\times\hat{x} = \vec{0}$ and $\hat{x}\times\hat{y} = \hat{z}$, by anti-commutativity and \equaref{eq: xhat cross yhat zhat}.  Therefore,
\begin{align*}
    \vec{u}\times\vec{v} = uv_y\hat{z}.
\end{align*}
To contrast this result with the dot product, the statement above means that the linearity of the cross product \textit{selects} the component of $\vec{v}$ that is perpendicular to $\vec{u}$ and totally \textit{ignores} the parallel component.  Furthermore, this property means that it does not matter how large the parallel component is | all that matters is the size of the perpendicular part!

\section{Concluding Remarks}
In this chapter we covered the basics of vector operations.  In the very beginning, we started the discussion with talking about mathematics as a game that has a set of rules (the proper name for \textquotedblleft rule\textquotedblright$\,$ is \textit{axiom}). Although the first few sections may have initially seemed a little unnecessarily abstract when they simply were talking about real numbers and functions, I kept them in because the abstractness is to illustrate how we can formulate familiar mathematics in terms of this game structure. From here, we moved to defining a vector geometrically as lengths in a particular direction. Using geometry, we saw that we can define a commutative function of vectors that we interpret as vector addition (\textbf{tip-to-tail!}).  We then found an anti-commutative function that in turn acts as vector subtraction (tail-to-tail).  When coupled to a system of coordinates, we saw that we must be able to describe vectors in terms of their lengths along each basis vector | we call these lengths the components of the vector. Later on, to be consistent between algebra and geometry, the we found rules for scalar multiplication and the dot product. Finally, using areas of parallelograms, we built the component form of the cross product AND inserted the idea of handedness into our coordinate systems by defining cross product relationships between our basis vectors. \tblref{tbl: Summary of Vector Equations} lists the equations that are the most important (and general) throughout this chapter.  I included the equation references to bring you back to the discussion where we derived them, just in case you need some context to refresh your memory about what each equation means.

Sprinkled throughout the chapter are several proofs that I recommended that you try to do.  All of them are practically identical to what I have already done in the chapter, although they may have more components or a couple more steps.  I sincerely suggest that you try to do a few of them just to get a feel for how these proofs are done on your own | in physics and math, one of the best ways to deeply understand derivations or proofs is by putting in the time to do it out for yourself (although in my experience, a physics and math educations leaves very little time for anything but perpetual confusion...).  

I know that for many of you reading this chapter, my lack of numerical values will be unsettling | I will only get more algebraic as the chapters progress.  I do this on purpose though.  My first reason is that it is frankly easier once you get used to it.  I do remember that the phase transition between needing numbers in math and exclusively using letters is not a smooth one.  It will take time to master | probably as much time as it did for you when you first started using the symbol $\pi$ instead of $3.14159$ back in the day.  The advantage to only using letters or symbols is that you eventually need not worry about how the intermediate numerical values affect the outcome of your mathematics. By extension, the necessary variables will be left in your physical models of the natural world giving a deeper insight into how the universe works.  Unfortunately, unless you already have a handle on the physics, injecting numerical values at intermediate steps will obfuscate this insight.  But again, it takes time (a.k.a. practice) to get used to doing everything algebraically.  Hopefully this chapter and the following will serve as an external perturbation to make your phase transition that much easier.

\begin{table}
    \centering
    \caption{A summary of the important and general equations derived for vector operations.  \label{tbl: Summary of Vector Equations}}
    \scalebox{0.83}{
    \begin{tabular}{ccc}
         \hline\hline
         \textbf{Equation Description}& \textbf{Equation Formula} & \textbf{Text Reference}  \\ \hline &&\\ 
         3-Dimensional Cartesian Vector & $\vec{v} = v_x\hat{x}+v_y\hat{y}+v_z\hat{z}$ & \equaref{eq: 3d Cartesian vector}
         \\ && \\
         $n$-Dimensional Cartesian Vector & $\vec{v} = v_1\hat{x}_1 + v_2\hat{x}_2 + \dots = \sum\limits_{j=1}^n v_j\hat{x}_j $ & \equaref{eq: nd Cartesian vector}
         \\ && \\
         Angle between $xy$ Components & $\tan\theta_{xy} = \dfrac{v_y}{v_x}$ & \equaref{eq: theta xy plane}
         \\&&\\
         Length/Magnitude of Vector & $ \vert \vec{v}\vert = v = \sqrt{\sum\limits_{j=1}^n v_j^2} $ & \equaref{eq: nd length}
         \\&&\\
         Sum of Two Vectors & $\vec{u} + \vec{v} = \sum\limits_{j=1}^n (u_j+v_j)\hat{x}_j$ & \equaref{eq: nd vector addition}
         \\&&\\
         Difference of Two Vectors & $\vec{u} - \vec{v} = \sum\limits_{j=1}^n (u_j-v_j)\hat{x}_j$ & \equaref{eq: nd vector subtraction}
         \\&&\\
         Scalar Multiplication & $a\vec{v} = \sum\limits_{j=1}^n (av_j)\hat{x}_j$ & \equaref{eq: nd vector scaling}
         \\&&\\
         Unit vector in Direction of $\vec{v}$ & $\hat{v} = \dfrac{\vec{v}}{\vert \vec{v}\vert} = \dfrac{\vec{v}}{v}$ & \equaref{eq: General Unit Vector}
         \\&&\\
         Dot Product & $\vec{u}\cdot\vec{v} = \sum\limits_{j=1}^n u_jv_j = uv\cos{\theta}$ & \equaref{eq: nd dot product}
         \\&&\\
         Projection of $\vec{v}$ onto $\vec{u}$ & $\textrm{proj}_{\hat{u}}(\vec{v}) = \dfrac{\vec{u}\cdot\vec{v}}{\vert \vec{u}\vert} = \dfrac{\vec{u}\cdot\vec{v}}{u}$ & \equaref{eq: proj v onto u}
         \\&&\\
         Length/Magnitude and Dot Product & $\vert\vec{v}\vert = v = \sqrt{\vec{v}\cdot\vec{v}}$ & \equaref{eq: dot product magnitude}
         \\&&\\
         Magnitude of Cross Product & $\vert\vec{v}\times\vec{u}\vert = uv\sin\theta$ & \equaref{eq: area of parallelogram}
         \\&&\\
         Right-Handed Cartesian Basis & $\begin{aligned}
            \hat{x} \times\hat{y} &= \hat{z}
            \\
            \hat{z}\times\hat{x} &= \hat{y}
            \\
            \hat{y}\times\hat{z} &= \hat{x}
         \end{aligned}$ & \equaref{eq: Right-Handed Cartesian Basis}
         \\&&\\
         3-Dimensional Cross Product & {$\begin{aligned}
             \vec{u}\times\vec{v} &= \;\;\;\; \left(v_yu_z -v_zu_y \right)\hat{x}
             \\
             &\;\;\;\;+ \left(v_zu_x - v_xu_z \right)\hat{y}
             \\
             &\;\;\;\;+ \left( v_xu_y - v_yu_x \right)\hat{z}
            \end{aligned}$} & \equaref{eq: 3d cross product}
         \\&&\\
         Definition of Linear Operator & $\mathcal{O}(a\vec{u} + b\vec{v}) = a\mathcal{O}(\vec{u}) + b\mathcal{O}(\vec{v})$ & \equaref{eq: def linearity}
         \\&&\\
         \hline\hline
    \end{tabular}
    }
\end{table}

\newpage


\setcounter{example}{0}
\setcounter{problem}{0}

\chapter{Complex Algebra}
In this chapter, we will start a perhaps unfamiliar form of mathematics for many of you, but it is crucial to not only our understanding of quantum phenomena, but also our understanding of differential equations, Fourier Analysis of signals, all kinds of waves, and various forms of data analysis. By the end of this chapter, we want to be able to add complex numbers to our set of \textquotedblleft game pieces\textquotedblright$\,$ that we have been developing, and then eventually get to the point that we can create functions of complex numbers.  This chapter should leave you in a pretty good position to understand most, if not all, the complex mathematics you will cover in your undergraduate physics curriculum. This chapter should also set you up to begin to learn the calculus of complex-valued functions later on in your mathematics career.

\section{The Lie of \textit{Imaginary} Numbers}
Before we get going, there is a common misconception that I want to clear up.  There is no such distinction between \textit{real} and \textit{imaginary} numbers in the colloquial sense; that is, there is no set of objects that are somehow tangible that we call the real numbers, $\mathds{R}$, versus the somehow intangible objects called the imaginary numbers, $\mathds{I}$.  These sets of objects are certainly distinct mathematically, but that is due to a rotation rather than some metaphysical and mystical separation that seems to exist by calling two things real and imaginary. 

The reason why I want to address this is because the term \textquotedblleft imaginary\textquotedblright$\,$ has a totally different connotation in normal life than it does in mathematics. To be clear, at some point in the development of our algebra system, some mathematicians like Rene Decartes did truly believe that imaginary numbers really were not a thing, but a rather convenient way out of an otherwise harder problem \cite{ComplexNumberHistory}. However, mathematicians today effectively only use the word as a label whose name bears no deep meaning.  We credit people like Gauss, Cauchy, Euler, and Riemann, among others, for changing the way we think about these objects mathematically. Gauss showed that imaginary numbers are truly just an extension of the more conventional real numbers, and had even tried to get them rebranded as \textit{lateral} numbers instead.  His quote on this subject is below \cite{Gauss_Quote}.

\begin{quote}
    \textit{That this subject [imaginary numbers] has hitherto been surrounded by mysterious obscurity, is to be attributed largely to an ill adapted notation. If, for example, $+1$, $-1$, and the square root of $-1$ had been called direct, inverse and lateral units, instead of positive, negative and imaginary (or even impossible), such an obscurity would have been out of the question.}
\end{quote}

The problem with taking the word \textquotedblleft imaginary\textquotedblright$\,$ too literally in physics makes the interpretation of certain natural phenomena appear as fake or something akin to pseudoscience. For example, when describing an electron's spin, the imaginary (lateral!) unit $i = \sqrt{-1}$ appears when talking about the projection of the spin vector along the $y$-axis. If we were to naively see the presence of $i$, we might be tempted to conclude that there is something mystical about this part of our physical world. Or even worse, we might conclude that we could never measure the $y$ component of the spin because it is imaginary!  But this interpretation is not true.  Gauss' idea of \textit{lateral} numbers can be used to more appropriately explain the appearance of $i$ in electron spins.  In this case, as you will learn later in your physics career, we can only ever know precisely the projection of an electron's spin along one axis in space; we denote the forward direction with a $+$ sign and the backward direction with a $-$ sign.  However, we can measure the statistical effects of the spin's vector components in the other two dimensions in 3D space.  What this means is we have a total of 3 sets of distinct pairs of basis vectors in this spin space, which is supposed to have physical meaning in all of 3D space. Without going into too much linear algebra, we essentially need 4 components to represent the final two perpendicular axes in 3D space | but this is impossible with only the real numbers! Real components can only ever give you the magnitude and direction that a vector points along a particular line, as we discussed in the chapter on Vectors.  Hence, our very \textit{real} measurements of the natural world force us to extend the real numbers to include their lateral counterparts in order to accurately describe electron spin.

Without further ado, I will quit my (legitimate) grumbling, and proceed with our introduction to the world of complex algebra.

\section{Some Important Definitions}
Before we move on with more algebra, we need to lay down some ground rules for these things that I'm calling \textit{complex numbers}.  The first time they are typically introduced (although not the first time they were ever contrived \cite{ComplexNumberHistory}) is with the standard defining equation
\begin{align}
    x^2 + 1 = 0,
\end{align}
where one solves for $x$ to find $x = \pm \sqrt{-1}$.  Since we showed in Section \ref{subsec: Vectors - Real Numbers} that for any real number\footnote{Remember that we use the symbol $\mathds{R}$ to represent the set of all reals and the symbol $\in$ to mean \textquotedblleft is an element of.\textquotedblright} $a\in\mathds{R}$, $a^2 > 0$, then we find that there can be no real number $x$ that satisfies $x^2 + 1 = 0$.  Thus we define the \textbf{imaginary unit} as
\begin{align}
    i = \sqrt{-1}.\label{eq: def sqrt -1}
\end{align}
Likewise, we could easily define the \textbf{real unit} with the following equation $x^2 - 1 = 0$, and so we obtain $x = \pm 1$. These ideas fit in with our understanding that $1$ represents \textquotedblleft unit\textquotedblright$\,$ length along a number line.  We will define any \textbf{imaginary number} $\alpha$ as $\alpha = ai$, where $a\in\mathds{R}$. Such an object would solve the quadratic $x^2 + a^2 = 0$ for $x$.  Likewise, a \textbf{real number} $a\in\mathds{R}$ would solve the quadratic $x^2 - a^2 = 0$ for $x$.

When we move to more complicated quadratics, we have an equation that looks something like $ax^2 + bx + c = 0$, whose solutions are given by the quadratic formula
\begin{align}
    x = -\frac{b}{2a} \pm \frac{1}{2a}\sqrt{b^2 - 4ac}. \label{eq: Quadratic Formuler}
\end{align}
If we pay attention to the quantity called the \textbf{discriminant}, $D = b^2 - 4ac$, we should take note there are exactly three cases for $D$ given by the ordering property of the reals.  They are given explicitly as 
\begin{align}
    D > 0 &\Rightarrow b^2 > 4ac, \nonumber
    \\
    D = 0 &\Rightarrow b^2 = 4ac, 
    \\
    D < 0 &\Rightarrow b^2 < 4ac. \nonumber
\end{align}
The third case, $D < 0$, implies that we will again have a negative number inside of a square-root, and so this implies that there will be an imaginary unit involved in some way.  Written explicitly, when $D < 0$, then it must be true that $-D > 0$. Then
\begin{align}
    x = -\frac{b}{2a} \pm \frac{1}{2a}\sqrt{D} = -\frac{b}{2a} \pm \frac{1}{2a}\sqrt{-(-D)} = -\frac{b}{2a} \pm \frac{i}{2a}\sqrt{-D} = -\frac{b}{2a} \pm \frac{i}{2a}\sqrt{4ac - b^2}
\end{align}
What we have now is something that hopefully is a little jarring to you, especially if you have never seen complex algebra before.  We have an expression that is somehow telling us to add a real number $-b/2a$ with the imaginary number $i\sqrt{4ac-b^2}/2a$. But can we? There is mathematically a pretty large distinction between real numbers and imaginary numbers; the square (and any other nonzero even power) of a real number is always positive.  Meanwhile, the square (and any other nonzero even power) of an imaginary number is always negative.  We \textit{proved} the former, whereas we had to \textit{define} the latter. So what gives? 

The way we deal with this conundrum is actually by defining something new. It turns out that if you were jarred before, you were right, because \textbf{there is no way to add a nonzero real number with a nonzero imaginary number and get either a purely real or imaginary number out}.  The sets of objects are just too different.  All of their differences can pretty much be reduced to the fact that
\begin{align}
    1^2 = 1 \textrm{ and } i^2 = -1,
\end{align}
and so we get around this issue by saying each of these numbers is totally distinct from one another, but they can be combined to form a greater set of numbers; just as the basis vectors $\hat{x}$ and $\hat{y}$ are totally distinct, but can be combined to form a plane.  Actually, if we take the real unit and the imaginary unit as basis vectors, we can create the \textbf{complex plane}, denoted by $\mathds{C}$. Furthermore, we define a \textbf{complex number} $z$ as the \textit{vector sum} of a \textbf{real part} $z = \mathrm{Re}(z)$ and an \textbf{imaginary part} $z = \mathrm{Im}(z)$ like the following
\begin{align}
    z = \mathrm{Re}(z) + i\,\mathrm{Im}(z).
\end{align}
It is important to see then that this definition constrains both the real part and the imaginary part to be real numbers! Please read that sentence again | it confuses a lot of people. Even though the imaginary part of a complex number is called the \textit{imaginary part}, it itself is real.  It represents the \textit{projection} (see \equaref{eq: proj v onto u}) of the complex number in the direction of $i$, whereas the real part is the projection of the complex number in the direction of $1$. 

Using this vector-like interpretation of complex numbers, then it must be true the so-called \textquotedblleft real axis\textquotedblright$\,$ and \textquotedblleft imaginary axis\textquotedblright$\,$ must together span a \textbf{complex plane}, just like the $x$-axis and $y$-axis span the $xy$-plane.  By convention, we denote $x =\mathrm{Re}(z)$ and $y = \mathrm{Im}(z)$, so we can write any complex number $z\in\mathds{C}$ as
\begin{align}
    z\in\mathds{C}\textrm{ if and only if } z = x + iy, \textrm{ where } i = \sqrt{-1}  \textrm{ and } x,y\in\mathds{R}.\label{eq: def complex number}
\end{align}
Since the complex numbers form a plane, then we can more easily see what Gauss was talking about when he claimed that the imaginary unit should be instead named the \textit{lateral unit}. The existence of this plane just means that the real numbers are accompanied by another \textit{orthogonal} axis that has had the misfortune of having us silly humans call them \textquotedblleft imaginary\textquotedblright.  The lateral numbers are present with or without us claiming they are figments of our imaginations | they are an extension of the real numbers into the generalized complex plane.  With that said, I will continue to refer to them as imaginary numbers just so you get accustomed to the vernacular. But please do not think any of this is some kind of fantasy.  If you are willing to accept the existence of the real number line, it is clear there must exist an accompanying imaginary (lateral!) number line.

\section{Conjugates and Magnitudes}
A question that we can now ask since we have defined complex numbers is whether there is a way to construct the real and imaginary parts of any complex number computationally.  In other words, if we know know any real number $x$ and any real number $y$, we can compute a complex number $z = x + iy$. That's old news.  The new question is whether we can find another complex number $z^\ast$ if we know already know a complex number $z$ such that we could calculate $x = \mathrm{Re}(z)$ and $y = \mathrm{Im}(z)$.  If we were to assume that this were possible, then we would have to start by clarifying that if $z^\ast$ were complex, then $z^\ast = u + iv$ for some combination of real numbers $u$ and $v$.  Thus,
\begin{align*}
    z &= x + iy
    \\
    z^\ast &= u + iv
\end{align*}
As of right now, we have two \textit{separate} equations with two unknowns. By \textquotedblleft separate\textquotedblright, I mean that as of right now, we don't have an equation to relate any of the variables.  Hence, we are free to \textit{choose} that $x = \mathrm{Re}(z) = (z + z^\ast)/2$.  Thus,
\begin{align*}
    x = \mathrm{Re}(z) = \frac{1}{2}\left( z + z^\ast \right) = \frac{1}{2}\left[ (x + u) + i (y + v) \right] = \frac{x + u}{2} + i\,\frac{y + v}{2}
\end{align*}
Since $x$ is purely real, then that means that it cannot be imaginary.  Thus the coefficient attached to $i$ must vanish; in other words, $y + v = 0 \Rightarrow v = -y$. This means that 
\begin{align*}
    x = \frac{x + u}{2} + i\,\frac{y + v}{2} = \frac{x + u}{2} + i0 =  \frac{x + u}{2}
\end{align*}
By multiplying both sides by 2 and subtracting over the remaining x we then find $x = u$. This means we have an expression for $z^\ast$, given $z$:
\begin{align}
    \textrm{if } z = x + iy \textrm{ then } z^\ast = x + i(-y) = x - iy.\label{eq: Complex Conjugate}
\end{align}

The complex number $z^\ast$ is a very helpful quantity | so helpful, in fact, that it is given the name of \textbf{the complex conjugate to} $\mathbf{z}$. Additionally, this complex conjugate is \textit{unique}, as based on our rules for the real numbers in Section \ref{subsec: Vectors - Real Numbers}, there exists only one $v = -y\in\mathds{R}$ if $y\in\mathds{R}$\footnote{This uniqueness property justifies our phrasing of \textit{the} complex conjugate instead of \textit{a} complex conjugate.}. Notice that $z^\ast = x -iy = x + (-i)y$.  This means that \textbf{whenever we want a complex conjugate of $z = x + iy$, then all we have to do is replace every $i$ with $-i$}. If it is not immediately clear why this is true, it's because we have defined complex numbers in such a way that any number may be written as a real part plus $i$ times an imaginary part, and then we defined the complex conjugate in terms of those arbitrary real and imaginary parts. This rule is particularly helpful for when you have a rather nasty function of complex variables, but you need the conjugate to compute something useful.  

Now what would be useful to compute with a complex conjugate? For starters, we know that we can find the real part of $z$ (or $z^\ast$) with the following formula.
\begin{align}
    \mathrm{Re}(z) = \frac{z + z^\ast}{2}. \label{eq: Real part conjugates}
\end{align}
We can actually calculate the imaginary part of $z$ as well.  I leave it to you to verify that 
\begin{align}
    \mathrm{Im}(z) = \frac{z - z^\ast}{2i}. \label{eq: Imaginary part conjugates}
\end{align}
To show this, set $y = \mathrm{Im}(z)$ and use \equaref{eq: Complex Conjugate} to solve for $y$.  If the division by $i$ is weird for you, then instead use $1/i = (1/i)(i/i) = i/(i^2) = i/(-1) = -i$. The third part of \exref{ex: real and imaginary} shows another helpful tool that the complex conjugate provides.

\vspace{0.15in}
\begin{example}[Real and Imaginary Parts]{ex: real and imaginary}
Consider the complex numbers $z = x + iy$ and $w = u + iv$, where $x,y,u,v\in\mathds{R}$.  We seek the real and imaginary parts of $z \pm w$, $zw$, and $z/w$.
\begin{enumerate}
    \item We start with $z\pm w$.
        \begin{align}
            z\pm w &= (x + iy) \pm (u + iv) \nonumber
                \\ &= (x \pm u) + i(y \pm v). \label{eq: complex z pm w}
        \end{align}
        Hence, $\mathrm{Re}(z\pm w) = x\pm u$ and $\mathrm{Im}(z\pm w) = y\pm v$.
    \item Next, we compute $zw$.
        \begin{align}
            zw &= (x + iy)(u + iv) \nonumber
            \\ &= xu + ixv + iyu + i^2yv \nonumber
            \\ &= (xu - yv) + i(xv + yu). \label{eq: complex zw}
        \end{align}
        Thus, $\mathrm{Re}(zw) = xu - yv$ and $\mathrm{Im}(zw) = xv + vu$.
    \item Lastly, we seek the quantity $z/w$ (this one is a little tricky).
        \begin{align}
            \frac{z}{w} &= \frac{x+iy}{u+iv} \nonumber
                \\ &= \frac{x+iy}{u+iv}\,\frac{w^\ast}{w^\ast} \nonumber
                \\ &= \left(\frac{x+iy}{u+iv}\right)\left(\frac{u - iv}{u - iv}\right), \textrm{ Using \equaref{eq: Complex Conjugate} for } w \nonumber
                \\ &= \frac{(x+iy)(u-iv)}{(u+iv)(u-iv)} \nonumber
                \\ &= \frac{(xu + yv) + i(xv - yu)}{u^2 + v^2} \nonumber
                \\ &= \frac{xu + yv}{u^2 + v^2} + i\,\frac{xv-yu}{u^2+v^2}. \label{eq: complex z/w}
        \end{align}
        Therefore we have
        \begin{align*}
            \mathrm{Re}\left(\frac{z}{w}\right) = \frac{xu + yv}{u^2 + v^2} \textrm{ and } \mathrm{Im}\left(\frac{z}{w}\right) = \frac{xv-yu}{u^2+v^2}
        \end{align*}
\end{enumerate}
It is important to note that since we can establish clear real and imaginary parts for each of the arithmetic operations above | addition, subtraction, multiplication, and division | these quantities are also complex numbers, by definition.  Thus, when we add, subtract, multiply, or divide complex numbers, we will always get complex numbers back!
\end{example}
\vspace{0.15in}

For a next calculation, let's find the product of $z$ with its conjugate $z^\ast$ which is often denoted as $z^\ast z$; we do so by direct substitution.
\begin{align*}
    z^\ast z &= (x - iy)(x + iy) 
    \\
    &= x^2 -iyx +ixy -i^2 y^2
    \\
    &= x^2 -(-1)^2y^2 + i(xy - yx) 
    \\
    &= x^2 + y^2 +i0 
    \\
    &= x^2 + y^2.
\end{align*}
Outright, this formula may not look very impressive, so let's try $\sqrt{z^\ast z}$:
\begin{align}
    \sqrt{z^\ast z} = \sqrt{x^2 + y^2} = \sqrt{[\textrm{Re}(z)]^2 + [\textrm{Im}(z)]^2}. \label{eq: Complex modulus}
\end{align}
Hopefully this catches your eye as the Pythagorean Theorem, or even more importantly, the equation for the \textit{length} of a two-dimensional vector \equaref{eq: 2d length}, where the $x$-component of the vector is just $x$ and the $y$-component is just $y$! Using \equaref{eq: Complex modulus}, called the \textbf{complex modulus} of $z$, we can calculate a \textit{real modulus} and and \textit{imaginary modulus} if we consider only a purely real number $x$ or purely imaginary number $\Upsilon = iy$.
\begin{align*}
    \sqrt{x^\ast x} &= \sqrt{x^2 + 0} = \vert x\vert, \textrm{ a real number has no imaginary part}
    \\
    \sqrt{\Upsilon^\ast \Upsilon} &= \sqrt{0 + y^2} = \vert y\vert, \textrm{ an imaginary number has no real part}
\end{align*}
where the absolute value bars are necessary because we implicitly took the positive square root, and so our moduli should be positive for arbitrary $x,y\in\mathds{R}$. But these expressions are just the magnitudes of the real numbers $x$ and $y$.  In other words, the complex modulus is the generalization of magnitude | a.k.a. absolute value | in the complex plane!  In other words, we may write 
\begin{align}
    \vert z\vert = \sqrt{z^\ast z} = \sqrt{[\textrm{Re}(z)]^2 + [\textrm{Im}(z)]^2}, \label{eq: complex modulus absolute value}
\end{align}
when talking about the magnitude of any complex number.

Before moving on, there are a couple of more things I want to talk about.  Firstly, what happens if we need to find the complex conjugate of a sum (or difference)? Suppose we have two complex numbers $z$ and $w$, as we do in the first part of \exref{ex: real and imaginary}, and we want to find $(z\pm w)^\ast$. We could do so directly as shown below using \equaref{eq: complex z pm w},
\begin{align}
    (z\pm w)^\ast = (x\pm u) + (-i)(y\pm v) = (x - iy) \pm (u - iv) = z^\ast \pm w^\ast. \label{eq: conjugate z pm w}
\end{align}
Notice that we replaced the $+i$ in \equaref{eq: complex z pm w} with the $-i$ to find the conjugate initially. Hence, when we have a sum (or difference) of two complex numbers, then the conjugate of the sum (or difference) is just the sum (or difference) of the conjugates. How about a product of two complex numbers? Here we use \equaref{eq: complex zw},
\begin{align}
    (zw)^\ast &= (xu - yv) + (-i)(xv + yu),\nonumber
    \\
    &= (xu + i^2yv) -i(xv + yu),\nonumber
    \\
    &= xu + (iy)(iv) - ixv - iyu, \textrm{ combine like-terms in $x$ and $-iy$}\nonumber
    \\
    &= x(u-iv) - iy(u-iv),\nonumber
    \\
    &= (x - iy)(u-iv),\nonumber
    \\
    &= z^\ast \, w^\ast. \label{eq: conjugate zw}
\end{align}
(Note derivation is a little tricky because I needed to remember that $-1 = i^2$ in the second step and then group one $i$ with $y$ and the other with $v$ in the third step.) This relation shows that the conjugate of the product is simply the product of the conjugates!  At this point, however, we have enough information to be certain that \textbf{complex conjugation is NOT a linear operation for complex numbers} (see \equaref{eq: def linearity} for the definition of a linear operator).  To see this more clearly, we first extend our definition of a linear operator to the complex plane
\begin{align}
    \mathcal{O}(a\vec{z} + b\vec{w}) = a\mathcal{O}(\vec{z}) + b\mathcal{O}(\vec{w}),\;\;\textrm{for all } a,b\in\mathds{C}. \label{eq: complex linearity def}
\end{align}
For right now, consider the \textit{complex vectors} $\vec{z}$ and $\vec{w}$ as being normal vectors whose components are complex-valued.  The specific details on these complex vectors are not totally necessary (we'll save that for linear algebra...) for right now.  What is important here is that if we have to compute $(a\vec{z} + b\vec{w})^\ast$, then we would have
\begin{align}
    (a\vec{z} + b\vec{w})^\ast &= (a\vec{z})^\ast \pm (b\vec{w})^\ast, \textrm{ by \equaref{eq: conjugate z pm w}}
    \\
    &= a^\ast\,\vec{z}^\ast \pm b^\ast\,\vec{w}^\ast. \textrm{ by \equaref{eq: conjugate zw}}
\end{align}
The only way this is equal to $a(\vec{z}^\ast) + b(\vec{w}^\ast)$ is if $a = a^\ast$ and $b = b^\ast$, implying that $a,b\in\mathds{R}$. But since our complex vectors have complex-valued components, and we have already shown that multiplication between complex numbers produced complex numbers, it follows that in general the condition that both $a$ and $b$ are real is NOT general.  Thus, in general, complex conjugation is not linear\footnote{This fact is specifically exploited in quantum mechanics and quantum field theory when dealing with time-reversal symmetry in real physical systems.}!

\vspace{0.15in}
\begin{problem}[Conjugate of a Quotient]{prob: conjugate of quotient}
We have shown that the conjugate of the sum is the sum of the conjugates (\equaref{eq: conjugate z pm w}) and the conjugate of the product is the product of the conjugates (\equaref{eq: conjugate zw}).  Show now that the conjugate of the \textit{quotient} is the quotient of the conjugates. In other words, show that $(z/w)^\ast = z^\ast/w^\ast$.

\vspace{0.15in}
\noindent (Hint: either start with $z^\ast/w^\ast$, substitute in $-i$ for the $+i$, and then show that it is the conjugate of \equaref{eq: complex z/w}, OR start with the conjugate of \equaref{eq: complex z/w} and manipulate it algebraically into $z^\ast/w^\ast$. I think both methods take similar amounts of algebra...)
\end{problem}

\vspace{0.15in}
\begin{problem}[Magnitude of the Conjugate]{prob: magnitude of conjugate}
With purely real numbers, it is true that $\vert a\vert = \vert -a\vert$ for all $a\in\mathds{R}$. Thus, magnitudes are not \textit{unique}; however, we can use a sign difference to order our real numbers from least to greatest.  This problem focuses on whether we can do a similar thing in the complex plane | if we could, then it would be true that only $z$ and $-z$ share the same magnitude, just like the reals.  We used $\sqrt{z^\ast z}$ as the definition of the magnitude of a complex number.  Show that the \textbf{complex numbers cannot be ordered like the reals} by finding at least one other complex number that shares the same magnitude of $z$ (and $-z$).  For example, show that $\vert z\vert = \vert z^\ast\vert$ for all $z = x + iy\in\mathds{C}$.

\vspace{0.15in}
\noindent (Hint: to avoid confusing yourself with all the $z$s and asterisks, define $w = z^\ast$ and then find $\vert w\vert$ using \equaref{eq: complex modulus absolute value}. Finally, compare your result with \equaref{eq: Complex modulus}.)
\end{problem}

\section{Cartesian versus Polar Representations}
\begin{figure}
    \centering
    \includegraphics[width = 4in, keepaspectratio]{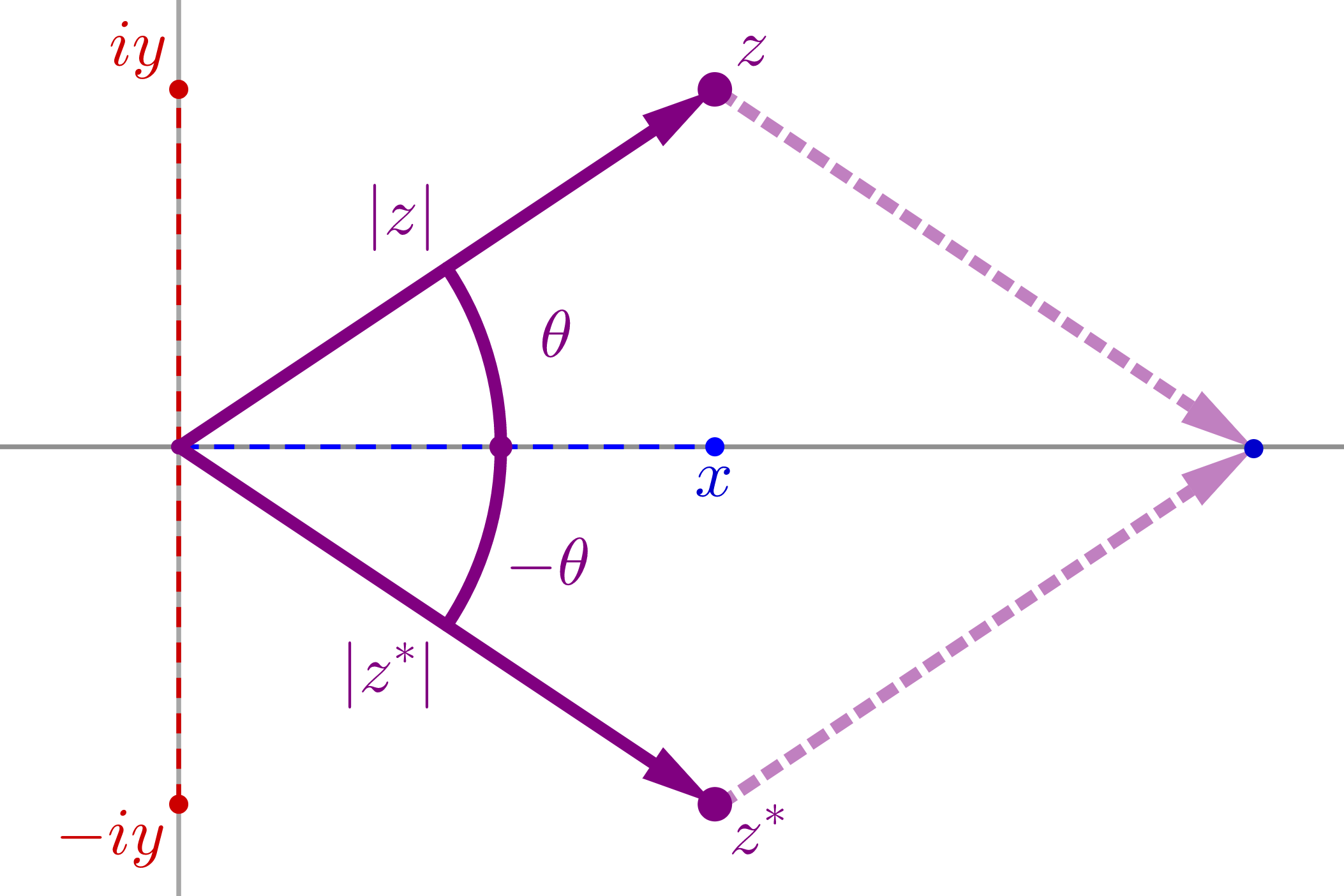}
    \caption{If we consider $z = x + iy$, then we can draw it and its conjugate, $z^\ast = x - iy$, in the complex plane. }
    \label{fig: complex vector}
\end{figure}

The fact that the real and imaginary axes form a complex plane implies that we should be able to draw any complex number in a geometric plane.  Further, since we have a Pythagorean Theorem-type relationship for the magnitude of the complex number (and $1$ is totally distinct from $i$), we are able to infer that the real and imaginary axes are orthogonal to one another | just like the $x$-axis and the $y$-axis. Figure \ref{fig: complex vector} shows a possible complex number $z = x + iy$ in the complex plane. Additionally, if we know the components $x$ and $y$, then we could easily draw the conjugate of $z$, namely $z^\ast = x - iy$. It is important to note that the points $z$ and $z^\ast$ are the complex numbers; meanwhile, based on our knowledge of vectors, we could easily draw an arrow from the origin to each complex number, where the length of each vector would the the magnitude (or length) of the complex number. Now for some trigonometry.

From the definition of sines, cosines, and tangents (see Section \ref{subsec: vector addition and coordinate systems}), we can use the geometrical right angle between the real and imaginary axes to define the angle of the the complex number, or the so-called \textbf{argument of a complex number}, denoted by $\mathrm{arg}(z) = \theta$.  As is shown in \figref{fig: complex vector}, $\theta$ is the angular elevation of the complex number above the $+x$-axis. Since we know that $\mathrm{Im}(z^\ast) = -\mathrm{Im}(z)$ while the real parts are identical, we know that $\mathrm{arg}(z^\ast) = -\mathrm{arg}(z)$. Using these angles we can then determine
\begin{align}
    \cos\theta &= \frac{\mathrm{adjacent}}{\mathrm{hypotenuse}} = \frac{x}{\vert z\vert} = \frac{x}{\sqrt{x^2 + y^2}},
    \\
    \sin\theta &= \frac{\mathrm{opposite}}{\mathrm{hypotenuse}} = \frac{y}{\vert z\vert} = \frac{y}{\sqrt{x^2 + y^2}}.
\end{align}
By solving the second equalities for $x$ and $y$ and then substituting these quantities in $z = x + iy$, we find
\begin{align}
    z = \vert z \vert \cos\theta + i\vert z\vert \sin\theta = \vert z\vert (\cos\theta + i\sin\theta). \label{eq: complex r cosine p i sine}
\end{align}
Note that this picture is consistent with just substituting in $-i$ for $+i$ to obtain a conjugate, because 
\begin{align}
    z^\ast = \vert z^\ast\vert [\cos\theta + (-i)\sin\theta] =\vert z\vert [\cos(-\theta) + i\sin(-\theta)].
\end{align}
where we have used the result of \probref{prob: magnitude of conjugate} to set $\vert z^\ast\vert = \vert z\vert$, and then we used the \textit{even} and \textit{odd} symmetry of the sinusoids: $\cos(-\theta) = \cos(\theta)$ and $\sin(-\theta) = -\sin\theta$.

Equation \ref{eq: complex r cosine p i sine} shows that any complex number $z = x + iy$ can be represented as the product of a \textbf{radial part}, and an \textbf{angular part}, as shown below
\begin{align}
    z = \underbrace{\vert z\vert}_{\textrm{Radial Part}} (\underbrace{\cos\theta + i\sin\theta}_{\textrm{Angular Part}} ),
\end{align}
as a complement to the real-and-imaginary-part representations. Again, for the sake of completeness, the radial part is the modulus of the complex number while the angular part is the argument of the complex number.  Take note that while the radial part is a real number like $\mathrm{Re}(z)$ and $\mathrm{Im}(z)$, the angular part is still complex!

Thus, we have written any complex number in either a real-and-imaginary representation and in a radial-and-angular representation.  These terms, although straight-to-the-point, are fairly clunky, so instead we name them the \textbf{Cartesian} and \textbf{Polar} representations of a complex number, respectively.  The \textquotedblleft pole\textquotedblright$\,$ in this case is the distance (magnitude/length) the complex number is from the origin.  A much more useful (and \textbf{e}l\textbf{e}gant) way to write a complex number $z = x + iy\in\mathds{C}$ and its conjugate in the polar representation is actually with an \textit{imaginary exponential} $\txte^{i\theta}$, given as
\begin{align}
    z &= r\txte^{i\theta}, \label{eq: complex polar representation}
    \\
    z^\ast &= r\txte^{-i\theta}, \label{eq: conjugate polar representation}
\end{align}
where
\begin{align}
    r &= \vert z\vert = \sqrt{x^2 + y^2}, \label{eq: complex radius}
    \\
    \theta &= \mathrm{arg}(z) = \arctan\left( \frac{y}{x}\right). \label{eq: complex argument}
\end{align}
Of course, to write such a thing, it would have to be true that
\begin{align}
    \txte^{i\theta} = \cos\theta + i\sin\theta, \label{eq: Euler's identity}
\end{align}
where $\txte \approx 2.71828182846$ is the base of the natural logarithm. Surprisingly, this result is true \textit{for all values of} $\theta$! Remarkably, this result is not too difficult to prove even though it took a supergenius like Leonhard Euler\footnote{Pronounced \textit{oiler}, unlike how they said it in \textit{The Imitation Game}, much to my chagrin...} to first do it\footnote{Hence, it is usually called the Euler identity.} | it can be done with an introductory understanding of \textit{Taylor Series} in a Calculus II course | although it is a bit beyond the scope of this chapter\footnote{Don't worry! We will come back to it later on and YOU will prove it in Problem \ref{prob: Euler's Identity} (I help you along though).}.

\vspace{0.15in}
\begin{problem}[Practice with the Polar Representation]{prob: practice with polar}
It is initially a little strange to go from the Cartesian representation to the polar representation, but ultimately using the polar representation is more convenient than the Cartesian representation (otherwise physicists wouldn't bother with it!).  So this problem is designed to have you practice the conversion for a few important numbers in physics.

\begin{enumerate}[(a)]
    \item Show that $i = \txte^{i\pi/2}$ by arguing $r(i) = 1$ and $\mathrm{arg}(i) = \pi/2$.
    
    \item Show that, in the Cartesian representation, $\sqrt{2}\,\txte^{-i\pi/4} = 1 - i $
    
    \item Consider the complex number $z = \txte^{i\pi}$. What is $z + 1$ in both Polar and Cartesian representation? 
\end{enumerate}

\end{problem}
\vspace{0.15in}

Using the polar representation, it is possible for us to construct two of the most widely used expressions in all of physics | we are going to rewrite sine and cosine in terms of exponentials. To start consider the complex number of \textit{unit magnitude} given by \equaref{eq: Euler's identity}.  Then it must be true
\begin{align}
    \mathrm{Re}\left(\txte^{i\theta}\right) = \cos\theta\in\mathds{R},
    \\
    \mathrm{Im}\left(\txte^{i\theta}\right) = \sin\theta\in\mathds{R}.
\end{align}
But by \equaref{eq: Real part conjugates} and \equaref{eq: Imaginary part conjugates}, we can write the real and imaginary parts of any complex number as a \textit{superposition} of it and its complex conjugate. Then,
\begin{align}
    \cos\theta &= \frac{\txte^{i\theta} + \txte^{-i\theta}}{2}, \label{eq: cosine in polar representation}
    \\
    \sin\theta &= \frac{\txte^{i\theta} - \txte^{-i\theta}}{2i}. \label{eq: sine in polar representation}
\end{align}
Part of the reason why these relationships are so useful, is it allows us to treat trigonometric functions | objects that are defined geometrically | as exponential functions instead, which then allow us to employ a slew of useful (and quick) multiplication, differentiation, and integration rules to otherwise algebraically cumbersome functions.  In physics, this polar representation of trigonometric functions helps us describe certain geometries and spaces in terms of essentially successive multiplications. As you can see in the chapter on Fourier Analysis, being able to convert sines and cosines into combinations of exponential functions makes otherwise impossible algebra much simpler.  As we continue through this book, I will highlight more locations where these can be immediately implemented to make your life easier because the sooner you begin to feel comfortable with Eqs. \ref{eq: cosine in polar representation} and \ref{eq: sine in polar representation}, the faster you will begin to see through the mathematics of difficult subjects like signals, optics, and quantum mechanics to understand the underlying phenomena in a much more precise way.  There is something potentially unsettling by writing geometric formulas as exponential formulas; it seems to imply there is a clear \textit{rotational} aspect of multiplying two numbers, rather than the simpler stretching-and-shrinking interpretation that was valid with the real numbers.  We study this more in the next section.

\section{Multiplication = Dilation + Rotation}
\begin{figure}
    \centering
    \includegraphics[width = 4in, keepaspectratio]{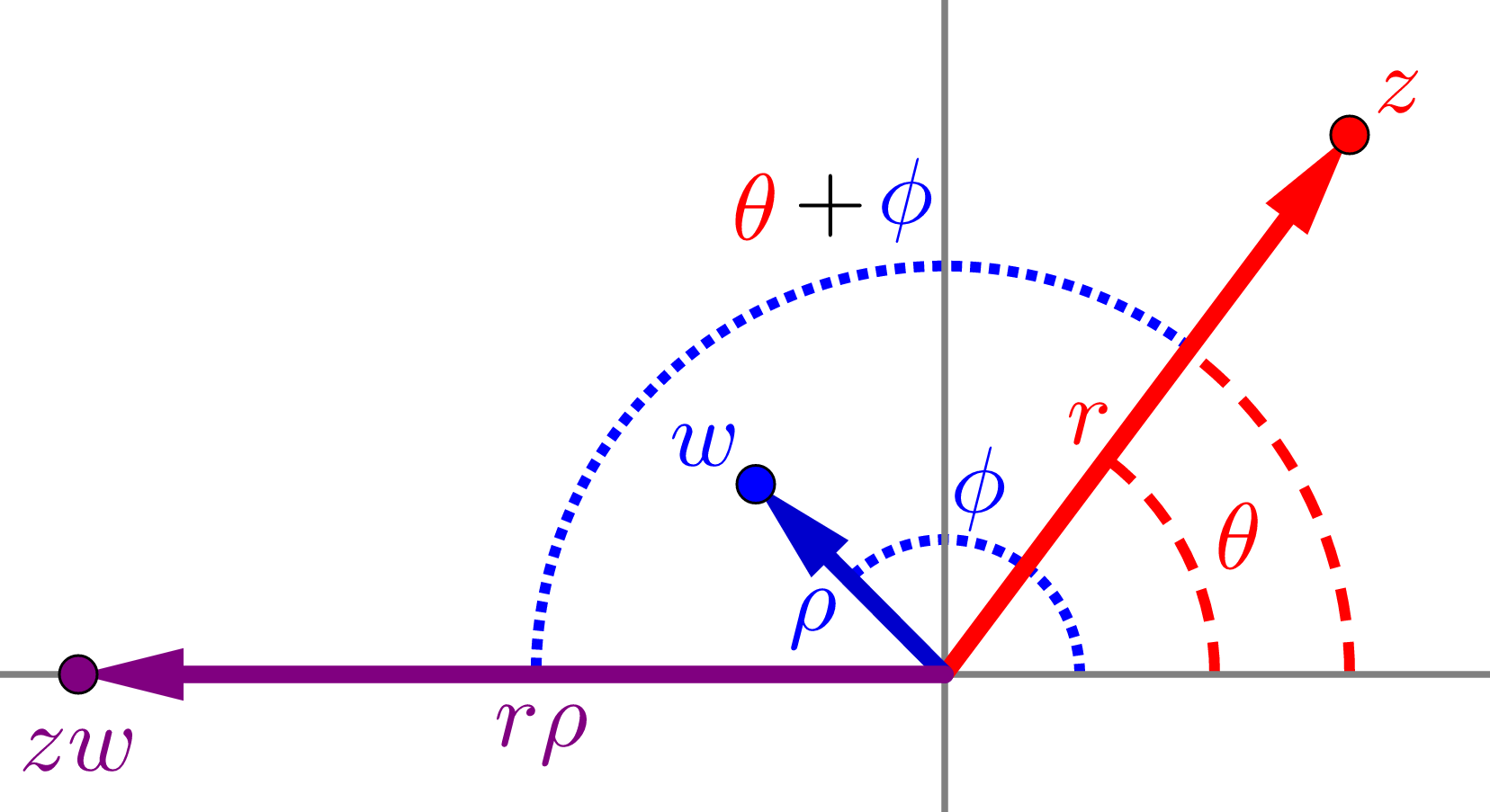}
    \caption{When we multiply the complex number $z = r\txte^{i\theta}$ by $w = \rho \txte^{i\phi}$, the magnitude of $z$ is \textit{dilated} by a factor of $\rho$, while the angle of $z$ is \textit{rotated} by $\mathrm{arg}(w) = \phi$.  Thus $\vert zw\vert = r\rho$ and $\mathrm{arg}(zw) = \theta + \phi$. \label{fig: multiplication dilation rotation}}
\end{figure}

Let's consider two complex numbers in polar representation, given by $z = r\txte^{i\theta}$ and $w = \rho\txte^{i\phi}$. We will calculate their product and difference in the polar representation.  
\begin{align}
    zw &= \left( r\txte^{i\theta} \right)\left( \rho\txte^{i\phi} \right) = r\rho \txte^{i\theta}\txte^{i\phi} = (r\rho)\, \txte^{i(\theta + \phi)}, \label{eq: multiplication dilation rotation}
    \\
    \frac{z}{w} &= \frac{ r\txte^{i\theta} }{ \rho\txte^{i\phi}} = \frac{r}{\rho}\,\txte^{i\theta}\txte^{-i\phi} = \left(\frac{r}{\rho}\right)\, \txte^{i(\theta - \phi)}, \label{eq: division dilation rotation}
\end{align}
Thus, by \equaref{eq: multiplication dilation rotation}, multiplying a complex number by another is equivalent to \textit{dilating} the magnitude of the first by the second and \textit{rotating} the first complex number \textbf{counter-clockwise} by the second's argument.   Figure \ref{fig: multiplication dilation rotation} shows the geometry behind multiplication. Likewise, since division is the multiplicative inverse, we should not be surprised by \equaref{eq: division dilation rotation} which says that dividing a complex number by another is equivalent to \textit{constricting}\footnote{My word choice for \textquotedblleft anti-dilating.\textquotedblright} the magnitude of the first by the second and rotating the first complex number \textbf{clockwise} by the second's argument.

\begin{figure}
    \centering
    \includegraphics[width = 3.5in, keepaspectratio]{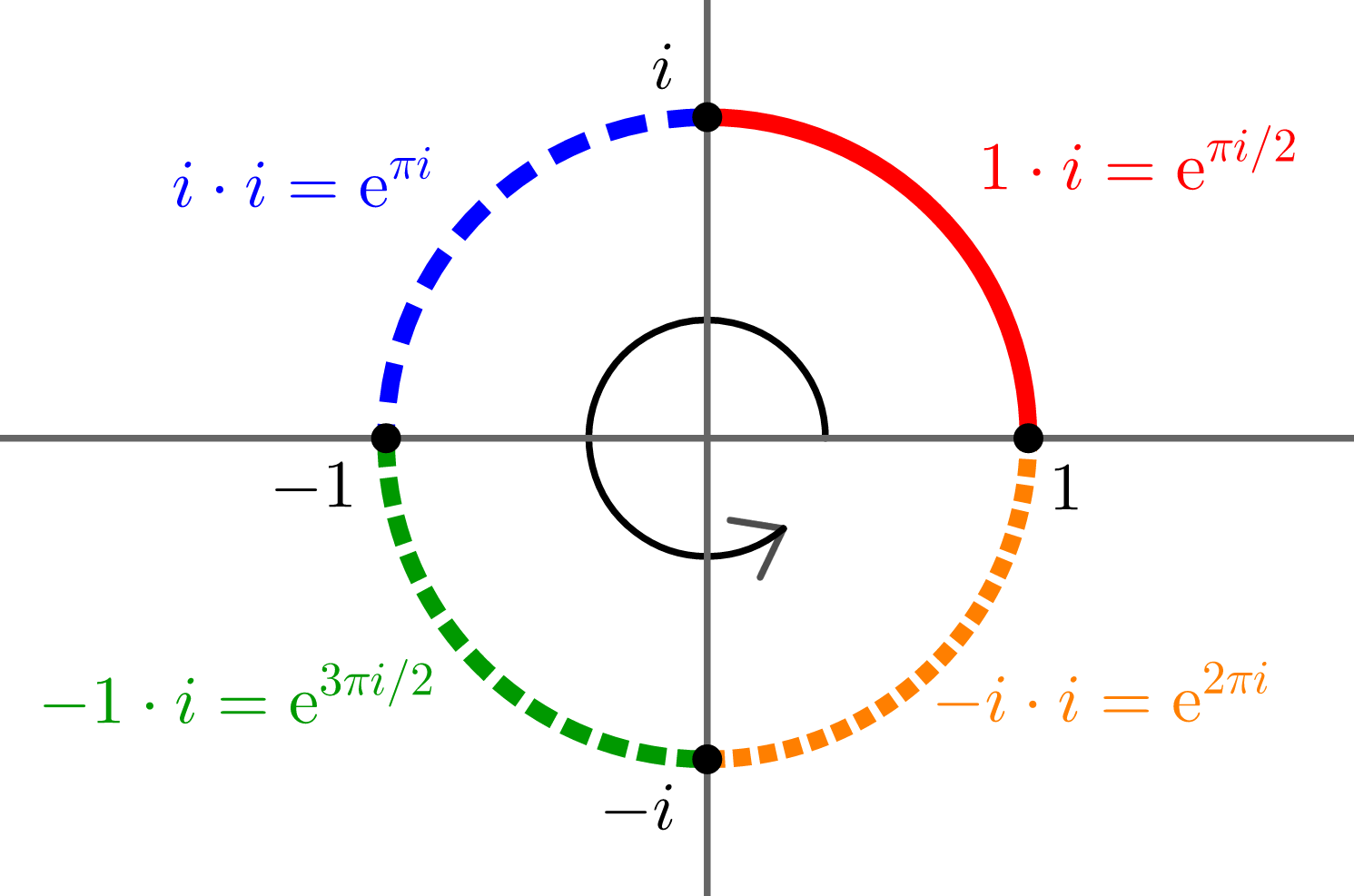}
    \caption{In this figure, we multiply each of the following points along the complex unit circle by the imaginary (lateral!) unit $i = \txte^{\pi i/2}$: $(1,0)$, $(0,i)$, $(-1,0)$, and $(0,-i)$. Starting with $(1,0)$, we have $1\cdot i = i = \txte^{\pi i/2}$ over the solid red arc. Hence, multiplying $1$ by $i$, \textit{rotates} $1$ by $\pi/2$ radians ($90^{\mathrm{o}}$). Next, we multiply $i$ by $i$ in the dashed blue arc. But $i^2 = -1 = \txte^{\pi i/2 + \pi i/2} = \txte^{\pi i}$. Thus, again, multiplying by $i$ rotates our starting complex number $i$ by $\pi/2$ radians.  I leave you to confirm the other two rotations for yourself.}
    \label{fig: multiply by i}
\end{figure}

What is often the case, at least in physics, when we deal with complex numbers, we are usually only taking about ones with \textit{unit magnitude} | that is, the complex modulus is equal to one.  Any complex number of the form in \equaref{eq: Euler's identity} fits this description.  The reason why these numbers are so important is that they do not dilate the modulus of any complex number they are multiplying.  Instead, they exclusively rotate the complex number they are multiplying.  Figure \ref{fig: multiply by i} serves to help you get some intuition to how the multiplication = rotation bit works with more concrete numerical examples like $\pm 1$ and $\pm i$. Very often, physicists will talk of the \textit{phase} of some quantity (such as in electromagnetic theory, optics, or quantum mechanics), and refer to the entire complex number $\txte^{i\theta}$ as this phase.  To be perfectly precise, the actual angle (argument) is the phase, NOT the entire complex number.  However, as we have seen by \equaref{eq: multiplication dilation rotation} where $r = 1$, multiplying by the entire number $z = \txte^{i\theta}$ only changes the \textit{total phase} of the product, since it rotates the original number by $\mathrm{arg}(z) = \theta$. Thus, in terms of things we can experimentally detect (and therefore know are truly there), we can only find the resultant change of phase due to the multiplication of a complex number by another of unit magnitude.  

Our ability to detect these so-called \textit{phase differences} is actually rather remarkable.  For example, in both classical and quantum mechanics, magnetic fields exert a torque on any charged object with angular momentum. This torque causes the charged object to \textit{precess} in a circle; in other words, the charged object will behave similarly to how a toy top begins to wobble in circles around its central axis before it falls over due to gravity.  In this case, the field that generates the wobble is the magnetic field instead of the gravitational field. Anyway, in quantum mechanics, a typical experiment to measure the phase difference goes something like this: generate a beam of identical particles, split the beam into two parts, do something to one beam and leave the other alone, then recombine the beams and see if anything happens.  So one such experiment deals with measuring the intensity of particles with inherent angular momentum (spin) after splitting up the beam and having one part travel through a magnetic field over some distance.  The amount those particles wobble due to the magnetic field then acts as the phase difference between the particles in the beam.  As you will eventually learn, the intensity of the recombined beam is a function of the phase difference, meaning we can experimentally vary either the strength of the magnetic field or increase the distance the beam travels through it, and then \textit{change the intensity of the resulting beam of particles!}  For an example of such an experiment, check out \cite{spin_interferometry}.

Before proceeding, let's take note of one special case of multiplication: exponentiation.  In other words, we can raise any real number $a\in\mathds{R}$ to the $n\in\mathds{R}$ power by successively multiplying $a$ by itself $n$ times.  For example, if $a =2$ and $n = 4$, then $a^n = 2^4 = 2\cdot 2\cdot 2\cdot 2 = 16$.  Likewise, if we have a negative exponent, then we divide successively, while if $n$ has a noninteger fractional part, we take the appropriate root $(2^{1/3} = \sqrt[3]{2})$. Using the polar representation of a complex number $z = r\txte^{i\theta}$, we can generalize exponentiation rather straightforwardly as 
\begin{align}
    z^n = r^n\,(\txte^{i\theta})^n = r^n\,\txte^{in\theta}.
\end{align}
By setting a new angle $\psi = n\theta$, and a new radial part as $s = r^n\in\mathds{R}$, we have
\begin{align}
    z^n = s\,\txte^{i\psi}.
\end{align}
Making use of Euler's Identity (\equaref{eq: Euler's identity}), we see that
\begin{align}
    z^n = s (\cos\psi + i\sin\psi ) = r^n(\cos n\theta + i\sin n\theta).\label{eq: complex z^n}
\end{align}
Since the quantity $z^n$ can be written in terms of a radial and angular part, it must be complex-valued, as those parts yield real and imaginary parts.  Thus $z^n\in\mathds{C}$ which means our rules for complex numbers lead to \textit{closure} under exponentiation! In other words, we cannot ever possibly end up with a non-complex-valued quantity purely by exponentiation\footnote{This is a very good thing, mind you, for it essentially gives us motivation for starting to see if more complicated algebraic and transcendental functions always return complex numbers.}.

There is, meanwhile, a much more subtle identity that we may not have initially noticed while showing $z^n\in\mathds{C}$.  We made use of a statement known as \textbf{De Moivre's Theorem} that says for any \textit{integer} $n$
\begin{align}
    \left(\cos\theta + i\sin\theta \right)^n = \cos(n\theta) + i\sin(n\theta). \label{eq: De Moivre's Theorem}
\end{align}
Looking closely at \equaref{eq: De Moivre's Theorem}, we see that this statement is actually incredibly complicated. It manages to relate something as easy to compute as raising a number to an integer power $n$ to the much more difficult trigonometric functions, cosine and sine. But more importantly De Moivre's Theorem puts the exponent $n$ \textit{inside} of the argument of the trig function!  Usually, there is no clear way to translate the argument of a trig function to anything outside of the function | for example $\cos(2x)\neq 2\cos x\neq (\cos x)^2$ for every value of $x$. Trigonometric functions just don't behave nicely like this.  However, by making use of the complex plane, De Moivre's Theorem gives us a way of applying algebraically simply operations to compute otherwise very difficult, if not outright impossible, quantities.

I do want to emphasize that De Moivre's Theorem only applies for integer exponents, whereas \equaref{eq: complex z^n} applies for all possible values of $n$ (even complex ones!). What gives? Where is there a difference?  We look into this question next.

\subsection{Roots of Unity}
Let's consider the case in De Moivre's Theorem where $n = 1/2$.  In this particular case, then if De Moivre's Theorem \textit{were to hold}, then it would be true (but it is not)
\begin{align*}
     \left(\cos\theta + i\sin\theta \right)^{1/2} = \cos\left(\frac{\theta}{2}\right) + i\sin\left(\frac{\theta}{2}\right).
\end{align*}
Let's choose the easier case of $\theta = 0$ and $\theta = 2\pi$ since on the left-hand side, we will have
\begin{align*}
    (\cos 0 + i\sin 0)^{1/2} = (\cos 2\pi + i\sin 2\pi)^{1/2} = 1^{1/2}.
\end{align*}
This follows since $\sin 0 =\sin2\pi = 0$ and $\cos 0 = \cos 2\pi = 1$. There is no problem yet.  The issue arises when we look at the right-hand side to find
\begin{align*}
     \cos\left(\frac{0}{2}\right) + i\sin\left(\frac{0}{2}\right) = \cos 0 &= 1
     \\
      \cos\left(\frac{2\pi}{2}\right) + i\sin\left(\frac{2\pi}{2}\right) = \cos\pi &= -1.
\end{align*}
(Remember that $\sin\pi = 0$.)  So, if De Moivre's Theorem were to be trusted, we would find that $1^{1/2} = 1 = -1$. In other words, we would find that our answer implies that $1=-1$ and this contradicts our rules for real numbers.

This result is not too surprising from an Algebra II point-of-view since we already know that both $(1)^2$ and $(-1)^2 =1$, which we normally write instead as $1^{1/2} = \pm 1$.  But this means that the square-root function is \textit{mutli-valued}, and in general the $n^{\mathrm{th}}$-root is also multi-valued.  The problem with De Moivre's Theorem is that it does not explicitly account for this phenomenon. We could easily generalize it to account for the multi-valuedness in exponentiation by rational exponents.

We start with the so-called \textbf{Roots of Unity} as they are both fundamental and something that we have already developed the motivation for. Let's start off easy, and find the cubic roots of unity.  We will proceed in the same way as we did before with the square-roots of unity.  Here we have
\begin{align*}
     \left(\cos\theta + i\sin\theta \right)^{1/3} = \cos\left(\frac{\theta}{3}\right) + i\sin\left(\frac{\theta}{3}\right).
\end{align*}
Like before, we choose $\theta = 0$ and $\theta = 2\pi$ because $\cos 0 = \cos 2\pi = 1$.  There is actually another case we can consider, too: $\theta = 4\pi \Rightarrow \cos 4\pi = 1$.  Calculating each of these cases, we have
\begin{align*}
    &\cos\left(\frac{0}{3}\right) + i\sin\left(\frac{0}{3}\right) = 1
     \\
     &\cos\left(\frac{2\pi}{3}\right) + i\sin\left(\frac{2\pi}{3}\right) = -\frac{1}{2} + i\frac{\sqrt{3}}{2}
     \\
     &\cos\left(\frac{4\pi}{3}\right) + i\sin\left(\frac{4\pi}{3}\right) = -\frac{1}{2} - i\frac{\sqrt{3}}{2}
\end{align*}
And so the cubic-roots of unity we found are
\begin{align*}
    1^{1/3} \in \left\lbrace 1, -\frac{1}{2} + i\frac{\sqrt{3}}{2} , -\frac{1}{2} - i\frac{\sqrt{3}}{2} \right\rbrace
\end{align*}

Hopefully, you aren't satisfied with this derivation of the cubic-roots of unity so far because I just arbitrarily decided to include the $4\pi$ part when I didn't include it for the square-roots. The reason why I was able to include $4\pi$ for the cubic-roots and not for the square-roots is because of the oscillatory behavior of sines and cosines | they repeat themselves every $2\pi$ radians.  So essentially, when looking for $\theta$ values that would show the roots of unity are indeed multi-valued, I needed to watch out for two criteria. First, I needed to make sure $\txte^{i\theta} = 1$ since we are taking about roots of \textit{unity}, AND I needed to make sure that the right-hand side did not repeat itself.  Let's analyze each criterion independently, and by doing so, we will generalize the square and cubic cases to the $n^\mathrm{th}$-root.

If we want $\txte^{i\theta} = 1$, then we need $\cos\theta + i\sin\theta = 1$.  But the value $1$ is totally real, hence the sine coefficient attached to the $i$ must vanish for our $\theta$ values.  This condition holds for $\theta \in\{0, \pm \pi, \pm 2\pi, \pm 3\pi, \dots\}$. Next, we are dealing with roots of \textit{unity}, not roots of negative-unity.  Thus, we can only allow for \textit{even multiplies} of $\pi$ so that $\cos(2\pi m) = 1$, where $m \in \{0, \pm 1, \pm 2, \pm 3, \dots\}$. So this explains why I kept choosing $\theta = 0, 2\pi$ and then $4\pi$.  But why not the negatives, too?  The answer to that comes again from the even symmetry of the cosine function:
\begin{align*}
    \cos(-\theta) = \cos(\theta).
\end{align*}
In other words, the cosine functions ignore the overall negative sign inside of their argument and produce the same result either way.  Hence, we \textit{could} include the negative values, $-2\pi, -4\pi,$ \textit{et cetera}, but we would ultimately always recover the same set of possible $\theta$ values.

Now we move onto the second criterion which says that $\theta$ cannot make the right-hand side repeat itself.  We use the first criterion that we had established where all of the roots of unity will have $\theta$ values that are an even multiple of $2\pi$. Let's call this multiple $m$ and write $1^{1/n}$ as 
\begin{align*}
    1^{1/n} = \cos \left( \frac{2\pi m}{n}\right) + i\sin\left( \frac{2\pi m}{n}\right),\;\; m\in\{0,1,2,\dots\}.
\end{align*}
If we start with $m = 0$, then we will always find $1^{1/n} = 1$, as we expect, since $1^n = 1$. Then, as we continue to check through all of the possible multiples we have, we find the first repeating value at $m = n$:
\begin{align*}
    1^{1/n} \underbrace{=}_{m=n} \cos \left( \frac{2\pi n}{n}\right) + i\sin\left( \frac{2\pi n}{n}\right) = \cos 2\pi = 1
\end{align*}
But this returns the exact same value as $m=0$. Furthermore, any multiple of $n$ will ALWAYS be the same as the $m=0$ case.  Meanwhile, all of the possible multiples up to $n$ are totally allowed because, in general,
\begin{align*}
    \frac{2\pi m}{n }\neq 2\pi,
\end{align*}
the only case where equality holds is when $m = n$.  Thus, our $n^{\mathrm{th}}$-roots of unity are given as
\begin{align}
    1^{1/n} = \cos \left( \frac{2\pi m}{n}\right) + i\sin\left( \frac{2\pi m}{n}\right),\;\; m\in\{0,1,\dots, n-1\}, \label{eq: n roots of unity Cartesian}
\end{align}
in the Cartesian representation.  In the polar representation, we pack the sinusoids into the exponential function to find
\begin{align}
    1^{1/n} = \mathrm{exp} \left( \frac{2\pi m i}{n} \right),\;\; m\in\{0,1,\dots, n-1\}. \label{eq: n roots of unity polar}
\end{align}
(Note that $\mathrm{exp}(x) = \txte^{x}$ for all $x$.  I used the exp representation because the equation would have looked gross if I used $\txte$ instead.) 

\begin{figure}
    \centering
    \includegraphics[width = 5.25in, keepaspectratio]{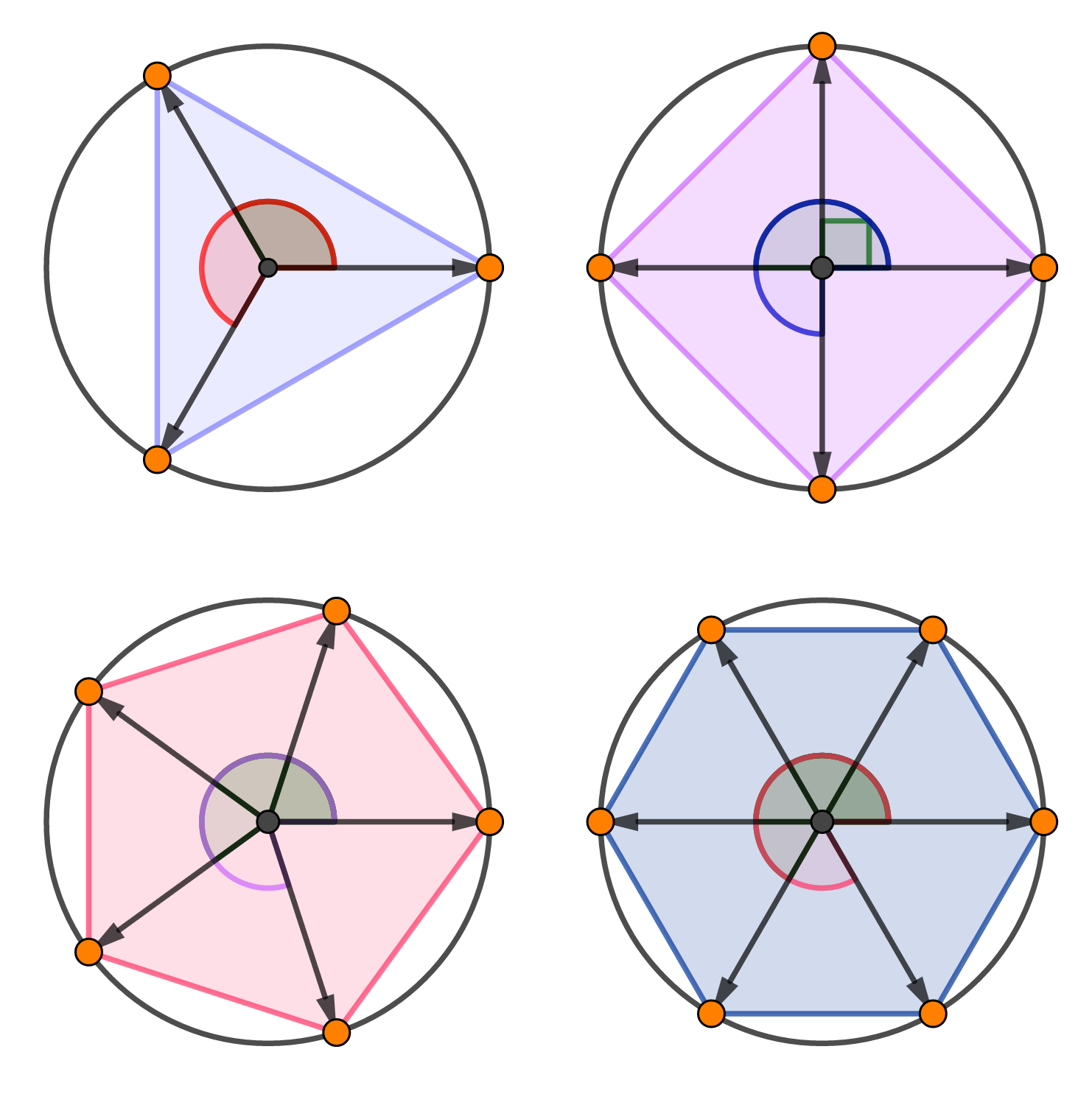}
    \caption{A few plots of the symmetry in the roots of unity.  The orange points are the complex numbers centered around their common origin. The vectors attached to each point is to help with our geometrical interpretation of the complex plane.  The regular polygons within each unit circle represent the internal symmetry within the set of the $n^\mathrm{th}$-roots of unity. }
    \label{fig: 3 to 6 Roots of Unity}
\end{figure}

The $n^\mathrm{th}$-roots of unity are quite nice geometrically because they chop up the unit circle ($2\pi$ radians) into regular $n$-gons by dividing the full $2\pi$ radians into equally-spaced $2\pi/n$ intervals.  A few of these are shown in \figref{fig: 3 to 6 Roots of Unity}. What is more beautiful from a physical point-of-view is that since the roots of unity form regular $n$-gons inscribed within a unit circle, their \textit{sum} must vanish.  This can be directly applied to study cylindrically (and polygonally) symmetrical systems in physics; for example, point like-charges (masses) arranged in some polygonal shape whose net electric (gravitational) fields must vanish at their geometrical center. In more computational applications in physics, this property allows us to compute discrete Fourier Transforms of signals which is a mathematical operation that allows us to understand the signal in terms of its frequency-dependence instead of its time-dependence\footnote{The Fourier Analysis chapter in this book does not deal with the discrete transform, but the tools developed in the chapter in the continuum can help you understand the intuition behind the discrete version.}.

To show how the roots sum to zero explicitly, recall the square-roots of unity: $1$ and $-1$.  Together they form a line (lame), but their sum is $1 + (-1) = 0$.  We can also do the cubic-roots of unity that form an equilateral triangle (less lame):
\begin{align*}
    1 -\frac{1}{2} + i\frac{\sqrt{3}}{2} -\frac{1}{2} - i\frac{\sqrt{3}}{2} = 1 - 1 + i0 = 0.
\end{align*}
Now let's generalize (disclaimer: this will be one of the more abstract things so far).

\vspace{0.15in}
\begin{problem}[The $4^\mathrm{th}$ and $5^\mathrm{th}$ Roots of Unity]{prob: 4th and 5th roots of unity}

Using \equaref{eq: n roots of unity Cartesian}, find the Cartesian representation of the $4^\mathrm{th}$ and $5^\mathrm{th}$ roots of unity and show that the sum of their roots is zero.
\end{problem}
\vspace{0.15in}

Consider the sum of the $n^\mathrm{th}$-roots of unity in polar representation, written as $S_n$, and given as
\begin{align*}
    S_n = 1 + \txte^{2\pi i/n} + \dots + \txte^{2\pi i(n-1)/n} = \sum_{m = 0}^{n-1} \txte^{2\pi i m/n}
\end{align*}
If we look closely at the summation, we will see that we actually have a finite \textit{geometric series}, defined generally as
\begin{align}
    S_n = 1 + a + a^2 + \dots + a^{n-1} = \sum_{m=0}^{n-1} a^m, \label{eq: finite geometric series}
\end{align}
for some value $a$. In this case, $a = \txte^{2\pi i/n}$ and
\begin{align*}
    S_n = \sum_{m=0}^{n-1} \left( \txte^{2\pi i/n} \right)^m
\end{align*}
To determine what this \textit{finite} sum is, we can employ a few mathemagical\footnote{Bad pun?} tricks that I \textbf{highly} recommend you work out with me.  Knowing how they work WILL come in handy later on in your physics career, at least when it comes to handling finite sums.  If you work these tricks out with me, then you are prone to remembering them in the future. Here we go.

We will calculate the quantity $1 - S_n$ using the definition of $S_n$ given in \equaref{eq: finite geometric series}, and at the end we will substitute in $a = \txte^{2\pi i/n}$.
\begin{align*}
    1 - S_n &= 1 - \sum_{m=0}^{n-1} a^m 
    \\
    &= 1 - (1 + a + a^2 + \dots a^{n-1})
    \\
    &= -(a + a^2 + \dots + a^{n-1})
    \\
    &= -a(1 + a + \dots + a^{n-2})
    \\
    &= -aS_{n-1}.
\end{align*}
In the last equality, I used the definition of the finite geometric series again, except since the sum only goes to $n-2$, then that means the proper subscript on $S$ is $n-1$ since $(n-1) -1 = n-2$. In this form, we cannot proceed because we needed $S_n$, NOT $S_{n-1}$. Thus we need to figure out a way to relate the two different finite sums.

Let's consider now $S_{n-1} = 1 + a + \dots + a^{n-2}$. If we look again at \equaref{eq: finite geometric series}, we will see that if we were to add $a^{n-1}$ to $S_{n-1}$, we would have
\begin{align*}
    S_{n-1} + a^{n-1} = 1 + a + \dots + a^{n-2} + a^{n-1} = 1 + a + \dots a^{n-1} = S_n.
\end{align*}
Hence, $S_{n} - a^{n-1} = S_{n-1}$.  Now we are going to substitute this expression in for $S_{n-1}$ inside the last equality for $1- S_n$ and isolate $S_n$.
\begin{align*}
    1 - S_n = -aS_{n-1} = -a\left(S_n - a^{n-1} \right) = -aS_n + a^n.
\end{align*}
By moving the $S_n$ on the left-hand side to the right-hand side we have
\begin{align*}
    1 = S_n - aS_n + a^n \Rightarrow 1-a^n = (1 - a)S_n,
\end{align*}
and finally we have
\begin{align}
    S_n = \frac{1 - a^n}{1 - a}, \label{eq: finite geometric series answer}
\end{align}
which is the exact way to calculate a \textit{finite} geometric series for any value of $a \neq 1$ (the formula blows up at $a = 1$ because if $a$ were 1 then we would have a $S_n - S_n = 0$ step in our derivation, thus we had to implicitly assume that $a \neq 1$). Finally, we substitute in $a = \txte^{2\pi i/n} \neq 1$:
\begin{align}
    S_n(\txte^{2\pi i/n}) =  \frac{1 - (\txte^{2\pi i/n})^n}{1 - \txte^{2\pi i/n}} = \frac{1 - \txte^{2\pi in/n}}{1 - \txte^{2\pi i/n}} = \frac{1 - \txte^{2\pi i}}{1 - \txte^{2\pi i/n}} = \frac{1 - 1}{1 - \txte^{2\pi i/n}} = 0. \label{eq: sum roots of unity}
\end{align}
And so it is true that the sum of all of the $n^{\mathrm{th}}$-roots of unity will always be identically zero! Furthermore, since a geometric series is built from multiplying the same object $a = \txte{2\pi i/n}$ by itself a bunch of time, and we know that multiplication by a complex number of unit magnitude is a pure rotation, then the complex number $\txte^{2\pi i/n}$ represents a \textit{symmetry} present in the set of the $n^\mathrm{th}$-roots of unity. The symmetry here is that our set of complex numbers are \textit{invariant} under a rotation of $2\pi/n$ radians | i.e. the number of points looks identical if we rotate all of them by the same $2\pi/n$ angle! For example, if we have the square-roots $\{1,-1\}$ and we rotate each number through the complex plane by $2\pi/2 = \pi$ radians, we have
\begin{align*}
    \{1,-1\} \xrightarrow{\textrm{rotate by } \pi \textrm{ radians}} \{-1,1\}, 
\end{align*}
as shown in Since the rotation $\txte^{\pi i}$ maps the set of square-roots onto themselves, this rotation is the \textit{symmetry} I was talking about that is directly embedded within the set of square-roots of unity. 

\begin{figure}
    \centering
    \includegraphics[width = 3in, keepaspectratio]{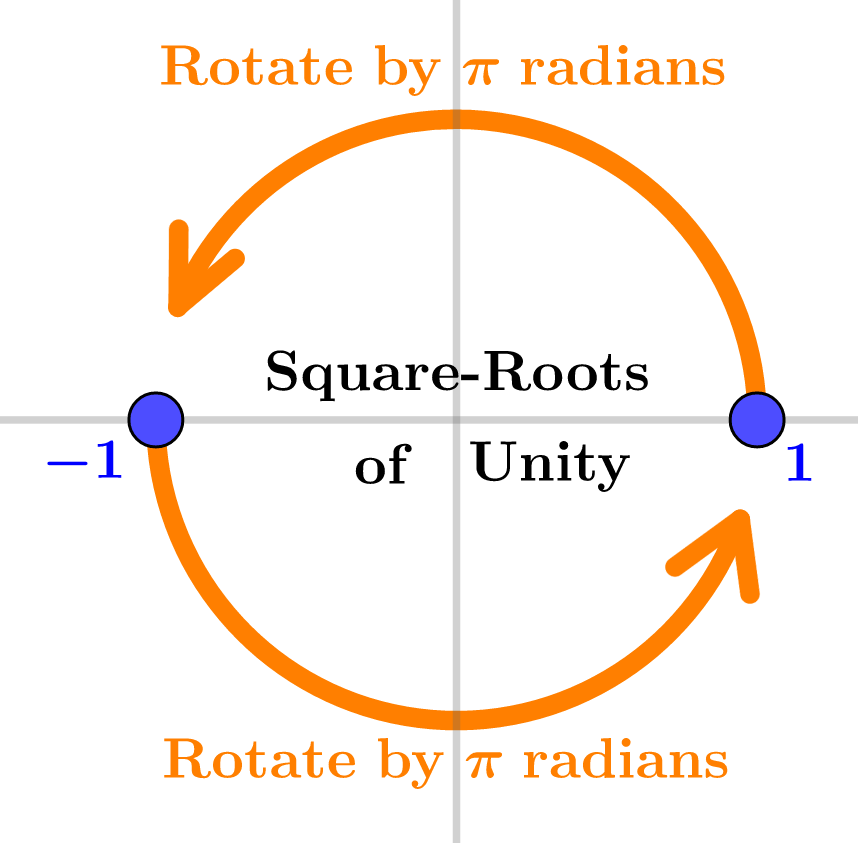}
    \caption{The symmetry in the square-roots of unity is a rotation by $\pi$ radians since the set of roots is identical under this rotation.}
    \label{fig: square roots of unity}
\end{figure}

Since the ambiguity in De Moivre's Theorem is taken care of, we can return to \equaref{eq: complex z^n} to account for all of the different possible roots of $z$.  Specifically, we can talk about any root $1/n$, we have
\begin{align}
    z^{1/n} &= \left(r\txte^{i\theta} \right)^{1/n} \nonumber
    \\
    &= r^{1/n}\cdot 1^{1/n}\cdot \txte^{i\theta/n} \nonumber
    \\
    &= r^{1/n}\,\txte^{2\pi im/n}\,\txte^{i\theta/n},\;\; m \in \{0,1,\dots, n-1\} \nonumber
    \\
    &= r^{1/n}\,\mathrm{exp}\left( i\frac{\theta + 2\pi m}{n} \right),\;\; m \in \{0,1,\dots, n-1\} \nonumber
    \\
    &= r^{1/n}\left[ \cos\left( \frac{\theta + 2\pi m}{n} \right) + i\sin\left( \frac{\theta + 2\pi m}{n} \right) \right],\;\; m \in \{0,1,\dots, n-1\}
\end{align}
Then, by extension, a complex number $z$ to any rational power $p/q$ (for example 4/3) can be written as 
\begin{align}
    z^{p/q} &= \left[r^{1/q}\,\mathrm{exp}\left( i\frac{\theta + 2\pi m}{q} \right)\right]^p,\;\; m \in \{0,1,\dots, q-1\} \nonumber
    \\
    &= r^{p/q}\,\mathrm{exp}\left( ip\frac{\theta + 2\pi m}{q} \right)^p,\;\; m \in \{0,1,\dots, q-1\} \nonumber
    \\
    &= r^{p/q}\left\lbrace \cos\left[ \frac{p}{q} (\theta + 2\pi m) \right] + i\sin\left[ \frac{p}{q} (\theta + 2\pi m) \right] \right\rbrace,\;\; m \in \{0,1,\dots, q-1\}, \label{eq: rational De Moivre}
\end{align}
The power with raising a complex number to a rational exponent is that we can then use this as a basis for raising a complex number to an \textit{irrational} exponent since we can always successively approximate an irrational number with a rational number.  For example, $\sqrt{2} \approx 1.414 = 1414/1000$.  But how about raising a complex number to a complex exponent? We will study this and more in the next section.

\section{Functions of a Complex Variable}
In this section, we will study a few algebraic and transcendental functions of a complex variable. Unfortunately, we will not have time to study these functions beyond just domain and range, but rest assured that understanding the inputs and outputs of complex functions is totally sufficient for a B.S. in physics. As one studies either more math or more physics, ideas from \textit{Complex Analysis} become relevant if not actually crucial, whether they be in the form of circuit analysis, fluid mechanics, or quantum field theory. I have some references for anyone interested at this point in their studies \cite{complex_analysis_wikipedia_2018, beck_marchesi_pixton_sabalka_2017, arnold_complexanalysis} (feel free to come back at a later date for them though!).

I have used the term \textquotedblleft map\textquotedblright$\,$ before when talking about functions, and perhaps you've heard others use that word in the same context. In more colloquial settings, people use maps to get from one place to another, and in this section the connection between a function and an everyday map will become clearer.  We will see that functions, particularly complex-valued functions, serve to get from one place in the complex plane to another.  To illustrate this idea, let's consider the following example where we see how regions of the complex plane are connected through the function, or \textit{mapping}, $w(z) = z^2$.

\begin{figure}
    \centering
    \includegraphics[width = 5.0in,keepaspectratio]{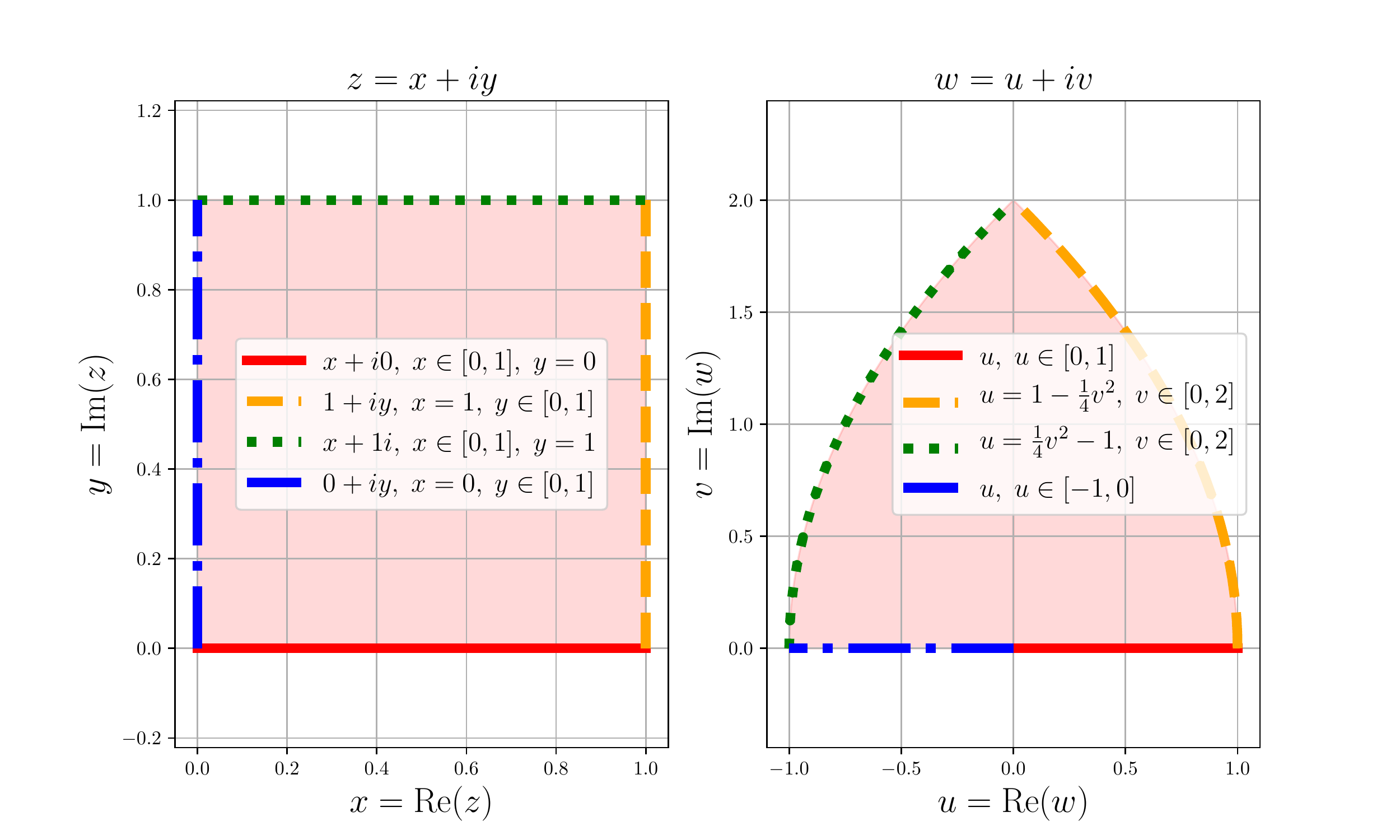}
    \caption{In the left plot, we define the unit square in the $z$-plane.  It is through the $w(z) = z^2$ mapping that we connect the unit square to the square-squared region in the $w$-plane in the right plot.  The colored curves represent sets of the complex plane that are connected through the $w(z) = z^2$ \textquotedblleft map.\textquotedblright}
    \label{fig: quadratic map}
\end{figure}

\vspace{0.15in}
\begin{example}[Squaring the Square]{ex: squaring the square}

Consider the complex-valued function $w(z) = z^2$.  We want to study $w$ over a unit square in the first quadrant of the complex plane.  To start, we consider $z = x+iy$ in the Cartesian representation.  Then,
\begin{align}
    w(z) = z^2 = (x+iy)^2 = x^2 - y^2 +2ixy. \label{eq: z squared}
\end{align}
If we define $w = u+iv$, then we have $u = x^2 - y^2$ and $v = 2xy$ for the real and imaginary parts of $w$, respectively. Now we define the boundary of the unit square for our input, or domain, of interest in the complex plane
\begin{align*}
   z = \begin{cases}
        x + i0,& x\in[0,1]
        \\
        1 + iy,& y\in[0,1]
        \\
        x + 1i,& x\in[0,1]
        \\
        0 + iy,& y\in[0,1]
   \end{cases}
\end{align*}
If we substitute these coordinates for $z$ into $w$, we have
\begin{align*}
   w(z) = \begin{cases}
        x^2,& x\in[0,1],\; y=0
        \\
        1-y^2 + 2iy,& x=1,\; y\in[0,1]
        \\
        x^2 - 1 + 2ix,& x\in[0,1], \; y=1
        \\
        -y^2, & x = 0, \;y\in[0,1]
   \end{cases}
\end{align*}
The \textquotedblleft mapping\textquotedblright, in this case, comes in when we rewrite $w(z)$ into its own set of real and imaginary parts, given by $w = u +iv$. By looking at the four same regions, we have
\begin{align*}
    w(\textrm{unit cube}) \in \begin{Bmatrix}
        u = x^2,\; &v = 0
        \\
        u = 1-y^2,\; &v = 2y
        \\
        u = x^2 - 1,\; &v = 2x
        \\
        u = -y^2,\; &v = 0
    \end{Bmatrix} = 
    \begin{Bmatrix}
        u, & u\in [0,1]
        \\
        u = 1 - \frac{1}{4}v^2,& v\in\left[0, 2 \right]
        \\
        u = \frac{1}{4}v^2 - 1,& v\in\left[0, 2\right]
        \\
        u, & u\in[-1,0]
    \end{Bmatrix}
\end{align*}
It is definitely difficult to visualize how the equation above defines a map, at least not in this representation.  Figure \ref{fig: quadratic map} shows how the equation above can serve as a set of instructions of how to turn one plot, or regions of the complex plane, is connected to another through the $w(z) = z^2$ function.
\end{example}
\vspace{0.15in}

From \exref{ex: squaring the square}, we can conclude a couple of things. First, complex-valued functions unite different regions of the complex plane. And second, specifically for this quadratic function, we \textit{chose} to only consider the unit square. We could have chosen a square of size 2, or $\pi$, or a gazillion. Then $w$ would also increase in size by \equaref{eq: z squared}. I will leave it to you to pick particular points on the square of side length $s$ and plug it into \equaref{eq: z squared} to see how large the region in the $w$-plane becomes (hint: try the points $(s,0),\; (s,s),$ and $(0,s)$). Since we can choose a square of any size, we can artificially stretch the region in the $w$-plane to infinity as we increase the domain in the $z$-plane. Notice though, that as long as our domain is only in the first quadrant in the $z$-plane, we will only ever be in the first or second quadrants in the $w$-plane. This region is sometimes referred to as the \textit{upper-half plane}.  As you may see after doing \probref{prob: lower half plane}, other regions will map to fill the \textit{lower-half plane}, and by extension, we can stretch regions out as well to fill the entire lower-half plane. Hence, the function $w(z) =z^2$ maps the entire complex plane into the entire complex plane!  Actually, functions such as these are specifically named \textbf{entire functions} since they are defined\footnote{Technically, functions are entire only if they \textit{converge} everywhere in the plane. However, convergence is beyond the scope of this chapter, so understanding entire functions as simply being \textit{defined} everywhere is good enough for right now. } everywhere in the plane.  Since we have begun to understand complex mappings in terms of complex domains and ranges, we seek to study a few more fairly commonly-used complex functions in physics.

\vspace{0.15in}
\begin{problem}[The Lower-Half Plane]{prob: lower half plane}

Use \equaref{eq: z squared} to show that the square defined by 
\begin{align*}
    z = \begin{cases}
        x + i0,& x\in[-1,0],\;y=0
        \\
        -1 + iy,& x=-1,\;y\in[0,1]
        \\
        x+1y,& x\in[-1,0],\;y=1
        \\
        0 + iy,&x = 0,\;y\in[0,1]
    \end{cases}
\end{align*}
\textit{maps} to the same region in \figref{fig: quadratic map}, except that it is reflected over the $u$-axis into the lower-half plane.
\end{problem}
\vspace{0.15in}

\subsection{Polynomials}
For the first type of function, we consider functions of the form
\begin{align}
    w(z) = P_n(z) = \sum_{j = 0}^n a_j z^j = a_0 + a_1 z + a_2 z^2 + \dots + a_n z^n, \label{eq: complex polynomial}
\end{align}
where all of the constant coefficients $\{a_j\}$ are arbitrary complex numbers, but each exponent is a nonnegative integer.  This class of functions, often denoted by the symbol $P_n$ for \textquotedblleft polynomial of $n^{\mathrm{th}}$-order,\textquotedblright$\,$ is the complex generalization of the real polynomials that you already know.  This means that instead of only allowing ourselves to \textit{input} real numbers into a polynomial, we now are allowing ourselves to insert a complex number $z = r\txte^{i\theta}$ as the argument of the function (in other words, an element from the domain).  With this value of $z$ and by \equaref{eq: complex z^n}, we have
\begin{align}
    P_n\left(z = r\txte^{i\theta}\right) = \sum_{j = 0}^n a_j r^j \txte^{ij\theta} = a_0 + a_1r\txte^{i\theta} + a_2r^2\txte^{2i\theta} + \dots  + a_n r^n \txte^{in\theta}.
\end{align}
It is important to note that we are entirely allowed to compute every single one of these terms using the polar representations of the coefficients, if we had them given. Since each term is defined and computable, then we are definitely allowed to add all of their corresponding real and imaginary parts. Thus, for any $z\in\mathds{C}$ that we throw at this thing, we will always be able to find its output as a complex number.  Thus a general polynomial maps the complex plane onto\footnote{To be clear, there is a difference between the words \textit{into} and \textit{onto}. A function from set $A$ to set $B$, $f:A \rightarrow B$, is said to be \textbf{onto} if every element of $B$ is in the range of $f$.  Pictorally, this means that there are no parts of $B$ that are not connected by $f$ to $A$.  Otherwise, mathematicians use the word \textit{into}.} the complex plane. Our conclusion is important because, just like with real numbers, complex polynomials are actually relatively easy functions to handle, whether it be in calculus, computational physics, mathematical modeling, \textit{et cetera}, because we can handle them term-wise.  Other functions, (like the transcendental functions) are not so nice.

\vspace{0.15in}
\begin{problem}[Practice with Complex Polynomials]{prob: practice with complex polynomials}
Consider the complex quadratic,
\begin{align*}
    P_2(z) = i + \txte^{i\pi/4}z - 2z^2.
\end{align*}
Use either the Cartesian or the polar representation to evalutate $P_2(z)$ at the following points: $z_1 = 1$, $z_2 = -i\pi$, $z_3 = 1-i$.  Also, write down the set of coefficients $\{a_j\}$ (this part is NOT a trick. I just want to make sure you know what the set of cofficients is).
\end{problem}
\vspace{0.15in}

\subsection{Complex Exponentiation Revisted}
The last time we dealt with complex exponentials was with deriving the rational generalization to De Moivre's Theorem in \equaref{eq: rational De Moivre}.  We were explicit, however, that the exponent itself had to be a rational number $p/q\in\mathds{R}$, where $p$ and $q\neq 0$ are integers along the real axis.  Now we have to ask the question of whether it is possible to raise a complex number $z$ to a constant complex exponent $\xi$.  In other words, we want to study the function $w(z) = z^\xi$.

Since $\xi\in\mathds{C}$, we choose to work with it in its Cartesian representation.  Let $\alpha = \mathrm{Re}(\xi)$ and $\beta = \mathrm{Im}(\xi)$ such that $\xi = \alpha + i\beta$.  Thus
\begin{align*}
    w(z) = z^\xi = z^{\alpha + i\beta} = z^{\alpha}\,z^{i\beta}.
\end{align*}
Now, we will make the substitution of $z = r\txte^{i\theta}$ into the equation above.
\begin{align*}
    w(z) &= \left( r\txte^{i\theta} \right)^\alpha \left( r\txte^{i\theta} \right)^{i\beta},
    \\
    &= \left( r^\alpha \txte^{i\theta\cdot\alpha} \right)\,\left( r^{i\beta}\,\txte^{i\theta\cdot i\beta}\right),
    \\
    &= r^{\alpha}\txte^{i\alpha\theta}\, r^{i\beta} \txte^{-\beta\theta},
    \\
    &= r^{\alpha+i\beta}\txte^{-\beta\theta}\,\txte^{i\alpha\theta}.
\end{align*}
Now, before proceeding further, since $r$ is a real number, that means it lies entirely along the real axis, and we can use our regular logarithms on it, since those are defined over the positive real axis.  Thus, we choose to write $r = \txte^{\ln r}$, as the (real) functions $\txte^{x}$ and $\ln x$ are inverse functions such that $\txte^{\ln x} = x = \ln( \txte^{x})$. When we make this substitution into $w$, we can derive our final result.
\begin{align}
    w(z) &= \txte^{\ln( r)\cdot (\alpha+i\beta)}\txte^{-\beta\theta}\,\txte^{i\alpha\theta}, \nonumber
    \\
    &= \txte^{\alpha \ln( r) + i\beta\ln(r)}\txte^{-\beta\theta}\,\txte^{i\alpha\theta}, \nonumber
    \\
    &=\txte^{\alpha\ln(r) - \beta\theta}\,\txte^{i[\beta\ln(r) + \alpha\theta ]},\nonumber
    \\
    &= r^\alpha\,\txte^{-\beta\theta}\,\txte^{i[\beta\ln(r) + \alpha\theta ]}. \label{eq: w(z) complex exponential}
\end{align}
It is important to note that the long string of symbols given in \equaref{eq: w(z) complex exponential} is something that is entirely computable.  What I mean by that is this expression is written in the polar representation of a complex number, where
\begin{align*}
    \vert w(z) \vert  &= r^{\alpha}\txte^{-\beta\theta} \in\mathds{R}^+,
    \\
    \mathrm{arg} [w(z)] &= \beta\ln(r) + \alpha\theta \in\mathds{R}.
\end{align*}
Hence \equaref{eq: w(z) complex exponential} is already in the form of a complex number, regardless of which values of $\alpha$ and $\beta$ we choose!  To be clear, if we were to convert back to a Cartesian representation then we would need to be careful with any ambiguities in the rational exponents, as is given in \equaref{eq: rational De Moivre}.  But otherwise, we are free to map the complex plane $z$ onto\footnote{Since there are many possible roots for complex exponents, the $z$-plane actually gets mapped onto the $w$-plane in multi-valued ways, which we won't go into.  These peculiarities (translated to \textquotedblleft headaches\textquotedblright in math-speak) are known as \textit{branch cuts}, and special note really should be paid to them from a mathematical point-of-view. However, in physics, they are only noted if necessary. } the complex plane $w$ via complex exponentiation. 

\vspace{0.15in}
\begin{problem}[$i$ to the Power of $i$]{prob: i to the power of i}
Use the polar representation and \equaref{eq: w(z) complex exponential} to show that $i^i = \txte^{-\pi/2}$. This example shows that, while complex exponentiation maps the complex plane onto the complex plane, exponentiation does NOT necessarily map imaginary numbers into other imaginary numbers. This is interesting because exponentiation will always map real numbers into other real numbers.
\end{problem}

\subsection{Complex Logarithms}
Since we have covered the case of exponentiation, it follows that we should cover \textit{inverse exponentiation}, that is, we need to discuss how to find the logarithm of a complex number.  To study this function, consider $w(z) = \ln (z)$, where $\ln$ denotes \textquotedblleft natural logarithm\footnote{In many advanced mathematics, the \textquotedblleft common log\textquotedblright,$\,$ denoted by $\log$, is used instead of $\ln$.  However, the \textquotedblleft common log\textquotedblright$\,$ of a number $x$ in physics is almost always given by $\ln(x)/\ln(10) = \log_{10}(x)$. }\textquotedblright.  Just as before, this problem is most easily tackled in the polar representation.
\begin{align}
    w(z) = w(r\txte^{i\theta}) = \ln (r\txte^{i\theta}) = \ln(r) + \ln(\txte^{i\theta}) = \ln(r) + i\theta. \label{eq: w(z) = ln(z)}
\end{align}
To be clear, in this derivation, we implicitly borrowed the logarithm of a product rule from the real numbers, $\ln(ab) = \ln(a) + \ln(b)$, and we implicitly enforce that the natural logarithm is the function-inverse of $\txte^{x}$. It is very important to note, however, that ANY complex number $\xi = r_{\xi} \txte^{i\theta_{\xi}}$ is invariant under a full rotation about the origin | therefore $\xi$ invariant under a rotation of $2\pi n$, where $n$ is an integer.  This is equivalent to saying $\xi\,\txte^{2\pi n i} = \xi \cdot 1 = \xi$. However, the exponents are additive, therefore
\begin{align*}
    \xi\txte^{2\pi n i} = r_\xi \txte^{i\theta_{\xi}}\txte^{2\pi n i} = r_\xi \txte^{i(\theta_\xi + 2\pi n)}
\end{align*}
But if we apply this idea to $z = r\txte^{i\theta}$, then we have $z = r\txte^{i(\theta + 2\pi n)}$. Thus, when we take a logarithm, we have
\begin{align}
    w(z) = \ln(r) + i(\theta + 2\pi n) = \ln(r) + i\theta + 2\pi n i.
\end{align}
But wait. From \equaref{eq: w(z) = ln(z)}, have 
\begin{align}
    w(z) = \ln(r) + i\theta = \ln(r) + i\theta + 2\pi n i,
\end{align}
which seems to imply that $2\pi ni = 0$ if this equation is to be true. Since $2\pi$ and $i$ are both nonzero quantities, this means that $n$ would have to be zero! However, we said $n$ can be \textit{any} integer, not just zero. What gives? Did we accidentally stumble on a contradiction within complex algebra?

So the truth is that, with our current system of complex numbers, we actually did find a contradiction, and it has to do with our interpretation of multiplication as rotation and $\txte^{2\pi n i} = 1$ in the polar representation. But this contraction is not a totally new idea | it is similar to the idea that rational exponents are multivalued.  It turns out that in the complex plane, the logarithm also accrues multiple values for the same input.  In this sense, the full logarithm is \textit{not} a function since for any input there are many outputs (so it does not pass the complex-version of the vertical line test).  How do we deal with this problem? We use the idea of \textbf{branch cuts} to save the day, where a \textquotedblleft branch cut\textquotedblright$\,$ is a fancy name for specifying a particular range of the logarithm over a particular domain. In fact, this approach is identical to the one you learned in trigonometry, where we would only specify the inverse cosine ($\arccos(x) = \cos^{-1}(x)$) or inverse sine ($\arcsin{}(x) = \sin^{-1}(x)$) on the interval $x\in[-1,1)$, since there are (infinitely) many values of $x$ such that $\cos(x) = \pm 1$ or $\sin(x) = \pm 1$.  Although this is (sort of) passing-the-buck on the issue of having the logarithm be mutlivalued, it DOES allow us to retain our interpretation of multiplication as rotations in the complex plane | we just need to be careful about which \textit{branch} we're talking about when using the logarithm.

So the problem with the logarithm is due to the multiplicity, or rotational symmetry, surrounding the $2\pi n$ angle.  Thus, it helps to define a \textbf{principal branch} of the logarithm, so there really isn't a lot of ambiguity surrounding our equations.  All we need to do is define the imaginary part of the logarithm over a $2\pi$-interval to keep things consistent, since it is the imaginary part in \equaref{eq: w(z) = ln(z)} that has the multiple values. The interval we choose may initially seem odd, but actually is more intuitive in physics, and is given by $\theta_{principal} \in [-\pi,\pi)$. Therefore, the \textbf{principal logarithm} is defined as
\begin{align}
    \mathrm{Ln}(z) = \mathrm{Ln}(r\txte^{i\theta}) = \ln(r) + i\theta, \; \theta \in [-\pi, \pi). \label{eq: principal ln}
\end{align}
Here, the capital $\mathrm{L}$ signifies the special branch of the multivalued logarithm given in \equaref{eq: w(z) = ln(z)}. Notice though that there is a lowercase $\mathrm{l}$ in the real part of \equaref{eq: principal ln}. This is because $r$ is a positive real number, and we can always take a normal, non-multivalued logarithm of any positive real number.  All of the multivaluedness, again, is only due to the imaginary part.

\vspace{0.15in}
\begin{problem}[Logarithms of Negative Numbers]{prob: ln negative numbers}

You have probably been told for your whole life that it is impossible to take a logarithm of a negative number. Unfortunately, that is just not true.  Use \equaref{eq: w(z) = ln(z)} defined for $\mathrm{arg}(z)\in[\pi,3\pi)$ to calculate $\ln(-1)$.  

To check whether your answer makes sense, suppose you compute $\ln(-1)$, and let us denote it by $w = \ln(-1)$.  This would imply that $\txte^{w} = -1$, or $\txte^{w} + 1 = 0$. Do you recognize this equation?

Next, compute $\ln(-a)$, where $a$ is a positive real number. Hence, $-a$ represents any point along the negative real axis with an absolute value of $a$. For example, if $a = 9$, then I would be asking you to evaluate $\ln(-9)$.  Since the number $-a$ represents \textit{any} negative real number, what can you conclude about $\ln(-a)$? Is it ever a purely real number (Hint: No it isn't.)? Is this why you've been taught that you cannot take the logarithm of a negative number (Hint: Yes it is.)?
\end{problem}
\vspace{0.15in}

\subsection{Hyperbolic Trigonometry}
\begin{figure}
    \centering
    \includegraphics[width = 5in, keepaspectratio]{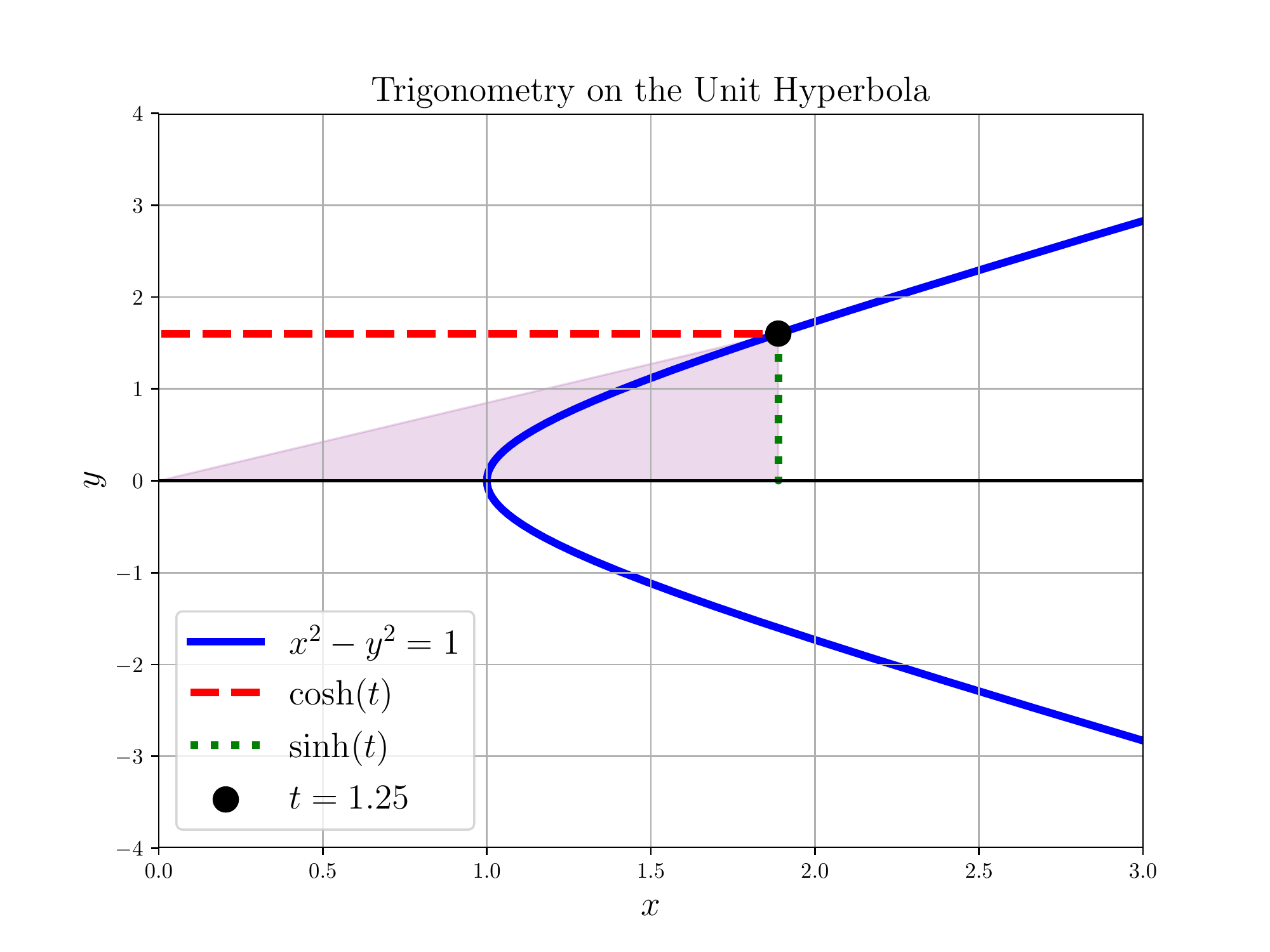}
    \caption{The unit hyperbola $x^2-y^2 = 1$ parameterized by the functions $x = \cosh t$ and $y = \sinh t$. Think of the parameter $t$ as the time it takes for the black point to move from the point $(x,y) = (1,0)$ to the point $(x,y) = (x(t),y(t))$ along the unit hyperbola.}
    \label{fig: unit hyperbola}
\end{figure}

The final class of functions that I want to discuss is within the set of functions that most people refer to as hyperbolic trigonometry. The geometry is actually remarkably simple to jump to from normal, or \textit{circular}, trigonometry.  The latter is governed by the unit circle, defined by $x^2 + y^2 = 1$.  We are instead going to be interested in the so-called \textit{unit hyperbola}, defined by $x^2 - y^2 = 1$.  Although the geometry may be interesting, I am going to be honest with you: the only time I've ever actually seen it be relevant is in orbital mechanics where trajectories matter, or general relativity, where the geometry of spacetime is the dynamical variable of interest (in other fields of physics, like Physics 1, the dynamical variables are position and velocity as a functions of time). With that said, the actual hyperbolic functions themselves are often incredibly useful tools to use in physics.  They, for example, are used to describe how the electrostatic potential changes in space from a source. I have included \figref{fig: unit hyperbola} to help you see what unit hyperbola looks like, and how the hyperbolic trigonometric functions relate triangles to it.

\begin{figure}
    \centering
    \includegraphics[width = 4.5in, keepaspectratio]{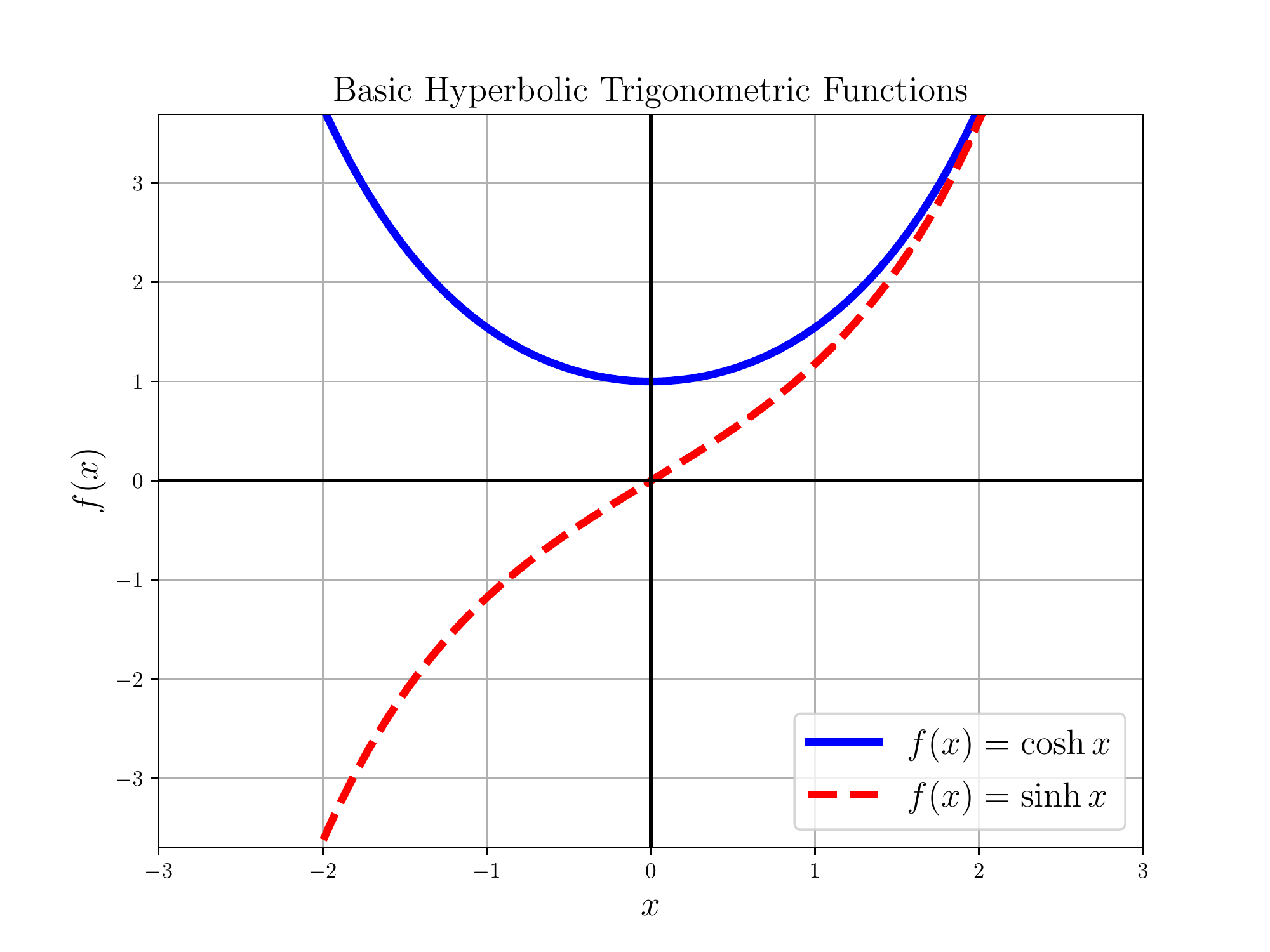}
    \caption{Plots of the hyperbolic cosine (\equaref{eq: cosh def}) and hyperbolic sine (\equaref{eq: sinh def}) functions over a few real numbers $x$.}
    \label{fig: cosh sinh sketch}
\end{figure}

The most commonly used hyperbolic functions are the \textbf{hyperbolic cosine} and the \textbf{hyperbolic sine} functions, defined by
\begin{align}
    \cosh{x} &= \frac{\txte^x + \txte^{-x}}{2}, \label{eq: cosh def}
    \\
    \sinh{x} &= \frac{\txte^x - \txte^{-x}}{2}, \label{eq: sinh def}
\end{align}
for any $x\in\mathds{R}$. Here the extra $\mathrm{h}$ that's attached is the \textit{hyperbolic} part, and the $\cosh$ notation is pronounced \textit{cah-sh} while the $\sinh$ part is often pronounced \textit{sin-ch}. I'm not 100\% sure why there is an added \textit{ch} in the $\sinh$ pronounciation, but my best guess is that \textit{sin-h} is hard to say. Since I know that these real-valued functions will be brand new to most of you reading, I included a plot of them in \figref{fig: cosh sinh sketch} so you can visualize them.  Notice that as $x \rightarrow +\infty$, the hyperbolic sine and hyperbolic cosine become equal, whereas on the other side of the real axis, the $\sinh$ and $\cosh$ approach values equal magnitudes but opposite signs. Note that if we set $x = \cosh(t)$ and $y = \sinh(t)$, then
\begin{align*}
    x^2 - y^2 &= (\cosh t)^2 - (\sinh t)^2,
    \\
    &= \left(\frac{\txte^t + \txte^{-t}}{2} \right)^2 - \left(\frac{\txte^t - \txte^{-t}}{2} \right)^2, 
    \\
    &= \frac{1}{4}(\txte^{2t} + \txte^{-2t} + 2) - \frac{1}{4}(\txte^{2t} + \txte^{-2t} - 2), 
    \\
    &= \frac{1}{4}(\txte^{2t} + \txte^{-2t} - \txte^{2t} + \txte^{-2t} +2 + 2), 
    \\
    &= 1
\end{align*}
which shows that these definitions of the $\cosh$ and $\sinh$ appropriately \textit{parameterize} the unit hyperbola with the parameter $t$.  To better understand the parameterization, imagine $t$ as a unit of time, and $x(t)$ and $y(t)$ are the $x,y$ positions of a point moving along the unit hyperbola.

It is also really important to notice that the $\cosh$ is an even function, whereas the $\sinh$ is an odd function, in the sense that an even function $f(x)$ has the property $f(-x) = f(x)$ while odd functions have the property $f(-x) = -f(x)$. I will prove these properties for the $\cosh$ and $\sinh$ to you below.

\begin{align*}
    \cosh(-x) &= \frac{\txte^{(-x)} + \txte^{-(-x)}}{2} = \frac{\txte^{-x} + \txte^{x}}{2} = \frac{\txte^x + \txte^{-x}}{2} = \cosh(x).
    \\
    \sinh(-x) &= \frac{\txte^{(-x)} - \txte^{-(-x)}}{2} = \frac{\txte^{-x} - \txte^{x}}{2} = \frac{-(\txte^x - \txte^{-x})}{2} = -\sinh(x).
\end{align*}

I wanted to point these properties out to you because they actually identically mimic those of the normal trigonometric functions, where the cosine is even and the sine is odd!  There are actually a slew of other analogous properties between circular trigonometry and hyperbolic trigonometry, but I will not address them here.  From here I am going to shift our attention to uniting circular trig with hyperbolic trig through the complex plane.

\vspace{0.15in}
\begin{problem}[The Hyperbolic Tangent]{prob: Hyperbolic Tangent}

We defined a hyperbolic cosine and a hyperbolic sine.  Recall that the normal (circular) tangent is the quotient of the sine and cosine.  Use this quotient to derive the hyperbolic tangent function over the real numbers $x$, denoted by $\tanh{x}$ and pronounced \textit{tan-ch}, and then sketch it so you see what it looks like.  Based on your expression for the $\tanh$, prove that it is an odd function over the reals.
\end{problem}
\vspace{0.15in}

Okay, back to complex numbers.  Consider the function $w(z) = \cosh{z}$.  This time, we will choose to work with $z$ in its Cartesian representation so that $z = x +iy$.  Then, using \equaref{eq: cosh def}, we have
\begin{align}
    w(z) &= \cosh(x+iy), \nonumber
    \\
    &= \frac{\txte^{x+iy} + \txte^{-(x+iy)}}{2},\nonumber
    \\
    &= \frac{\txte^{x}\txte^{iy} + \txte^{-x}\txte^{-iy}}{2},\nonumber
    \\
    &= \frac{\txte^{x}\left( \cos y + i\sin y \right) + \txte^{-x}\left(\cos y -i\sin y \right)}{2},\nonumber
    \\
    &= \frac{\left( \txte^{x} + \txte^{-x}\right)\cos y + i\left(\txte^{x} - \txte^{-x}\right)\sin y}{2},\nonumber
    \\
    &= \frac{ \txte^{x} + \txte^{-x}}{2}\,\cos y + i\frac{\txte^{x} - \txte^{-x}}{2}\,\sin y,\nonumber
    \\
    &=\cosh(x)\cos(y) + i\sinh(x)\sin(y). \label{eq: w(z) = cosh(z)}
\end{align}
Thus, we have shown that the $\cosh$ of a complex argument mixes circular and hyperbolic trigonometry. Furthermore, let's assume that $z = iy$, meaning it is a purely imaginary (lateral) number.  Then
\begin{align}
    \cosh(iy) = \cos(y). \label{eq: cosh(iy) = cos(y)}
\end{align}
This relationship goes the other way as well.  If we consider $\cos(iy)$ then we have from \equaref{eq: cosine in polar representation}

\begin{align}
    \cos(iy) = \frac{\txte^{i(iy)} + \txte^{-i(iy)}}{2} = \frac{\txte^{-y} + \txte^{y}}{2} = \frac{\txte^{y} + \txte^{-y}}{2} = \cosh(y). \label{eq: cos(iy) = cosh(y)}
\end{align}

Since the imaginary arguments can be used to convert between circular and hyperbolic trig using these equations, we conclude that hyperbolic triangles are really just a lateral/adjacent form of straight-line versions. This conclusion could have actually been reached with the unit circle itself, $x^2 + y^2 = 1$, if we changed the $y$ part to an imaginary number $y\rightarrow iy$, thus converting the unit circle to $x^2+(iy)^2 = x^2 - y^2 = 1$, the unit hyperbola! Perhaps if we had not given $i=\sqrt{-1}$ the name of \textquotedblleft imaginary\textquotedblright, then we could have known more about geometry, in general, immediately from the start.

\vspace{0.15in}
\begin{problem}[Hyperbolas to Circles and Back]{prob: circular sines to hyperbolas}

I derived the relationships between the hyperbolic cosine and the circular cosine.  Using an almost identical treatment, derive the following equations:
\begin{align}
    \sinh(x+iy) &= \sinh(x)\cos(y) + i\cosh(x)\sin(y),\label{eq: w(z) = sinh(z)}
    \\
    \sinh(iy) &= i\sin(y),\label{eq: sinh(iy) = isin(y)}
    \\
    \sin(iy) &= i\sinh(y).\label{eq: sin(iy) = isinh(y)}
\end{align}
\end{problem}
\vspace{0.15in}

\section{Concluding Remarks}
In this chapter, the fundamentals of complex algebra were introduced and used to generalize a few commonly used real functions into the complex plane.  Knowing how to use complex numbers to study physical systems is particularly helpful.  For example, it is pretty common to treat all electromagnetic waves as traveling complex exponentials given by $\txte^{i(\vec{k}\cdot\vec{r} - \omega t)}$, where $\vec{k}$ is the wavenumber of the wave, $\vec{r}$ is the position of the wavefront, $\omega$ is the frequency of the wave, and $t$ is the time. To be clear, the electromagnetic waves are purely real objects meaning that all of their physical properties would be quantities that we would associate with the real line (they are not \textit{lateral} quantities, whatever those may be...). So why are these imaginary/lateral parts relevant in physics? It turns out that exponentials have a plethora of useful properties in calculus and they are much more manageable than the trigonometric waves defined by sines and cosines.  Thus, we usually only need to worry about the real part of the electric and magnetic fields.  Meanwhile, whenever the waves move through a material, it turns out that the $\vec{k}$ can be identified as having both a real part and a lateral part, where the latter actually plays a role in the heat-loss of the wave in that medium (this property is called \textit{attenuation}).  In other words, knowing how to split up complex numbers into real and imaginary (lateral) parts and more can tell us very important things about how physical objects interact with one another.  And this case is one where we use complex numbers to make our mathematical modeling simpler | it does not include situations in physics where the whole complex plane is absolutely necessary, such as anything where the word \textquotedblleft quantum\textquotedblright$\,$ is involved.

There is still a lot of topics within the topic of complex numbers at large, but this chapter hopefully gives you a solid foundation in something that, once you get the hang of it, will truly make that math you use in physics much simpler, and by extension, your life much easier.  Not to mention, complex algebra at least makes things like numbers a little prettier to look at.

I have included a table of helpful equations (\tblref{tab: complex algebra table}) for you to reference for whenever you need them later on in your career.  They also have a reference to the text where they were described.

\begin{table}
    \centering\small
    \caption{A summary of the important and general equations in Complex Algebra.}
    \scalebox{0.85}{
    \begin{tabular}{ccc}
        \hline\hline
        \textbf{Equation Description} & \textbf{Equation Formula} & \textbf{Text Reference} \\
        \hline \\
        Definition of Complex Numbers & $z = x + iy,\;\; x,y\in\mathds{R}$ & \equaref{eq: def complex number}
        \\ & & \\
        Definition of Complex Conjugate & $z = x - iy,\;\; x,y\in\mathds{R}$ & \equaref{eq: Complex Conjugate}
        \\ & & \\
        Real Part of a Complex Number & $\mathrm{Re}(z) = \dfrac{z + z^\ast}{2}\in\mathds{R}$ & \equaref{eq: Real part conjugates}
        \\ & & \\
        Imaginary Part of a Complex Number & $\mathrm{Im}(z) = \dfrac{z - z^\ast}{2i}\in\mathds{R}$ & \equaref{eq: Imaginary part conjugates}
        \\ & & \\
        Addition of Complex Numbers & $z\pm w = (x \pm u) + i(y \pm v)$ & \equaref{eq: complex z pm w}
        \\ & & \\
        Product of Complex Numbers & $zw = (xu - yv) + i(xv + yu)$ & \equaref{eq: complex zw}
        \\ & & \\
        Quotient of Complex Numbers & $\dfrac{z}{w} = \dfrac{xu + yv}{u^2 + v^2} + i\,\dfrac{xv-yu}{u^2+v^2}$ & \equaref{eq: complex z/w}
        \\ & & \\
        Modulus of a Complex Number & $\vert z\vert = \sqrt{z^\ast z} = \sqrt{[\textrm{Re}(z)]^2 + [\textrm{Im}(z)]^2}$ & \equaref{eq: complex modulus absolute value}
        \\ & & \\
        Polar Representation  & $z = \vert z\vert \txte^{i\,\mathrm{arg}(z)}$ & \equaref{eq: complex polar representation}
        \\ & & \\
        Argument of a Complex Number & $\mathrm{arg}(z) = \arctan\left[ \dfrac{\mathrm{Im}(z)}{\mathrm{Re}(z)}\right]$ & \equaref{eq: complex argument}
        \\ & & \\
        Euler's Identity & $\txte^{i\theta} = \cos\theta + i\,\sin\theta$ & \equaref{eq: Euler's identity}
        \\ & & \\
        Cosine Formula & $\cos\theta = \dfrac{\txte^{i\theta} + \txte^{-i\theta}}{2}$ & \equaref{eq: cosine in polar representation}
        \\ & & \\
        Sine Formula & $\sin\theta = \dfrac{\txte^{i\theta} - \txte^{-i\theta}}{2i}$ & \equaref{eq: sine in polar representation}
        \\ & & \\
        Multiplication as Dilation and Rotation & $zw = (r\rho)\,\txte^{i(\theta + \phi)}$ & \equaref{eq: multiplication dilation rotation}
        \\ & & \\
        De Moivre' Theorem (for Integer $n$) & $(\cos\theta + i\sin\theta)^n = \cos(n\theta) + i\sin(n\theta)$ & \equaref{eq: De Moivre's Theorem}
        \\ & & \\
        $n^{\mathrm{th}}$ Roots of Unity & $1^{1/n} = \exp\left( \dfrac{2\pi m i}{n} \right),\;\, m \in \{0, 1,\dots,n-1\}$ & \equaref{eq: n roots of unity polar}
        \\ & & \\
        $n$-Term Geometric Series & $S_n = \sum\limits_{m=0}^{n-1} a^m = \dfrac{1 - a^n}{1 - a},\; a\neq 1$ & \equaref{eq: finite geometric series answer}
        \\ & & \\
        Sum of $n^{\mathrm{th}}$ Roots of Unity & $\sum\limits_{m=0}^{n-1} \left(\txte^{2\pi i/n} \right)^{m} = 0$ & \equaref{eq: sum roots of unity}
        \\ & & \\
        General Complex Exponentiation & $z^{\alpha + i\beta} = \left( r\txte^{i\theta} \right)^{\alpha + i\beta} = r^n\txte^{-\beta\theta} \,\txte^{i[\beta\ln(r) +\alpha\theta]} $ & \equaref{eq: w(z) complex exponential}
        \\ & & \\
        Principal Logarithm & $\mathrm{Ln}(z) = \mathrm{Ln}(r\txte^{i\theta}) = \ln(r) + i\theta,\;\; \theta\in[-\pi,\pi)$ &\equaref{eq: principal ln} 
        \\ & & \\
        \hline\hline
    \end{tabular}
    }
    \label{tab: complex algebra table}
\end{table}

\newpage

\setcounter{example}{0}
\setcounter{problem}{0}

\chapter{Calculus of a Single Variable}

Up until now, I have not assumed that you have any prior knowledge of calculus, mostly because it is frankly unnecessary to understand vectors and complex algebra. Thus, everything that we have covered is totally \textit{static}. It, in no way, describes anything that can possibly change at all.  For example, if the position vector of a particular object is known at a particular time, we would thus far only have the tools to describe the object's vector at that time. Sure, we could resolve the vector into its components, we could talk about the direction of the vector, we could talk about it being orthogonal to other vectors, but we would have NO WAY of talking about how that vector is changing at that time.  Hence could not speak of how the object will move in time. 

You might be wondering why we could not just reference the object's velocity vector at that time, since surely the velocity will tell us about how the position vector changes.  As we will see, this is absolutely true. However, none of the mathematical tools we've discussed so far allow us to describe this change.  Meanwhile, you can definitely \textit{intuit} how position will change because you have experienced changes in your everyday life.  Our goal in this chapter is to develop your natural intuition into something mathematically precise and immutable | our goal is to develop calculus.

As a heads up, chapter is going to be dense.  Like, degenerate matter dense\footnote{A classical model of the degeneracy will probably suffice though.}. We are going to cover essentially the same amount of material that one would see as they progressed from a Calculus I and II sequence of courses.  Although we will cover a lot of material, it will be condensed to a form that is sufficient for implementation in physics.  Truth be told, once you understand derivatives and integrals from a geometric/changey perspective, then the number of dimensions rarely makes a difference to a physicist.  Sure, it might add a few extra terms here and there, but the actual modeling of the natural world using rates of change is what matters to us, and it is almost always easy, if not \textit{trivial}, to discuss the laws of physics using calculus.  That is not to say that actually solving problems in physics is trivial, \textbf{it isn't}, but the nontriviality of the problem solving in physics is partly due to the inclusion of our clunky form of algebra with calculus. The other part is usually just because nature is smarter than us.

I plan to introduce many calculus topics at are relevant for physics in ways that are easy to visualize. I am not going to get too bogged down in the details about which functions have which derivatives or antiderivatives, although I will talk about a few important ones.  Calculus really should not be about the memorization of a bunch of special-case formulas; when it is, it is usually very hard for students to see its value or applicability. Thus, I will reference outside sources for specific formulas if they are not quick and easy to derive (tables of derivatives and integrals are ubiquitous and freely available on the Internet).  Although the transition can be hard, once you are able to see beyond the mess of special cases, using calculus to describe the universe will be effortless, and you will probably wonder how you didn't already describe it is such a fluent way\footnote{Okay, so maybe this is a little anecdotal and waxing poetic, but, hey, if using calculus were worse than what we had before, why would we still be using it?}.

\section{Position, Velocity, and Acceleration}

People usually have an almost instinctive understanding of how things move (with the exception of feathers and bowling balls in free fall). Unfortunately, it took an incredible amount of time for us to develop a means of communicating our understanding with others in a precise way. We can thank people like Galileo and Newton for figuring that last bit out for us.  In historical hindsight though, developing the crucial ideas of calculus are pretty straightforward.

\begin{figure}
    \centering
    \includegraphics[width = 5.25in, keepaspectratio]{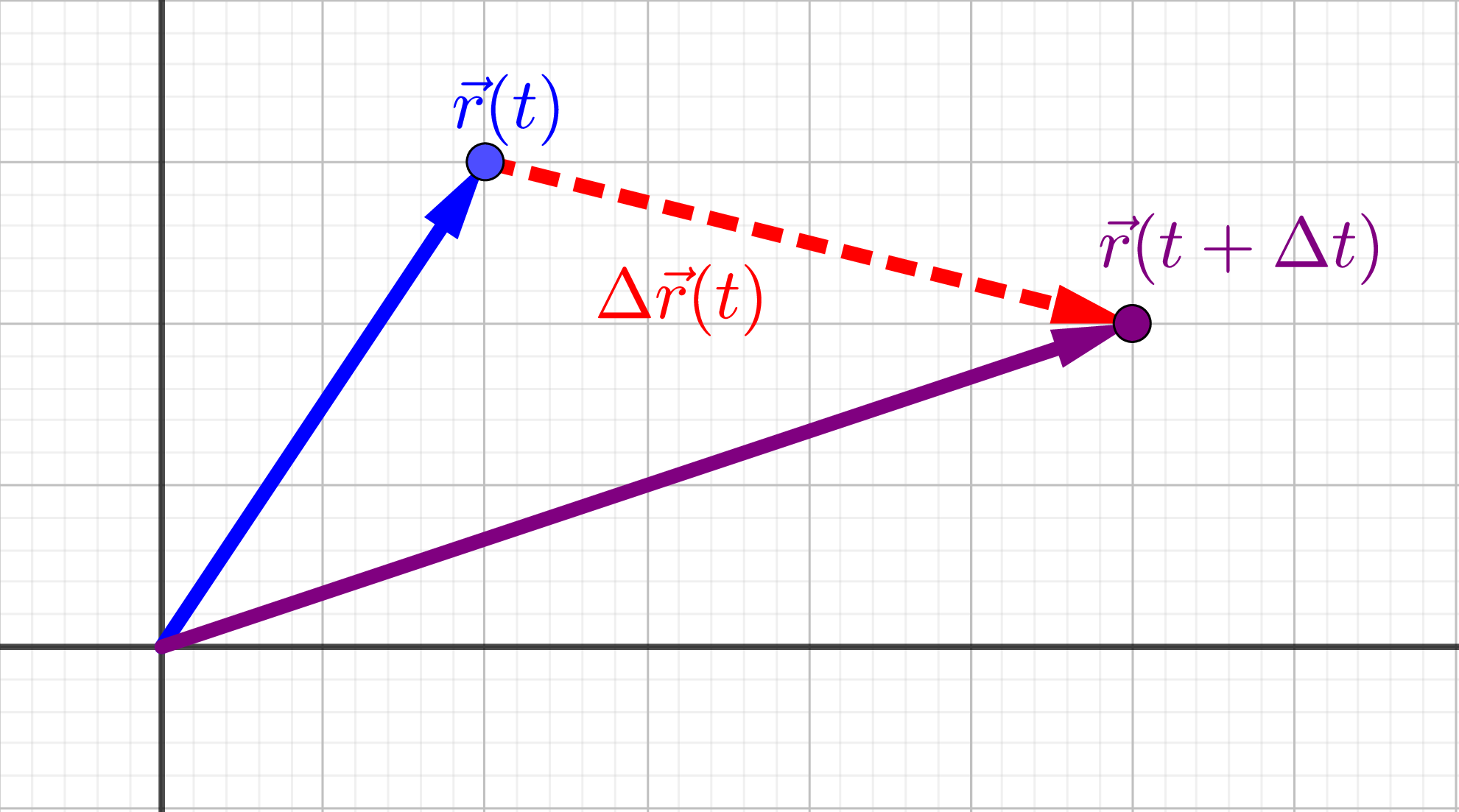}
    \caption{If the position of an object is known at two times, perhaps at $t$ and $t+\Delta t$, it is possible to describe the change in position from $\vec{r}(t)$ to $\vec{r}(t+\Delta t)$ as $\Delta\vec{r}(t) = \vec{r}(t+\Delta t) - \vec{r}(t)$.}
    \label{fig: position two points}
\end{figure}

We are going to start by exploiting the position, velocity, and acceleration of some object (maybe a car or volleyball or bug) and we will be careful to remember that each is a vector.  Furthermore, let us assume that we know the position vectors at a few different times, and let us denote these vectors by $\vec{r}(t)$.  Using vector addition (recall \textbf{tip-to-tail}), we can then define a vector that \textit{changes} the position at one time $t$ to one at a later time $t + \Delta t$.  Here this $\Delta t$ represents an increment in time, for example, it could represent a year, a century, a second, a femtosecond, a unit of Planck time, \textit{et cetera}. Physically, this increment in time would be whatever time difference it takes to get the object from one position to the next. This change vector at time $t$ is going to be denoted as $\Delta \vec{r}(t)$ and \textit{connects} the position at $t$, $\vec{r}(t)$, to the position at $t+\Delta t$, $\Delta \vec{r}(t + \Delta t)$, as is shown in \figref{fig: position two points}.

A question that follows from this set up would be something like, \textquotedblleft Is it possible to describe how quickly the object moved from $\vec{r}(t)$ to $\vec{r}(t+\Delta t)$?\textquotedblright$\,$ Intuitively, we might propose that the change in position, or the position increment, $\Delta\vec{r}(t)$, is proportional to the increment in time, $\Delta t$.  For example, if we were running at a specific pace for a longer period of time, we will travel farther. However, time is \textit{scalar}, whereas the change in position is definitely a vector by \figref{fig: position two points}. Thus, we are left to define the \textit{velocity}\footnote{For anyone who already knows calculus, this is the average velocity.} of the object at time $t$ as the vector responsible for the change in the object's position. This is denoted as $\vec{v}(t)$ and can be written mathematically as 
\begin{align}
    \Delta \vec{r}(t) = \vec{r}(t + \Delta t) - \vec{r}(t) = \vec{v}(t)\Delta t,
\end{align}
or if we divide both sides by the scalar time increment, we have
\begin{align}
    \vec{v}(t) = \frac{\Delta \vec{r}(t)}{\Delta t} = \frac{\vec{r}(t + \Delta t) - \vec{r}(t)}{\Delta t}, \label{eq: definition of average velocity}
\end{align}
It is important to remember that at this point the $(t)$ things everywhere are meant to symbolize that the letter immediately before is a \textit{function} of $t$; it does not mean multiply everything by $t$. For example, $\vec{v}(t)$ means that the velocity is a function of time $t$ rather than multiply $\vec{v}$ and the scalar $t$.

By looking at \equaref{eq: definition of average velocity}, we see that the equation has the form of a \textbf{slope}, in that it is a function that has a change divided by a change, or more colloquially, it has a \textit{rise-over-run} type form. Furthermore, we can see where the units of miles-per-hour or meters-per-second come from, since we multiply the position vector measured in units of length by the scalar $1/\Delta t$ which has units of per-time. Also, using this definition, since the time increment takes on the value of whatever time-recording device we have allows, if there is no change in position, we conclude that there is zero velocity.  Likewise, if there is ever a time where the velocity is zero, then there will be no resulting change in position. Alright, now that the relationship between changes in position and velocity is established between two points in time, what happens when we introduce more points?  Or from a more scientific perspective, what happens when we measure the position of an object at a greater number of points in time?

\begin{figure}
    \centering
    \includegraphics[width = 5.5in, keepaspectratio]{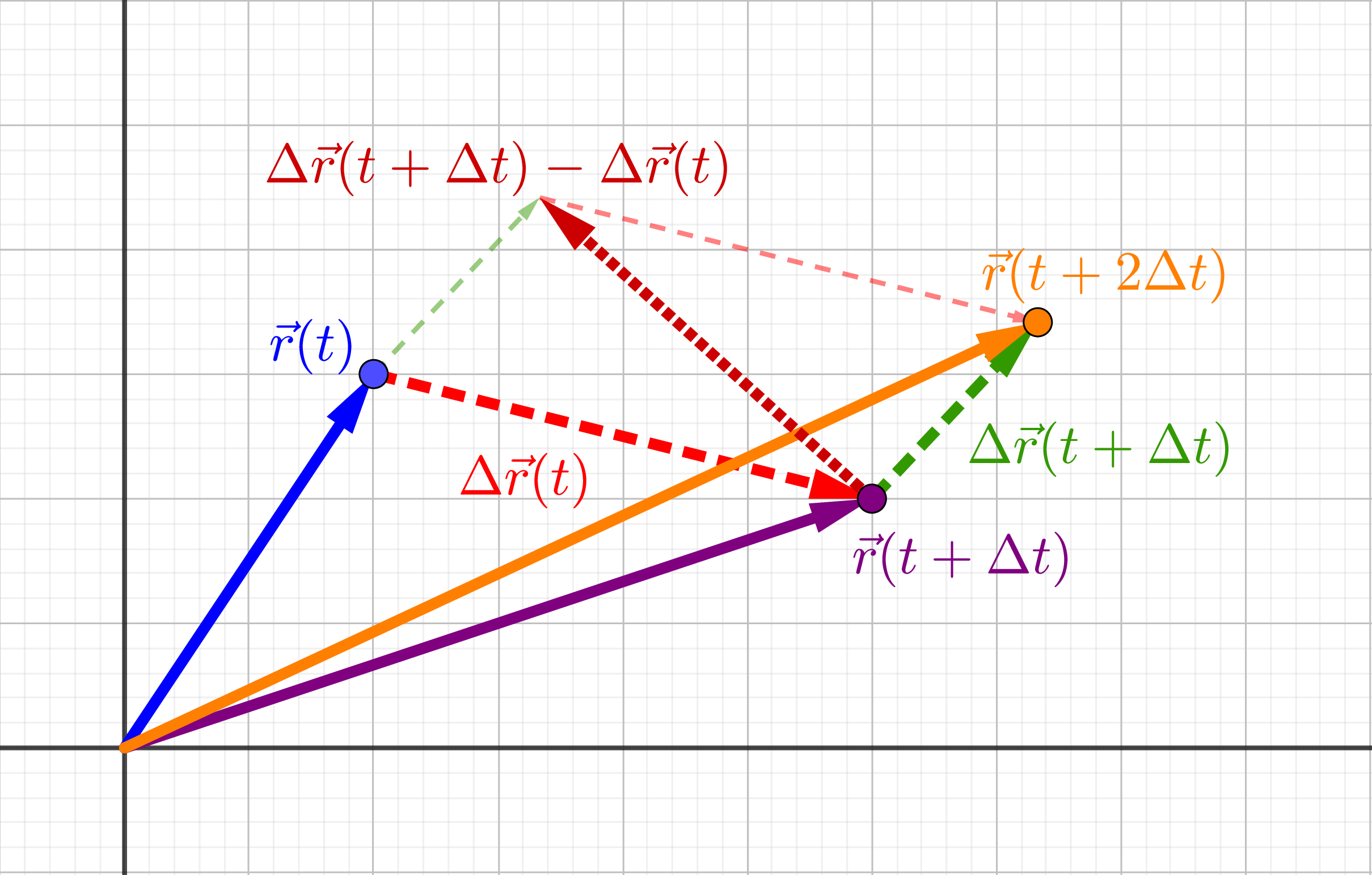}
    \caption{If the position of an object is known at three times, perhaps at $t$, $t+\Delta t$, and $t + 2\Delta t$, it is possible to describe the change in the change in the position from $\vec{r}(t)$ to $\vec{r}(t+2\Delta t)$ as $\Delta \Delta\vec{r}(t+\Delta t) = \Delta \vec{r}(t+\Delta t) - \Delta \vec{r}(t)$. Since the change in position is a measure of velocity by \equaref{eq: definition of average velocity}, then this figure shows the existence in the \textit{change in velocity}, also known as the acceleration.}
    \label{fig: position three points}
\end{figure}

For simplicity, let's assume we know the position of a particle at three points in time: $t$, $t+\Delta t$,  and $t+2\Delta t$.  Then, just as we did before in \figref{fig: position two points}, it is possible for us to connect the position at $t$ with that at $t+\Delta t$, and we can now connect the position at $t+\Delta t$ with that at $t + 2\Delta t$. A picture of this scenario is shown in \figref{fig: position three points}, and in this figure, it is evident that there are really two changes in position:
\begin{align*}
   \Delta\vec{r}(t) &= \vec{r}(t+\Delta t) - \vec{r}(t),
   \\
   \Delta\vec{r}(t+\Delta t) &= \vec{r}(t+2\Delta t) - \vec{r}(t + \Delta t).
\end{align*}
Thus, we may ask \textquotedblleft what is the change in the change in position?\textquotedblright$\,$ Let us denote this change of change in position as $\Delta\Delta \vec{r}(t+\Delta t)$, since the change of change position occurs at $t+\Delta t$. Let's denote the change of change as $\Delta\Delta$, and evaluate it directly
\begin{align*}
    \Delta\Delta\vec{r}(t+\Delta t) = \Delta\vec{r}(t+\Delta t) - \Delta\vec{r}(t)
\end{align*}
If we now assume that this change is proportional to some vector rate, just like we did when we came up with the notion of velocity, and denote this rate as $\vec{a}(t+\Delta t)$, then we can posit that 
\begin{align*}
    \Delta\Delta\vec{r}(t+\Delta t) = \vec{a}(t+\Delta t)\Delta t^2,
\end{align*}
Here $\Delta t^2 = (\Delta t)^2$. Solving for $\vec{a}(t+\Delta t)$ directly, we have
\begin{align*}
    \vec{a}(t+\Delta t) = \frac{\Delta\vec{r}(t+\Delta t) - \Delta\vec{r}(t)}{\Delta t^2} = \frac{ \frac{\Delta\vec{r}(t+\Delta t)}{\Delta t} - \frac{\Delta\vec{r}(t)}{\Delta t}  }{\Delta t}.
\end{align*}
In the second equality, I moved one of the $\Delta t$s into the numerator to make two fractions because these fractions are defined explicitly in \equaref{eq: definition of average velocity}.  Using this definition, we have 
\begin{align}
   \vec{a}(t + \Delta t) = \frac{\vec{v}(t + \Delta t) - \vec{v}(t)}{\Delta t}, \label{eq: definition of average acceleration}
\end{align}
which we recognize as the definition of \textit{acceleration}\footnote{Average acceleration to be precise.} Again, this equation is one that has the form of a slope. Further, we would say that if there is a change in \textit{velocity}, then there must have been an acceleration to cause it.  However, since the acceleration is related to the change of change in position, we could also say that a \textquotedblleft second-order\textquotedblright$\,$ change in position is due to the existence of an acceleration.

So what have we learned by measuring more points? Well, it appears that each time we measure a new point, we can define a new \textit{rate} associated with the addition of a new position measurement.  This is generally true for position measurements.  Actually, the names for a few of the even higher order changes ($3^{rd} - 6^{th}$) are the jerk, snap, crackle, and pop \cite{thompson_systems_technology}. We also know that the little time increment $\Delta t$ is some measurement, but its actual value depends on the time-measuring device we have (for example, atomic clocks have smaller $\Delta t$s than grandfather clocks). In principle, we could have exactly one position measurement for every single time measurement we have, and let's assume we have $N$ total of these position measurements. Then we could write them down in a table of sorts as

\begin{align*}
\begin{array}{c|c|ccccc}
    \textbf{Measure:} & \textrm{Time } t & t & t + \Delta t & t + 2\Delta t & \dots & t + (N-1)\Delta t  \\\hline
    \textbf{Measure:} & \textrm{Position }\vec{r} & \vec{r}(t) & \vec{r}(t + \Delta t) & \vec{r}(t + 2\Delta t) & \dots & \vec{r}(t + (N-1)\Delta t) 
\end{array}
\end{align*}

And so each time there is a change in position, we would presumably have a corresponding velocity measurement, and additionally, each time we have a change in velocity we would have an acceleration. So we would have two more rows in that table

\begin{align*}
\begin{array}{c|c|ccccc}
    \textbf{Measure:} & \textrm{Time } t & t & t + \Delta t & t + 2\Delta t & \dots & t + (N-1)\Delta t  \\\hline
    \textbf{Measure:} & \textrm{Position }\vec{r} & \vec{r}(t) & \vec{r}(t + \Delta t) & \vec{r}(t + 2\Delta t) & \dots & \vec{r}(t + (N-1)\Delta t)  \\\hline
    \textbf{Calculate:} & \textrm{Velocity } \vec{v} & - & \vec{v}(t + \Delta t) & \vec{v}(t + 2\Delta t) & \dots & \vec{v}(t + (N-1)\Delta t)  \\\hline
    \textbf{Calculate:} & \textrm{Acceleration } \vec{a} & - & \vec{a}(t + \Delta t) & \vec{a}(t + 2\Delta t) & \dots & -
\end{array}
\end{align*}

\begin{figure}
    \centering
    \includegraphics[width = 5.5in, keepaspectratio]{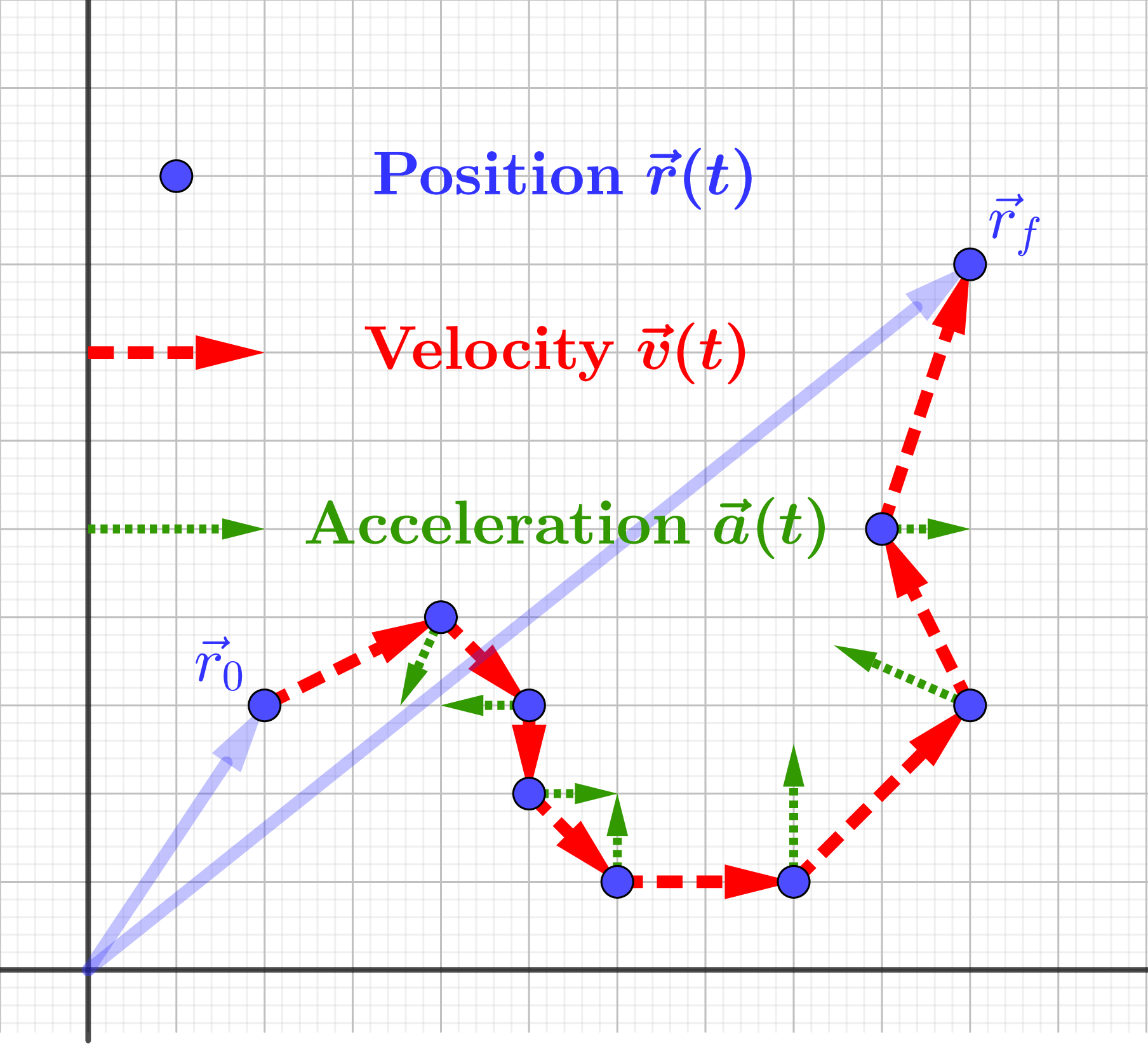}
    \caption{The position, velocity, and acceleration of particle as a function of time $t$ and it meanders from an initial position $\vec{r}_0$ to a final position $\vec{r}_f$. During the trip, it can have many different velocities and accelerations, whose directions are represented by the red dashed arrows and the green dotted arrows, respectively.}
    \label{fig: position bunch of points}
\end{figure}

Here, the dashed marks appear since there are no changes initially for either the position or velocity; thus there are no velocities or accelerations at this time. To better help visualize this table, imagine that all of these measurements came from an object's trajectory as that shown in \figref{fig: position bunch of points}, where the object starts at an initial position of $\vec{r}_0$ and moves to a final position of $\vec{r}_f$. In the figure, the blue points represent the positions as functions of time (remember that these are all vectors, hence the ghostly vectors to $\vec{r}_0$ and $\vec{r}_f$), the red dashed vectors represent the velocities as functions of time, and the green dotted vectors represent the accelerations as functions of time.

An important highlight from drawing out all of these position measurements is that the velocity vectors can really be seen to connect all of the position measurements together.  More precisely, if we wanted to know the position of the object at any time measurement, let's say at the $j^{\mathrm{th}}$ time increment, then we could find it by starting at $\vec{r}_0$ and then \textit{adding} all of the changes in position up until the $j^{\mathrm{th}}$ increment.  In other terms,
\begin{align*}
    \vec{r}(t+ j\Delta t) &= \vec{r}_0 + \Delta\vec{r}(t) + \Delta\vec{r}(t + \Delta t) + \Delta\vec{r}(t+2\Delta t) + \dots \Delta\vec{r}(t + (j-1)\Delta t),
    \\
    &= \vec{r}_0 + \vec{v}(t)\Delta t + \vec{v}(t + \Delta t)\Delta t + \vec{v}(t + 2\Delta t)\Delta t + \dots + \vec{v}(t + (j-1)\Delta t)\Delta t.
\end{align*}
All of the $\Delta t$s here show up from the implicit definition of the velocity at a point. The equation above can be written much more succinctly with summation notation, and so we have
\begin{align}
    \vec{r}(t + j\Delta t) = \vec{r}_0 + \sum_{k = 0}^{j-1}\vec{v}(t + k\Delta t)\Delta t. \label{eq: position sum of velocity}
\end{align}
Although it is harder to visualize from \figref{fig: position bunch of points} (but not impossible!), there is actually an analogous summation relationship that exists to relate the velocity at $t + j\Delta t$ to all of the intermediate accelerations:
\begin{align}
    \vec{v}(t + j\Delta t) = \vec{v}_0 + \sum_{k = 0}^{j-1}\vec{a}(t + k\Delta t)\Delta t. \label{eq: velocity sum of acceleration}
\end{align}
These equations have the form of the \textbf{area of a rectangle}, where $\Delta t$ is the width of the rectangle, and the velocity or acceleration would be the height of the rectangle.  Furthermore, we actually add up all of the intermediate areas to find something new.

In short, this is all that calculus is: either evaluating slopes or adding up a bunch of boxes. If we want to describe how quickly something changes in terms of another variable | for example, if we want to find how fast position changes as a function of time | we would use a slope.  If, on the other hand, we wanted to describe the aggregate effects of many successive actions | for example, if we want to find out where our final position is after moving with a bunch of different velocities | we would add boxes. Even further, we can \textit{undo} the effect of taking a slope by adding a bunch of areas together; an operational inversion analogous to how multiplying something by $4$ and then dividing the result by $4$ returns the original something. And that is truly all of calculus.  There really is nothing else to conceptualize.

However, like most things, this conceptualization is easier said than done. And that's okay.  As we progress through this chapter, we will study many more applications of these ideas and refine them into some theorems.  Fortunately for us, the context of this chapter will largely be the natural world, and so if the mathematics ever gets too messy, we can always find real-world analogies to bolster our intuition and understand which message the math is trying to convey.

\section{Continuity}
We have discussed how slopes and box-addition represent function-transitions between position, velocity, and acceleration. But everything we did was built on the assumption that we were taking measurements of time with some device that only measured in increments of $\Delta t$. While this picture is accurate from a numerical perspective, and definitely from an experimental point-of-view, it is not necessarily physical and certainly not mathematical.  

Physically, objects have positions at intermediate values between the points we measure.  We unfortunately are limited by how quickly we can measure something's position; however, this is no way implies that objects only move in discrete ways. In principle, if we were able to \textquotedblleft take more measurements\textquotedblright$\,$ then we could describe an object's position, velocity, or acceleration at \textit{any time} $t$ that we wanted. 

To better illustrate this idea, we assume that something only moves over a fixed time interval, perhaps $t_f - t_0$ and that we have made a total of $N$ measurements.  Then the time increment would be
\begin{align*}
    \Delta t = \frac{t_f - t_0}{N}.
\end{align*}
Notice that this time increment gets \textit{smaller} when the number of samples $N$ gets larger.  If this is not clear, choose $t_f - t_0$ to be your favorite number (physicists really like $1$) and then divide by successively larger numbers | for example, $1/1 = 1$, $1/2 = 0.5$, $1/3 = 0.33\dots$, ..., $1/100 = 0.01$, ... So as long as the $t_f - t_0$ is a fixed number, $\Delta t$ will get very small as $N$ gets very large. 

The small time increment is helpful, because it allows us to probe more time-values.  For example, if $t_0 = 1$, $t_f = 2$, and $N = 10$, then we can talk about times like
\begin{align*}
    t_0 + \Delta t &= 1 + 1/10 = 1.1,
    \\
    t_0 + 2\Delta t &= 1 + 2/10 = 1.2,
    \\
    &\vdots 
    \\
    t_0 + (N-1)\Delta t &= 1 + 9/10 = 1.9
\end{align*}
But we could never reach a time value like $1.11$, since our time increment is larger than that extra $0.01$. In other words, our \textit{resolution} is not good enough to \textit{see} a time increment of $0.01$. So what can we do? Instead of only taking $10$ measurements, how about we take $100$? Then we could have
\begin{align*}
    t_0 + \Delta t &= 1 + 1/100 = 1.01,
    \\
    t_0 + 2\Delta t &= 1 + 2/100 = 1.02,
    \\
    &\vdots
    \\
    t_0 + (N-1)\Delta t &= 1 + 99/100 = 1.99
\end{align*}
By the same argument, we could jump down to a resolution of 0.001 by taking 1000 measurements to have
\begin{align*}
    t_0 + \Delta t &= 1 + 1/1000 = 1.001,
    \\
    t_0 + 2\Delta t &= 1 + 2/1000 = 1.002,
    \\
    &\vdots
    \\
    t_0 + (N-1)\Delta t &= 1 + 999/1000 = 1.999
\end{align*}
Hence, it is clear that by increasing the number of samples, we can know the position over smaller time increments and therefore observe what happens at those intermittent steps that we would have been incapable of observing before.  So now the question is: \textquotedblleft how many samples do we need to get a time increment to probe time up to any arbitrary level of precision $\epsilon$?\textquotedblright

To solve this problem, we just need the time increment to be less than this arbitrary precision $\epsilon$ (before $\epsilon \in \{ 0.1, 0.01, 0.001\}$).  Thus, what we do is posit that $\Delta t < \epsilon$, or 
\begin{align}
    \frac{t_f-t_0}{N} < \epsilon \Rightarrow  \frac{t_f-t_0}{\epsilon} < N, \label{eq: Delta t N limit}
\end{align}
The second inequality\footnote{This inequality rule can be tricky for a lot of people the first time they see it. Most people know that if you multiply an inequality by a negative sign, then you must flip the inequality. The same is true for division. For example, it is true that $1/2 < 1/1$, but it is not true that $2 < 1$, when we invert the inequality.} is our answer. I recommend you test it for yourself.  

The beautiful thing about this idea is that $\epsilon$ can literally be \textit{anything}, as long as it's positive. That means we could increase it or decrease it at will | this is the math equivalent of increasing or decreasing the volume of your phone\footnote{Or whatever the cool sound-emitting technological marvel you future people use.}, except there is no upper or lower bound! Thus, we could, at least in principle, probe time measurements to \textit{any} level of precision we want.  All we would need to do is take enough measurements.

In mathematics, we give this idea the name, \textit{limit of a sequence}.  In this case, the sequence would be $\Delta t$, and the elements of the sequence would be each $\Delta t$ with a incrementally higher $N$ value substituted in.  In this case, we would say that the limit of $\Delta t$ would be zero since we can definitively get within a precision of $\epsilon$ of zero using the inequalities above. More formally, we would say that the limit of some sequence $a_N$, denoted by $L$, exists by the following criterion:
\begin{quote}
    The sequence $a_N$ is said to \textbf{converge} to the \textbf{limit} $L$ if there exists positive integers $n$ and $M$ such that $\vert a_n - L\vert < \epsilon$ for every $\epsilon > 0$, whenever $n > M$.  
\end{quote}
When the limit does in fact exist, we say that
\begin{align}
    \lim_{N\rightarrow\infty} a_N = L, \label{eq: definition of sequence limit}
\end{align}
where the calculation of how many measurements are required to reach a precision of $\epsilon$ is given by the positive integer $M$ and the \textquotedblleft taking enough measurements\textquotedblright$\,$ is taken into account when $N\rightarrow\infty$ in \equaref{eq: definition of sequence limit}.

Knowing more and more precise time measurements is great, but what does it get us? How can we be sure that there will truly be a position measurement for us to take if we \textit{zoom} in on the temporal axis with smaller and smaller time increments?  To the best of my knowledge, in physics, we assume that this is always true for moving objects.  But this would imply that as we more and more precise time measurements, we could get more and more precise position measurements. Or at least, if the position at one point in time is known, then the position as $t + \Delta t$ should be nearby that at $t$.  Otherwise, the object would somehow jump/teleport to another location randomly. And that would be weird/not anything we have ever observed, so in physics we assume that we can always zoom in infinitely far for all spatial and temporal variables ($N\rightarrow \infty$).

But how would we represent this mathematically? For this case, we will assume that we have taken infinitely many temporal measurements so that we can zoom in arbitrarily around some special point in time, let's call it $t^\prime$.  If we have zoomed in to some arbitrary level of precision, now denoted by $\delta t$\footnote{The lowercase delta is because our precision is tiny, so I figured a little variable would be appropriate.}, then we can probe any time values within what's called an \textbf{open ball of radius} $\delta t$. The ball itself is defined as the set of all $t$-values that are within the radius $\delta t$ of the central point $t^\prime$, and for those who are interested it is written as
\begin{align*}
    \mathcal{B}(t^\prime,\delta t) = \{t\in\mathds{R}: \vert t - t^\prime \vert < \delta t \}.
\end{align*}
In a single dimension though, this ball is just the open interval $t \in (t^\prime -\delta t, t^\prime +\delta t)$ along the real line. It is a circular disk in two dimensions, and actually is a ball (sphere) in three dimensions. In four and higher dimensions, it is harder to visualize\footnote{Yes, this is a dare to prove me wrong.}.  

Anyway, we want to be sure that an object doesn't suddenly teleport elsewhere, so we need the position of an object at time $t$ to be nearby the position at time $t^\prime$.  Therefore, if we can find a position ball whose radius $\vert \delta \vec{r}\vert$ is somehow the result of the $\delta t$ precision we chose, then we can \textit{guarantee} that the position of the object at time $t$ will \textit{always} be within the vicinity of the position at time $t^\prime$.  And most importantly, means
\begin{align}
    \vert \vec{r}(t) - \vec{r}(t^\prime) \vert < \vert \delta \vec{r}\vert,
\end{align}
where $\vert \delta \vec{r}\vert$ is a finite number, of again, arbitrary precision.

This whole idea that zooming in far on the independent variable consequently leads to us freely zooming on the dependent variable is what is known as \textbf{continuity}.  To be more mathematically precise:
\begin{quote}
    A function $f:\mathds{R}\rightarrow\mathds{R}$ is said to be \textbf{continuous} at $x^\prime$ if there exists a $\delta > 0$ for every $\epsilon >0$ such that for every $x$ where $\vert x - x^\prime\vert < \delta$ then implies $\vert f(x) - f(x^\prime)\vert < \epsilon$.
\end{quote}
When the above statement holds, we call the output $f(x^\prime)$ the \textit{limit} of $f(x)$ as $x\rightarrow x^\prime$ which is written as
\begin{align}
    \lim_{x\rightarrow x^\prime} f(x) = f(x^\prime). \label{eq: definition continuity limit}
\end{align}
Sometimes, the limit is instead defined as $F$ so we can study functions where $f(x^\prime)$ need not be defined.

The definition above is what you may have seen/heard of before | it is the $\epsilon$-$\delta$ definition of continuity | but it is strictly in \textit{one-dimension}, whereas actually everything else up until this point was in an arbitrarily high number of spatial dimensions; so maybe we were taking about an object only moving along a single axis (one-dimensional), or on a curve on a surface (two-dimensional), or even in a trajectory through space (three-dimensional). Here we condensed back to one-dimension with the $\epsilon$-$\delta$ definition of continuity, because the mathematical precision in higher dimensions is actually a little too nuanced to jump in with immediately. Actually, to be completely honest, I don't think the definition, as it is written above, has ever come up in a physics class that I've taken. However, the idea of being able to zoom in as far as we like does come up constantly; that is why I spent so much time developing the intuition behind the $\epsilon$-$\delta$ definition of continuity. Nevertheless, I do want to unpack that definition a little, because it will be extremely relevant for things like derivatives later on.

First things first, the statement $f:\mathds{R}\rightarrow\mathds{R}$ translates to \textquotedblleft function $f$ that has real number inputs (first $\mathds{R}$) and then outputs real numbers (second $\mathds{R}$).\textquotedblright $\,$ This means we are only dealing with the good-ol' one-dimensional functions that you learned in high school.  The next thing that is listed is that this definition of continuity only applies for functions at a \textit{single} point $x^\prime$ in the domain.  It says nothing about the continuity of $f$ at nearby points. The actual requirement for continuity is that there has to be an $\delta>0$, a.k.a. an arbitrarily small number, to zoom in on $x$ nearby $x^\prime$ when we zoom in on $f(x^\prime)$ itself to a chosen precision of $\epsilon$.  Therefore, this definition posits that the existence of the $\epsilon$ precision guarantees the existence of the $\delta $ precision in the independent variable. This means that continuity allows us to continue taking smaller and smaller independent-variable increments to measure smaller and smaller dependent-variable increments, just like our intuition with an object's position informed us.  So this definition, although I agree that it is pretty intangible the first time you see it, incorporates all of what we've discussed in the section so far, but it does it in a much more concise manner.

Example \ref{ex: continuity of simple functions} shows how we might use the definition of continuity to show simple functions are continuous.  The truth is that it usually takes a different algebra tricks to establish the continuity of individual functions, and oftentimes these tricks are NOT obvious. If you ever take a course in mathematical analysis, you will see it. Therefore, it usually helps to establish the continuity of general classes of functions, like that of sums, products, or compositions, because we can then argue that whatever messy function we have is truly just a combination of continuous functions.  Thus, I am going to prove a few theorems to you regarding the continuity of sums, products, and compositions. 

\textbf{Disclaimer:} the proofs will be pretty abstract, and so if it is too hard to follow at first, that's okay.  The fundamentally important continuity rules in physics will be the numbered limit equations that follow each proof, but I wouldn't be able to live with myself if I just told you what they were without justification. I mean, this is supposed to be science after all, right?

\vspace{0.15in}
\begin{example}[The Continuity of Simple Functions]{ex: continuity of simple functions}
    Definitions are great, but their utility really comes in applying them. In this example, I will prove to you that constant and linear functions are continuous.
    
    We first start with a constant function.  Let's assume our constant function looks like $f(x) = A$ for all $x\in\mathds{R}$. If you don't like the $A$ here, replace it with the number $\pi$ and then, at the end, replace all the $\pi$s with $A$s.  Okay, so the proof goes something like this: 
    \begin{quote}
        Consider the point $x^\prime\in\mathds{R}$ and the open ball $\mathcal{B}(f(x^\prime), \epsilon)$ centered at $f(x^\prime)$ with radius $\epsilon>0$.  More visually, this means we have zoomed in on the dependent variable $f$ to a precision of $\epsilon$. We want to show that there exists a $\delta>0$ such that $\vert x - x^\prime\vert < \delta$, no matter what $\epsilon$ precision we choose.
        
        Now consider the absolute difference $\vert f(x)-f(x^\prime)\vert$, where I will substitute in the values of $x$ and $x^\prime$ into $f(x) = A$. 
        \begin{align*}
            \vert f(x)-f(x^\prime)\vert = \vert A - A\vert = 0
        \end{align*}
        But by definition, $\epsilon > 0$.  If we let $\delta = \epsilon >0$, then for every $x$ such that $\vert x - x^\prime\vert < \delta$, we have
        \begin{align*}
            \vert f(x)-f(x^\prime)\vert = \vert A - A\vert = 0 < \delta = \epsilon
        \end{align*}
        This shows that $f$ is continuous at $x^\prime$.  However, since this argument holds for all $x^\prime\in\mathds{R}$, we conclude that \textit{constants functions are continuous over all of the real numbers}.
    \end{quote}
    
    For the next proof, we look at the linear function $f(x) = bx$ for all $x\in\mathds{R}$, where $b\neq 0$ is a constant coefficient.  The proof is similar in set up to the one above, but its execution will probably look very different to you the first time you see it. Remember that the key is to show the existence of any $\delta >0$ for every conceivable $\epsilon > 0$.
    \begin{quote}
        Consider the point $x^\prime\in\mathds{R}$ and the open ball $\mathcal{B}(f(x^\prime), \epsilon)$ centered at $f(x^\prime)$ with radius $\epsilon>0$. 
        
        We will argue that for every $\epsilon>0$, we can find a $\delta = \epsilon/\vert b\vert >0$ such that when $\vert x - x^\prime\vert < \delta$, we have $\vert f(x) - f(x^\prime)\vert < \epsilon$.
        \begin{align*}
             \vert f(x)-f(x^\prime)\vert &= \vert bx - bx^\prime\vert
             \\
             &= \vert b(x-x^\prime)\vert 
             \\ 
             &= \vert b\vert \,\underbrace{\vert x - x^\prime \vert}_{ < \delta}
             \\
             &< \vert b \vert\,\frac{\epsilon}{\vert b\vert} 
             \\
             &= \epsilon
        \end{align*}
        Thus, $\vert f(x)-f(x^\prime)\vert < \epsilon$ when $\vert x-x^\prime \vert < \delta$, for all $\delta > 0$; thereby completing the proof that $f(x) = bx$ is continuous at $x^\prime$. Further, since $x^\prime$ is arbitrary, again, this argument holds for all of the real numbers.  We conclude that \textit{linear functions are continuous over all of the real numbers}.
    \end{quote}
\end{example}
\vspace{0.15in}

Before we proceed to this point, we need one very important theorem concerning the absolute function: the so-called \textbf{triangle inequality}.  It states simply that for any $x,y\in\mathds{R}$, we have
\begin{align}
    \vert x + y \vert \leq \vert x \vert  + \vert y \vert. \label{eq: triangle inequality}
\end{align}
A proof of this statement is straightforward, and follows from the fact that $x^2 = (\vert x\vert)^2$ for all real numbers.  It goes as something like this,
\begin{align*}
    (\vert x\vert + \vert y\vert)^2 &= (\vert x\vert)^2 + (\vert y\vert)^2 + 2\vert x\vert\,\vert y\vert
    \\
    &\geq (\vert x\vert)^2 + (\vert y\vert)^2 + 2xy, \textrm{ since either $x$ or $y$ could possibly be negative}
    \\
    &= x^2 + y^2 + 2xy
    \\
    &= (x + y)^2
    \\
    &= (\vert x + y\vert)^2.
\end{align*}
Thus, $(\vert x + y\vert)^2 \leq (\vert x\vert + \vert y\vert)^2 $. By taking the square root of both sides, we have the triangle inequality; a theorem that gets its name because it essentially says that if a triangle is made up of three sides, one side cannot have a magnitude greater than the sum of the other two. This inequality will help us establish that sums of continuous functions are continuous.

Now for continuity in summation.  To prove this, consider the function $h(x) = f(x) + g(x)$, where $f(x)$ and $g(x)$ are both continuous at $x^\prime$.  Since both $f$ and $g$ are continuous at $x^\prime$, then there must exist $\delta$-precisions for both functions for every $\epsilon$-precision, namely $\delta_f$ and $\delta_g$, respectively, such that 
\begin{align*}
    \vert x - x^\prime\vert < \delta_f \textrm{ implies }\vert f(x) - f(x^\prime) \vert &< \frac{\epsilon}{2},
    \\
    \vert x - x^\prime\vert < \delta_g \textrm{ implies }\vert g(x) - g(x^\prime) \vert &< \frac{\epsilon}{2},
\end{align*}
Then, by substitution into $h$, we have
\begin{align*}
    \vert h(x) - h(x^\prime) \vert &= \vert f(x) + g(x) - f(x^\prime) - g(x^\prime) \vert 
    \\
    &= \vert f(x) - f(x^\prime) + g(x) - g(x^\prime) \vert 
    \\
    &\leq \vert f(x) - f(x^\prime)\vert  + \vert g(x) - g(x^\prime)\vert, \textrm{ thanks to the triangle inequality} 
    \\
    &< \frac{\epsilon}{2} + \frac{\epsilon}{2}, \textrm{ whenever } \vert x-x^\prime\vert < \min\{\delta_f,\;\delta_g\},
    \\
    &= \epsilon
\end{align*}
Thus, when we define $\delta = \min\{\delta_f,\;\delta_g\}$, where the $\min$ function outputs the smaller value of $\delta_f$ and $\delta_g$, then we have that $\vert x-x^\prime\vert < \delta$ implies $\vert h(x) - h(x^\prime) \vert < \epsilon$  for every $\epsilon>0$. This means that when $f$ and $g$ are continuous at $x^\prime$, so is the sum of the two. I want to emphasize that this conclusion implies the following limit rule:
\begin{align}
    \lim_{x\rightarrow x^\prime} \left[ f(x) + g(x) \right] = \left[\lim_{x\rightarrow x^\prime} f(x)\right] + \left[\lim_{x\rightarrow x^\prime} g(x)\right]. \label{eq: limit of sum}
\end{align}
Hence, \textbf{the limit of the sum is the sum of the limits}.  Furthermore, if (and only if) \textit{both} $f$ and $g$ are continuous over all of the real numbers, then so is their sum $h$.

We next want to prove that the product of two functions that are continuous at $x^\prime$ is also continuous at that point when two functions are defined\footnote{The defined condition guarantees that I can write something like $f(x^\prime)$ and not have is accidentally blow up to infinity.} at $x^\prime$. Now let $h(x) = f(x)g(x)$, and let the definitions of $\delta_f$ and $\delta_g$ be modified a little from what is given above as
\begin{align*}
    \vert x - x^\prime\vert < \delta_f \textrm{ implies }\vert f(x) - f(x^\prime) \vert &< \frac{\epsilon}{2(1 + \vert g(x^\prime)\vert)},
    \\
    \vert x - x^\prime\vert < \delta_g \textrm{ implies }\vert g(x) - g(x^\prime) \vert &< \frac{\epsilon}{2(1 + \vert f(x^\prime)\vert)}.
\end{align*}
These odd looking definitions will come in handy later on so that we will find $\vert h(x) - h(x^\prime)\vert < \epsilon$ and not some function of $\epsilon$. Then we have by substitution,
\begin{align*}
     \vert h(x) - h(x^\prime) \vert &= \vert f(x)g(x) - f(x^\prime)g(x^\prime) \vert, 
    \\
    &= \vert f(x)g(x) + 0 - f(x^\prime)g(x^\prime) \vert, 
    \\
    &= \vert f(x)g(x) - f(x^\prime)g(x) + f(x^\prime)g(x) - f(x^\prime)g(x^\prime) \vert,
    \\
    &= \left\vert \left[f(x)g(x) - f(x^\prime)g(x)\right] + \left[f(x^\prime)g(x) - f(x^\prime)g(x^\prime)\right] \right\vert,
    \\
    &\leq \vert f(x) - f(x^\prime)\vert\,\vert g(x)\vert + \vert f(x^\prime)\vert\, \vert g(x) - g(x^\prime)\vert, \textrm{ via the triangle inequality}
    \\
    &< \frac{\epsilon}{2(1 + \vert g(x^\prime)\vert)}\, \vert g(x)\vert + \vert f(x^\prime)\vert\,\frac{\epsilon}{2(1 + \vert f(x^\prime)\vert)}, \textrm{ when } \vert x-x^\prime\vert < \min\{\delta_f,\;\delta_g\}.
\end{align*}
Before continuing, we must figure out a way to deal with the $\vert g(x)\vert$ because we do not know the specific value of $g$ at $x$ | we only know that $g(x^\prime)$ is defined and that $g$ is continuous at $x^\prime$. But since $g$ is continuous at $x^\prime$, we are free to zoom in as far as we like on $g(x^\prime)$ and we are \textit{guaranteed} to be able to find an $x$-interval that corresponds to our $g$-precision.  Thus, what we will do is choose to limit our $g$-precision to be less than a set number, for example, $1$.  Then, by the continuity of $g$, there exists a $\delta_1$ such that
\begin{align*}
    \vert x - x^\prime\vert < \delta_1 \textrm{ implies }\vert g(x) - g(x^\prime) \vert &< 1.
\end{align*}
Then we can say using the triangle inequality that
\begin{align*}
    \vert g(x)\vert  = \vert g(x) + 0\vert = \vert g(x) - g(x^\prime) + g(x^\prime)\vert \leq \vert g(x)-g(x^\prime)\vert + \vert g(x^\prime)\vert < 1 + \vert g(x^\prime)\vert,
\end{align*}
where the last inequality holds for every $x$ such that $\vert x-x^\prime\vert < \delta_1$. Okay, so now that we have these conditions in place, and if we remember that $\vert f(x^\prime)\vert \leq 1 + \vert f(x^\prime)\vert$, we can define our $\delta$-precision such that $\delta = \min\{\delta_f,\; \delta_g,\; \delta_1 \}$. Then for every $x$ such that $\vert x - x^\prime \vert < \delta$, we have
\begin{align*}
    \vert h(x) - h(x^\prime) \vert &< \frac{\epsilon}{2(1 + \vert g(x^\prime)\vert)}\, \vert g(x)\vert + \vert f(x^\prime)\vert\,\frac{\epsilon}{2(1 + \vert f(x^\prime)\vert)},
    \\
    &\leq \frac{\epsilon\, (1 + \vert g(x^\prime)\vert )}{2(1 + \vert g(x^\prime)\vert)}  + \frac{\epsilon\,(1 + \vert f(x^\prime)\vert )}{2(1 + \vert f(x^\prime)\vert)},
    \\
    &= \frac{\epsilon}{2} + \frac{\epsilon}{2}
    \\
    &= \epsilon.
\end{align*}
Since we were able to find an $\delta >0$ for every $\epsilon >0$ such that$\vert x - x^\prime \vert < \delta$ implies/leads to $\vert h(x) - h(x^\prime) \vert < \epsilon$, then we can conclude that $h(x) = f(x)g(x)$ is continuous at $x^\prime$.  This time, the corresponding limit equation would look like
\begin{align}
    \lim_{x\rightarrow x^\prime} \left[ f(x)g(x) \right] = \left[\lim_{x\rightarrow x^\prime} f(x) \right]\, \left[\lim_{x\rightarrow x^\prime} g(x) \right], \label{eq: limit of a product}
\end{align}
or the \textbf{limit of a product is the product of limits}. Just as was the case for the the sum of continuous functions, if \textit{both} $f$ and $g$ are continuous over all of the reals, so is their product.  However, if only one function is continuous everywhere, then the product is NOT continuous everywhere | it, too, is limited\footnote{Nice pun, right?} to only be continuous for the values $x^\prime$ where \textit{both} of its factors are.

The last continuity rule deals with compositions of functions.  As a brief review compositions are sometimes written as $g(f(x))$ or $g\circ f(x)$. They are functions that look something like this: if $g(x) = x^2$ and $f(x) = 4\sin x$, then $g(f(x)) = (4\sin x)^2$.  Thus, the notation of $g(f(x))$ really just means put whatever the output of $f$ is into $g$ as the input. Although this class of functions may seem like more of a special case compared to sums or products (it did to me the first time I learned of it), compositions of functions appear everywhere in both calculus and physics.  Thus, knowing how they behave is imperative.

Our goal in particular is to show that compositions of continuous functions are also continuous. To do this, we then need to show that a function $h(x) = g(f(x))$ is continuous at the point $x^\prime$, given that $f$ is continuous at $x^\prime$ and $g$ is continuous at $f(x^\prime)$.  To make this proof a little clearer, define 
\begin{align*}
    u = f(x),\; u^\prime = f(x^\prime).
\end{align*}
By the continuity of $g$ at $u^\prime$, we know that for every $\epsilon>0$, there exists a $\delta_u>0$ such that 
\begin{align*}
    \vert u-u^\prime\vert = \vert f(x) - f(x^\prime) \vert < \delta_u \textrm{ implies } \vert g(u) - g(u^\prime) \vert  < \epsilon.
\end{align*}
Then by the continuity of $f$ at $x^\prime$, we know that for every $\epsilon_f$-precision, there exists a $\delta >0$ such that 
\begin{align*}
    \vert x-x^\prime\vert < \delta \textrm{ implies } \vert f(x) - f(x^\prime) \vert  = \vert u - u^\prime \vert < \epsilon_f.
\end{align*}
Therefore, since we want something that looks like
\begin{align*}
    \vert x-x^\prime\vert < \delta \textrm{ implies } \vert h(x) - h(x^\prime) \vert  = \vert g(f(x)) - g(f(x^\prime)) \vert < \epsilon,
\end{align*}
then we choose $\epsilon_f = \delta_u$ such that our final statement reads as
\begin{align*}
    \vert x-x^\prime\vert < \delta \textrm{ implies } \vert f(x) - f(x^\prime) \vert = \vert u - u^\prime \vert < \delta_u \textrm{ implies } \vert g(u) - g(u^\prime) \vert = \vert h(x) - h(x^\prime) \vert  < \epsilon. 
\end{align*}
for every $\epsilon > 0$.  As long as both $f$ and $g$ are continuous everywhere, then the composite function will be continuous everywhere.  If there are any points where $g$ is not continuous, then the composite function will not be continuous at those points\footnote{There is a huge caveat here: sometimes functions that are not specifically defined at a point can still look continuous at that point.  These types of function discontinuities are called \textit{removable discontinuities} or \textit{holes}.}.  The corresponding limit equation for this composition property can be written as
\begin{align}
    \lim_{x\rightarrow x^\prime} \left[ g(f(x)) \right] = g\left( \lim_{x\rightarrow x^\prime} f(x) \right), \label{eq: limit of composition of functions}
\end{align}
which can be summarized as: \textbf{the limit of the composition is the composition of the limits}.

Again, these continuity rules (Eqs. \ref{eq: limit of sum}, \ref{eq: limit of a product}, and \ref{eq: limit of composition of functions}) are incredibly useful tools to determine if any function is continuous, or at least continuous at a specified point.  

\vspace{0.15in}
\begin{problem}[Applications of Continuity Rules]{prob: application cont rules}

This problem is designed to show you how to use the continuity rules defined by Eqs. \ref{eq: limit of sum}, \ref{eq: limit of a product}, and \ref{eq: limit of composition of functions}, given that any constant functions or linear functions are continuous from \exref{ex: continuity of simple functions}.

As a quick example of how to use these continuity rules, by \ref{eq: limit of sum}, it follows that the line $f(x) = bx + A$, where $b$ and $A$ are constants, is continuous for every single real number input, $x$.  This is because \textit{both} $bx$ and $A$ are continuous everywhere from \exref{ex: continuity of simple functions}, therefore their sum is also continuous everywhere. 

\begin{enumerate}[(a)]
    \item For the first part of the problem, use \equaref{eq: limit of a product} to argue that $f(x) = bx^2$ is continuous everywhere.
    \item Then, using \equaref{eq: limit of sum}, argue that \textit{any} parabola $f(x) = ax^2 + bx + c$ is continuous everywhere.
    \item Now argue that \textit{any} cubic polynomial $P_3(x) = a_3 x^3 + a_2 x^2 + a_1 x + a_0$ is continuous everywhere, where every element of the set of coefficients $\{a_n\} = \{a_0,a_1,a_2,a_3\}$ is constant.
    \item Finally, use the same argument to show that \textit{any} $n^\mathrm{th}$ order polynomial, denoted by $P_n$, and written as
        $$ P_n(x) = a_n x^n + a_{n-1}x^{n-1} + \dots + a_2 x^2 + a_1x + a_0 $$
    and accompanied by the set of constant coefficients $\{a_n\} = \{a_0,a_1,a_2\dots, a_{n-1},a_{n}\}$ is continuous everywhere.
\end{enumerate}
\end{problem}
\vspace{0.15in}

Although the ability to potentially take an infinite amount of measurements is great, it is honestly more helpful to know, for example, how an object is moving because we can then predict where it will be in the next instant in time. In physics, equations like these are called \textit{equations of motion}, and the pursuit of these equations is pretty much all that physicists want to find for they all us to predict the future. Anyway, in order to be able to actually obtain the position of an object at every position, we need to know the velocity \textit{exactly} at every time.  So that means our $\Delta t$ increment would have to go to zero in \equaref{eq: definition of average velocity}! Furthermore, if we  were to predict the object's position, then the number of terms we would have to add in \equaref{eq: position sum of velocity} would go to infinity! In summary, the goal of calculus is to obtain \textit{finite} numbers when either dividing by zero or adding an infinite number of terms together.  The amazing part is that this ability stems directly from the idea of continuity.

\section{Differentiability and Integrability}
Without further ado, I've been building up the idea that there are two major fields in calculus: one that deals with ratios (slopes) and one that deals adding terms together (areas). When we refer to the study of the slopes or areas of functions as \textbf{differentiation} and \textbf{integration}, respectively, and we give the exact slope the name of \textquotedblleft\textbf{derivative}\textquotedblright$\,$ while we call the exact area \textquotedblleft\textbf{integral}\textquotedblright. 

It is very common for math courses to present the derivative an integral as two separate objects and then combine them later on.  I will instead try to introduce them as simultaneously as possible so that they are inextricably paired in your mind just as addition and subtraction or multiplication and division are.  We will then jump back to higher dimensions because our job as scientists is to explain the world, and the world is not just one-dimensional\footnote{Some people think it is 11-dimensional.}, and so we will have to deal with the higher dimensions almost immediately to explain the phenomena we observe.  The nice thing is that the ideas of calculus generalize quite easily | some of the applications of calculus do not generalize nicely, but those harder applications are largely the topics of other courses and are outside the scope of this book. Again, I will not list a bunch of specific derivatives or integrals because they are all over the place on the internet or in other books.  Our goal is to get a better understanding of how they work and then if you need the specific formulas outside of this book, you can use them fluidly as they arise. 

\subsection{Definition of Derivative: Rise Over Run}
We start with derivatives.  What we do here follows from what we have built so far when we combine the idea of rates and continuity.  We seek an object's velocity at every single instant in time, and the only way to do this experimentally is to take infinitely many time measurements so our time increment ($\Delta t = (t_f-t_0)/N$) is small enough for us to connect any instant in time with another instant.  Here, when we say \textquotedblleft instant\textquotedblright, we mean a literal point along the temporal axis with absolutely no uncertainty.  So experimental measurements are out of the question because it is impossible for anyone to measure something an infinite amount of times, but luckily, the continuity framework we developed allows us to look at the overall behavior of the slopes of functions as our time increment gets infinitely small, and if the slopes seem to converge to some finite value, then we extract this value as the \textbf{instantaneous slope}, or the derivative of our function.

We start by defining the derivative of a function $f:\mathds{R}\rightarrow\mathds{R}$ at a specified point $x^\prime$ as the following
\begin{align}
    \left. \tderiv[x]{f} \right\vert_{x = x^\prime} = \lim_{x\rightarrow x^\prime} \frac{f(x) - f(x^\prime)}{x - x^\prime}. \label{eq: definition of derivative}
\end{align}
The notation here of the big vertical line with the subscript of $x = x^\prime$ translates to \textquotedblleft evaluated at the value of $x = x^\prime$.\textquotedblright$\,$ Notice then, that the function $f$ \textbf{must be continuous at} $x^\prime$ for the limit definition above to actually be defined.  We say then that if this limit can be computed, that is to say if the \textbf{difference quotient},
\begin{align}
    \textrm{Difference Quotient: } \frac{f(x) - f(x^\prime)}{x - x^\prime}, \label{eq: difference quotient}
\end{align}
is continuous at $x^\prime$, then the function $f$ is \textit{differentiable} at $x^\prime$. Here, if we define the change in the input variable $\Delta x$ as $\Delta x = x - x^\prime$, and we define the associated change in the dependent variable as $\Delta f = f(x)-f(x^\prime)$, then the equation above becomes
\begin{align}
    \tderiv[x]{f} = \lim_{\Delta x^\rightarrow 0} \frac{\Delta f}{\Delta x}, \label{eq: derivative as slope}
\end{align}
and so it is a little clearer that the derivative truly is an \textit{instantaneous} slope of $f$, as the fraction inside the limit is just the \textit{rise-over-run} formula that you learned in algebra as the slope of a line. In this formula, I have omitted the vertical line telling us where to evaluate the derivative just to make it a little aesthete. Make sure you remember that the derivative is evaluated at a point though, and if a formula in general does not include the vertical line, it is implying that the equation for the derivative holds FOR ALL inputs.

A lot of non-pure-math people take \equaref{eq: derivative as slope} as justification of saying something along the lines of \textquotedblleft the derivative is when we take the limit as $\Delta \rightarrow \mathrm{d}$.\textquotedblright $\,$ This is only true in the most literal interpretation possible, because, as you may have noticed, the $\mathrm{d}$s in the left-hand side of \equaref{eq: definition of derivative} and \equaref{eq: derivative as slope} just appeared as soon as I defined the derivative; thus their appearance is purely a \textbf{notation}, and not any physically or mathematically meaningful.  With that said, the reason why this $\mathrm{d}$-notation is helpful is because it does remind us that a derivative is a slope, but its calculation comes from a \textit{limiting process}; that is, we must study the continuity of the function $f$ at $x^\prime$ to evaluate its derivative at that point.

For the sake of completeness, before I move on to showing you how to calculate a couple of simple derivatives, I need to let you know that, unfortunately, there are about as many notations for derivatives as there are baryonic particles in the universe. Okay, that's not true. But there are a bunch of different notations that people use, and the context kind of drives which one is used where.  My personal favorite is the fraction of the $\mathrm{d}$s, known as \textit{Leibniz's notation}, because of its aforementioned relation to slopes. But some other ones that you might see are boxed below.
\begin{align*}
    \boxed{ \tderiv[x]{f} = f^\prime (x) = \mathrm{D}_x f(x) = \textrm{d}_x f(x) = \dot{f}(x) = \dots }
\end{align*}
And there are a few more nightmares that come about for slopes of slopes (think acceleration) that you can find here \cite{deriv_notations}.  I will almost exclusively use the Leibniz notation throughout this book, but for your own future reference, it is always a very good idea to make sure you understand the context; for example, I and many other physics books use $f^\prime$ to denote a different value than $f$, but math courses almost always use the $f^\prime$ to denote a derivative.

\vspace{0.15in}
\begin{example}[Simple Derivatives]{ex: simple derivatives}
    In this example, we will find the derivatives for some more simple functions.  We start with a constant function $f(x) = C$ for all $x$, where $C\in\mathds{R}$.  To do so, we pick a special point $x^\prime$ and measure deviations from it in accordance with \equaref{eq: definition of derivative}.
    \begin{align}
        \left. \tderiv[x]{f} \right\vert_{x = x^\prime} &= \lim_{x \rightarrow x^\prime} \frac{f(x) - f(x^\prime)}{x - x^\prime}, \textrm{ by \equaref{eq: definition of derivative}} \nonumber
        \\
        &= \lim_{x \rightarrow x^\prime} \frac{C - C}{x - x^\prime},\nonumber
        \\
        &= \lim_{x\rightarrow x^\prime} \frac{0}{x - x^\prime},\nonumber
        \\
        &= 0. \label{eq: derivative of constant at point}
    \end{align}
    Since $x^\prime$ is arbitrary, this derivative must hold for every value of $x^\prime \in \mathds{R}$ we choose. Therefore, we would write
    \begin{align}
        \frac{\mathrm{d}}{\mathrm{d}x}\left[ C \right] = 0. \label{eq: derivative of C}
    \end{align}
    as the derivative of $C$ at any input $x$. This result is satisfying because we know that constants never change, and therefore their derivatives | their instantaneous rates of changes | must never change.
    
    We now consider the relationship $f(x) = Ax$ for all $x$, where $A$ is some constant coefficient.  We proceed in the same way as before, where we choose a special point $x^\prime$ and consider deviations in $f$ away from $f(x^\prime)$, yet again.
    \begin{align}
        \left. \tderiv[x]{f} \right\vert_{x = x^\prime} &= \lim_{x \rightarrow x^\prime} \frac{f(x) - f(x^\prime)}{x - x^\prime}, \textrm{ by \equaref{eq: definition of derivative}} \nonumber
        \\
        &= \lim_{x \rightarrow x^\prime} \frac{ Ax - Ax^\prime }{ x - x^\prime}, \nonumber
        \\
        &= \lim_{x \rightarrow x^\prime} \frac{A(x-x^\prime)}{x - x^\prime}, \nonumber \textrm{ divide through by }x-x^\prime
        \\
        &= \lim_{x \rightarrow x^\prime} A, \nonumber
        \\
        &= A. \label{eq: derivative of line at point}
    \end{align}
    Again, since this result holds for any arbitrary $x^\prime$ we choose, it must hold for all $x^\prime\in \mathds{R}$. We would therefore write
    \begin{align}
        \frac{\mathrm{d}}{\mathrm{d}x}\left[ Ax \right] = A. \label{eq: derivative of Ax}
    \end{align}
    when talking about the derivative of $Ax$ at any input $x$. Thus, we have found that the instantaneous slope of the line $f(x) = Ax$ is $A$ for every single input $x$.
    
    Finally, we consider a parabola given by $f(x) = Bx^2$ for all $x\in\mathds{R}$, where $B$ is just some constant coefficient. We want to measure the instantaneous rate of change of $f(x)$, just as we did with the other functions, and so 
    \begin{align}
        \left. \tderiv[x]{f} \right\vert_{x = x^\prime} &= \lim_{x \rightarrow x^\prime} \frac{f(x) - f(x^\prime)}{x - x^\prime}, \textrm{ by \equaref{eq: definition of derivative}} \nonumber
        \\
        &= \lim_{x\rightarrow x^\prime} \frac{Bx^2 - B{x^\prime}^2}{x - x^\prime}, \textrm{ factor by difference to two squares} \nonumber
        \\
        &= \lim_{x \rightarrow x^\prime} \frac{B(x - x^\prime)(x+x^\prime)}{x - x^\prime}, \textrm{ divide through by }x - x^\prime \nonumber
        \\
        &= \lim_{x \rightarrow x^\prime} B(x + x^\prime), \nonumber
        \\
        &= 2Bx^\prime. \label{eq: derivative of parabola at point}
    \end{align}
    This result holds for any $x^\prime$, and so we would write
    \begin{align}
        \frac{\mathrm{d}}{\mathrm{d}x}\left[ Bx^2 \right] = 2Bx, \label{eq: derivative of Bx2}
    \end{align}
    for arbitrary and arbitrary $x$ input. Remarkably, it shows that the instantaneous rate of change of a parabola is also changing. In fact, by \equaref{eq: derivative of line at point}, the derivative of the derivative is $2B$. Since parabolas are the simplest non-straight curve, then we can associate the \textit{curviness} with a nonzero \textit{second derivative}, or a \textit{changing first derivative}.
    
    The final function we will show is a simple hyperbola, given by $f(x) = a/x$ for all $x\neq 0$, and where $a\in\mathds{R}$.  So we take any point $x^\prime \neq 0$, as $1/x$ is continuous for all points where $x \neq 0$, and we consider deviations in $a/x$ nearby $x^\prime$.
    \begin{align}
        \left. \tderiv[x]{f}\right\vert_{x = x^\prime \neq 0} &= \lim_{x \rightarrow x^\prime \neq 0} \frac{\frac{a}{x} - \frac{a}{x^\prime}}{x - x^\prime}, \nonumber
        \\
        &=  \lim_{x \rightarrow x^\prime \neq 0} \frac{\frac{ax^\prime - ax}{xx^\prime}}{x - x^\prime}, \nonumber
        \\
        &= \lim_{x \rightarrow x^\prime \neq 0} \frac{a}{xx^\prime}\,\frac{x^\prime - x}{x - x^\prime}, \nonumber \textrm{ divide by } x - x^\prime
        \\
        &= \lim_{x \rightarrow x^\prime \neq 0} \frac{-a}{xx^\prime}, \nonumber
        \\
        &= -\frac{a}{{x^\prime}^2}, \label{eq: derivative of hyperbola at point}
    \end{align}
    Since this too holds for any arbitrary $x^\prime \neq 0$, then we have
    \begin{align}
        \frac{\mathrm{d}}{\mathrm{d}x}\left[ \frac{a}{x} \right] = -\frac{a}{x^2}, \, x\neq 0. \label{eq: derivative of a/x}
    \end{align}
    This derivative then shows that the derivative is always negative when $a > 0$, thus the \textit{change in the function}, is always negative for $a > 0$, as we move from left to right along the real number line. If $a < 0$, then the word \textquotedblleft negative \textquotedblright$\,$ becomes \textquotedblleft positive.\textquotedblright$\,$
\end{example}
\vspace{0.15in}

Example \ref{ex: simple derivatives} shows how one would use the limit definition to find the derivative of a constant function, a simple linear function, a parabola, and a hyperbola.  What is conventional at this point in many calculus references is to just list out a bunch of other derivatives because there are a bunch more that can be computed. A simple Google search would return any such list, but I've included a couple here for your reference \cite{Schaums, pauls_derivative_table}. Then, in practice, one would just use the functional expression (like Eqs. \ref{eq: derivative of C}, \ref{eq: derivative of Ax}, \ref{eq: derivative of Bx2}, and \ref{eq: derivative of a/x}) whenever one is needed.  It is truly a rare thing in physics for one to use the limit definition to ever compute a derivative\footnote{However, the limit definition is used constantly in fields of computational physics to evaluate the numerical derivative | in essence, one chooses a point $x^\prime$ in the computer and then finds a number with the difference quotient (\equaref{eq: difference quotient}).}, although we do use the limit definition often to say there exists a derivative within our equations.  Nevertheless, there are three rules that we must establish with the limit definition before we can move on. These are the linearity of differentiation, the product rule, and the chain rule.

\subsubsection{Linearity of the Derivative}
An extremely useful and important property of derivatives is that they are \textit{linear operators}.  Recall from \equaref{eq: def linearity} that a linear operator $\mathcal{O}$ has the following property:
\begin{align}
    \mathcal{O}\left[ af(x) + bg(x) \right] = a\mathcal{O}\left[ f(x) \right] +  b\mathcal{O}\left[ g(x) \right], \textrm{ for all } a,b\in\mathds{R}, \label{eq: definition of linear operator on function}
\end{align}
where we have replaced the vectors in \equaref{eq: def linearity} with the functions $f$ and $g$.  Thus, our mission here is to prove that the following statement holds
\begin{align*}
    \frac{\mathrm{d}}{\mathrm{d}x}\left[ af(x) + bg(x) \right] = a\frac{\mathrm{d}}{\mathrm{d}x}\left[ f(x) \right] +  b\frac{\mathrm{d}}{\mathrm{d}x}\left[ g(x) \right] = a\tderiv[x]{f} + b\tderiv[x]{g}, \textrm{ for all } a,b\in\mathds{R}
\end{align*}
In other words, we seek to prove \textit{two} things: first, \textbf{derivatives ignore constant coefficients}, and two, \textbf{the derivative of the sum is the sum of the derivatives}.

Okay so now we start with the following function: $f(x) = cg(x)$, where $c\in\mathds{R}$.  Let's assume that $g$ is \textit{differentiable} at the point $x^\prime$. Then we have
\begin{align}
    \left. \tderiv[x]{f}\right\vert_{x = x^\prime} &= \lim_{x\rightarrow x^\prime} \frac{cg(x) - cg(x^\prime)}{x - x^\prime}, \nonumber
    \\
    &= \lim_{x \rightarrow x^\prime} c \left[ \frac{g(x) - g(x^\prime)}{x - x^\prime} \right], \nonumber
\end{align}
But $c$ is a constant function which is continuous everywhere and converges to $c$, and by assumption the difference quotient for $g$ is continuous at $x^\prime$ and converges to the derivative of $g$, therefore we can use the limit of the product is the limit of the product rule in \equaref{eq: limit of a product}.  Thus,
\begin{align}
    \left. \tderiv[x]{f}\right\vert_{x = x^\prime} &= \lim_{x \rightarrow x^\prime} c \left[ \frac{g(x) - g(x^\prime)}{x - x^\prime} \right], \nonumber
    \\
    &= \left( \lim_{x \rightarrow x^\prime} c \right) \left[ \lim_{x \rightarrow x^\prime} \frac{g(x) - g(x^\prime)}{x - x^\prime} \right], \nonumber
    \\
    &= c \left. \tderiv[x]{g}\right\vert_{x = x^\prime}. 
\end{align}
And so we find that, as long as $g$ is differentiable at any arbitrary input $x$, then
\begin{align}
    \frac{\mathrm{d}}{\mathrm{d}x}\left[ cg(x) \right] = c\frac{\mathrm{d}}{\mathrm{d}x}\left[ g(x) \right] = c\tderiv[x]{g}, \label{eq: derivative ignores constant coefficients}
\end{align}
that is, the derivative operator indeed totally ignores any constant coefficients.

Now we must talk about sums of functions.  So suppose we have a function $h(x) = f(x) + g(x)$, where both $f$ and $g$ are differentiable at $x^\prime$.  Then, we simply have
\begin{align}
    \left. \tderiv[x]{h}\right\vert_{x = x^\prime} &= \lim_{x \rightarrow x^\prime} \frac{h(x) - h(x^\prime)}{x - x^\prime}, \nonumber
    \\
    &= \lim_{x \rightarrow x^\prime} \frac{f(x) + g(x) - f(x^\prime) - g(x^\prime)}{x - x^\prime}, \nonumber
    \\
    &= \lim_{x \rightarrow x^\prime} \frac{f(x) -f(x^\prime) + g(x) - g(x^\prime)}{x - x^\prime}, \nonumber
    \\
    &= \lim_{x\rightarrow x^\prime} \left[ \frac{f(x) - f(x^\prime)}{x - x^\prime} + \frac{g(x) - g(x^\prime)}{x - x^\prime} \right], \nonumber
\end{align}
By assumption, both difference quotients are continuous at $x^\prime$ and converge to the derivatives of $f$ and $g$ at $x^\prime$, therefore we can use the limit of the sum rule in \equaref{eq: limit of sum}. Thus,
\begin{align}
    \left. \tderiv[x]{h}\right\vert_{x = x^\prime} &= \lim_{x\rightarrow x^\prime} \left[ \frac{f(x) - f(x^\prime)}{x - x^\prime} + \frac{g(x) - g(x^\prime)}{x - x^\prime} \right], \nonumber
    \\
    &= \lim_{x\rightarrow x^\prime} \left[ \frac{f(x) - f(x^\prime)}{x - x^\prime} \right] + \lim_{x\rightarrow x^\prime} \left[ \frac{g(x) - g(x^\prime)}{x - x^\prime} \right], \nonumber
    \\
    &= \left. \tderiv[x]{f}\right\vert_{x = x^\prime} + \left. \tderiv[x]{g}\right\vert_{x = x^\prime}.
\end{align}
As long as $f$ and $g$ are continuous at any arbitrary input $x$, then we can write
\begin{align}
    \frac{\mathrm{d}}{\mathrm{d}x}\left[ f(x) + g(x) \right] = \frac{\mathrm{d}}{\mathrm{d}x}\left[ f(x) \right] + \frac{\mathrm{d}}{\mathrm{d}x}\left[ g(x) \right] = \tderiv[x]{f} + \tderiv[x]{g}. \label{eq: derivative of sum}
\end{align}
Therefore, we do see that the derivative of the sum is the sum of the derivatives.

Finally, we combine these two proofs together into our statement of linearity.  Take the function $h(x) = af(x) + bg(x)$, where $f$ and $g$ are at least differentiable at $x^\prime$. Then,
\begin{align}
    \left. \tderiv[x]{h} \right\vert_{x = x^\prime} &= \left. \frac{\mathrm{d}}{\mathrm{d}x}\left[ af(x) + bg(x) \right]\right\vert_{x = x^\prime}, \textrm{ use \equaref{eq: derivative of sum}} \nonumber
    \\
    &= \left. \frac{\mathrm{d}}{\mathrm{d}x} \left[ af(x) \right]\right\vert_{x = x^\prime} + \left. \frac{\mathrm{d}}{\mathrm{d}x} \left[ bg(x) \right]\right\vert_{x = x^\prime}, \textrm{use \equaref{eq: derivative ignores constant coefficients}} \nonumber
    \\
    &= a\left. \frac{\mathrm{d}}{\mathrm{d}x} \left[ f(x) \right]\right\vert_{x = x^\prime} + b\left. \frac{\mathrm{d}}{\mathrm{d}x} \left[ g(x) \right]\right\vert_{x = x^\prime}. \nonumber
\end{align}
When we combine the first and the last equality, we indeed have our statement of the linearity of the derivative, given explicitly as 
\begin{align}
    \left. \frac{\mathrm{d}}{\mathrm{d}x}\left[ af(x) + bg(x) \right]\right\vert_{x = x^\prime} = a\left. \frac{\mathrm{d}}{\mathrm{d}x} \left[ f(x) \right]\right\vert_{x = x^\prime} + b\left. \frac{\mathrm{d}}{\mathrm{d}x} \left[ g(x) \right]\right\vert_{x = x^\prime}, \textrm{ for all } a,b\in\mathds{R}. \label{eq: derivative is linear at point}
\end{align}
and again, whenever both $f$ and $g$ are differentiable everywhere, so is their sum. In this case, the equation above would shed its vertical lines as
\begin{align}
    \frac{\mathrm{d}}{\mathrm{d}x}\left[ af(x) + bg(x) \right] = a\frac{\mathrm{d}}{\mathrm{d}x}\left[ f(x) \right] +  b\frac{\mathrm{d}}{\mathrm{d}x}\left[ g(x) \right] = a\tderiv[x]{f} + b\tderiv[x]{g}, \textrm{ for all } a,b\in\mathds{R}, \label{eq: derivative is linear everywhere}
\end{align}
as we had as our claim.

\subsubsection{The Product Rule}
Now that we have established the way in which derivatives behave with sums (\equaref{eq: derivative of sum}), how about the derivative of a product? This proof will take a little more work, but combining it with sums will allow us to calculate many individual derivatives whenever they arise. So without further ado, consider the function $h(x) = f(x)g(x)$ where $f(x)$ and $g(x)$ are both differentiable at $x^\prime$.  Now we want to determine what the derivative of $h$ is at that same point.
\begin{align}
    \left. \tderiv[x]{h} \right\vert_{x = x^\prime} &= \lim_{x \rightarrow x^\prime} \frac{h(x) - h(x^\prime)}{x - x^\prime}, \nonumber
    \\
    &= \lim_{x \rightarrow x^\prime} \frac{f(x)g(x) - f(x^\prime)g(x^\prime)}{x - x^\prime}, \textrm{ add and subtract zero}\nonumber
    \\
    &= \lim_{x \rightarrow x^\prime} \frac{f(x)g(x) - f(x^\prime)g(x) + f(x^\prime)g(x) - f(x^\prime)g(x^\prime)}{x - x^\prime}, \nonumber
    \\
    &= \lim_{x \rightarrow x^\prime} \frac{\left[ f(x) - f(x^\prime) \right]g(x) + f(x^\prime)\left[ g(x) - g(x^\prime) \right]}{x - x^\prime}, \nonumber
    \\
    &= \lim_{x \rightarrow x^\prime} \left\lbrace \left[ \frac{f(x) - f(x^\prime)}{x - x^\prime}  \right] g(x) + f(x^\prime) \left[ \frac{g(x) - g(x^\prime)}{x - x^\prime}  \right] \right\rbrace. \nonumber
\end{align}
Since both $f$ and $g$ are differentiable at $x^\prime$, they are both defined and continuous at $x^\prime$.  Additionally, their difference quotients are also continuous at $x^\prime$.  Therefore each term in the limit is continuous at $x^\prime$ and so we can use \equaref{eq: limit of sum} to split up the sum, and then we can use \equaref{eq: limit of a product} to split up each of the factors in each term.  Thus, we have
\begin{align}
    \left. \tderiv[x]{h} \right\vert_{x = x^\prime} &= \lim_{x \rightarrow x^\prime} \left\lbrace \left[ \frac{f(x) - f(x^\prime)}{x - x^\prime}  \right] g(x) + f(x^\prime) \left[ \frac{g(x) - g(x^\prime)}{x - x^\prime}  \right] \right\rbrace, \nonumber
    \\
    &= \left[ \lim_{x \rightarrow x^\prime} \frac{f(x) - f(x^\prime)}{x - x^\prime} \right]\, \left[ \lim_{x \rightarrow x^\prime} g(x) \right] + \left[ \lim_{x \rightarrow x^\prime} f(x^\prime) \right]\,\left[ \lim_{x \rightarrow x^\prime} \frac{g(x) - g(x^\prime)}{x - x^\prime} \right], \nonumber
    \\
    &= \left[ \left. \tderiv[x]{f}\right\vert_{x = x^\prime} \right]\,g(x^\prime) + f(x^\prime)\, \left[ \left. \tderiv[x]{g}\right\vert_{x = x^\prime} \right]. \nonumber
\end{align}
When we are free to omit the vertical lines, then the equation above has the more common form of
\begin{align}
    \frac{\mathrm{d}}{\mathrm{d}x} \left[ f(x)g(x) \right] = \left(\tderiv[x]{f} \right)\,g(x) + f(x)\left( \tderiv[x]{g} \right), \label{eq: product rule}
\end{align}
which is known as the \textit{product rule for differentiation}.  It essentially says that the derivative of the product is more complicated than simply the product of the derivatives. However, nice little mnemonic phrases exist for the product rule, such as \textquotedblleft the derivative of the product is the derivative of the first times the second plus the first times the derivative of the second.\textquotedblright$\,$ Okay, so maybe the mnemonic isn't that great. But the formula itself is not that difficult to remember, so that's good, right?

\begin{problem}[Derivation of the Power Rule for Differentiation]{prob: derivation of power rule for derivs}
Now that we have established how to take the derivative of a product, let's use it to calculate the derivative of \textit{any} integer power function.  

To start, let's begin with the function $x^2$.  From \equaref{eq: derivative of parabola at point}, we would set $B = 1$ and find the derivative is $2x$ everywhere.  We could also use the product rule instead when we recognize that $x^2 = x\cdot x$.  By \equaref{eq: derivative of Ax}, we set $A = 1$ and we have that the derivative of $x$ is 1.  Then, by product rule,
\begin{align*}
    \frac{\mathrm{d}}{\mathrm{d}x} \left[ x^2 \right] = \frac{\mathrm{d}}{\mathrm{d}x} \left[ x\cdot x \right] =  \left(\tderiv[x]{x} \right)\,x + x\left( \tderiv[x]{x} \right) = 1\cdot x + x\cdot 1 = 2x.
\end{align*}
Using this same logic, show that the derivative of $x^3 = x\cdot x^2$ is $3x^2$. Then, show that the derivative of $x^4 = x\cdot x^3$ is $4x^3$.  Based on this line of reasoning, show that
\begin{align}
    \frac{\mathrm{d}}{\mathrm{d}x} \left[ x^m \right] = \frac{\mathrm{d}}{\mathrm{d}x} \left[ x\cdot x^{m-1} \right] = mx^{m-1}, \nonumber
\end{align}
where $m$ is a positive integer.

Now we can equivalently calculate the negative integers using the same reasoning.  I'll start with the derivative of $1/x^2 = (1/x)\cdot (1/x)$ using \equaref{eq: derivative of a/x} with $a = 1$.
\begin{align*}
    \frac{\mathrm{d}}{\mathrm{d}x} \left[ \frac{1}{x^2} \right] = \frac{\mathrm{d}}{\mathrm{d}x} \left[ \frac{1}{x}\,\cdot\, \frac{1}{x} \right] = \left[  \frac{\mathrm{d}}{\mathrm{d}x} \left( \frac{1}{x} \right) \right]\, \frac{1}{x} + \frac{1}{x}\, \left[  \frac{\mathrm{d}}{\mathrm{d}x} \left( \frac{1}{x} \right) \right] = \left(-\frac{1}{x^2}\right)\,\frac{1}{x} + \frac{1}{x}\,\left(-\frac{1}{x^2} \right)
\end{align*}
so we have 
\begin{align*}
    \frac{\mathrm{d}}{\mathrm{d}x} \left[ \frac{1}{x^2} \right] = -\frac{2}{x^3}.
\end{align*}
Again, using the same logic show that the derivative of $1/x^3 = (1/x)\cdot (1/x^2)$ and $1/x^4 = (1/x)\cdot (1/x^3)$ is $-3/x^4$ and $-4/x^5$, respectively.  Then, use the same idea to show that
\begin{align}
    \frac{\mathrm{d}}{\mathrm{d}x} \left[ \frac{1}{x^m} \right] = \frac{\mathrm{d}}{\mathrm{d}x} \left[ \frac{1}{x}\cdot \frac{1}{x^{m-1}} \right] = -\frac{m}{x^{m+1}}, \nonumber
\end{align}
where $m$ is again a positive integer (hint: $1/x^m = x^{-m}$).

Finally, argue that when we let $n$ be any integer (either positive or negative), we have
\begin{align}
     \frac{\mathrm{d}}{\mathrm{d}x} \left[ x^n \right] = \frac{\mathrm{d}}{\mathrm{d}x} \left[ x\cdot x^{n-1} \right] = nx^{n-1}. \label{eq: derivative power rule} 
\end{align}
To be clear, by argue, I mean connect the negative integer case with the positive integer case in the tidy little formula above that we will call the \textbf{Power Rule of Differentiation}.
\end{problem}
\vspace{0.15in}

\subsubsection{The Chain Rule}
The final rule that needs to be covered is what happens when we have a composition of functions, as was our last case for continuity. Again, compositions of functions arise all the time in physics and math, so we better know how to handle them when they do.  So suppose we have a function $h(x) = g(f(x))$, where $f$ is differentiable at $x^\prime$ and $g$ is differentiable at $f(x^\prime)$.  Now we employ the limit definition of differentiation to study the tiny variations in $h$ associated with the variations in $x$ nearby $x^\prime$.
\begin{align}
    \left. \tderiv[x]{h} \right\vert_{x = x^\prime} &= \lim_{x \rightarrow x^\prime} \frac{g(f(x)) - g(f(x^\prime))}{x - x^\prime}, \textrm{ multiply by one} \nonumber
    \\
    &= \lim_{x \rightarrow x^\prime} \left[\frac{f(x) - f(x^\prime)}{f(x) - f(x^\prime)}\frac{g(f(x)) - g(f(x^\prime))}{x - x^\prime} \right], \nonumber \textrm{ interchange the denominators}
    \\
    &= \lim_{x \rightarrow x^\prime} \left[\frac{f(x) - f(x^\prime)}{x - x^\prime}\,\frac{g(f(x)) - g(f(x^\prime))}{f(x) - f(x^\prime)} \right]. \nonumber
\end{align}
The above equation might look a little gross because of all the parentheticals.  Therefore, let's define $u = f(x)$ and $u^\prime = f(x^\prime)$, where $u$ is continuous at $x^\prime$, as we assumed $f$ was.  Then we have a limit of continuous difference quotients, and so we can split up the product according to \equaref{eq: limit of a product}.
\begin{align}
    \left. \tderiv[x]{h} \right\vert_{x = x^\prime} &= \lim_{x \rightarrow x^\prime} \left[\frac{f(x) - f(x^\prime)}{x - x^\prime}\,\frac{g(f(x)) - g(f(x^\prime))}{f(x) - f(x^\prime)} \right]. \nonumber
    \\
    &= \left[ \lim_{x \rightarrow x^\prime} \frac{f(x) - f(x^\prime)}{x - x^\prime} \right]\,\left[ \lim_{u \rightarrow u^\prime} \frac{g(u) - g(u^\prime)}{u - u^\prime} \right], \nonumber
    \\
    &= \left[ \left. \tderiv[x]{f}\right\vert_{x = x^\prime} \right]\, \left[ \left. \tderiv[u]{g}\right\vert_{u = u^\prime} \right], \nonumber \textrm{ re-substitute in } u = f(x)
    \\
    &= \left[ \left. \tderiv[x]{f}\right\vert_{x = x^\prime} \right]\, \left[ \left. \tderiv[f]{g}\right\vert_{f(x) = f(x^\prime)} \right]. \nonumber
\end{align}
When $f$ and $g$ are differentiable everywhere, then the formula above becomes what is more widely recognized as the \textbf{Chain Rule for Differentiation}, given by
\begin{align}
    \frac{\mathrm{d}}{\mathrm{d}x}\left[ g(f(x)) \right] = \tderiv[x]{f}\,\tderiv[f]{g}, \label{eq: Chain Rule differentiation}
\end{align}
The formula above is pretty easy to remember because it essentially shows that we can change the independent variable in $g$ from $x$ to $f$ by multiplying and dividing by the differential quantity $\mathrm{d}f$ so that they \textquotedblleft cancel\textquotedblright$\,$ to one. Now, as you saw above, in order to multiply by 1, we had to do it in a continuous limiting fashion, so it was a little more complicated than multiplying by, for example, $4/4$. But the spirit is the same, mathematically. In physics though, most of the time we treat differential quantities, a. k. a. differentials, as single numerical values like $4$, with the somewhat obvious implication that they are only defined in the sense of the limits shown above. 

\vspace{0.15in}
\begin{problem}[Examples of the Handiness of the Chain Rule]{prob: handiness of the chain rule}
This problem is more brute force rather than something more elegant, but it is ultimately to help convince you of how useful the Chain Rule can be. Consider the function $f(x)$ defined for all $x$ as 
\begin{align*}
    f(x) = (ax+b)^2 = a^2x^2 + 2abx + b^2
\end{align*}
where $a,b\in\mathds{R}$. Using the Chain Rule calculate the derivative of $(ax+b)^2$ where the inner function is $ax+b$. Then use the linearity of the derivative and the Power Rule to find the same derivative.

Next, use the Chain Rule to show that the derivative of $(ax^2 +bx^3)^n$ is $n(2ax + 3bx^2)(ax^2 + bx^3)^{n-1}$, where $n$ is a number.
\end{problem}
\vspace{0.15in}

\begin{example}[Implicit Differentiation]{ex: implicit differentiation}
    In this example, we will show how to compute the a quantity called the \textit{implicit derivative} of a function using the Chain Rule. This quantity can be used to calculate instantaneous slopes but cast them in terms of other variables. For example, in classical mechanics, the implicit derivative allows us to rewrite Newton's Second Law | something that normally describes how velocities change with time given a \textit{time-dependent} net force $F(t)$ | as an equation describing how velocities change with position, given a \textit{position-dependent} force field $F(x)$. 
    
    To show some of its more mathematical utility,  how it can be used to generalize Power Rule to rational powers.  To start, let's assume $g(u) = g(u(x))$, and let's say we want to find the derivative of $u$ with respect to $x$.  Then, when we take a derivative of $g$ and we have
    \begin{align*}
        \tderiv[x]{g} = \underbrace{\tderiv[x]{u}}_{\textrm{Implict Derivative}}\,\tderiv[u]{g}.
    \end{align*}
    In the case where $u(x) = x^{1/n}$, where $n$ is an integer, we can raise both sides to the $n$-power to get $u^n(x) = x$.  Now we set $g(x) = x$ and $g(u) = u^n$. Then the derivative of $g$ is
    \begin{align*}
        \tderiv[x]{g} = \frac{\mathrm{d}}{\mathrm{d}x}[x] = 1.
    \end{align*}
    Meanwhile, by Chain Rule
    \begin{align}
        \tderiv[x]{u} &= \left(\tderiv[u]{g} \right)^{-1}\, \tderiv[x]{g}, \nonumber
        \\
        &= \left[ \frac{\mathrm{d}}{\mathrm{d}u} \left( u^n \right) \right]^{-1}\cdot 1, \nonumber
        \\
        &= \left( nu^{n-1} \right)^{-1}, \nonumber
        \\
        &= \frac{1}{nu^{n-1}}, \nonumber
        \\
        &= \frac{1}{n (x^{1/n})^{n-1}}, \nonumber
        \\
        &= \frac{1}{n x^{(n-1)/n} }, \nonumber
        \\
        &= \frac{1}{n x^{1-1/n}}, \nonumber
        \\
        &= \frac{1}{n}x^{1/n - 1}. \nonumber
    \end{align}
    But this is the exact form of Power Rule with integer powers, except now we can include any fractional powers, where fraction is $1/n$.  By Chain Rule, we could generalize this expression further to any rational power $p/q$, where $p$ and $q$ are integers:
    \begin{align}
        u(x) = x^{p/q} = (x^p)^{1/q}. \nonumber
    \end{align}
    Then the derivative of $u$ is
    \begin{align}
        \tderiv[x]{u} &= \left( \frac{\mathrm{d}}{\mathrm{d}x}[x^p] \right)\, \left. \frac{\mathrm{d}}{\mathrm{d}v}[v^{1/q}]\right\vert_{v = x^p}, \nonumber
        \\
        &= \left( px^{p-1} \right)\, \left[ \frac{1}{q}(x^p)^{1/q-1} \right], \nonumber
        \\
        &= \frac{p}{q}\,x^{p-1}\,x^{p/q - p}, \nonumber
        \\
        &= \frac{p}{q} x^{p/q - 1}. \label{eq: generalized power rule}
    \end{align}
    Hence, for \textit{any} number $p/q$, we have a way to calculate the derivative of $x^{p/q}$; all thanks to the implicit derivative. Thus whenever we have a function that is $x^\alpha$, where $\alpha$ is representable by any rational number, then we can compute its derivative using the Power Rule. Since any real number can be approximated iteratively by rational numbers, then this must hold for \textit{all} real numbers.
\end{example}
\vspace{0.15in}

\begin{example}[Slopes of Inverse Functions]{ex: inverse functions}
    Suppose that we have a pair of inverse functions $f$ and $f^{-1}$ such that $f^{-1}(f(x)) = x$ and $f(f^{-1}(x)) = x$.  An example of such a pair is the exponential function and the natural logarithm, $\txte^{x}$ and $\ln x$, such that $\ln(\txte^x) = x$ and $\txte^{\ln x} = x$.  We can use the Chain Rule to find the slope of an inverse function given slope of any function.
    
    To start, define $g(x) = f(f^{-1}(x)) = x$, and assume that $f$ is differentiable and nonzero at some point $f^{-1}(x^\prime)$.  Then,
    \begin{align}
        \left. \tderiv[x]{g}\right\vert_{x = x^\prime} = \frac{\mathrm{d}}{\mathrm{d}x}\left[ f(f^{-1}(x)) \right] = \left. \tderiv[x]{f^{-1}}\right\vert_{x = x^\prime}\,\left. \tderiv[f^{-1}]{f} \right\vert_{f^{-1} = f^{-1}(x^\prime)}. \nonumber
    \end{align}
    But the derivative of $g$ with respect to $x$ is 1. Therefore, we can solve for the derivative of the inverse function,
    \begin{align}
        \left. \tderiv[x]{f^{-1}}\right\vert_{x = x^\prime} &= \left. \tderiv[x]{g}\right\vert_{x = x^\prime}\,\left[ \left. \tderiv[f^{-1}]{f} \right\vert_{f^{-1} = f^{-1}(x^\prime)} \right]^{-1}, \nonumber
        \\
        &= 1\cdot \left[ \left. \tderiv[f^{-1}]{f} \right\vert_{f^{-1} = f^{-1}(x^\prime)} \right]^{-1}, \nonumber
        \\
        &= \left[ \left. \tderiv[f^{-1}]{f} \right\vert_{f^{-1} = f^{-1}(x^\prime)} \right]^{-1}. \label{eq: derivative of inverse}
    \end{align}
    Thus the slope of the inverse function at a point $x^\prime$ is the reciprocal of the slope of the function at the point $f^{-1}(x^\prime)$.
    
    For example, consider the function $f(x) = x^2$. Its inverse function is $f^{-1}(x) = x^{1/2}$. Using \equaref{eq: derivative of inverse}, we have
    \begin{align*}
        \left. \frac{\mathrm{d}}{\mathrm{d}x}[x^{1/2}]\right\vert_{x = x} = \left[ \left. \frac{\mathrm{d}}{\mathrm{d}x}[x^{2}]\right\vert_{x = x^{1/2}} \right]^{-1} = \left. \frac{1}{2x}\right\vert_{x = x^{1/2}} = \frac{1}{2}\,x^{-1/2} = \frac{1}{2}\,x^{1/2-1}
    \end{align*}
    which is the exact derivative we would have expected from the Power Rule, \equaref{eq: generalized power rule}.
\end{example}

\subsubsection{Removable Discontinuities in the Difference Quotient}
\textbf{Disclaimer:} this subsection is really only to appease the mathematician in me. Not-as-mathy people need not read this part too seriously because it won't influence the veracity of the derivative in almost every setting imaginable.  

For almost all of the derivatives section, I've been dividing things by zero. At any time when we have been taking the limit of something like $1/(x-x^\prime)$ as $x\rightarrow x^\prime$, we have been violating the main premise of continuity in that the function \textit{must be defined at the point-of-interest}. So what gives? Why have I been seemingly contradicting myself this whole time? It deals with the whole difference quotient business, actually. So let me give it a name, $\mathcal{Q}(f,x,x^\prime)$, and define it formally below
\begin{align}
    \mathcal{Q}(f,x,x^\prime) = \begin{cases}
        \dfrac{f(x) - f(x^\prime)}{x - x^\prime}, & x \neq x^\prime
        \\
        \left. \dfrac{\mathrm{d}f}{\mathrm{d}x}\right\vert_{x = x^\prime}, & x = x^\prime
    \end{cases}. \label{eq: formal difference quotient}
\end{align}
The three-variable inputs in $\mathcal{Q}$ simply mean that the difference quotient takes in a function $f$ and evaluates it at $x$ and $x^\prime$.  So the only problem with the difference quotient defined in \equaref{eq: difference quotient} is that, numerically speaking, it is not strictly defined at $x = x^\prime$ because that is where we run into the issue of dividing by zero.  However, sometimes we have functions where we can \textit{remove} this discontinuity, as is the case with the the derivatives of the parabola and hyperbola in \equaref{eq: derivative of parabola at point} and \equaref{eq: derivative of hyperbola at point}, respectively. Recall that in both of those cases, we were able to divide out the $x - x^\prime$ and we had some leftover bits that I called the derivative.  That's actually the equivalent definition to what's happening in \equaref{eq: formal difference quotient} | it evaluates the local slope of $f$ nearby $x^\prime$ where it can, and then when it can't because $x = x^\prime$, then we take the value of the derivative to be what the slope was otherwise approaching.  So in the case of the parabola, the slope was had the value of $x + x^\prime$ for all values of $x\neq x^\prime$. For all values of $x \neq x^\prime$, this slope seems to approach $2x^\prime$ and so we removed the discontinuity at $x = x^\prime$ by replacing the value of the slope at $x = x^\prime$ with $2x^\prime$. Therefore, we never actually deal with division by zero.  We just get \textit{really really} close to dividing by zero.  

Are we allowed to do this? Won't there be some kind of spike when we suddenly switch the value of $\mathcal{Q}$? The answer to this question is \textquotedblleft no\textquotedblright$\,$ as long as the slope is continuous! Thus the difference quotient $\mathcal{Q}$ must be continuous for the derivative itself to have instantaneous slope as its interpretation, and so whenever I would say that the difference quotient was continuous, this is what I really meant. Okay, I'm glad I got that off my chest. It was killing me.

\subsection{Fundamental Theorems of Calculus}
So we have discovered how we might talk about instantaneous slopes of functions.  To do so, we ultimately need to analyze the difference quotient of any function nearby some special point and study what happens to the difference quotient when we get closer and closer to that point.  Geometrically, this represents taking a secant line to a tangent line | thus, at the actual point-of-interest, the line approximating the slope of our function at that point intercepts the function at only that point.  This approximating line actually comes straight out of the limit definition of the derivative.  If we consider the derivative at some point $x^\prime$ of a differentiable function $f$, then it must be true that
\begin{align}
    \left. \tderiv[x]{f}\right\vert_{x = x^\prime} \approx \frac{f(x) - f(x^\prime)}{x - x^\prime}, \label{eq: approximate slope}
\end{align}
by continuity in the difference quotient. This function can be rewritten as
\begin{align}
    f(x) \approx f(x^\prime) + \left. \tderiv[x]{f}\right\vert_{x = x^\prime} \left(x - x^\prime\right). \label{eq: linear approximation}
\end{align}
Since we know our function and our point $x^\prime$, then $f(x^\prime)$ is just a number that we can compute. Likewise, the derivative itself is evaluated at the specific point $x^\prime$, so it also just takes on a numerical value.  Therefore, the only \textquotedblleft variables\textquotedblright$\,$ in the equation above are $x$ and $f(x)$ while the rest are constants.  This equation itself then has the form $y = y_0 + m(x-x_0)$, which is the equation of a line from algebra.  We therefore call \equaref{eq: linear approximation} the \textbf{linear approximation} of the function $f$.  

The reason why I bring up the linear approximation in this section rather than in one above is because it actually represents something much deeper in calculus.  Specifically, it shows that if we multiply a derivative by a tiny width $x-x^\prime$, then we recover can recover the difference in the function $f$.  Or, at least this holds within a tiny little vicinity around the point $(x^\prime, f(x^\prime))$. But what happens when the linear approximation doesn't hold that well? 

To consider this case, I will change up my notation a little bit.  Let me define the following quantity
\begin{align}
    \Delta x_N = \frac{x - x^\prime }{N},
\end{align}
where $N$ is a positive integer | note that this definition is that of a sequence whose limit is zero.  The linear approximation above holds for $\Delta x_1$.  Now let's suppose that the approximation is really only holds for the intervals $(x^\prime, x^\prime + \Delta x_2)$ and $(x^\prime + \Delta x_2, x)$. Then by \equaref{eq: approximate slope}, we have
\begin{align}
    \left. \tderiv[x]{f}\right\vert_{x = x^\prime} \approx \frac{f(x^\prime + \Delta x_2) - f(x^\prime)}{\Delta x_2}\; \textrm{ and }\; \left. \tderiv[x]{f}\right\vert_{x = x^\prime + \Delta x_2} \approx \frac{f(x) - f(x^\prime + \Delta x_2)}{\Delta x_2}
\end{align}
Remember, just as before, we wanted to find the difference in $f$ between two points using the product of the derivative and some width in $x$.  For fun, let's try adding these two derivatives together and try to simplify over the common denominator of $\Delta x_2$.
\begin{align}
    \left. \tderiv[x]{f}\right\vert_{x = x^\prime} + \left. \tderiv[x]{f}\right\vert_{x = x^\prime + \Delta x_2} &\approx \frac{f(x^\prime + \Delta x_2) - f(x^\prime)}{\Delta x_2} + \frac{f(x) - f(x^\prime + \Delta x_2)}{\Delta x_2}, \nonumber
    \\
    &= \frac{f(x^\prime + \Delta x_2) - f(x^\prime) + f(x) - f(x^\prime + \Delta x_2)}{\Delta x_2}, \nonumber
    \\
    &= \frac{f(x) + f(x^\prime + \Delta x_2) - f(x^\prime + \Delta x_2) - f(x^\prime)}{\Delta x_2}, \nonumber
    \\
    &= \frac{f(x) + 0 - f(x^\prime)}{\Delta x_2}, \nonumber
    \\
    &= \frac{f(x) - f(x^\prime)}{\Delta x_2}.
\end{align}
Therefore, we find that the difference in $f$ between $x$ and $x^\prime$ is
\begin{align}
    f(x) - f(x^\prime) \approx \Delta x_2 \left( \left. \tderiv[x]{f}\right\vert_{x = x^\prime} + \left. \tderiv[x]{f}\right\vert_{x = x^\prime + \Delta x_2} \right).
\end{align}
If we were to do this process again with $\Delta x_3$ instead of $\Delta x_2$, then we would need three derivatives to connect $x^\prime$ to $x$ and the expression would look like
\begin{align}
     f(x) - f(x^\prime) \approx \Delta x_3 \left( \left. \tderiv[x]{f}\right\vert_{x = x^\prime} + \left. \tderiv[x]{f}\right\vert_{x = x^\prime + \Delta x_3} + \left. \tderiv[x]{f}\right\vert_{x = x^\prime + 2\Delta x_3} \right),
\end{align}
and with $\Delta x_4$ and four connection points,
\begin{align}
     f(x) - f(x^\prime) \approx \Delta x_4 \left( \left. \tderiv[x]{f}\right\vert_{x = x^\prime} + \left. \tderiv[x]{f}\right\vert_{x = x^\prime + \Delta x_4} + \left. \tderiv[x]{f}\right\vert_{x = x^\prime + 2\Delta x_4} + \left. \tderiv[x]{f}\right\vert_{x = x^\prime + 3\Delta x_4} \right),
\end{align}
and then with a total of $N$ terms, we would have
\begin{align}
    f(x) - f(x^\prime) \approx \Delta x_N \sum_{j = 0}^{N-1} \left. \tderiv[x]{f}\right\vert_{x = x^\prime + j\Delta x_N}. \label{eq: approximate integral}
\end{align}
The final question we must ask is when does the approximation become an exact equality?  Well, this all started from \equaref{eq: approximate slope} where we assumed that the difference quotient was about equal to the instantaneous slope. However, the only time that it is truly equal for differentiable functions is at the point-of-interest.  Therefore, we must squeeze our $\Delta x_N$ down to zero to get the true value of the derivative.  It just so happens\footnote{It's almost as if I had a plan while writing this...} that we are totally allowed to take our $\Delta x_N$ to be as arbitrarily small as we like by taking $N$ to be as large as we need, as is done in \equaref{eq: Delta t N limit}.  

All that is required of $f$ is that it be \textit{continuous} along the interval $(x^\prime, x)$, so that we can guarantee there is some little $\delta$-interval in $x$ for our arbitrarily small $\Delta x_N = \epsilon$-interval such that $\vert f(x^\prime + j\Delta x_N) - f(x^\prime (j-1)\Delta x_N)\vert < \epsilon$.  All this part does is really allow us to continue to take smaller and smaller $\Delta x_N$ and still find locations where we can take the difference between adjacent points in $f$.  Of course, we need the derivative to be defined at all points in the interval $(x^\prime, x)$, or else the right-hand side of \equaref{eq: approximate integral} is undefined; however, we actually only ever needed the difference quotient to approximate the slope and then only added and subtracted values of $f$ along the interval. Hence, we really only ever need the difference quotient, but more on that in the next section.

Now for the first of two fundamental theorems of calculus.  As my high school physics teacher would say, when we take the limit where $\Delta x_N\rightarrow 0$, then we \textit{yank} on the top and bottom of the $\Sigma$ until it looks like an $S$ | or at least a very fancy looking $S$. Then we can set the approximation in \equaref{eq: approximate integral} to be an equality as
\begin{align}
    f(x) - f(x^\prime) = \lim_{N\rightarrow \infty} \sum_{j = 1}^{N-1} \Delta x_N \left. \tderiv[x]{f}\right\vert_{x = x^\prime  + j\Delta x} = \int_{x^\prime}^x \mathrm{d}s\,\tderiv[s]{f}.
\end{align}
The $\int$ symbol is the fancy $S$ my physics teacher was talking about (it's an old script $S$ first written by Leibniz \cite{bbc_1986}).  That quantity on the right is called a \textbf{definite integral} and it says to add up \textit{all} of the values of the derivative of $f$ between $x^\prime$ and $x$ and then multiply them each by the infinitesimal width $\mathrm{d}s$.  The lowercase $s$ here is called a \textit{dummy variable} meaning that it really is only a placeholder in our notation but bears no weight in the final expression.  When we write that equation a little clearer, we arrive at the first fundamental theorem of calculus:
\begin{align}
    f(x) - f(x^\prime) = \int_{x^\prime}^x \mathrm{d}s\,\tderiv[s]{f}, \label{eq: first fundamental theorem of calculus}
\end{align}
which says that \textbf{the integral is the inverse operator of the derivative}. To put it another way, you can undo the derivative by taking an integral. 

We can use \equaref{eq: first fundamental theorem of calculus} to establish the second fundamental theorem.  We do so by differentiating $f$ with respect to $x$ while $x^\prime$ is a fixed constant making $f(x^\prime)$ a fixed constant.
\begin{align}
    \tderiv[x]{f} = \frac{\mathrm{d}}{\mathrm{d}x}\left( f(x^\prime) + \int_{x^\prime}^x \mathrm{d}s \tderiv[s]{f} \right).
\end{align}
By the linearity of the derivative operator, we can take the derivative of both terms individually.
\begin{align}
    \tderiv[x]{f} = \frac{\mathrm{d}}{\mathrm{d}x}\left[ f(x^\prime) \right] + \frac{\mathrm{d}}{\mathrm{d}x}\left (\int_{x^\prime}^x \mathrm{d}s \tderiv[s]{f} \right).
\end{align}
But $f(x^\prime)$ is just a constant, and so its instantaneous slope is zero everywhere (\equaref{eq: derivative of C}).  Then we are left with
\begin{align}
    \tderiv[x]{f} = \frac{\mathrm{d}}{\mathrm{d}x}\int_{x^\prime}^x \mathrm{d}s \tderiv[s]{f}, \label{eq: second fundamental theorem of calculus}
\end{align}
as our second fundamental theorem of calculus. In English, the equation says that \textbf{the derivative is the inverse operator of the integral}, and so whenever we want to undo any integral we have performed, we simply take a derivative.

I know that \equaref{eq: first fundamental theorem of calculus} and \equaref{eq: second fundamental theorem of calculus} are probably not the exact replicas of the ones you may have learned in your calculus course. And that's okay.  The ones I learned back in the day were 
\begin{align*}
    &F(b) - F(a) = \int_{a}^b f(x)\,\mathrm{d}x, \textrm{ where } f(x) = \tderiv[x]{F}
    \\
    &\frac{\mathrm{d}}{\mathrm{d}t}\int_a^{x(t)} \mathrm{d}t \, f(t) = \tderiv[t]{x}\,f(x(t))
\end{align*}
as the first and second fundamental theorems, respectively. But the first one hides all of the inverse properties of the integral and derivative, while the second is honestly the same thing that we have but with some extra Chain Rule sprinkled in for some reason.  Either way, the math content is the same: these fundamental theorems show that derivatives and integrals act just like addition and subtraction or multiplication and division. One will undo the other.

\subsection{Definition of (Riemann) Integration: Adding Boxes}
We have talked so far about differentiation and the fundamental theorems of calculus in a single dimension.  There is still one major topic that we must cover before we can jump to applying calculus to solve some problems. We need to address a few important facets of integration.

\begin{figure}
    \centering
    \includegraphics[width = 5.25in]{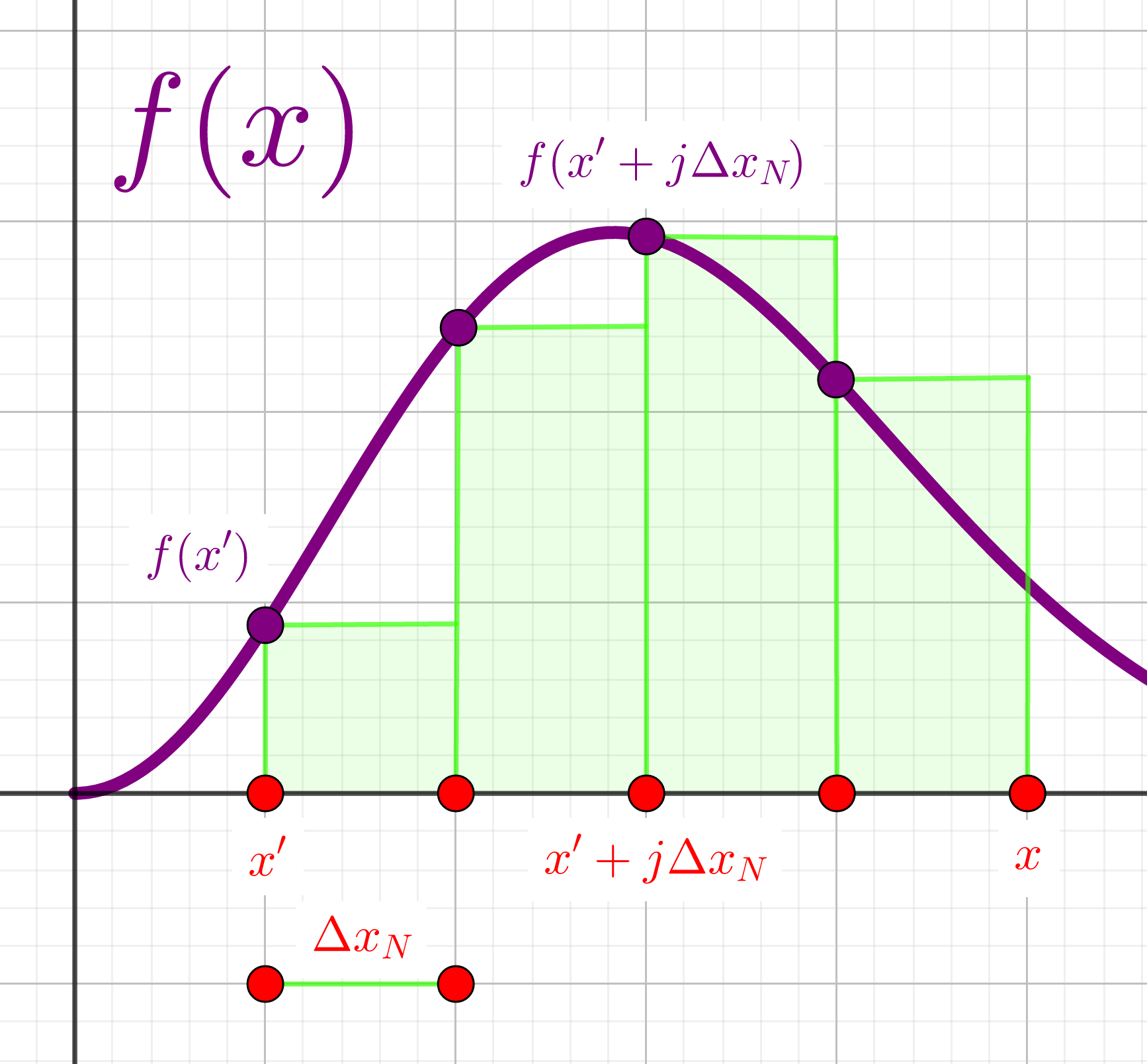}
    \caption{A sketch of a function $f$ (purple solid curve) and its corresponding Riemann sum  between $x^\prime$ and $x$ (green shaded boxes).  Here the function has a particular value at each point $x^\prime + j\Delta x_N$, and that value is multiplied by $\Delta x_N$. This is geometrically equivalent of multiplying a base and a height, and thus we have an approximation for the area under a little segment of the function $f(x)$. By adding up all of the little boxes we can then approximate the total area under the curve.}
    \label{fig: riemann sum}
\end{figure}

First off, we need to formally define this thing called integration.  As hinted at before, integration is really just taking function values and multiplying by some tiny little width | hence we say that integration is a formal way of evaluating the area under the curve.  A picture of such an area is given in \figref{fig: riemann sum}.  Specifically, we take a given function $f(x)$ and approximate the area under it between two \textit{bounds} $x^\prime$ and $x$ with the following equation
\begin{align}
    \textrm{area under curve} \approx \sum_{j=0}^{N-1} \Delta x_N f(x^\prime + j\Delta x_N), \textrm{ where } \Delta x_N = \frac{x - x^\prime}{N}.
\end{align}
The summation in this equation is called the \textbf{Riemann Sum} over a partition of the $x$-axis given by
\begin{align}
    \textrm{Partition } P_N = \Set{x_k: x^\prime < x^\prime + \Delta x_N < x^\prime + 2\Delta x_N < \dots < x^\prime + (N-1)\Delta x_N},
\end{align}
where the partition itself just chops up the $(x^\prime, x)$-interval into equally spaced sections as is shown in \figref{fig: riemann sum}.  At this point I want to clarify something: the Riemann Sum approximates the \textbf{signed area}, not just an area which is conventionally taken as a nonnegative value.  In other terms, the Riemann Sum is allowed to be negative and is whenever a significant portion of the function is below the $x$-axis.

Now that we have the Riemann Sum defined, we can now define Riemann Integration as 
\begin{align}
    \underbrace{\int_{x^\prime}^x \mathrm{d}s\, f(s)}_{\textrm{Integral}} = \lim_{N\rightarrow \infty} \underbrace{\sum_{j=0}^{N-1} \Delta x_N\,f(x^\prime + j\Delta x_N)}_{\textrm{Riemann Sum}}, \label{eq: definition of riemann integration}
\end{align}
where $s$ is simply a dummy variable. Since the limiting process makes $\Delta x_N$ smaller and smaller, then we must have continuity in $f(x)$ over this whole interval. The (definite) integral on the left-hand side is read as \textquotedblleft the definite integral of $f$ with respect to $s$ from $x^\prime$ to $x$.\textquotedblright$\,$ Notice that in this limit, just as was true with the definition of differentiation, the $\Delta$ becomes a $\mathrm{d}$ as the finite width $\Delta x_N$ becomes an infinitesimal width $\mathrm{d}x$. Since this $\mathrm{d}x$ represents the limiting width of the box in the Riemann Sum, it can be multiplied on either side of $f(x^\prime  + j\Delta x_N)$. In other words, since
\begin{align*}
    \lim_{N\rightarrow \infty} \sum_{j=0}^{N-1} \Delta x_N\,f(x^\prime + j\Delta x_N) = \lim_{N\rightarrow \infty} \sum_{j=0}^{N-1} f(x^\prime + j\Delta x_N)\,\Delta x_N,
\end{align*}
then it must be true that
\begin{align*}
    \int_{x^\prime}^x \mathrm{d}s\,f(s) = \int_{x^\prime}^x f(s)\,\mathrm{d}s.
\end{align*}
The left-hand side is pretty common notation in physics, particularly in quantum mechanics, whereas the right-hand side is a more common notation in mathematics. Either way, the same story is told: add up the areas of infinitely many infinitesimal boxes.  The only thing that MUST NOT be forgotten is the $\mathrm{d}s$ inside the integral.  It is \textit{crucial} to help one remember that the definite integral represents the (signed) area under the curve. More importantly though, in multiple dimensions it is called the \textit{volume-element} and forgetting about it can actually mess up your math by an unbelievable amount.  But more of that when you get there. For right now, keep your mathematical hygiene up and write the $\mathrm{d}s$ inside the integral.

There are a few properties of integrals that we need to discuss, namely antiderivatives, integration in the opposite direction, linearity, integrating products, and integration compositions of functions.  But before that, I know some of you may not be familiar with the idea of dummy variables, and so I will use the definition of Riemann Integration to try and make it a little clearer of how they work. Alright, so if we start with the definition, then we have
\begin{align*}
    \int_{x^\prime}^x \mathrm{d}s\,f(s) &= \lim_{N\rightarrow \infty} \sum_{j=0}^{N-1} \Delta x_N\,f(x^\prime + j\Delta x_N),
    \\
    &= \lim_{N\rightarrow\infty} \frac{x - x^\prime}{N}\,\left[f(x^\prime) + f\left(x^\prime + \frac{x - x^\prime}{N} \right) + \dots + f\left(x^\prime + (N-1)\,\frac{x - x^\prime}{N} \right)  \right].
\end{align*}
So $s$ does not appear anywhere within the actual Riemann Sum. Hence, it really is just a placeholder to represent the partition of the domain of the function (independent variable).  It would have made honestly just as much sense to use $x^{\prime\prime}$, $\xi$, $\psi$, $\aleph$, \Coffeecup, \textit{et cetera}. Thus, all of the following expressions are true:
\begin{align*}
    \int_{x^\prime}^x \mathrm{d}s\,f(s) = \int_{x^\prime}^x \mathrm{d}x^{\prime\prime}\,f(x^{\prime\prime}) = \int_{x^\prime}^x \mathrm{d}\xi\,f(\xi) = \int_{x^\prime}^x \mathrm{d}\psi\,f(\psi) = \int_{x^\prime}^x \mathrm{d}\textrm{\Coffeecup}\,f(\textrm{\Coffeecup}).
\end{align*}
Again, all of these equalities are true because none of those symbols appear in the Riemann Sum, and so the choice of which symbol to use is arbitrary. So if you ever get a little confused on what the dummy variables represent, just remember to expand out the Riemann Sum again to recall what the integral notation represents.

\subsubsection{Antiderivatives}
The idea of antiderivatives springs from the first of our fundamental theorems of calculus given by \equaref{eq: first fundamental theorem of calculus} where the integral can undo the derivative.  It goes something like this: assume that there exists a continuous function $f$ | that is, a function that is continuous at every point.  Then there must exist a function $F$ such that
\begin{align}
    F(x) - F(x^\prime) = \int_{x^\prime}^x \mathrm{d}s\,f(s). \label{eq: definition of antiderivative}
\end{align}
Then by the first fundamental theorem, the integral undoes the derivative, and so we must have
\begin{align}
    f(x) = \tderiv[x]{F}.
\end{align}
We can check that this is true by substitution to get
\begin{align}
    F(x) - F(x^\prime) = \int_{x^\prime}^x \mathrm{d}s\, \tderiv[s]{F},
\end{align}
which is indeed the exact form of the first fundamental theorem in \equaref{eq: first fundamental theorem of calculus}, except we have exchanged the lowercase $f$ in that equation for a captial $F$ in this one.  We define the function captial $F$ as the \textbf{antiderivative} of lowercase $f$ and note that its difference over some interval is the exact area underneath the lowercase $f$ curve. Example \ref{ex: power rule for integration} shows how one might determine an antiderivative for the general power function.

\vspace{0.15in}
\begin{example}[Power Rule for Antiderivatives]{ex: power rule for integration}
    Based on \exref{ex: implicit differentiation}, we have a generalized Power Rule for differentiation given by \equaref{eq: generalized power rule} and written as
    \begin{align*}
        \frac{\mathrm{d}}{\mathrm{d}x}\left[ x^\alpha \right] = \alpha x^{\alpha - 1},\textrm{ where } \alpha\in\mathds{R}.
    \end{align*}
    Our goal in this example is to use the first fundamental theorem to find an antiderivative $F(x)$ for any power function $f(x) = x^\alpha$.  Recall that if $F$ is the antiderivative of $f$, then
    \begin{align*}
        f(x) = x^\alpha = \tderiv[x]{F}.
    \end{align*}
    But we know that derivatives of power functions are also power functions, so we propose an \textit{ansatz}\footnote{An ansatz is a fancy word for mathematical guess.} given by $F(x) = Ax^\beta$, where $A$ and $\beta$ are fixed real numbers. Now we use this ansatz and the definition to $f(x)$ to see what the values of $A$ and $\beta$ are to make this the true antiderivative.
    
    We start with a derivative
    \begin{align}
        \tderiv[x]{F} = \frac{\mathrm{d}}{\mathrm{d}x}\left[ Ax^\beta \right] = A \frac{\mathrm{d}}{\mathrm{d}x}\left[ x^\beta \right] = A\beta x^{\beta - 1}. \nonumber
    \end{align}
    But this derivative must be equal to $f(x) = x^\alpha$. Therefore,
    \begin{align*}
        A\beta x^{\beta - 1} = x^\alpha.
    \end{align*}
    By dividing both sides by $x^\alpha$, we have
    \begin{align*}
        A\beta x^{\beta - (\alpha + 1)} = 1, \textrm{ for all }x\neq 0
    \end{align*}
    So to be clear, even though the left-hand side varies with $x$, it is always equal to one, therefore $x^{\beta - (\alpha - 1)}$ must be a constant.  But the only way to reduce a power function to a constant is by making the power equal to zero because $x^0 = 1$ for all $x\neq 0$.  Therefore, $\beta - (\alpha + 1) = 0$ implying that $\beta = \alpha + 1$.  This leaves us with
    \begin{align*}
        A\beta x^{\beta - (\alpha + 1)} = A\beta x^0 = A\beta = 1. 
    \end{align*}
    Hence, $A = 1/\beta = 1/(\alpha + 1)$. Thus, we have closed-form expression for our antiderivative $F(x)$ as 
    \begin{align}
        F(x) = \frac{1}{\alpha + 1}\,x^{\alpha + 1}, \textrm{ where } \alpha\neq -1 \textrm{ and if } \alpha \leq 0,\; x\neq 0 . \label{eq: power rule antiderivatives}
    \end{align}
    To check that this solution is the correct one, we take again take the derivative to see if we get $f(x)$ back
    \begin{align*}
        \tderiv[x]{F} = \frac{\mathrm{d}}{\mathrm{d}x}\left[ \frac{1}{\alpha + 1}\,x^{\alpha + 1} \right] = \frac{1}{\alpha + 1} \frac{\mathrm{d}}{\mathrm{d}x}\left[ x^{\alpha + 1} \right] = \frac{\alpha + 1}{\alpha + 1} x^{\alpha + 1 - 1} = x^\alpha = f(x)
    \end{align*}
    Since this antiderivative yields the expected behavior in $f(x)$, and so we're done (fun fact: we just solved a differential equation).
\end{example}
\vspace{0.15in}

If we go back to \equaref{eq: definition of antiderivative}, then we can also solve for $F(x)$ in terms of the fixed number $F(x^\prime)$ explicitly. 
\begin{align}
    F(x) = F(x^\prime) + \int_{x^\prime}^x \mathrm{d}s\,f(s).
\end{align}
But $x^\prime$ is just some fixed number and so is $F(x^\prime)$.  Presumably, when $x^\prime$ changes, so will the value of $F(x^\prime)$, but the overall structure of $F(x)$ will not.  And most importantly, $f(x)$ will still be the $x$-derivative of $F(x)$, since the derivative of any constant number is zero.  What this means is that we are actually free to add any constant number to $F(x)$ and retain the nature of the integral undoing the derivative.  Therefore, we come to define the \textbf{indefinite integral} as a bound-independent integral given by
\begin{align}
    \int \mathrm{d}x\,f(x) = F(x) + C, \textrm{ where } f(x) = \tderiv[x]{F} \textrm{ and } C\in\mathds{R}. \label{eq: definition of indefinite integral}
\end{align}
Notice there are no bounds on the integral, meaning this quantity is purely a function of $x$.  For that reason, the integration variable $x$ inside the integral is no longer a dummy variable, but is used to tell us what we are integrating with respect to (this time it is not a dummy variable because it shows up again on the right-hand side). Since $C$ is literally any constant, we are free to move it over to the other side while keeping it positive as 
\begin{align*}
    F(x) = \int \mathrm{d}x\, f(x) + C.
\end{align*}
This equation shows that technically speaking, there is no such thing as \textit{the} antiderivative, for it is only unique up to some additive constant $C$.  There are only infinitely many quantities that we would refer to as \textit{an} antiderivative.  However, it is commonplace to package all of the ambiguity in the antiderivative totally within the indefinite integral, as is done in \equaref{eq: definition of indefinite integral}.  Then it is fairly easy to write down that 
\begin{align*}
    \textrm{Indefinite Integral } = \textrm{ Antiderivative } + \textrm{ Constant}
\end{align*}
where we would use the fundamental theorems of calculus to find the antiderivative of any function and add an arbitrary constant to it to find the indefinite integral.  Otherwise we would need some other scheme to find the indefinite integral to then find the antiderivative | this would ultimately reduce to evaluating the Riemann Sum directly which is without a doubt extremely difficult to do.  If we do just evaluate the antiderivative separately from the arbitrary constant, then we can use the ease of derivatives to find areas instead; in practice, evaluating derivatives is very easy while evaluating sums is not at all.

\subsubsection{Integrating Backwards and Integrating Nothing}
I need to clarify something before we get too deep into integration: this whole time I have been implicitly assuming that our integration interval is $(x^\prime, x)$ and that \textbf{we integrate from left to right}.  This is the convention in most courses and books, but it actually is not a strict requirement.  All it essentially does is keeps the value $x-x^\prime > 0$. However, by the ordering property of the real numbers (see Section \ref{subsec: Vectors - Real Numbers}), there are actually two other cases we may consider: $x - x^\prime < 0$ and $x - x^\prime = 0$.  

We will start with the first case.  If we assume that $x - x^\prime < 0$, then by the definition of Riemann Integration (\equaref{eq: definition of riemann integration}) and the fact that $\vert y\vert = -y$ for $y<0$, we have 
\begin{align*}
    \int_{x^\prime}^x \mathrm{d}s\,f(s) &= \lim_{N\rightarrow \infty} \sum_{j=0}^{N-1} \Delta x_N\,f(x^\prime + j\Delta x_N),
    \\
    &= \lim_{N\rightarrow \infty} \sum_{j=0}^{N-1} \frac{x - x^\prime}{N}\,f\left(x^\prime + j\frac{x-x^\prime}{N}\right),
    \\
    &= \lim_{N\rightarrow \infty} \sum_{j=0}^{N-1} -\frac{\vert x - x^\prime\vert }{N}\,f\left(x^\prime - j\frac{\vert x-x^\prime\vert}{N}\right),
    \\
    &= -\lim_{N\rightarrow \infty} \sum_{j=0}^{N-1} \frac{\vert x - x^\prime\vert }{N}\,f\left(x^\prime - j\frac{\vert x-x^\prime\vert}{N}\right)
    \\
    &= -\lim_{N\rightarrow \infty} \sum_{j=0}^{N-1} \vert\Delta x_N\vert \,f\left(x^\prime - j\vert \Delta x_N\vert\right).
\end{align*}
By dividing through by the negative sign we recover 
\begin{align}
    -\int_{x^\prime}^x \mathrm{d}s\,f(s) = \lim_{N\rightarrow \infty} \sum_{j=0}^{N-1} \vert\Delta x_N\vert \,f\left(x^\prime - j\vert \Delta x_N\vert\right). \label{eq: integrate backwards}
\end{align}
What this shows is that if we moving along the $x$-axis in the opposite direction, that is we move backwards along it, then the Riemann Sum spits out a negative sign.  In this case, our partition would be
\begin{align*}
    P_N = \Set{x_k: x^\prime > x^\prime - \vert\Delta x_N\vert >  x^\prime - 2\vert\Delta x_N\vert > \dots >  x^\prime - (N-1)\vert\Delta x_N\vert},
\end{align*}
which says that we are getting smaller input values by integrating \textit{from right to left}. There actually is another way to achieve the same effect as integrating backwards though!  Suppose for a moment that instead we had $x - x^\prime >0$, again.  If we had instead simply \textit{switched the bounds} on the definite integral then
\begin{align*}
    \int_{x}^{x^\prime} \mathrm{d}s\,f(s) &= \lim_{N\rightarrow \infty} \sum_{j=0}^{N-1} \frac{x^\prime - x}{N}\,f\left(x +  j\frac{x^\prime-x}{N}\right),
    \\
    &= -\lim_{N\rightarrow \infty} \sum_{j=0}^{N-1} \frac{x - x^\prime}{N}\,f\left(x - j\frac{x-x^\prime}{N}\right),
    \\
    &= - \lim_{N\rightarrow \infty} \sum_{j=0}^{N-1} \vert\Delta x_N\vert \,f\left(x - j\vert \Delta x_N\vert\right)
\end{align*}
And so we see that we again increment down the $x$-axis again while starting at the high-end of the interval.  We conclude from this that \textbf{integrating backwards is equivalent to switching the bounds on the definite integral}. Furthermore, the integral picks up a negative sign by integrating backwards, so, in math terms,
\begin{align}
    \int_x^{x^\prime}\mathrm{d}s\, f(s) = - \int_{x^\prime}^x \mathrm{d}s\,f(s). \label{eq: integrate switch bounds}
\end{align}

\subsubsection{Linearity of the Integral}
Since the derivative is a linear operator, it follows that its inverse should be linear operator as well, right? This was always a statement that I would hear in class and I frankly didn't see it to be too obvious. For me, there needed to be a stronger proof, and so this is what this subsection is about.  Although it is possible to establish the linearity with the fundamental theorems, I prefer to show it with the Riemann Sum because it is not too difficult to do, for starters, but also it is a helpful reminder of the Riemann Sum being there (unfortunately too many students forget about it).

We will proceed with a very similar proof as we did with differentiation.  Assume that a function $h(x) = af(x) + bg(x)$, where $a,b\in\mathds{R},$ is continuous on the $(x^\prime, x)$ interval, where of course $f$ and $g$ must be continuous on that same interval.  So we simply integrate $h$
\begin{align*}
    \int_{x^\prime}^x \mathrm{d}s\,h(s) &= \lim_{N\rightarrow \infty} \sum_{j=0}^{N-1} \Delta x_N\,h(x^\prime + j\Delta x_N),
    \\
    &= \lim_{N\rightarrow \infty} \sum_{j=0}^{N-1} \Delta x_N\,\left[af(x^\prime + j\Delta x_N) + bg(x^\prime + j\Delta x_N)\right].
\end{align*}
Then we use can split up this limit into two terms since $f$ and $g$ are continuous on this interval by \equaref{eq: limit of sum}. Further, we can take the constants $a$ and $b$ out of the limits since constants are continuous functions that converge exactly to those constants. Thus,
\begin{align*}
    \int_{x^\prime}^x \mathrm{d}s\,h(s) &= \lim_{N\rightarrow \infty} \sum_{j=0}^{N-1} \Delta x_N\,\left[af(x^\prime + j\Delta x_N) + bg(x^\prime + j\Delta x_N)\right],
    \\
    &= a\left[\lim_{N\rightarrow \infty} \sum_{j=0}^{N-1} \Delta x_N\, f(x^\prime + j\Delta x_N) \right] + b\left[\lim_{N\rightarrow \infty} \sum_{j=0}^{N-1} \Delta x_N\, g(x^\prime + j\Delta x_N)\right],
    \\
    &=a\int_{x^\prime}^x \mathrm{d}s\,f(s) + b\int_{x^\prime}^x \mathrm{d}s\,g(s).
\end{align*}
Therefore, we have
\begin{align}
    \int_{x^\prime}^x \mathrm{d}s\,\left[af(s) + bg(s) \right] = a\int_{x^\prime}^x \mathrm{d}s\,f(s) + b\int_{x^\prime}^x \mathrm{d}s\,g(s), \textrm{ for all } a,b\in\mathds{R}. \label{eq: linear integrals} 
\end{align}
But this is the exact form of a linear operator acting on a function, given in \equaref{eq: definition of linear operator on function}, where 
\begin{align*}
    \mathcal{O} = \int_{x^\prime}^x \mathrm{d}s,
\end{align*}
and so by definition of Riemann Integration, we have established that the (definite) integral is a linear operator.  I leave it to you to wrap your mind around how the indefinite integral is also linear (the difference is you need to figure out how the arbitrary constants behave when you add and subtract them (hint: $C + C + C = C$ for any combination of arbitrary constants.)).

\subsubsection{Integration by Parts}
Just as there is a Product Rule for differentiation, there is an equivalent product rule for integration.  To establish this relationship, we will use the first fundamental theorem of calculus, because in this proof the Riemann Sums are a little too clunky to be helpful.  We start by assuming that a function $h(x) = f(x)g(x)$ that is differentiable on the interval $(x^\prime,x)$ (again, since $h$ is differentiable on this interval then both $f$ and $g$ are also differentiable on this interval).  We \textit{differentiate} $h$ using Product Rule (\equaref{eq: product rule}), to find
\begin{align}
    \tderiv[x]{h} = g(x)\tderiv[x]{f} + f(x)\tderiv[x]{g}\nonumber.
\end{align}
Now we will use the first fundamental theorem to undo this differentiation
\begin{align}
    h(x) - h(x^\prime) = \int_{x^\prime}^x \mathrm{d}s\,\tderiv[s]{h} = \int_{x^\prime}^x \mathrm{d}s\,g(s)\tderiv[s]{f} + \int_{x^\prime}^x\mathrm{d}s\,f(s)\tderiv[s]{g}. \label{eq: integrate product rule}
\end{align}
Then, we isolate the second term to find
\begin{align}
    \int_{x^\prime}^x\mathrm{d}s\,f(s)\tderiv[s]{g} = f(x)g(x) - f(x^\prime)g(x^\prime) - \int_{x^\prime}^x\mathrm{d}s\,g(s)\tderiv[s]{f},
\end{align}
where I have substituted in $h(x) = f(x)g(x)$.  To make this expression a little nice, people use the evaluation vertical line again
\begin{align}
    \int_{x^\prime}^x\mathrm{d}s\,f(s)\tderiv[s]{g} = f(s)g(s)\bigg{\vert}_{s = x^\prime}^{s = x} - \int_{x^\prime}^x\mathrm{d}s\,g(s)\tderiv[s]{f}, \label{eq: definite integration by parts}
\end{align}
where the vertical bar notation is translated as 
\begin{align*}
    h(s)\bigg{\vert}_{s = x^\prime}^{s = x} = h(x) - h(x^\prime).
\end{align*}
Equation \ref{eq: definite integration by parts} is commonly referred to as \textbf{integration by parts}. By writing $u = f(x)$ and $v = g(x)$, then this equation is typically written without bounds as
\begin{align}
    \int u\,\mathrm{d}v = uv - \int v\,\mathrm{d}u. \label{eq: indefinite integration by parts} 
\end{align}
In physics, integration by parts is used CONSTANTLY, although it is usually only used in higher dimensions (such as in electromagnetic theory). The reason why is it directly gives us a way of relating functions, their derivatives, and the boundary values to each other in one need little equation.  For example, it can relate electric potential, the electric field, and the electric potential on the boundary to each other.  However, using either \equaref{eq: definite integration by parts} or \equaref{eq: indefinite integration by parts} is initially pretty tricky. Example \ref{ex: practice with integration by parts} is here to show you how to get started with a not-so-obvious case of how integration by parts can be used with meaningful results.

\vspace{0.15in}
\begin{example}[Practice with Integration by Parts]{ex: practice with integration by parts}
    In this example, we see how to use integration by parts to evaluate otherwise impossible integrals.  For this example, we will integrate $\ln x$.  Note that the natural logarithm is \textit{defined} as the following integral
    \begin{align}
        \ln x = \int_1^x \frac{\mathrm{d}s}{s}, \label{eq: definition of logarithm}
    \end{align}
    Therefore, by the first fundamental theorem,
    \begin{align}
        \frac{\mathrm{d}}{\mathrm{d}x} \left( \ln x\right) = \frac{1}{x}. \label{eq: derivative of ln}
    \end{align}
    We use \equaref{eq: indefinite integration by parts} to find $\int \mathrm{d}x\,\ln x$. The key with all of these types of integrals is to look at the integrand and try to see if you can recognize any either easily differentiable functions or any easily integrable functions. And then use
    \begin{align}
        \int u(x)\,\tderiv[x]{v}\,\mathrm{d}x = u(x)v(x) - \int v(x)\,\tderiv[x]{u}\,\mathrm{d}x. \label{eq: indefinite integration by parts explicit derivatives}
    \end{align}
    The equation above is identical to \equaref{eq: indefinite integration by parts}, but where we \textquotedblleft multiplied and divided by $\mathrm{d}x$\textquotedblright.$\,$ Here, we only know the derivative of $\ln x$, so we must set $u(x) = \ln x$.  But then what is $v(x)$?  To find it we must integrate the other function multiplying $\ln x$.
    \begin{align*}
        \int \ln x\,\mathrm{d}x = \int \ln x\,\tderiv[x]{v}\,\mathrm{d}x
    \end{align*}
    By comparing these expressions, we see that
    \begin{align*}
        \tderiv[x]{v} = 1 = x^{0} \Rightarrow v(x) = \int \mathrm{d}x\, x^{0}.
    \end{align*}
    By Power Rule, \equaref{eq: power rule antiderivatives}, we can evaluate the antiderivative as 
    \begin{align*}
        v(x) = \frac{1}{0 + 1}\,x^{0 + 1} = x^1 = x.
    \end{align*}
    With this quantity known, we now just plug stuff into \equaref{eq: indefinite integration by parts explicit derivatives}. 
    \begin{align*}
        \int \ln x\cdot 1\,\mathrm{d}x &= \ln (x)\cdot x - \int x \, \frac{\mathrm{d}}{\mathrm{d}x} \left( \ln x\right)\,\mathrm{d}x,
        \\
        &= x\ln x - \int \frac{x\mathrm{d}x}{x},
        \\
        &= x\ln x - \int 1\,\mathrm{d}x.
    \end{align*}
    But the second term has already been evaluated as $x$.  Therefore
    \begin{align}
        \int \ln x\,\mathrm{d}x = x\ln x - x + C = x\left( \ln x - 1\right) + C. \label{eq: integral of logarithm}
    \end{align}
    In the final form I inserted the $C$'s as the indefinite integral is incomplete with only the antiderivative (again the addition of any arbitrary constant keeps the derivative of \equaref{eq: integral of logarithm} the same function of $\ln x$).
    \end{example}
\vspace{0.15in}

Although integration by parts can be a really useful tool, it can also be a little cranky.  There is usually a bit of an ambiguity regarding which function should be taken as the derivative and which function should be taken as the antiderivative that we should then differentiate to make our lives easier.  This is all due to the fact that I chose to isolate the second term in \equaref{eq: integrate product rule} instead of the first term. Sometimes, we have to integrate by parts multiple times to actually arrive at something that whose antiderivative is known, and this too can make using it a little nightmarish. As a good rule of thumb, when you can, choose the antiderivative to be one who eventually has a terminating derivative.  By this I mean essentially polynomial functions.  The reason why is if we continually take derivatives of positive integer powers, we eventually differentiate until we get a constant back, and then the next derivative is zero.  This is because we always subtract one from the power whenever we differentiate, therefore, if we have an $n^{\mathrm{th}}$-order polynomial, $n+1$ derivatives of it will return zero.  If we eventually have a terminating derivative, then we know that at the penultimate step we have a constant function times some other function which may be integrable.  But just like a lot of other things in math, the cases of weird quirks are far outnumbered by cases that do actually work.

\subsubsection{Change of Variables}
Lastly, we must talk about how to deal with compositions of functions.  This set up will be very similar to that of integration by parts where we will make use of the first fundamental theorem, although this time we are going to use it for the sake of brevity.  The Riemann Sum proof is not as difficult to do as integration by parts, and hey, if you're bored and want to give it a go, more power to you.  Anyway, let's assume that we have a differentiable composition of functions $h(x) = g(f(x))$ over the interval $(x^\prime, x)$.  Then, when we \textit{differentiate}, we must use the Chain Rule (\equaref{eq: Chain Rule differentiation}), given by
\begin{align}
    \tderiv[x]{h} = \tderiv[x]{f}\,\tderiv[f]{g}.\nonumber
\end{align}
Then, by the first fundamental theorem, we can undo this differentiation as 
\begin{align}
    h(x) - h(x^\prime) = \int_{x^\prime}^x \mathrm{d}s\,\tderiv[s]{f}\,\tderiv[f]{g}
\end{align}
Although it is probably pretty difficult to see in this form, the equation above tells us that the product of one function's derivative times another function's derivative with respect to the first has a closed-form antiderivative given by the composition.  A lot of introductory calculus courses instead choose to write $u = f(x)$ such that $h(x) = g(u)$ and then call this class of integrals problems \textquotedblleft solvable with a $u$-substitution\textquotedblright. With this insertion, then equation above becomes
\begin{align}
    h(x) - h(x^\prime) = \int_{x^\prime}^x \mathrm{d}s\,\tderiv[s]{u}\,\tderiv[u]{g} = g(u) - g(u^\prime).
\end{align}
But the second equality is, by definition, the definite integral of $g$ over the $u^\prime = u(x^\prime)$, $u(x)$ interval instead of over the $(x,x^\prime)$-interval. Thus, 
\begin{align}
    \int_{x^\prime}^x \mathrm{d}s\,\tderiv[s]{u}\,\tderiv[u]{g} = g(u) - g(u^\prime) = \int_{u^\prime}^u \mathrm{d}t\,\tderiv[t]{g}.
\end{align}
Or, written without bounds, we have
\begin{align}
    \int \mathrm{d}x\,\tderiv[x]{u}\,\tderiv[u]{g} = g(u(x)) + C = \int \mathrm{d}u\, \tderiv[u]{g}, \label{eq: indefinite u substitution}
\end{align}
The crux of this method of integrating comes from the change of integration variable as we read the equation above from left to right.  We start integrating with respect to $x$ and then transition to integrating with respect to $u$. Example \ref{ex: a u-substitution} shows how one might use a Change of Variables with a $u$-substitution to evaluate a typical integral.

\begin{example}[A Prototypical $u$-Substitution]{ex: a u-substitution}
    This example is to show you how we might use a $u$-substitution in practice.  Let's suppose we wanted to find the indefinite integral of the function $f(x) = x^2\ln(3x^3)$. Specifically, we want to find the integral $I(x)$ such that
    \begin{align*}
        I(x) = \int x^2\ln(3x^3)\,\mathrm{d}x.
    \end{align*}
    We would use Change of Variables in this case because we know that $x^2$ is related to the derivative of $3x^3$, since 
    \begin{align*}
        \frac{\mathrm{d}}{\mathrm{d}x}\left[ 3x^3\right] = 3\cdot 3x^2 = 9x^2,
    \end{align*}
    using Power Rule for differentiation.  Thus, 
    \begin{align*}
        x^2 = \frac{1}{9}\frac{\mathrm{d}}{\mathrm{d}x}\left[ 3x^3\right],
    \end{align*}
    and by substitution into the integral $I(x)$, we have
    \begin{align*}
        I(x) = \int x^2\ln(3x^3)\,\mathrm{d}x = \int \mathrm{d}x\, \frac{1}{9}\frac{\mathrm{d}}{\mathrm{d}x}\left[ 3x^3\right]\,\ln(3x^3).
    \end{align*}
    Now we use the power of the Change of Variables by setting $u(x) = 3x^3$. Then the $\mathrm{d}x$'s in the integral \textquotedblleft cancel\textquotedblright$\,$ and we have
    \begin{align*}
        I(u) = \int \frac{1}{9}\mathrm{d}u\,\ln u = \frac{1}{9}\int \mathrm{d}u\,\ln u = \frac{1}{9}\,u\left( \ln u - 1\right) + C
    \end{align*}
    The second-to-last equality holds by the linearity of the integral and the last equality holds from \equaref{eq: integral of logarithm}.  Finally, since $u(x) = 3x^3$, we substitute this expression back into $I(u)$ to find
    \begin{align*}
        I(x) = \frac{1}{9}\,(3x^3)\left[ \ln(3x^3) - 1\right] + C = \frac{1}{3}\, x^3\left[ \ln(3x^3) - 1 \right] + C
    \end{align*}
    which is our final result.
\end{example}

\begin{problem}[Your Turn with a $u$-Substitution]{prob: your turn with u-substitution}
Now that we have evaluated a $u$-substitution problem together, you should try one out on your own just to be sure you have the hang of it.  For this problem, find the indefinite integral $I(x)$ such that
\begin{align*}
    I(x) = \int \mathrm{d}x\, \frac{\ln(\pi\ln x)}{x}.
\end{align*}
\begin{enumerate}[(a)]
    \item What is the derivative of $\pi \ln x$?
    \item Solve for $1/x$ in terms of the derivative of $\pi \ln x$.
    \item Substitute this value into the integrand and cancel the $\mathrm{d}x$'s.
    \item Define $u(x) = \pi \ln x$. Show that 
    $$ I(u) = \int \mathrm{d}u\, \ln u. $$
    \item Evaluate the integral with respect to $u$ and substitute in $u(x)$ to show that
    \begin{align*}
        I(x) = \ln(x)\left[ \ln(\pi \ln x) - 1 \right] + C.
    \end{align*}
\end{enumerate}
\end{problem}

Although most students now know them as $u$-substitutions, I want to be clear that $u$-subs truly do belong to the much larger class of solvable problems called \textit{Change of Variables}.  Oftentimes, problems are impossible to solve without using some kind of change of variable, or in physics, some functions (like Bessel functions) may only have integrals known in one particular variable, and so it will be your job to massage the equation into something more tangible.  And this is totally okay. You are more than welcome to do such a thing as long as you remember to change the integration variable in accordance with \equaref{eq: indefinite u substitution}.

\section{Applications of One-Dimensional Calculus}
We now have enough tools to start talking about how we might apply calculus to solve problems in mathematics, physics, and beyond.  It is pretty disappointing to me that I don't have more time to talk about many more advanced or niche topics.  If it is not clear yet, the topic of calculus is a bit of a sinc-hole. There are simply too many things for us to talk about and almost all of them are directly relevant for physics (maybe that's why there is a physics major in the first place?). So I will focus our attention to a few absolutely crucial applications of one-dimensional calculus.

\subsection{Local Extrema}
One of the primary examples of how useful calculus can be is in finding local extrema; points on a function that represent the maximum or minimum value within a certain neighborhood.  The power of this type of mathematical analysis is that we if we have a particular situation that can be modeled as a function and we want to know where any maxima or minima are, then we do not need to just input values into a calculator until we somehow get lucky and find them. Instead we can use the analytical properties of calculus to find them exactly and without nearly as much trouble. Furthermore, if we were just to plug numbers into some calculator or computer, we would only ever be able to scan a finite number of points and so there is always some trepidation over whether we scanned enough points to be sure of any extrema we may find. Meanwhile, calculus allows to scan \textit{infinitely} many points meaning there is no uncertainty in our results.

Allow me to set the stage within the context of our moving object. Let's assume our object takes the form of a car moving on a single-lane straight road. Under these assumptions, it is fair for us to take its position along the road as points along the $x$-axis measured as functions of time, $t$. So in this scenario, the time $t$ is the independent variable while the position $x$ is the dependent variable.  If, for example, we measured the positions $x(t)$ of the car along the road from some starting point $x(0) = 0$, then we could ask the question, \textquotedblleft what is the furthest distance the car is from that starting point?\textquotedblright$\,$ Intuitively, this question only really makes sense if the car ever stops and start to move backwards; otherwise there is no \textquotedblleft furthest distance\textquotedblright$\,$ from the start.  Thus the act of stopping is what defines a maximum, where in this case the maximum is a maximum of distance. But if something is to stop moving, then its velocity must vanish.  Therefore, if we were able to find the $t$-values at which the \textit{derivative} of the position goes to zero, then we could then plug those values into the $x(t)$ formula to find the maximum distance from the starting point. These special $t$-values where the first derivative vanish are called \textbf{critical points}.  

\begin{figure}
    \centering
    \includegraphics[width = 4.5in, keepaspectratio]{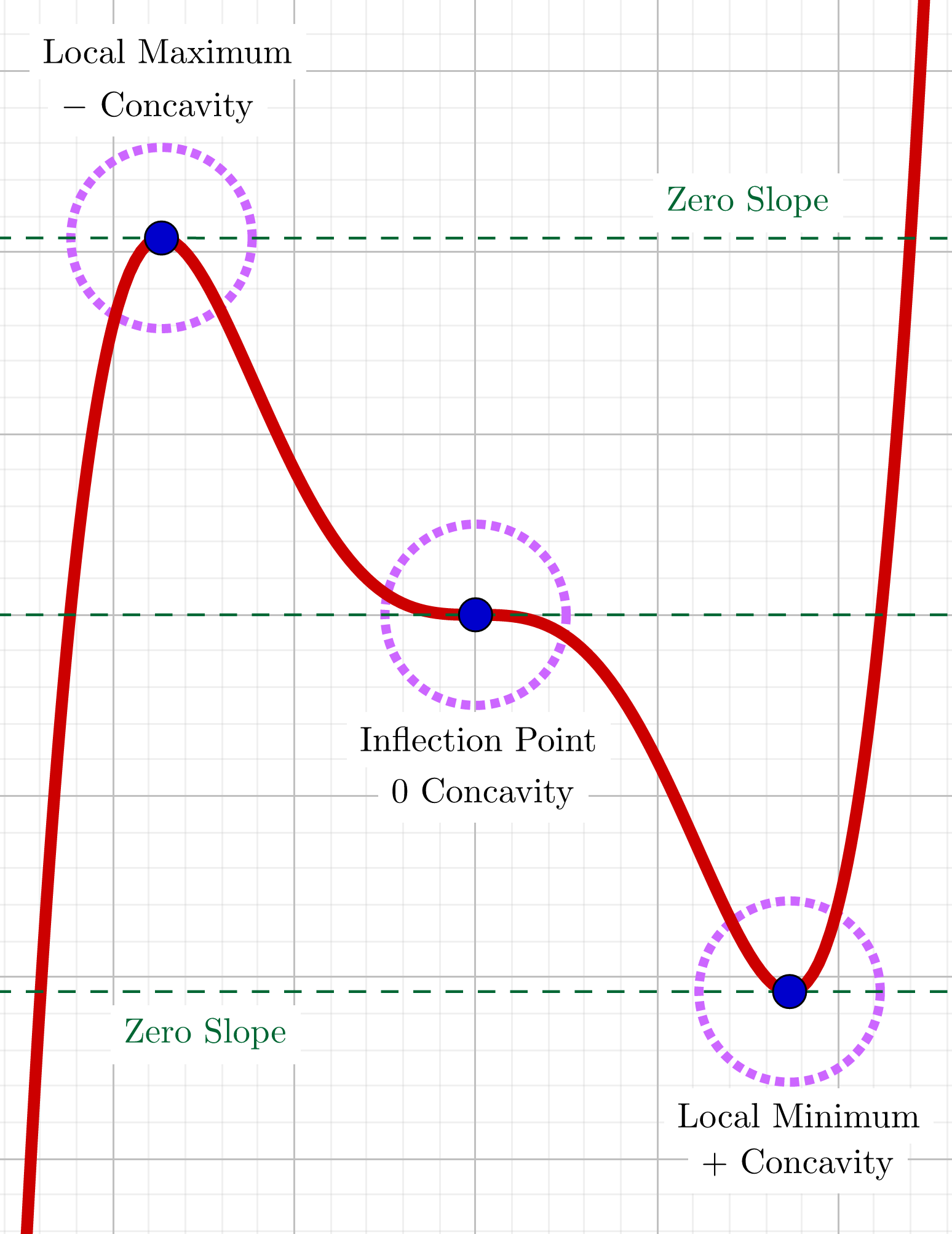}
    \caption{A function and its local extrema.  When there is a local maximum, the nearby points are lower than it, creating an upside-down cup shape.  When there is a local minimum, the nearby points are all above it, creating a rightside-up shape. When there is an inflection point, the nearby points are both above and below it and the concavity changes sign. Notice that these are all local extrema, but they are not global.}
    \label{fig: concavity}
\end{figure}

Unfortunately not all critical points lead to extrema. For example, if the car where to stop and then continue moving in the same direction, then there is also no maximum distance from the starting point.  So how do we tell if a critical point leads to an extremum or if it leads to nothing?  Well, for the car to actually travel to some maximum distance, then the car must be \textit{slowing down} before it stops and then \textit{speeding up} in the \textit{opposite direction} after it stops. Thus, its velocity must be getting \textit{more negative} nearby a critical point for a maximum distance to make sense.  Since there is a negative change in velocity, then the first derivative of the velocity (acceleration) must be negative, or the second derivative of the position must be negative, for the critical point to lead to a maximum in position. The sign of the second derivative denotes \textbf{concavity}. When any function reaches a maximum, then all nearby points must be lower than it. This makes a local upside-down cup shape which is called negative concavity, whereas a local minima has all nearby points higher than it, and so the cup is rightside-up. In the case where the car stops and then continues in the same direction, we have a negative acceleration (slowing down) and then a positive acceleration (speeding up in the \textit{same} direction).  Since the acceleration changes sign, we say that the function has zero concavity at this critical point and call the point and \textbf{inflection point}. These three cases are shown in \figref{fig: concavity}.

This reasoning is essentially the same outside of the context of moving particles. If we ever want to find the local extrema of a particular function, then we differentiate it and find points in its domain where the first derivative vanishes.  The reason why is that at these locations in the domain, the function is neither increasing nor decreasing since the instantaneous slope (derivative) is zero. Thus, it is only at these locations that we may expect to find the function to be \textit{extremized}. 

There is a bit of a caveat here. Many courses talk of the \textit{global} extrema at the same time that they talk about \textit{local} extrema.  However, calculus cannot easily account for global extrema, where the term \textquotedblleft global\textquotedblright$\,$ refers to our consideration of the entire function itself.  Sometimes calculus will tell us about a global extremum | for instance, the function $f(x) = x^2$ has a global minimum at $x = 0$ because the function goes to positive infinity on either side of the vertical axis | but it will in no way identify it to us as a global extremum versus a local extremum.  Personally, I cannot stand it when other courses introduce local and global extrema together in a unit as if they are some tidy little package for this very reason. The calculus with the local extrema is fine, but the global extrema is no longer a calculus thing; it's an algebra visualization thing.

One nice thing about my internal debate over local and global extrema is its connection to the notion of locality in physics.  By this point, I have tried to emphasize the ideas of intervals and continuity in most of the calculus we have done because it is fundamental to calculus from a mathematical point-of-view.  From a physical point-of-view, since all, or at least almost all, of our laws of physics or based on derivatives and integrals, our laws of physics are said to be \textit{local}.  This means that tiny changes in one little neighborhood of space-time may affect its neighborhood, but it takes some finite propagation in either time or space to affect regions outside of such a neighborhood. For regions very, very far outside of the neighborhood, the original tiny changes will essentially never be propagate due to some kind of energy dissipation. In the case of my debate with myself over local and global extrema, the derivative function only scans points within some infinitesimally small interval to evaluate the limit of the difference quotient.  Therefore, it cannot propagate its information outside of this region to tell us more about global properties of functions far outside this tiny neighborhood.  The best candidate that we have talked about thus far to do such a thing is the integral, for it takes into account all the points over some interval.  But there is rarely some global extrema test for integration where we would use the integral to find all the points where the derivative vanishes, because it actually isn't ever any different than just simply plotting our original function by the first fundamental theorem of calculus.

Alright so the formal steps behind finding local extrema are as follows: 
\begin{enumerate}
    \item Use the \textbf{first derivative} test.
        \begin{enumerate}
            \item Take the first derivative of the function.
            \item Set the derivative equal to zero.
            \item Find all points in the function's domain where the first derivative is zero.
            \item Substitute these critical points into the actual function to find possible extrema.
        \end{enumerate}
    \item Use the \textbf{second derivative} test.
        \begin{enumerate}
            \item Take the second derivative of the function.
            \item Substitute in the critical points.
            \begin{enumerate}
                \item If the sign of the second derivative is negative, then critical point leads to a local maximum.
                \item If the second derivative is zero, then the critical point is an inflection point.
                \item If the sign of the second derivative is positive, then the critical point leads to a local minimum.
            \end{enumerate}
        \end{enumerate}
\end{enumerate}

\subsection{Taylor Series}
The next very useful application of one dimensional calculus, especially for physicists, is the Taylor Series.  The main idea behind Taylor Series is to recast a known function in terms of much easier ones, or at least much easier ones to handle with calculus.  One thing that you may have noticed so far is we established the Power Rule for differentiation and integration (Eqs. \ref{eq: generalized power rule} and \ref{eq: power rule antiderivatives}), but I've stayed away from a lot of other derivatives and integrals, such as those that can be found in \cite{pauls_derivative_table, Schaums}. The reason why is that there are a couple of other functions in physics whose derivatives that we care about, such as exponentials and sinusoids, but otherwise the mathematics may be too clunky to use effectively.  Instead we turn to some type of infinite polynomial because the derivatives and integrals of such objects are known to use via linearity and the Power Rule.  Thus, our goal is to write a function $f(x)$ as a polynomial given by
\begin{align}
    f(x) = \sum_{n = 0}^\infty a_n x^n
\end{align}
where the $\set{a_n}$ are a set to to-be-determined coefficients. It helps to have functions that are infinitely differentiable at $x = 0$, and so we will assume we are dealing with these types of functions.  So we now need to find out what these coefficients are.

To start, let's just substitute in $x = 0$. Then
\begin{align}
    f(0) = a_0 + \sum_{n = 1}^\infty a_n\cdot x^n = a_0 +\sum_{n=0}^\infty 0 = a_0.
\end{align}
Therefore $a_0 = f(0)$. Great so we have one coefficient. Now we need the rest. Notice that if we take a single derivative, then the $a_0$ term vanished because it is a constant, and then we have
\begin{align}
    \tderiv[x]{f} = \sum_{n = 1}^\infty a_n\cdot nx^{n-1} = a_1 + \sum_{n=2}^\infty a_n\cdot nx^{n-1}
\end{align}
Therefore, if we evaluate the derivative at $x = 0$, then 
\begin{align}
    \left. \tderiv[x]{f}\right\vert_{x = 0} = a_1 + \sum_{n=2}^\infty a_n\cdot n\cdot 0^{n-1} = a_1 + \sum_{n = 2}^\infty 0 = a_1.
\end{align}
Now we have another coefficient, and so we just continue this pattern to find the other coefficients.
\begin{align}
    \left. \frac{\mathrm{d}^kf}{\mathrm{d}x^k}\right\vert_{x = 0} = k\cdot (k-1)\cdot (k-2)\cdot \dots 2\cdot 1\cdot a_k + \sum_{n=k+1}^\infty 0 = k!\cdot a_k.
\end{align}
Therefore, we have for any general coefficient $a_k$, we have
\begin{align}
    a_k = \frac{1}{k!}\,\left. \frac{\mathrm{d}^kf}{\mathrm{d}x^k}\right\vert_{x = 0}. \label{eq: taylor series coefficient}
\end{align}
Then finally we have a formula for recasting an infinitely differentiable function $f$ as 
\begin{align}
    f(x) = \sum_{n = 0}^\infty \frac{1}{n!}\, \left. \frac{\mathrm{d}^kf}{\mathrm{d}x^k}\right\vert_{x = 0} x^n.\label{eq: maclaurin series}
\end{align}
The series above is referred to in math classes as a \textbf{Maclaurin Series} because it is \textit{centered at} $x = 0$.  We could very easily choose instead to center the series at another point, let's say $x = c$, if it suits our needs better.  To do so, we let $x \rightarrow x - c$, and so the derivative is the same by Chain Rule.  Then we substitute $x -c$ into the equation above to find
\begin{align}
     f(x) = \sum_{n = 0}^\infty \frac{1}{n!}\, \left. \frac{\mathrm{d}^kf}{\mathrm{d}x^k}\right\vert_{x = c} (x-c)^n,\label{eq: taylor series}
\end{align}
which is what mathematicians would call a Taylor Series. An important thing to note right now is that physicists call both series Taylor Series.  Example \ref{ex: taylor series of e and cos and sin} shows how one might compute the full Taylor Series for a few functions.

\vspace{0.15in}
\begin{example}[Some Key Taylor Series]{ex: taylor series of e and cos and sin}
Now it is important to address exponentials and sinusoids.  The these three functions have the following derivatives
\begin{align}
    \frac{\mathrm{d}}{\mathrm{d}x}\left[ \txte^{x} \right] &= \txte^x, \label{eq: derivative of e^x}
    \\
    \frac{\mathrm{d}}{\mathrm{d}x}\left[ \cos x \right] &= -\sin x, \label{eq: derivative of cos x}
    \\
    \frac{\mathrm{d}}{\mathrm{d}x}\left[ \sin x \right] &= \cos x, \label{eq: derivative of sin x}
\end{align}
It is of particular importance to find out what these functions look like in terms of their Taylor Series, because when we do, you can see how they actually are all related to each other.

We start with the exponential and we choose to center our series at $x = 0$. Then $a_0 = \txte^{0} = 1$.  Next, we find $a_1$.
\begin{align*}
    a_1 = \left. \frac{\mathrm{d}}{\mathrm{d}x}\left[ \txte^{x} \right] \right\vert_{x=0} = \txte^{x}\bigg{\vert}_{x=0} = 1.
\end{align*}
Then for $a_2$,
\begin{align*}
    a_2 = \frac{1}{2!}\left. \frac{\mathrm{d}^2}{\mathrm{d}x^2}\left[ \txte^{x} \right] \right\vert_{x=0} = \frac{1}{2}\left. \frac{\mathrm{d}}{\mathrm{d}x}\left[ \txte^{x} \right] \right\vert_{x=0} = \frac{1}{2}\txte^{x}\bigg{\vert}_{x=0} = \frac{1}{2}.
\end{align*}
And we can continue through, but since the derivative of $e^x$ is just $e^x$, then all the derivatives of $e^x$ evaluated at $x = 0$ will all be 1. Hence, for general $k$,
\begin{align*}
    a_k = \frac{1}{k!}\left. \frac{\mathrm{d}^k}{\mathrm{d}x^k}\left[ \txte^{x} \right] \right\vert_{x=0} = \frac{1}{k!}
\end{align*}
and therefore 
\begin{align}
    \txte^x = \sum_{n = 0}^\infty \frac{x^n}{n!} = 1 + x + \frac{1}{2}x^2 + \frac{1}{3!}x^3 + \dots, \label{eq: Taylor series e^x}
\end{align}

We do the same thing now for cosine. First, $a_0 = \cos 0 = 1$.  Next, we take a derivative.
\begin{align*}
    a_1 = \left. \frac{\mathrm{d}}{\mathrm{d}x}\left[ \cos x \right] \right\vert_{x=0} = -\sin x\bigg{\vert}_{x = 0} = 0.
\end{align*}
Now we take the second derivative,
\begin{align*}
    a_2 = \frac{1}{2!}\left. \frac{\mathrm{d}^2}{\mathrm{d}x^2}\left[ \cos x \right] \right\vert_{x=0} = \frac{1}{2}\left. \frac{\mathrm{d}}{\mathrm{d}x}\left[ -\sin x \right] \right\vert_{x=0} = \frac{1}{2}(-\cos x)\bigg{\vert}_{x = 0} = -\frac{1}{2}.
\end{align*}
Now with the third derivative,
\begin{align*}
    a_3 = \frac{1}{3!}\left. \frac{\mathrm{d}^3}{\mathrm{d}x^3}\left[ \cos x \right] \right\vert_{x=0} = \frac{1}{3!} \left. \frac{\mathrm{d}}{\mathrm{d}x}\left[ -\cos x \right] \right\vert_{x=0} = \frac{1}{3!} (\sin x)\bigg{\vert}_{x = 0} = 0
\end{align*}
And the fourth,
\begin{align*}
    a_4 = \frac{1}{4!} \left. \frac{\mathrm{d}^4}{\mathrm{d}x^4}\left[ \cos x \right] \right\vert_{x=0} = \frac{1}{4!}\left. \frac{\mathrm{d}}{\mathrm{d}x}\left[ \sin x \right] \right\vert_{x=0} = \frac{1}{4!}(\cos x)\bigg{\vert}_{x = 0} = \frac{1}{4!}
\end{align*}
And so the pattern in the coefficients alternates between being positive and negative for only even powers of $x^n$ and then totally vanish for all of the odd powers. Therefore, we can write the cosine Taylor Series as
\begin{align}
    \cos x = \sum_{n=0}^{\infty} \frac{(-1)^nx^{2n}}{(2n)!} = 1 - \frac{1}{2}x^2 + \frac{1}{4!}x^4 - \frac{1}{6!}x^6 + \dots, \label{eq: cosine Taylor Series}
\end{align}
where the $2n$ was introduced to just to make sure that all of the powers are even.

For the sine, we could go back through and find the sine coefficients in the same way, or we can exploit the fact that the derivative of the cosine function is the negative sine.  In other words,
\begin{align*}
    \sin x = - \frac{\mathrm{d}}{\mathrm{d}x}\left[ \cos x \right].
\end{align*}
Then we substitute the Taylor Series for cosine into the equation above to get
\begin{align*}
    \sin x &= - \frac{\mathrm{d}}{\mathrm{d}x}\left[ 1 - \frac{1}{2}x^2 + \frac{1}{4!}x^4 - \frac{1}{6!}x^5 + \dots \right]
    \\
    &= -\left[ - \frac{1}{2}\cdot 2x + \frac{1}{4!}\cdot 4x^3 - \frac{1}{6!}\cdot 6x^5 + \dots \right]
    \\
    &= x - \frac{1}{3!}x^3 + \frac{1}{5!}x^5 - \dots
\end{align*}
Thus, only the odd powers of $x^n$ survive and there is no constant term in the beginning.  Then, we can finalize the Taylor Series of the sine function as
\begin{align}
    \sin x = \sum_{n = 0}^\infty \frac{(-1)^n x^{2n+1}}{(2n+1)!} = x - \frac{1}{3!}x^3 + \frac{1}{5!}x^5 - \frac{1}{7!}x^7 + \dots, \label{eq: sine Taylor Series}.
\end{align}
\end{example}
\vspace{0.15in}

Normally we need to talk about the convergence of the Taylor Series, but unfortunately we don't have time to talk about that either. Convergence just guarantees that if we keep adding infinitely many terms together that we will eventually get a finite number. Luckily, in physics, when we need series to converge, they do, or something weird happens like we pass through a conducting interface.  The Taylor Series for exponentials and sinusoids do converge everywhere though, so no need to worry about that for right now. 

Typically, in physics, we almost always terminate the infinite series anyway, either because the other terms are negligible (as in multipole expansions), or because normalization makes us terminate them, \textit{et cetera}.  It really is rare to have to compute \textit{all} the coefficients for any given function in physics.  But knowing how to do it can lead to some pretty incredible results.  For example, one can \textit{prove} that $\txte^{i\theta} = \cos\theta + i\sin\theta$ using Taylor Series, as is done in Problem \ref{prob: Euler's Identity}.

\vspace{0.15in}
\begin{problem}[Euler's Identity]{prob: Euler's Identity}
In this problem, you will show that $\txte^{i\theta} = \cos\theta + i\sin\theta$ holds for all inputs $\theta$. We will split it up so it is a little easier to handle.
\begin{enumerate}[(a)]
    \item Start with $i^n$, where $i =\sqrt{-1}$.  Then $i^2 = -1$.  What is $i^3$? How about $i^4$? What is $i^5$? Detect a pattern when you substitute in $i^n$ for both even and odd powers $n$.
    \item Using \equaref{eq: Taylor series e^x}, substitute in $x = i\theta$ into $\txte^x$ and write out the first few terms.
    \item Combine all of the terms into ones that are multiplied by $i$ and those that are not.  Factor out all the $i$'s from the term with them in it.
    \item Compare this expression with the Taylor Series for cosine and sine, given by \equaref{eq: cosine Taylor Series} and \equaref{eq: sine Taylor Series}, respectively.
    \item Conclude $\txte^{i\theta} = \cos\theta + i\sin\theta$.
\end{enumerate}
\end{problem}
\vspace{0.15in}

\subsection{A Couple of Key (Linear) Differential Equations in Physics}
The final topic we are going to cover is some very, very basic ordinary differential equations.  However, the differential equations we will cover are ubiquitous in physics.  We will first show how to use calculus to model the number of radioactive particles in a sample as a function of time. Then, we will show how one particular differential equation implies oscillatory motion. Even if the context is not the same in one area of physics, the mathematics will be the same, and so the form of the solutions will be the same in all areas of physics.

The study of differential equations turns out to be the study of which guesses work best.  That is the truth of the matter: we start with a particular differential equation (an equation containing derivatives). Sometimes, some guesses work for some types of differential equations, while these same methods may not work for other types of equations.  Thus, being able to classify differential equations is usually key to getting closer to the solution. Here we will only discuss linear and homogeneous differential equations. The first term means that only all dependent variables and their derivatives have powers of one in the equation, while the second term means that the differential equations are equal to zero.  There are other types of differential equations and you will cover them when you take your first semester of introductory differential equations. 

\subsubsection{First Order: Radioactive Decay}
We start with a \textit{first order linear homoegeneous} differential equation of the form
\begin{align}
    \tderiv[t]{N} + kN(t) = 0 \Rightarrow \tderiv[t]{N} = -kN(t), \label{eq: radioactive decay equation}
\end{align}
where $k\in\mathds{R}$. Here, $N(t)$ may represent the number of radioactive particles left in some sample within a lab.  The equation it self says that the rate of change of the number of particles is proportional to the number of particles there are, while the negative sign denotes that the number of particles is decreasing.  To solve this equation, we use a method known as \textbf{separation of variables} which is one where we collect all of the dependent variables on one side of the equation and then collect all of the independent variables on the other. Thus,
\begin{align*}
    \frac{\mathrm{d}N}{N} = k\mathrm{d}t.
\end{align*}
Now we integrate both sides
\begin{align*}
    \int_{N(t_0)}^{N(t)} \frac{\mathrm{d}N^\prime}{N^\prime} = \int_{t_0}^t k\mathrm{d}t^\prime
\end{align*}
where the upper and lower bounds on either side of the equal side correspond to the same point.  The integral on the left can be computed using the logarithm function, while the integral on the right can be evaluated using Power Rule.  Then
\begin{align*}
    \ln N(t) - \ln N(t_0) = \ln \frac{N(t)}{N(t_0)}= kt - kt_0 = k(t - t_0)
\end{align*}
By exponentiating and then multiplying by $N(t_0)$, we arrive at our solution
\begin{align}
    N(t) = N(t_0)\txte^{k(t- t_0)}. \label{eq: solution radioactive decay}
\end{align}
Here, quantity $N(t_0)$ represents the total number of particles at the initial time $t_0$.  Thus, one question we may ask is \textquotedblleft how long does it take before half of the particles have decayed?\textquotedblright$\,$

To solve this problem, we simply say $N(t) = N(t_0)/2$ and solve for the length of time $t - t_0$ such that 
\begin{align*}
    \frac{1}{2}N(t_0) = N(t_0)\txte^{-k(t-t_0)} \Rightarrow -k(t-t_0) = \ln \frac{1}{2} \Rightarrow t-t_0 = \frac{\ln 2}{k}.
\end{align*}
This length of time $t-t_0$ is called the \textit{half-life} of the sample and then $k$ is called the \textit{decay rate}.  There is another quantity that we can define as well: $\tau = 1/k$.  Depending on the context, $\tau$ is either called the \textit{lifetime} of the radioactive particles, or a \textit{time constant} for the sample.  Either way, if we substitute in $\tau = 1/k$ into the solution, we have
\begin{align*}
    N(t) = N(t_0)\txte^{-\frac{t - t_0}{\tau}}.
\end{align*}
Thus, when $t - t_0 = \tau$, we have 
\begin{align*}
    N(\tau + t_0) = N(t_0)\txte^{-\frac{\tau + t_0 - t_0}{\tau}} = N(t_0)\txte^{-1}.
\end{align*}
Hence the lifetime represents the amount of the independent variable that can pass before the dependent variable decreases to $\txte^{-1} \approx 37$\% of its original value.

Another example of how this equation appears in physics is when one seeks the electric charge $Q$ on a discharging capacitor in an RC circuit as a function of time $t$.  In this case, the differential equation is 
\begin{align*}
    R\tderiv[t]{Q} + \frac{1}{C}Q(t) = 0 \Rightarrow \tderiv[t]{Q} = -\frac{1}{RC}Q(t),
\end{align*}
where $R$ is the resistance in the circuit and $C$ is the capacitance of the capacitor. Here the derivative represents the change in charge per unit time which is the current in the circuit.  Anyway, there is an \textit{isomorphism} between the equation governing radioactive decay and the discharging capacitor in an RC circuit | \textit{iso-} for \textquotedblleft same\textquotedblright$\,$ and \textit{-morphism} for \textquotedblleft structure.\textquotedblright$\,$ This isomorphism is in \textit{changing the letters} $N(t) \rightarrow Q(t)$ and $k \rightarrow 1/RC$. Since all we did is literally just change the symbols mathematically, \textit{nothing} happened to the differential equations and therefore \textit{nothing} happened to the solution.  Thus, the charge on the discharging capacitor is
\begin{align*}
    Q(t) = Q(t_0)\txte^{-\frac{t - t_0}{RC}}.
\end{align*}
Hence, the time constant for this circuit is $\tau = RC$ which is the time it takes for the capacitor to hold about $37\%$ of its original charge. But there are even more isomorphisms in physics such as current-delay in an RL circuit ($L$ is for self-inductance), attenuation of photons through an absorbing material, relaxation of bulk magnetization of a material that had just undergone nuclear magnetic resonance, \textit{et cetera}. But in all of these cases we change the letters and the meaning of the quantities, while keeping the mathematical structure the same.

\subsubsection{Second Order: Simple Harmonic Motion}
The next equation that we must cover is one that governs all simple harmonic motion. It is a \textit{second order linear homogeneous} differential equation, meaning that it has a second-order derivative in it rather than a first-order derivative.  This solution is of the form
\begin{align}
    \frac{\mathrm{d}^2\psi}{\mathrm{d}t^2} + \omega^2\psi(t) = 0 \Rightarrow \frac{\mathrm{d}^2\psi}{\mathrm{d}t^2} = - \omega^2\psi, \label{eq: simple harmonic motion DE}
\end{align}
where $\omega$ is a positive real number. What this equation says is that whenever acceleration (second derivative) of a function opposes that function, then the solution will abide by simple harmonic motion.  So why do we call it simple harmonic motion?  To see, we must solve the differential equation.

We start with our \textit{ansatz} (guess) of $\psi(t) = A\txte^{rt}$, where $A$ and $r$ are to-be-determined constant coefficients.  The reason why we guess an exponential is because its derivatives are fairly straightforward to compute.  All we need to remember is that this is actually a composite function, $g(f(t))$, where $f(t) = rt$ and $g(f) = \txte^{f}$.  Then by Chain Rule,
\begin{align*}
    \frac{\mathrm{d}}{\mathrm{d}t}\left[ Ag(f(t)) \right] = A\tderiv[t]{f}\tderiv[f]{g} = A(r)(\txte^{f}) = Ar\txte^{rt}.
\end{align*}
By taking another derivative, then we pick up another multiplicative $r$.  Hence,
\begin{align*}
    \frac{\mathrm{d}^2}{\mathrm{d}t^2}\left[ A\txte^{rt} \right] = Ar^2\txte^{rt}.
\end{align*}
Now we plug this result into \equaref{eq: simple harmonic motion DE} to see what values $A$ and $r$ have to be so that our guess was correct.
\begin{align*}
    Ar^2\txte^{rt} = -\omega^2A\txte^{rt}.
\end{align*}
Notice that both the $A$'s and $\txte^{rt}$ are on both sides of the equation, and so they can be divided out.  This means that $A$ can have \textit{any} value we want and every single one \textit{will} be a solution to the differential equation.  After dividing everything out, we have
\begin{align*}
    r^2 = -\omega^2 \Rightarrow r = \pm \sqrt{-\omega^2} = \pm \omega\sqrt{-1} = -\pm i\omega.
\end{align*}
Thus, when we find the \textit{characteristic roots}, $r$, we have \textit{two} solutions. So which one do we pick? It turns out that we have to choose \textit{both} since \textit{both} roots satisfy the differential equation.

The way in which we choose both roots is in a linear fashion, since derivatives are linear operators.  Thus, our guess solution becomes
\begin{align*}
    \psi (t) = A\txte^{i\omega t} + B\txte^{-i\omega t},
\end{align*}
where $A$ and $B$ are any coefficients that would have to be determined with some kind of \textbf{initial condition}. Now I am going to pull a little trick to get rid of the imaginary numbers.  Let 
\begin{align*}
    A = \frac{C - iD}{2},\;\; B = \frac{C + iD}{2},
\end{align*}
where $C$ and $D$ are real numbers. Then, using Euler's Identity, 
\begin{align*}
    \psi(t) &= A\txte^{i\omega t} + B\txte^{-i\omega t},
    \\
    &= \left( \frac{C - iD}{2}\right) \left( \cos \omega t + i\sin \omega t\right) + \left( \frac{C + iD}{2} \right)\left( \cos \omega t - i\sin \omega t\right)
    \\
    &= \left( \frac{C + C}{2} \right)\cos\omega t + \left(\frac{D+D}{2} \right)\sin\omega t + \left( \frac{-iD + iD}{2} \right)\cos\omega t + \left( \frac{iC - iC}{2} \right)\sin\omega t
    \\
    &= C\cos\omega t + D\sin\omega t.
\end{align*}
Thus our guess solution is 
\begin{align}
    \psi (t) = C\cos\omega t + D\sin\omega t. \label{eq: simple harmonic motion solution}
\end{align}
I leave it to you to plug this solution back into \equaref{eq: simple harmonic motion DE} to prove to yourself that this is indeed our solution. Hence, regardless of what the physical quantity $\psi$ represents, if it abides by any differential equation of the form \equaref{eq: simple harmonic motion DE}, then its solution will always look like \equaref{eq: simple harmonic motion solution}, which says that the solution will oscillate (wiggle) at an angular frequency given by $\omega$.

\section{Concluding Remarks}
In this chapter, we have thoroughly discussed the concepts behind one-dimensional calculus. It is my hope that all of the build-up at the beginning of the chapter helped you see calculus for being a rather intuitive form of mathematics and is a way for us to translate our observations of reality into a precise language.  Yes, calculus may get a little tricky every once in a while, but it was ultimately developed by Newton and Leibniz to explain the natural world. Thus, whenever things get a little unclear, try your best to visualize what might being going on | whether it be an object in motion or something that you prefer more. However, in my experience, it is almost always the algebra that gets to be untenable, not the calculus itself.

I have included a table with a few of the more general equations for your future reference in \tblref{tbl: calculus}.

\begin{table}
    \centering\small
    \caption{A summary of the important and general equations in one dimensional calculus. \label{tbl: calculus}}
    \scalebox{0.67}{
    \begin{tabular}{ccc}
        \hline\hline
        \textbf{Equation Description} & \textbf{Equation Formula} & \textbf{Text Reference}\\ \hline & & \\
        Definition of Average Velocity & $\vec{v}(t) = \dfrac{\vec{r}(t + \Delta t) - \vec{r}(t)}{\Delta t}$ & \equaref{eq: definition of average velocity}
        \\ && \\
        Definition of Average Acceleration & $\vec{a}(t) = \dfrac{\vec{v}(t + \Delta t) - \vec{v}(t)}{\Delta t}$ & \equaref{eq: definition of average acceleration}
        \\ && \\
        Position from Velocity & $\vec{r}(t + j\Delta t) = \vec{r}_0 + \sum\limits_{k = 0}^{j-1}\vec{v}(t + k\Delta t)\Delta t$ & \equaref{eq: position sum of velocity}
        \\ && \\
         Velocity from Acceleration & $\vec{v}(t + j\Delta t) = \vec{v}_0 + \sum\limits_{k = 0}^{j-1}\vec{a}(t + k\Delta t)\Delta t$ & \equaref{eq: velocity sum of acceleration}
         \\ && \\
         Sufficiently Large $N$ & $\dfrac{t_f - t_0}{N} < \epsilon \Rightarrow \dfrac{t_f - t_0}{\epsilon} < N$ & \equaref{eq: Delta t N limit}
         \\ && \\
         Limit of a Continuous Function & $\lim\limits_{x \rightarrow x^\prime} f(x) = f(x^\prime)$ & \equaref{eq: definition continuity limit}
         \\ && \\
         Limit of a Sum & $\lim\limits_{x\rightarrow x^\prime} \left[f(x) + g(x) \right] = \left[ \lim\limits_{x\rightarrow x^\prime}f(x) \right] + \left[ \lim\limits_{x\rightarrow x^\prime} g(x) \right]$ & \equaref{eq: limit of sum}
         \\ && \\
         Limit of a Product & $\lim\limits_{x\rightarrow x^\prime} \left[f(x)g(x) \right] = \left[\lim\limits_{x\rightarrow x^\prime} f(x) \right] \,\left[\lim\limits_{x\rightarrow x^\prime} g(x) \right]$ & \equaref{eq: limit of a product}
         \\ && \\
         Limit of a Composition & $\lim\limits_{x\rightarrow x^\prime} \left[ g(f(x)) \right] = g\left( \lim\limits_{x\rightarrow x^\prime} f(x)\right) $  & \equaref{eq: limit of composition of functions}
         \\ && \\
         Definition of Derivative & $\left. \dfrac{\mathrm{d}f}{\mathrm{d}x} \right\vert_{x = x^\prime} = \lim_{x\rightarrow x^\prime} \dfrac{f(x) - f(x^\prime)}{x - x^\prime}$ & \equaref{eq: definition of derivative}
        \\ && \\
         Linearity of the Derivative & $\dfrac{\mathrm{d}}{\mathrm{d}x}\left[ af(x) + bg(x) \right] = a\dfrac{\mathrm{d}f}{\mathrm{d}x} + b\dfrac{\mathrm{d}g}{\mathrm{d}x},\; \forall a,b\in\mathds{R}$ & \equaref{eq: derivative is linear everywhere}
         \\ && \\
         Product Rule for Differentiation & $\dfrac{\mathrm{d}}{\mathrm{d}x}\left[ f(x)g(x) \right] = \left( \dfrac{\mathrm{d}f}{\mathrm{d}x} \right)\,g(x) + f(x)\,\left( \dfrac{\mathrm{d}g}{\mathrm{d}x} \right)$ & \equaref{eq: product rule}
         \\ && \\
         Chain Rule for Differentiation & $\dfrac{\mathrm{d}}{\mathrm{d}x}\left[ g(f(x)) \right] =  \dfrac{\mathrm{d}f}{\mathrm{d}x}\, \dfrac{\mathrm{d}f}{\mathrm{d}f}$ & \equaref{eq: Chain Rule differentiation}
         \\ && \\
         First Fundamental Theorem of Calculus & $f(x) - f(x^\prime) = \int\limits_{x^\prime}^x \mathrm{d}s\, \dfrac{\mathrm{d}f}{\mathrm{d}s}$ & \equaref{eq: first fundamental theorem of calculus} 
         \\ && \\
         Second Fundamental Theorem of Calculus & $ \dfrac{\mathrm{d}f}{\mathrm{d}x} =  \dfrac{\mathrm{d}}{\mathrm{d}x} \int\limits_{x^\prime}^x \mathrm{d}s\, \dfrac{\mathrm{d}f}{\mathrm{d}s}$ & \equaref{eq: second fundamental theorem of calculus}
         \\ && \\
         Definition of Riemann Integration & $\int\limits_{x^\prime}^x \mathrm{d}s\, f(s) = \lim\limits_{N\rightarrow \infty} \sum\limits_{j=0}^{N-1} \Delta x_N\, f(x^\prime + j\Delta x_N)$ & \equaref{eq: definition of riemann integration}
         \\ && \\
         Switching Bounds = Integrating Backwards & $\int\limits_{x^\prime}^x \mathrm{d}s\, f(s) = - \int\limits_x^{x^\prime} \mathrm{d}s\, f(s)$ & \equaref{eq: integrate switch bounds}
         \\ && \\
         Linearity of Integral & $\int\limits_{x^\prime}^x \mathrm{d}s\, \left[ af(s) + bg(s)\right] = a\int\limits_{x^\prime}^x \mathrm{d}s\, f(s) + b\int\limits_{x^\prime}^x \mathrm{d}s\, g(s),\; \forall a,b\in\mathds{R}$ & \equaref{eq: linear integrals}
         \\ && \\
         Integration by Parts & $\int u\,\mathrm{d}v  = uv - \int v\,\mathrm{d}u$ & \equaref{eq: indefinite integration by parts}
         \\ && \\
         Change of Variables & $\int \mathrm{d}x\,\dfrac{\mathrm{d}u}{\mathrm{d}x}\,\dfrac{\mathrm{d}g}{\mathrm{d}u} = g(u(x)) + C = \int \mathrm{d}u\, \dfrac{\mathrm{d}g}{\mathrm{d}u}$ & \equaref{eq: indefinite u substitution}
         \\ && \\
         Taylor Series Expansion & $f(x) = \sum\limits_{n = 0}^\infty \dfrac{1}{n!}\, \left. \dfrac{\mathrm{d}^kf}{\mathrm{d}x^k}\right\vert_{x = c} (x-c)^n$ & \equaref{eq: taylor series}
         \\
         & &
         \\
         \hline\hline
         & &
    \end{tabular}
    }
\end{table}








\newpage



\setcounter{example}{0}
\setcounter{problem}{0}

\chapter{Fourier Analysis}

Fourier analysis is one of the most important tools available for physicists.  It is used everywhere from signal processing, to image construction, to finding solutions to the Schr\"odinger equation.  The reason why it is so useful is that the Fourier Transform represents a \emph{duality} transformation between different representations (a real-space representation and a reciprocal space representation).  In other words, it retains all information but delivers it from a new angle that is usually much easier to understand. For example, if we were studying a time-dependent signal from a spring system, we can decompose it in terms of its fundamental frequencies (its wiggly parts) by taking the Fourier Transform.  Thus we could talk about the signal in terms of its time-dependence OR we could, equally, talk about the signal in terms of the ways in which the springs are oscillating | in fact, one way of talking about a system implies the other.  

My goal for this chapter is to introduce you to introductory Fourier Analysis in one dimension.  I will build the Fourier Transform from a Fourier Series, but I work to develop your intuition by using analogies with how to describe a vector components from its components. It is not possible to talk about many of the interesting properties of Fourier Analysis in a single chapter, so I picked a few that have immediate physical significance.  I have included some hyperlinked references for any readers interested in learning more than I have space for below. I have been pretty explicit with the algebraic manipulations I perform in the Examples, but I do this so you can use these examples to see how to apply the mathematical framework of Fourier Analysis to deduce interesting mathematics and physics.  Usually these examples have \textquotedblleft corollaries\textquotedblright $\;$ and so I left these for you to practice (although I did give many hints to help you). It is my hope that this chapter to helps you see Fourier Analysis as an indispensible tool for you as you study the behavior of the universe.

\section{Frequencies versus Wavenumbers}
Throughout this chapter, I will talk about frequencies and wavenumbers, but I will use the term \textquotedblleft frequency\textquotedblright$\;$ a lot more than I will use \textquotedblleft wavenumber\textquotedblright.  I do this intentionally, as the word \textquotedblleft wavenumber\textquotedblright $\;$ is cumbersome and just sounds weird (at least to me). Some people use \textquotedblleft spatial frequency\textquotedblright $\;$ instead of \textquotedblleft wavenumber\textquotedblright, but I personally think that is just as sludgy and verbose. The reason why we must talk about either is that Fourier Analysis occurs in the \emph{reciprocal space} of whatever real-space thing you are talking about, and the units we use to talk about the reciprocal space depend on the real space. This is due to, as you will see, the sinusoidal functions that are used in Fourier Analysis because the argument of a sinusoid must be unitless like the quantities $\omega t$, $\kappa x$, etc, that you may find in $\sin\omega t$ or $\cos\kappa x$.  example, if our real space were time | as is the case when we have a time-dependent signal from an oscilloscope measured in seconds | the reciprocal, or Fourier, space would have units of frequency: Hertz.  Meanwhile, if our real space signal has units of length in meters, then the reciprocal space units would be wavenumbers measured in meters$^{-1}$. Furthermore, unless I am absolutely explicit about it, we will treat radians as unitless measures (because they are\footnote{In radians, an angle is used as a proportionality constant to convert between a circle's radius and an arc's length. Since both have units of length, then an angle has no units. In other words, a radian is not a real unit like time, length, mass, temperature, etc.}). So when I say frequency or wavenumber, I really mean angular frequency or angular wavenumber.

I choose to use \textquotedblleft frequency\textquotedblright$\;$ (\textquotedblleft angular\textquotedblright$\;$ not included) simply because it sounds like an actual English word, and I use it to help with the interpretation of the mathematics at the risk of being dimensionally sloppy. It is usually easier to understand the math without having to repeatedly reread a sentence that is trying explain it because the sentence has impenetrable wording.  In the future though, whenever you do Fourier Analysis in the field, I strongly recommend quickly figuring out the units of your reciprocal space on your own so you don't get confused later on in your data analysis.  There may be situations when you need to explicitly differentiate between frequency and angular frequency or wavenumber and angular wavenumber, but those are typically special cases in physics. 

\section{Fourier Series as a Linear Combination}
In this section, we will derive the coefficients for a Fourier Series using the same tools as we would to find the components of a vector.  This is done to try to help you develop an intuition about how Fourier Analysis really does just take a known function and sees how strongly each oscillating mode is responsible for its makeup, just like how you would talk about which axes contribute most to the direction and magnitude of a vector.  Then, we will use the Fourier Series as a stepping stone to the Fourier Transform.

\subsection{Review of Linear Combinations}
Recall that any vector $\vec{v}$ in an $n$-dimensional vector space $\mathcal{V}_n$ can be expressed as 
\begin{align}
    \vec{v} = \sum_{j = 1}^n v_j \Hat{u}_j = v_1\Hat{u}_1 + \dots + v_n\Hat{u}_n \;, \label{LinearCombination}
\end{align}
where each $\Hat{u}_j$ is a basis vector of $\mathcal{V}_n$.  For example, if $\mathcal{V}_n$ were the $xy$-plane $\mathds{R}^2$, then we could write $\vec{v}\in \mathcal{V}_2 = \mathds{R}^2$ as
\begin{align*}
    \vec{v} = v_1\Hat{x} + v_2\Hat{y} \; .
\end{align*}
Each coefficient $v_j$ in Eq. \ref{LinearCombination} can then be interpreted geometrically as the length of $\textbf{v}$ pointing in the direction of the $\Hat{u}_j$ basis vector.  We can then compute each component $v_j$ by exploiting the orthogonality of the basis vectors, namely
\begin{align}
    \Hat{u}_j \cdot \Hat{u}_k = \begin{cases}
    1, & \textrm{if } j = k 
    \\
    0, & \textrm{if } j \neq k
    \end{cases} \;. \label{Orthonormality of u}
\end{align}
If we then dot $\vec{v}$ from Eq. \ref{LinearCombination} into $\Hat{u}_k$ we find
\begin{align}
    \vec{v}\cdot\Hat{u}_k = \left( \sum_{j = 1}^n v_j \Hat{u}_j \right) \cdot\Hat{u}_k = \sum_{j = 1}^n v_j \left(\Hat{u}_j\cdot\Hat{u}_k\right)
\end{align}
But by Eq. \ref{Orthonormality of u}, the dot product in the parenthesis in the second equality above is zero unless the counting index $j$ is equal to the index-in-question $k$. Thus 
\begin{align}
    \vec{v}\cdot \Hat{u}_k = 0 + \dots + v_k + \dots + 0 = v_k \;.
\end{align}
If we further recall that a unit vector $\Hat{u}_j$ is a normalized vector $\vec{u}_j$, then 
\begin{align}
    v_k = \frac{\vec{v}\cdot\vec{u}_k}{\Vert \vec{u}_k \Vert} \;, \label{Value of vk}
\end{align}
which shows algebraically that $v_k$ is the length of $\vec{v}$ pointing in the direction of $\Hat{u}_k$.  

In the next section, we abstract Eq. \ref{Value of vk} for application in Fourier analysis. An important takeaway from Eq. \ref{Orthonormality of u} is that when $j\neq k$, the dot product between $\Hat{u}_j$ or $\Hat{u}_k$ vanishes. If we recall that $\vec{a}\cdot \vec{b} = ab\cos\theta$, for any two vectors $\vec{a}$ and $\vec{b}$, then if the dot product is zero, we conclude $\theta = 90^{\textrm{o}}$, where $\theta$ is the angle between the two vectors\footnote{Of course either $\vec{a}$ or $\vec{b}$ could be the zero vector for $\vec{a}\cdot\vec{b} = 0$, but that would be boring.}. Hence, the \emph{orthogonality condition}, Eq. \ref{Orthonormality of u}, shows that none of the basis vectors point in the same direction as they are all perpendicular to each other. So, geometrically, all orthogonal basis vectors are totally distinct | there is no way to make one from the others\footnote{This property actually comes from the fact that the set of $\Hat{u}_j$ is a basis for $\mathcal{V}_n$, but the geometric picture here is really important so I chose to emphasize it. }.

\subsection{The Fourier Series}
Consider some function\footnote{A LOT of people also require that $f(x)$ be periodic. However, this condition is not strictly necessary for physical signals, such as those measured from an oscilloscope in a nuclear magnetic resonance experiment. Thus, I do not require it for this section. In general, the periodicity in $f$ does pose an issue, and this is covered in Section \ref{subsec: FT infinite domain}.} $f(x)$ over the interval $0 \leq x \leq L$.  We will assume that this function is integrable over this interval, which is generally true for most physical functions of space and time.  Oftentimes though, physical signals do not come in nice little algebraic expressions. They are typically more complicated, unfortunately. To deal with this problem, we will attempt to write these functions as nice little algebraic expressions anyway.  But to do this, we trade the complicated expression for a potentially infinite sum.

The easiest elementary functions that wiggle are sinusoids, i.e. sines and cosines.  We choose to analyze $f(x)$ in terms of wiggly parts because we in physics typically analyze systems that have oscillatory behavior | springs and waves are kind of our jam. Hence, we choose to write $f(x)$ as a linear combination of totally distinct sinusoids.
\begin{align}
    f(x) = \sum_{\kappa}  a_\kappa \cos(\kappa x) + b_\kappa \sin(\kappa x) \;. \label{FourierLinearCombination}
\end{align}
Here, the summation is taken over all possible values of $j$ such that the \emph{functions} $\cos(\kappa x)$ and $\sin(\kappa x)$ act as our basis vectors as in Eq. \ref{LinearCombination}.  We assume that the coefficients $a_\kappa$ and $b_\kappa$ are, in general, functions of the label $\kappa$. So we are left to ask: what can $\kappa$ be?

To answer this question, we recall that our basis vectors need to satisfy some kind of an orthogonality condition so that we may eventually determine the coefficients $a_\kappa$ and $b_\kappa$. However, each \textquotedblleft vector\textquotedblright $\,$ takes on a different length at any particular $x$ since the sine and cosine are functions of $x$. Thus we cannot initially proceed with the calculations as given by Eq. \ref{Value of vk} for basis vectors $\vec{u}_k$ with fixed lengths $\Vert\vec{u}_k\Vert$. What we decide to do instead is to integrate over the interval $0\leq x\leq L$ to account for all possible vector lengths\footnote{We are allowed to do this since $f(x)$ is integrable by assumption. Also, fun fact, this differs from the average of a function over $0\leq x\leq L$ by a factor of $1/L$.}  The domain for integration and the orthogonality condition then gives us some intuition about what $\kappa$ should be from the harmonic modes of sine and cosine:
\begin{align}
    \kappa_j = \frac{j\pi}{L} \;,\; j = 0,1,2,\dots\;.
\end{align}
We also note that our generalized orthogonality conditions for our basis vectors are then 
\begin{align}
    \frac{2}{L}\int_0^L \cos\left(\frac{j\pi x}{L}\right)\cos\left(\frac{k\pi x}{L}\right)\,\textrm{d}x &= \begin{cases}
    1, & \textrm{if } j = k \neq 0
    \\
    2, & \textrm{if } j = k = 0
    \\
    0, & \textrm{if } j \neq k
    \end{cases} \;, \label{Cosine Orthogonality}
    \\
     \frac{2}{L}\int_0^L \sin\left(\frac{j\pi x}{L}\right)\sin\left(\frac{k\pi x}{L}\right)\,\textrm{d}x &= \begin{cases}
    1, & \textrm{if } j = k 
    \\
    0, & \textrm{if } j \neq k
    \end{cases} \;, \label{Sine Orthogonality}
    \\
     \frac{2}{L}\int_0^L \cos\left(\frac{j\pi x}{L}\right)\sin\left(\frac{k\pi x}{L}\right)\,\textrm{d}x &= 0 \;. \label{Sine Cosine Orthogonality}
\end{align}
These integrals are fairly straightforward to evaluate, and I strongly recommend that you do evaluate them on your own at least once. It will help you see why there are those factors of $2/L$ in front of the integrals. To evaluate them, use the product-to-sum formulas given in \cite{wikipedia_Product2Sum}. 

\vspace{0.15in}
\begin{problem}[Orthogonality Relations]{prob: orthogonality relations}
Using either integration by parts (\equaref{eq: definite integration by parts}) or a product-to-sum formula \cite{wikipedia_Product2Sum}, verify each of the orgthogonality relations Eqs. \ref{Cosine Orthogonality}, \ref{Sine Orthogonality}, \ref{Sine Cosine Orthogonality}.
\end{problem}
\vspace{0.15in}

Thus the linear combination for $f(x)$ becomes
\begin{align}
   f(x) = a_0 + \sum_{j=1}^\infty a_j\cos\left(\frac{j\pi x}{L}\right) + b_j\sin\left(\frac{j\pi x}{L}\right)\,,\; 0\leq x\leq L  \label{Full Fourier Series}
\end{align}
which is the definition of the Fourier Series for $f(x)$ (note that the $b_0$ term vanishes because $\sin(0) = 0$). From the orthogonality conditions given by Eqs. \ref{Cosine Orthogonality}, \ref{Sine Orthogonality}, and \ref{Sine Cosine Orthogonality}, we can determine the coefficients $a_j$ and $b_j$ using a similar technique that we used for Eq. \ref{Value of vk}.

First, we will find $b_j$ explicitly, and then I will leave you to derive the expressions for $a_0$ and $a_j$ on your own. We then take our generalized \textquotedblleft dot product\textquotedblright of $f(x)$ with $\sin(\kappa_j x)$:
\begin{align}
    \frac{2}{L} \int_0^L f(x) \sin\left(\frac{k\pi x}{L}\right)\,\textrm{d}x &= \frac{2}{L}\int_0^L\left\lbrace a_0 + \sum_{j=1}^\infty a_j\cos\left(\frac{j\pi x}{L}\right) + b_j\sin\left(\frac{j\pi x}{L}\right) \right\rbrace \sin\left(\frac{k\pi x}{L}\right)\,\textrm{d}x \nonumber 
    \\
    &= a_0 \cdot\frac{2}{L}\int_0^L \sin\left(\frac{k\pi x}{L}\right)\,\textrm{d}x \nonumber \\
    &\;\;\;\;+ \sum_{j=1}^\infty a_j \cdot \frac{2}{L}\int_0^L \cos\left(\frac{j\pi x}{L}\right)\sin\left(\frac{k\pi x}{L}\right)\,\textrm{d}x \nonumber \\
    &\;\;\;\;+ \sum_{j=1}^\infty b_j \cdot \frac{2}{L}\int_0^L sin\left(\frac{j\pi x}{L}\right)\sin\left(\frac{k\pi x}{L}\right)\,\textrm{d}x \nonumber 
    \\ 
    &= a_0 \cdot 0 + \sum_{j=1}^\infty a_j \cdot 0 + b_k +
    \sum_{j\neq k} b_j \cdot 0 \nonumber
    \\
    &= b_k, \;\; 0\leq x\leq L. \label{Value for b_k}
\end{align}
The formulas for $a_0$ and $a_j$ are then 
\begin{align}
    a_0 &= \frac{1}{L} \int_0^L f(x)\,\textrm{d}x,\;\; 0\leq x\leq L, \label{Value for a_0}
    \\
    a_{j\geq 1} &= \frac{2}{L} \int_0^L f(x)\cos\left(\frac{j\pi x}{L}\right)\,\textrm{d}x, \;\; 0\leq x\leq L. \label{Value for a_j}
\end{align}
I would like to emphasize that the derivations for $a_0$ and $a_j$ are very similar to the one that found $b_k$.  Also, note that $j$ and $k$ are just indices that represent some whole number, so to find $b_j$ all you need to do is replace $k$ with $j$ in Eq. \ref{Value for b_k}.

\vspace{0.15in}
\begin{problem}[Fourier Coefficients]{prob: Fourier Coefficients}
Using the same approach as I did to find $b_k$ with \equaref{Value for b_k}, derive \equaref{Value for a_0} and \equaref{Value for a_j}.  Note there are two cases because of the extra case $(j = k = 0)$ in the cosine orthogonality relation. 
\end{problem}
\vspace{0.15in}

Once these coefficients are found, the Fourier Series of $f(x)$ is known exactly, even though this is usually easier said than done. There are many more things to be said about Fourier Series, such as the coefficients for intervals other than $0\leq x\leq L$, rules for convergence, Gibbs' sine integrals, and differentiability, but these are all beyond the scope of this Chapter.  However, if you are interested, you can find more of this information at the following references \cite{POMN_ConvergenceFourierSeries,wikipedia_GibbsFunctions}. Although the formulas above hold generally, it helps to see a couple examples done out explicitly to gain more intuition as to how the Fourier Series is related to oscillations.

\subsection{Calculating the Fourier Series}

\subsection*{The Fourier Series of Sinusoids}
For the first example, consider the function $f(x) = A\cos(6\pi x) + B\sin(\pi x)$ over the interval $0\leq x\leq 1$, where $A$ and $B$ are known constants. 

\vspace{0.15in}
\begin{example}[A Tale of Two Frequencies]{ex: Tale of Two Frequencies}
We seek the Fourier Series of $f(x)$ given by Eq. \ref{Full Fourier Series} where $L = 1$.  We first evaluate $b_j$.
\begin{align*}
    b_j &= 2\int_0^1 f(x)\sin\left(j\pi x\right)\,\textrm{d}x 
    \\
    &= 2 \int_0^1 \left[ A\cos(6\pi x) + B\sin(\pi x) \right]\sin\left(j\pi x\right)\,\textrm{d}x 
    \\
    &=2\int_0^1 A\cos(6\pi x)\sin\left(j\pi x\right)\,\textrm{d}x + 2\int_0^1 B\sin(\pi x) \sin\left(j\pi x\right)\,\textrm{d}x\;.
\end{align*}
By Eq. \ref{Sine Cosine Orthogonality}, the $A$ term vanishes completely, while from Eq. \ref{Sine Orthogonality}, the only $B$ term that is nonzero is the one for which $j = 1$. Thus,
\begin{align*}
    &b_1 = B\cdot 2\int_0^1 \sin^2 \left(\pi x\right)\, \textrm{d}x = B,
    \\
    &b_{j\neq 1} = 0.
\end{align*}
By similar reasoning for $a_j$, all $B$ terms vanish and the only nonvanishing $A$ term is the one where $j = 6$. Thus,
\begin{align*}
    &a_6 = A\cdot 2\int_0^1 \cos^2 \left(6\pi x\right)\, \textrm{d}x = A,
    \\
    &a_{j\neq 6} = 0.
\end{align*}
By substitution into Eq. \ref{Full Fourier Series}, we obtain
\begin{align*}
    f(x) &=  a_0 + \sum_{j=1}^\infty a_j\cos\left(j\pi x\right) + b_j\sin\left(j\pi x\right)
    \\
    &= 0 + A\cos(6\pi x) + \sum_{j\neq 6} 0\cdot\cos(j\pi x) + B\sin(\pi x) + \sum_{j\neq 1} 0\cdot \sin(j\pi x)
    \\
    &= A\cos(6\pi x) + B\sin(\pi x),
\end{align*}
as the Fourier Series.
\end{example}
\vspace{0.15in}

Notice that the Fourier Series of THIS function is the function itself! This is actually reassuring, though.  We started with a function that we knew had very specific angular frequencies that built it up | $6\pi$ in the cosine and $\pi$ in the sine | and then the Fourier Series told us that $f(x)$ is made up of exactly those frequencies and no others! Also, suppose that $A\gg B$ (so maybe $A = 1000$ and $B = 0.001$). Then $f(x) \approx A\cos(6\pi x)$. Notice then that the coefficient $\vert a_6\vert \gg \vert b_1\vert > a_{j\neq 6} = b_{j\neq 1}$. \textbf{We conclude then that the relative size of the Fourier coefficients} $a_j$ \textbf{and} $b_j$ \textbf{tells us how strongly one frequency dominates the rest.} Hence we can generalize this idea though to more complicated functions to say that if the Fourier coefficients have a peak, then the frequency at which the maximum is achieved is the dominating frequency of our function.  In practice, physicists refer to these as characteristic frequencies or wavelengths.

\subsubsection*{Quantum Mechanics or Fourier Analysis?}
Consider the function $f(x) = \delta(x-x_0),\; 0 < x_0 < L$, on the interval $0\leq x < L$, where the Dirac-$\delta$ function has the following properties:
\begin{align*}
    \delta(x-x_0) &= \begin{cases}
    0,\;&\textrm{if } x\neq x_0
    \\
    \infty,\;&\textrm{if }x = x_0
    \end{cases},
    \\
    \int_{-\infty}^{\infty} \delta(x-x_0)\,\textrm{d}x &= 1,
    \\
    \int_{-\infty}^{\infty} g(x)\delta(x-x_0)\,\textrm{d}x &= g(x_0)
\end{align*}
By the first two properties, this function then is integrable and only has one point where it is nonzero, namely at $x = x_0$. The third property is called the \emph{Sifting Property} of the Dirac-$\delta$ function and holds as long as $g(x)$ is continuous at $x_0$ (try to prove it using the first two properties and the idea that an integral is just a sum of areas of rectangles). This function is \textquotedblleft perfectly localized\textquotedblright, meaning it is only nonzero at $x = x_0$. In physics, if a particle or wave were exactly (out to infinite decimal places) at a point $x_0$, then a Dirac-$\delta$ would be involved in some way, and we would say that there is no error or uncertainty in our measurement of $x_0$.  We seek the Fourier Series representation of $f(x)$.

\vspace{0.15in}
\begin{example}[Fourier Series of a Localized Function]{ex: Fourier series localized function}
This time we will start with the $a_j$ coefficients because we did the $b_j$ out for the last example.  We begin with $a_0$ via Eq. \ref{Value for a_0}.
\begin{align*}
    a_0 &= \frac{1}{L}\int_0^L f(x)\,\textrm{d}x
    \\
    &= \frac{1}{L}\int_0^L \delta(x-x_0)\,\textrm{d}x
    \\
    &= 0 + \frac{1}{L}\int_0^L \delta(x-x_0)\,\textrm{d}x + 0
    \\
    &= \frac{1}{L}\int_{-\infty}^0 0\,\textrm{d}x + \frac{1}{L}\int_0^L \delta(x-x_0)\,\textrm{d}x + \frac{1}{L}\int_L^{\infty} 0\,\textrm{d}x
    \\
    &= \frac{1}{L}\int_{-\infty}^0 \delta(x-x_0)\,\textrm{d}x + \frac{1}{L}\int_0^L \delta(x-x_0)\,\textrm{d}x + \frac{1}{L}\int_L^{\infty} \delta(x-x_0)\,\textrm{d}x
    \\
    &= \frac{1}{L}\int_{-\infty}^{\infty} \delta(x-x_0)\,\textrm{d}x
    \\
    &= \frac{1}{L}.
\end{align*}
The way in which we used the integrability property of the Dirac-$\delta$ function with the addition of zeros is a useful technique to help us evaluate Dirac-$\delta$ integrals straight from the defining properties above. Now we evaluate $a_{j\geq 1}$ with Eq. \ref{Value for a_j}.
\begin{align*}
    a_{j\geq} &= \frac{2}{L}\int_0^L f(x)\cos\left(\frac{j\pi x}{L}\right)\,\textrm{d}x
    \\
    &= \frac{2}{L}\int_0^L \delta(x-x_0)\cos\left(\frac{j\pi x}{L}\right)\,\textrm{d}x
    \\
    &= \frac{2}{L}\int_{-\infty}^{\infty} \delta(x-x_0)\cos\left(\frac{j\pi x}{L}\right)\,\textrm{d}x
    \\
    &= \frac{2}{L}\,\cos\left(\frac{j\pi x_0}{L}\right).
\end{align*}
In the last step we made use of the Sifting Property. It follows then by this same method that
\begin{align*}
    b_j = \frac{2}{L}\,\sin\left(\frac{j\pi x_0}{L}\right).
\end{align*}
(You should do it out for the practice.) And therefore the Fourier Series representation of $f(x)$ is
\begin{align*}
    f(x) &= \frac{1}{L}\left\lbrace1 + 2\sum_{j=1}^{\infty} \left[ \cos\left(\frac{j\pi x_0}{L}\right)\cos\left(\frac{j\pi x}{L}\right) + \sin\left(\frac{j\pi x_0}{L}\right)\sin\left(\frac{j\pi x}{L}\right)\right]\right\rbrace
    \\
    &= \frac{1}{L}\left\lbrace 1 + 2\sum_{j=1}^{\infty} \cos\left[\frac{j\pi (x-x_0)}{L} \right] \right\rbrace
    \\
    &= \frac{1}{L}\left\lbrace -1 + 2\sum_{j=0}^{\infty} \cos\left[\frac{j\pi (x-x_0)}{L} \right] \right\rbrace .
\end{align*}
Here we have made use of the cosine-of-a-sum formula $$\cos(a\pm b) = \cos a\cos b \mp \sin a\sin b.$$
\end{example}
\vspace{0.15in}

The important thing to notice about this example is that we took a representation that was perfectly localized, $\delta(x-x_0)$, but we got a representation out that does not look to be localized in the slightest, especially not when it comes to which angular wavenumbers, $\kappa_j = j\pi/L$, are preferred. In fact, when we take the limit $x_0 \rightarrow 0^+$, we find
\begin{align*}
    \lim_{x_0 \rightarrow 0^+} f(x) = \frac{1}{L}\left[ -1 + 2\sum_{j=0}^{\infty} \cos\left(\frac{j\pi x}{L} \right) \right] = \frac{1}{L}\left[ -1 + 2\sum_{j=0}^{\infty} 1\cdot\cos\left(\frac{j\pi x}{L} \right) \right].
\end{align*}
I chose to emphasize that 1 in front of the $\cos$ in the last equality to show you that every single coefficient in the sum is weighted equally (with the exception of $a_0$ which is only $1/2$ of the others).  Remember that if this were a particle with a position $x$, then our perfectly localized particle has zero uncertainty associated with where it is located. Meanwhile, if we recall from quantum mechanics that the momentum is $p = \hbar \kappa$, then we conclude that we have an infinite uncertainty associated with what momentum our particle has, since it seems to require all possible momenta equally ($0\leq j< \infty$) to perfectly localize it at $x_0\rightarrow 0^+$. In other words, if we know position with perfect precision, we know nothing about the momentum | i.e. we just found an example of the Heisenberg Uncertainty Principle!

Now is this really the uncertainty principle, or is it simply a coincidence? It turns out that this \textquotedblleft loss of perfect localization\textquotedblright$\;$ is a mathematical consequence of having non-commuting linear operators which essentially just means that perfect localization cannot exist for both operator representations\footnote{For anybody interested, this means that the \emph{eigenstates} of the non-commuting operators are not identical}. Since linear operators are not strictly physical things, but they are still mathematical objects, then we conclude that it is the property of the mathematics we use which can produce the Heisenberg Uncertainty Principle. However, as soon as we drop in the $\hbar$ in front of all the $\kappa$, we find quantum mechanics | an absolutely experimentally-verified physical thing.  So this relationship is definitely non-accidental, at least in terms of how we currently know the universe behaves.  The beauty of this example is it shows that this behavior can be seen mathematically as well as experimentally!

\section{Fourier Transforms}
Now that we have the tools to talk about reciprocal representations of any integrable function, we will try to generalize them further so that we may call upon Fourier Analysis often when we study the laws that govern the universe.  This section is build based a lot on the framework developed in the last section, so be sure to refer back to it if you get stuck.

\subsection{The Case of the Infinite Domain \label{subsec: FT infinite domain}}
The great thing about the Fourier Series is that it converges exactly to any piece-wise continuous function $f(x)$ over its domain of definition \cite{POMN_ConvergenceFourierSeries}\footnote{The Fourier Series converges to the average of the left and right limits at each discontinuity \cite{POMN_ConvergenceFourierSeries}.}.  However, the Fourier Series is a function of sines and cosines | both of those are defined over an infinite domain.  Thus, the Fourier Series itself is defined over an infinite domain, even if our original function $f(x)$ is not. Furthermore, the fact that sines and cosines are $2\pi$-periodic, then our Fourier Series will also necessarily be periodic; the Fourier Series over $0\leq x\leq L$ in Eq. \ref{Full Fourier Series} is $2L$-period in $x$ as $j\pi(x+2L)/L = j\pi x/L + 2\pi j$.  Sometimes though, we need to talk about $f(x)$ without these periodic extensions. In other words, sometimes we must talk about $f(x)$ over an infinite domain otherwise we might misrepresent our data or include non-physical elements in our work. Hence we seek a linear combination of oscillating functions similar to Eq. \ref{Full Fourier Series} to represent $f(x)$ but over $-\infty \leq x \leq \infty$ instead of just $0\leq x\leq L$.

\subsection{The Complex Fourier Series}
Let's start again with Eq. \ref{FourierLinearCombination}, but this time let's assume that we can have negative frequencies.  We actually could have used negative frequencies before, but based on the integrals for the coefficients $a_j$ and $b_j$, the negative frequencies would not yield \emph{linearly independent} coefficients, so nothing is lost by ignoring them in the full Fourier Series in Eq. \ref{Full Fourier Series}.  Unfortunately, as they stand on their own, the sine and cosine function hide negative frequencies as they are functions with odd and even symmetry, respectively.  In other words,
\begin{align}
    \sin(-\theta) &= -\sin(\theta),
    \\
    \cos(-\theta) &= \cos(\theta),
\end{align}
which is a property that can be seen directly from comparing clockwise ($-\theta$) and counter-clockwise ($+\theta$) rotations around a unit circle.  To handle the negative frequencies explicitly, we make use of the Euler identity for imaginary exponentials:
\begin{align}
    \txte^{i\beta} = \cos\beta + i\sin\beta.
\end{align}
(Quick little exercise: use the symmetry properties of the sinusoids to show that $\txte^{-i\beta} = \cos\beta - i\sin\beta$.)  We can use $\txte^{i\beta}$ and its complex conjugate $\txte^{-i\beta}$ to derive an expression for cosine and sine (its real and imaginary parts, respectively).  We will do cosine out and then I recommend you follow the same steps to find the sine.
\begin{align}
    \txte^{i\beta} + \txte^{-i\beta} = \cos\beta + i\sin\beta +  \cos\beta - i\sin\beta = 2\cos\beta \Rightarrow \cos\beta = \frac{\txte^{i\beta} + \txte^{-i\beta}}{2}.
\end{align}
By almost identical reasoning,
\begin{align}
    \sin\beta = \frac{\txte^{i\beta} - \txte^{-i\beta}}{2i}.
\end{align}
We now substitute these into Eq. \ref{FourierLinearCombination} to find a suitable linear combination for $f(x)$:
\begin{align}
    f(x) = \sum_{\kappa} a_\kappa \left( \frac{\txte^{i\kappa x} + \txte^{-i\kappa x}}{2} \right) + b_\kappa \left( \frac{\txte^{i\kappa x} - \txte^{-i\kappa x}}{2i} \right).
\end{align}
Combining like terms in the positive frequency term $\txte^{i\kappa x}$ and the negative frequency term $\txte^{-i\kappa x}$, we have
\begin{align}
    f(x) = \sum_{\kappa} \left(\frac{a_\kappa - ib_\kappa}{2}\right)\txte^{i\kappa x} + \left(\frac{a_\kappa + ib_\kappa}{2}\right)\txte^{-i\kappa x}. \label{Complex Fourier Linear Combination}
\end{align}
So now the Fourier coefficients for the imaginary exponentials are linear combinations of those for the sinusoids. Recall that any complex number $z$ can be written as $z = x + iy$, where $x$ and $y$ are real numbers.  Likewise, the complex conjugate $z^{\ast} = x - iy$.  We will then define the complex Fourier coefficient $c_\kappa$ as 
\begin{align}
    c_\kappa = \lambda \left(\frac{a_\kappa - ib_\kappa}{2}\right),
\end{align}
where $\lambda$ is a real, positive number. I included $\lambda$ here because I will soon take its limit to infinity.  When we divide $c_\kappa$ by $\lambda$, we have the Fourier coefficients in Eq. \ref{Complex Fourier Linear Combination}.  Thus we can write $f(x)$ as 
\begin{align}
    f(x) &= \frac{1}{\lambda}\sum_{\kappa} c_\kappa\txte^{i\kappa x} + c_\kappa^\ast \txte^{-i\kappa x} \nonumber
    \\
    &= \frac{1}{\lambda}\sum_{+\kappa} c_{+\kappa}\txte^{i(+\kappa) x} + \frac{1}{\lambda}\sum_{-\kappa} c_{-\kappa}\txte^{i(-\kappa) x} \nonumber
    \\
    &= \frac{1}{\lambda}\sum_{\pm\kappa} c_{\kappa}\txte^{i\kappa x}. \label{Complex sum over kappa}
\end{align}
Here the sum means that we must account for both positive and negative $\kappa$, so we have recovered a linear combination that does explicitly show the negative frequencies. Before we dive in to the case where $-\infty < x < \infty$, let's first consider the case when $-\lambda/2 < x < \lambda/2$ since the length of the entire domain is $\lambda$.

Suppose $f(x)$ is integrable on $-\lambda/2 < x < \lambda/2$.  Then we choose to construct $f(x)$ from its harmonic modes, where $L = \lambda/2$.  Then
\begin{align}
    +\kappa_j = \frac{2\pi j}{\lambda},\;\; -\kappa_j = -\frac{2\pi j}{\lambda},\;\; j = 0, 1, 2, 3, \dots
\end{align}
But the expression for $-\kappa_j$ above is mathematically clunky, so we decide to absorb the negative sign into $j$ itself.  Thus,
\begin{align*}
    \kappa_j = \frac{2\pi j}{\lambda},\;\; j = 0, \pm 1, \pm 2, \pm 3, \dots
\end{align*}
That means we need to sum over all integers for $j$. Remember that the exponential function $\txte^\alpha = \exp{\alpha}$. Then Eq. \ref{Complex sum over kappa} becomes
\begin{align}
    f(x) = \frac{1}{\lambda}\sum_{j = -\infty}^{\infty} c_j \exp \left( \frac{2\pi i j x}{\lambda} \right), \;\; -\frac{\lambda}{2} < x < \frac{\lambda}{2}. \label{Complex Fourier Series}
\end{align}
This representation is called the Complex Fourier Series of $f(x)$.  To find $c_j$, we use the orthogonality relationship:
\begin{align}
    \int_{-\lambda/2}^{\lambda/2} \exp \left( \frac{2\pi i j x}{\lambda} \right) \exp \left( -\frac{2\pi i k x}{\lambda} \right)\,\textrm{d}x &= \int_{-\lambda/2}^{\lambda/2} \exp \left[ \frac{2\pi i (j-k) x}{\lambda} \right]\,\textrm{d}x \nonumber
    \\
    &= \frac{\lambda}{2\pi i(j-k)} \left[] \txte^{i\pi(j-k)} - \txte^{-i\pi(j-k)}  \right] \nonumber
    \\
    &= \lambda \frac{\sin[\pi(j-k)]}{\pi(j-k)} \nonumber
    \\
    &= \begin{cases}
    0,\;\;\textrm{if } j\neq k
    \\
    \lambda,\;\;\textrm{if } j = k
    \end{cases} \label{Discrete Complex Orthogonality}.
\end{align}
This orthogonality relationship holds since $j$ and $k$ are two integers, and so $j-k$ is an integer and the sine of any integer multiple of $\pi$ vanishes, meanwhile
\begin{align}
    \lim_{t\rightarrow 0} \frac{\sin t}{t} = 1,
\end{align}
which can be derived from L'H\^opital's rule for evaluating indeterminate limits. We could have also derived this orthogonality relationship from the ones between sines and cosines, but that has more steps and is less pretty to look at. Then, to find $c_j$, we have
\begin{align}
    c_j = \frac{1}{\lambda} \int_{-\lambda/2}^{\lambda/2} f(x)\exp\left(-\frac{2\pi i j x}{\lambda}\right)\,\textrm{d}x, \label{eq: complex fourier series coefficients}
\end{align}
where we used the same techniques as we did to find Eqs. \ref{Value for b_k}, \ref{Value for a_0}, and \ref{Value for a_j}.  Now, with this background in place, we take $\lambda\rightarrow\infty$.

\subsection{The Fourier Transform}
Consider again the Complex Fourier Series given by Eq. \ref{Complex Fourier Series}.  If we have function that is defined on $-\infty < x < \infty$, then we must allow $\lambda \rightarrow \infty$.  Before we do that though, let's rewrite Eq. \ref{Complex Fourier Series} so it is a little uglier.
\begin{align}
    f(x) &= \frac{1}{\lambda}\sum_{j = -\infty}^{\infty} c_j \exp \left( \frac{2\pi i j x}{\lambda} \right) \nonumber
    \\
    &= \frac{1}{2\pi}\sum_{j = -\infty}^{\infty} \frac{2\pi}{\lambda} c_j \exp \left( \frac{2\pi i j x}{\lambda} \right) \nonumber
    \\
    &= \frac{1}{2\pi}\sum_{j = -\infty}^{\infty} \frac{2\pi}{\lambda} c_j \exp \left( \frac{2\pi i j x}{\lambda} \right) \nonumber
    \\
    &= \frac{1}{2\pi}\sum_{j = 0}^{\infty} \frac{2\pi}{\lambda} c_j \exp \left( \frac{2\pi i j x}{\lambda} \right) + \frac{1}{2\pi}\sum_{j = 1}^{\infty} \frac{2\pi}{\lambda} c_{-j} \exp \left( -\frac{2\pi i j x}{\lambda} \right).
\end{align}
Now let's define $\Delta\kappa_\lambda = 2\pi/\lambda$.  Then we have 
\begin{align}
    f(x) &= \frac{1}{2\pi}\sum_{j = 0}^{\infty} \Delta\kappa_\lambda c_j \exp \left[i(0 + j\Delta\kappa_\lambda)x \right] + \frac{1}{2\pi}\sum_{j = 1}^{\infty} \Delta\kappa_\lambda c_{-j} \exp  \left[i(0 - j\Delta\kappa_\lambda)x \right]
\end{align}
But these are precisely the expression for Riemann sums as long as the function $c(\kappa)$ is integrable.  Here the integration variable is $\kappa$ and $\Delta\kappa_{\lambda\rightarrow \infty} = \textrm{d}\kappa$. Thus,
\begin{align}
    f(x) = \frac{1}{2\pi}\int_0^{\infty} c(\kappa)\txte^{i\kappa x}\,\textrm{d}\kappa + \frac{1}{2\pi}\int_{-\infty}^{0} c(\kappa)\txte^{i\kappa x}\,\textrm{d}\kappa,
\end{align}
or to be cleaner,
\begin{align}
    f(x) = \frac{1}{2\pi}\int_{-\infty}^{\infty} F(\kappa)\txte^{i\kappa x}\,\textrm{d}\kappa,\;\; -\infty < x < \infty. \label{Inverse Fourier Transform}
\end{align}
This is called the Inverse Fourier Transform, and what it says is that a function $f(x)$ can be written as a continuous, linear combination of angular frequencies or angular wavenumbers described by its Fourier Transform $F(\kappa)$.  It is conventional to write the transform of a function $g(x)$ as $G(x)$, so that's why I switched from $c(\kappa)$ to $F(\kappa)$.  The only requirements that $f(x)$ has to meet is that it must be piece-wise continuous, and it must at least be bound at $\pm\infty$, but it should preferably be zero at $\pm\infty$.  The latter may seem strict, but it turns out that most\footnote{As far as I know, all physical signals are finite, but I am giving myself room for error.} physical signals are finite, so the Fourier Transform is actually incredibly useful and it is a go-to tool for physicists. Now we must find out what the coefficients $F(\kappa)$ are.

What do you know, we employ another orthogonality relationship! In fact, we use the same one we already have derived, Eq. \ref{Discrete Complex Orthogonality}, but we again must take the limit as $\lambda\rightarrow \infty$.
\begin{align}
    \int_{-\lambda/2}^{\lambda/2} \exp \left( \frac{2\pi i j x}{\lambda} \right) \exp \left( -\frac{2\pi i k x}{\lambda} \right)\,\textrm{d}x &= \int_{-\lambda/2}^{\lambda/2} \exp \left( ij\Delta\kappa_\lambda x\right) \exp \left(  -ik\Delta\kappa_\lambda x\right)\,\textrm{d}x \nonumber
    \\
    &= \int_{-\lambda/2}^{\lambda/2} \exp \left[ i(j\Delta\kappa_\lambda - k\Delta\kappa_\lambda)x\right] \,\textrm{d}x \nonumber
    \\
    &= \begin{cases}
    0,\;\;\textrm{if } j\neq k
    \\
    \lambda,\;\;\textrm{if } j = k
    \end{cases}.
\end{align}
Define $\kappa = j\Delta\kappa_{\lambda\rightarrow\infty}$ and $\kappa_0 = k\Delta\kappa_{\lambda\rightarrow\infty}$.  Then we obtain the continuous version of the orthogonality relationship
\begin{align}
    \int_{-\infty}^{\infty} \txte^{i(\kappa - \kappa_0)x}\,\textrm{d}x  = \begin{cases}
    0,\;\;\;\;\textrm{if } \kappa\neq \kappa_0
    \\
    \infty,\;\;\textrm{if } \kappa = \kappa_0
    \end{cases}.
\end{align}
Hopefully this type of expression looks familiar, but just to be certain, we will integrate the above expression over all possible $\kappa$.
\begin{align}
    \int_{-\infty}^{\infty}\int_{-\infty}^{\infty} \txte^{i(\kappa - \kappa_0)x}\,\textrm{d}x\,\textrm{d}\kappa &= \int_{-\infty}^{\infty}\int_{-\infty}^{\infty} \txte^{i(\kappa - \kappa_0)x}\,\textrm{d}\kappa\,\textrm{d}x \nonumber
    \\
    &= \int_{-\infty}^{\infty}\lim_{K\rightarrow\infty}\int_{-K}^{K} \txte^{i(\kappa - \kappa_0)x}\,\textrm{d}(\kappa-\kappa_0)\,\textrm{d}x \nonumber
    \\
    &=\int_{-\infty}^{\infty}\lim_{K\rightarrow\infty} \frac{\txte^{iKx} - \txte^{-iKx}}{ix}\,\textrm{d}x \nonumber
    \\
    &= 2\lim_{K\rightarrow\infty}\int_{-\infty}^{\infty} \frac{\sin Kx}{x}\, \textrm{d}x \nonumber 
    \\
    &= 2\lim_{K\rightarrow\infty} \pi\nonumber
    \\
    &= 2\pi.
\end{align}
To be clear, the evaluation of the Sine Integral above is not straightforward at all, and it will distract us from our job here.  However, it is known to be $\pi K/\vert K\vert = \pi$, since $K>0$. For anyone interested in knowing how to do it out, follow a similar procedure given in \cite{stackSincFunction}. Anyway, what we have found is that 
\begin{align}
\frac{1}{2\pi}\int_{-\infty}^{\infty} \txte^{i(\kappa - \kappa_0)x}\,\textrm{d}x  &= \begin{cases}
    0,\;\;&\textrm{if } \kappa\neq \kappa_0
    \\
    \infty,\;\;&\textrm{if } \kappa = \kappa_0
    \end{cases}, \label{Continuous Complex Orthogonality}
    \\
    \frac{1}{2\pi}\int_{-\infty}^{\infty}\int_{-\infty}^{\infty} \txte^{i(\kappa - \kappa_0)x}\,\textrm{d}x\,\textrm{d}\kappa &= 1.
\end{align}
Thus, we have found that 
\begin{align}
    \frac{1}{2\pi}\int_{-\infty}^{\infty} \txte^{i(\kappa - \kappa_0)x}\,\textrm{d}x = \delta(\kappa - \kappa_0).
\end{align}
To find $F(\kappa)$ then, we (finally) multiply both sides of Eq. \ref{Inverse Fourier Transform} by $\txte^{-i\kappa_0 x}\,\textrm{d}x$ and integrate.
\begin{align}
    \int_{-\infty}^{\infty} f(x)\txte^{-i\kappa_0 x}\,\textrm{d}x &= \frac{1}{2\pi}\int_{-\infty}^{\infty}\int_{-\infty}^{\infty} F(\kappa)\txte^{i\kappa x}\,\textrm{d}\kappa\,\txte^{-i\kappa_0 x}\,\textrm{d}x \nonumber
    \\
    &= \frac{1}{2\pi}\int_{-\infty}^{\infty}\int_{-\infty}^{\infty} F(\kappa)\txte^{i(\kappa - \kappa_0) x}\,\textrm{d}x\,\textrm{d}\kappa \nonumber
    \\
    &= \int_{-\infty}^{\infty}F(\kappa)\left[\frac{1}{2\pi}\int_{-\infty}^{\infty} \txte^{i(\kappa - \kappa_0) x}\,\textrm{d}x\right]\,\textrm{d}\kappa \nonumber
    \\
    &= \int_{-\infty}^{\infty}F(\kappa)\delta(\kappa-\kappa_0)\,\textrm{d}\kappa \nonumber
    \\
    &= F(\kappa_0).
\end{align}
In the last equality we used the Sifting Property of the Dirac-$\delta$.  By relabeling the points of interest from $\kappa_0$ back to $\kappa$, we recover the Fourier Transform
\begin{align}
    F(\kappa) = \int_{-\infty}^{\infty} f(x)\txte^{-i\kappa x}\,\textrm{d}x,\;\; -\infty < \kappa < \infty. \label{Fourier Transform}
\end{align}
Sometimes people prefer to \textquotedblleft split up\textquotedblright$\;$ the factor of $2\pi$ equally to create the \textquotedblleft Normalized\textquotedblright$\;$ Fourier Transform and \textquotedblleft Normalized\textquotedblright$\;$ Inverse Fourier Transform, given below.

\begin{align}
   F(\kappa) &= \frac{1}{\sqrt{2\pi}}\int_{-\infty}^{\infty} f(x)\txte^{-i\kappa x}\,\textrm{d}x,\;\; -\infty < \kappa < \infty,\label{Normalized Fourier Transform}
    \\
    f(x) &= \frac{1}{\sqrt{2\pi}}\int_{-\infty}^{\infty} F(\kappa)\txte^{i\kappa x}\,\textrm{d}\kappa,\;\; -\infty < x < \infty. \label{Normalized Inverse Fourier Transform}
\end{align}
These functions have the same shape though as the non-normalized versions, they just have different scaling. In physics, we are almost always concerned with the characteristic frequencies or wavelengths of a particular signal which are completely unaffected by the rescaling by $\sqrt{2\pi}$. In other words, the functional form of the transform are what we physicists usually find important, not necessarily the numerical values of the tranforms themselves.  So it usually is just a matter of personal preference.  However, in quantum mechanics, the normalization is very important as the Fourier transform represents a continuous transformation between position representation and momentum representation.  The $\sqrt{2\pi}$ makes sure that probability is conserved between the two representations.  But more on that when you get to it in your first quantum mechanics class.

Fourier Transforms have some really interesting properties, most of which we will be unable to cover in this chapter. However, the examples below are designed to go over a few basic and important properties of the Fourier Transform for applications in physics.

\subsection{Some Important Examples}

\subsubsection{Uniqueness and Duality}
In this first example, we want to find out if there are multiple reciprocal space representations of the same function $f(x)$ defined on an infinite interval.  Let's start by assuming there are at least two Inverse Fourier Transforms that exist for $f(x)$, namely $F_1(\kappa)$ and $F_2(\kappa)$.  By the definition given by Eq. \ref{Inverse Fourier Transform} for the Inverse Fourier Transform, then
\begin{align*}
    f(x) &= \frac{1}{2\pi}\int_{-\infty}^{\infty} F_1(\kappa)\txte^{i\kappa x}\,\textrm{d}\kappa,\;\; -\infty < x < \infty,
    \\
    f(x) &= \frac{1}{2\pi}\int_{-\infty}^{\infty} F_2(\kappa)\txte^{i\kappa x}\,\textrm{d}\kappa,\;\; -\infty < x < \infty.
\end{align*}
If we subtract the first equation from the second, then the left-hand side is zero since $f(x) - f(x) = 0$. Therefore, right-hand side must also vanish.
\begin{align*}
    0 = \frac{1}{2\pi}\int_{-\infty}^{\infty} \left[F_2(\kappa) - F_1(\kappa)\right]\txte^{i\kappa x}\,\textrm{d}\kappa,\;\; -\infty < x < \infty.
\end{align*}
In this step I brought the subtraction inside of the integral and factored out the common $\txte^{i\kappa x}$.  For those of you who may already see how to proceed, continue ahead.  For those who do not, please allow me to get a little more mathy. Remember that we can write the integral above as a Riemann sum
\begin{align*}
    0 = \frac{1}{2\pi} \lim_{\lambda \rightarrow \infty} \sum_{j = -\infty}^\infty \left[F_2\left( \frac{2\pi j}{\lambda} \right) - F_1\left( \frac{2\pi j}{\lambda}\right)\right]\exp\left( \frac{2\pi i j x}{\lambda} \right) \frac{2\pi j}{\lambda},\;\; -\infty < x < \infty.
\end{align*}
What this statement says is that this summation is zero for every single value of $x$ while $\lambda$ gets larger and larger.  While it is true that $2\pi j/\lambda$ does approach zero, it never explicitly reaches it for any finite $\lambda$.  So the only possible \textquotedblleft source\textquotedblright $\;$ of the zero is the difference between the functions $F_2$ and $F_1$.  It is true that the whole sum must vanish, but we found this out without specifying any relationship at all between the individual values of $F_2$ and $F_1$.  For example, we do not know if $F_2(0) = -F_2(1)$ or any other possible combination.  The only thing we do know is that the whole summation must be zero as $\lambda$ gets larger and larger.  Thus, we conclude that 
\begin{align*}
    0 = \lim_{\lambda \rightarrow \infty} \left[ F_2\left( \frac{2\pi j}{\lambda} \right) - F_1\left( \frac{2\pi j}{\lambda}\right) \right],
\end{align*}
for every single value of $j$ as $\lambda$ gets larger and larger.  In other words,
\begin{align*}
    0 = F_2(\kappa) - F_1(\kappa) \Rightarrow F_2(\kappa) = F_1(\kappa),\;\; -\infty < \kappa < \infty.
\end{align*}
Thus these two reciprocal space representations of $f(x)$ are exactly identical for every single value of $\kappa$. We conclude that there really only is one reciprocal space representation of $f(x)$ | i.e. the Inverse Fourier Transform of $f(x)$ is unique.  

I leave it to you to show that for any $F(\kappa)$, there exists a unique real-space representation of it given by $f(x)$ using an almost identical argument to the one above. The reason why this is important is that it shows that the Fourier Transform and Inverse Fourier Transform change one function into exactly one other function and vice versa.  Thus, if we know the transform of one function in some context, then the transform of the same function in another context is exactly the same.  Furthermore, we can show that the  Fourier Transform of $F(x)$ is $2\pi f(-\kappa)$ if  
\begin{align*}
    f(x) &= \frac{1}{2\pi}\int_{-\infty}^{\infty} F(\kappa)\txte^{i\kappa x}\,\textrm{d}\kappa,\;\; -\infty < x < \infty,
    \\
    F(\kappa) &= \int_{-\infty}^{\infty} f(x)\txte^{-i\kappa x}\,\textrm{d}x,\;\; -\infty < \kappa < \infty.
\end{align*}
Let's take the first equation, and multiply it by $2\pi$.  We are also going to change $x\rightarrow -x$.
\begin{align*}
    2\pi f(-x) &= \int_{-\infty}^{\infty} F(\kappa)\txte^{i\kappa (-x)}\,\textrm{d}\kappa,\;\; -\infty < x < \infty,
    \\
    &= \int_{-\infty}^{\infty} F(\kappa)\txte^{-i\kappa x}\,\textrm{d}\kappa,\;\; -\infty < x < \infty.
\end{align*}
Now we are going to change the labels of the variables $x\rightarrow\kappa$ and $\kappa\rightarrow x$.
\begin{align*}
    2\pi f(-\kappa) &= \int_{-\infty}^{\infty} F(x)\txte^{-ix  \kappa}\,\textrm{d}x,\;\; -\infty < \kappa < \infty,
    \\
    &= \int_{-\infty}^{\infty} F(x)\txte^{-i\kappa x}\,\textrm{d}x,\;\; -\infty < \kappa < \infty.
\end{align*}
But this last expression fits the original definition of the Fourier Transform of a function, but in this case, $F(x)$.  We note that there is no extra factor of $2\pi$ for the Normalized Transforms (prove this for yourself if you do not see it at first glance). This property is called the \textit{Duality Property of Fourier Transform Pairs}.  Thus a function and its Fourier Transform are bound together through the Fourier Transform mapping. 

\vspace{0.15in}
\begin{example}[Dashes and Dots]{ex: dashes and dots}
Let's consider a time-dependent signal this time, so all $x\rightarrow t$ and all $\kappa\rightarrow \omega$.  We will consider a fairly basic signal for practice, but this signal also has some pretty interesting physics we can study.  So without further ado, consider a Morse code style dash, given by 
\begin{align*}
    f(t) = \begin{cases}
    0,\;\;&\textrm{if } -\infty < t < -\tau
    \\
    A,\;\;&\textrm{if } -\tau \leq t \leq \tau
    \\
    0,\;\;&\textrm{if } \tau < t < \infty
    \end{cases}
\end{align*}
where $A$ is a fixed value and $\tau > 0$. This signal could be, for example, a pulse of DC current of value $A$ that lasts for $2\tau$ time.  We seek the Fourier Transform of $f(t)$ given by Eq. \ref{Fourier Transform}.
\begin{align*}
    F(\omega) &= \int_{-\infty}^{\infty} f(t)\txte^{-i\omega t}\,\textrm{d}t,\;\; -\infty < \omega < \infty
    \\
    &= \int_{-\infty}^{-\tau} 0\txte^{-i\omega t}\,\textrm{d}t + \int_{-\tau}^{\tau} A\txte^{-i\omega t}\,\textrm{d}t + \int_{\tau}^{\infty} 0\txte^{-i\omega t}\,\textrm{d}t
    \\
    &= \left. \frac{A\txte^{-i\omega t}}{-i\omega}\right\vert_{-\tau}^\tau
    \\
    &= -\frac{A}{i\omega}\left( \txte^{-i\omega\tau} - \txte^{i\omega \tau}
    \right)
    \\
    &=\frac{2A}{\omega} \left( \frac{\txte^{i\omega\tau} - \txte^{-i\omega \tau}}{2i} \right)
    \\
    &= \frac{2A}{\omega} \sin\omega\tau
    \\
    &= 2A\tau\,\left(\frac{\sin\omega\tau}{\omega\tau}\right).
\end{align*}
The $\sin(u)/u$ function is pretty common in science and engineering and is called the \emph{sine cardinal} and written as $\textrm{sinc}(u) = \sin(u)/u$ (pronounced \textit{sink}).
\end{example}
\vspace{0.15in}

\begin{figure}
    \centering
    \includegraphics[width = 5.0in, keepaspectratio]{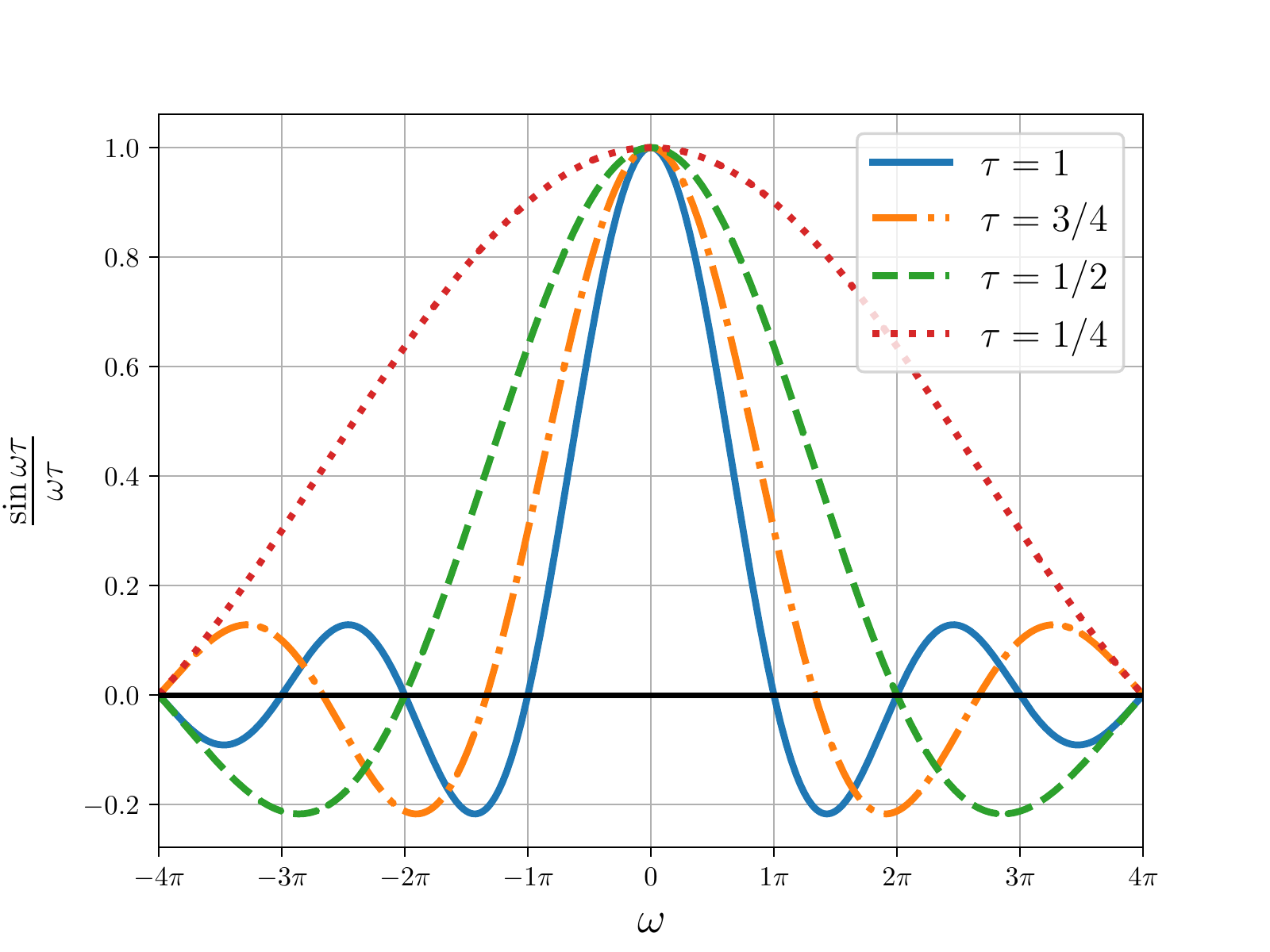}
    \caption{A few capital sines  $\mathrm{sinc}(\omega\tau) = \sin \omega\tau/\omega\tau$ plotted as functions of $\omega$ for varying values of $\tau$. Notice that as $\tau\rightarrow 0$, the whole function $\mathrm{sinc}(\omega\tau)$ gets wider, i.e. the Fourier Transform needs more values of $\omega$ to account for the smaller pulse.\label{fig: Sinc Plots}}
\end{figure}

What we want to look at now is what happens to $F(\omega)$ as we change the value of $\tau$.  We will focus on the sine part mostly. Recall that 
\begin{align*}
    \lim_{u\rightarrow 0} \frac{\sin u}{u} = 1,
\end{align*}
After this point, all the zeros of the sine capital function are the zeros of the sine function.  Thus we look for when
\begin{align*}
    \omega\tau = \pm j\pi \Rightarrow \omega = \pm\frac{j\pi}{\tau},\;\; j = 1,2,3,\dots
\end{align*}
This statement says that if $\tau$ is small, then the $\omega$ for which $F(\omega) = 0$ are farther apart, but if $\tau$ is large, then the zeros of $F(\omega)$ are closer together, as can be seen in \figref{fig: Sinc Plots}. Physically, these cases correspond to a shorter signal pulse and a longer signal pulse, respectively. Thus, shorter signals need a wider range of frequencies to make them up, while longer wider signals need shorter ranges of frequencies to make them up!  Again, we see behavior that is reminiscent of the Heisenberg Uncertainty Principle, but these behaviors occur outside of quantum mechanics, too!

Before we move onto the next example, I want to show how to use the Duality Property to evaluate potentially difficult integrals.  For example, let's say we want to find the Fourier Transform of $g(t) = \sin(t)/t,\; -\infty < t < \infty$.  Thus we would need to evaluate
\begin{align*}
    G(\omega) = \int_{-\infty}^{\infty} \frac{\sin t}{t} \txte^{-i\omega t}\,\textrm{d}t,\;\; -\infty < \omega < \infty.
\end{align*}
This is doable, although it is gross. Instead let's use what we found above.  We know that the Fourier Transform of the DC pulse of length $\tau$, $f(t)$, defined above is
\begin{align*}
    F(\omega) = 2A\tau\,\left(\frac{\sin\omega\tau}{\omega\tau}\right) = F(\omega\tau).
\end{align*}
Therefore, $g(\omega\tau)$ is 
\begin{align*}
    g(\omega\tau) = \frac{\sin\omega\tau}{\omega\tau} = \frac{F(\omega\tau)}{2A\tau}.
\end{align*}
So if we make the substitution, $\theta = \omega\tau$, then
\begin{align*}
    g(\theta) =  \frac{\sin\theta}{\theta} = \frac{F\left(\theta\right)}{2A\tau}.
\end{align*}
Then the expression for $G(\omega)$ becomes
\begin{align*}
    G(\omega) &= \int_{-\infty}^{\infty} \frac{\sin \theta }{\theta } \txte^{-i\omega \theta}\,\textrm{d}\theta,\;\; -\infty < \omega < \infty,
    \\
    &= \frac{1}{2A\tau}\int_{-\infty}^{\infty} F\left(\theta\right) \txte^{-i\omega \theta}\,\textrm{d}\theta,\;\; -\infty < \omega < \infty.
\end{align*}
But this integral (not the fraction in front) has the same form of the duality property when we now switch the label $\theta\rightarrow t$.  Hence,
\begin{align*}
    G(\omega) = \frac{1}{2A\tau} \cdot 2\pi f(-\omega) = \frac{\pi f(-\omega)}{A\tau},\;\; -\infty < \omega < \infty.
\end{align*}
Now we substitute in the definition of $f(t)$, but change every $t\rightarrow -\omega$.
\begin{align*}
    G(\omega) &= \frac{\pi}{A\tau} \begin{cases}
    0,\;\;&\textrm{if } -\infty < -\omega < -\tau
    \\
    A,\;\;&\textrm{if } -\tau \leq -\omega \leq \tau
    \\
    0,\;\;&\textrm{if } \tau < -\omega < \infty
    \end{cases}
    \\
    &= \frac{\pi}{\tau}\begin{cases}
    0,\;\;&\textrm{if } \infty > \omega > \tau
    \\
    1,\;\;&\textrm{if } \tau \geq \omega \geq -\tau
    \\
    0,\;\;&\textrm{if } \tau > \omega > -\infty
    \end{cases}
    \\
    &= \begin{cases}
    0,\;\;&\textrm{if } \infty > \omega > 1
    \\
    \pi,\;\;&\textrm{if } 1 \geq \omega \geq -1
    \\
    0,\;\;&\textrm{if } 1 > \omega > -\infty
    \end{cases}
\end{align*}
We note that this could have also been evaluated using Dirac-$\delta$ functions by evaluating the integral directly, but the Duality Property converts an otherwise advanced calculus problem into a relatively simple algebra problem, where we really just needed to be careful with the labels we chose as our variable names.

\vspace{0.15in}

\begin{example}[A Sinusoid of Finite Length]{ex: Finite Sinusoid}
In this example, we will check out how wiggly functions behave under Fourier Transformations.  Remember, a Fourier Transform is a linear combination of wiggly bits, so the Transform should tell us that the sinusoid $f(t) = A\cos(\omega_0 t + \phi)$ is wiggly too.  Let us consider a real signal though, perhaps from an LC-circuit, given by
\begin{align*}
    f(t) = \begin{cases}
    0,\;\;&\textrm{if } -\infty < t < -\tau
    \\
    A\cos(\omega_0 t + \phi),\;\;&\textrm{if } -\tau \leq t \leq \tau
    \\
    0,\;\;&\textrm{if } \tau < t < \infty
    \end{cases},
\end{align*}
where $A$ and $\phi$ are just numbers, and $\tau > 0$ as in the last example.  We seek the Fourier Transform of $f(t)$ from Eq. \ref{Fourier Transform}, where $x\rightarrow t$ and $\kappa\rightarrow \omega$.
\begin{align*}
    F(\omega) &= \int_{-\infty}^{\infty} f(t)\txte^{-i\omega t}\,\textrm{d}t,\;\; -\infty < \omega < \infty
    \\
    &= \int_{-\infty}^{-\tau} 0\txte^{-i\omega t}\,\textrm{d}t + \int_{-\tau}^{\tau} A\cos(\omega_0 t+\phi)\txte^{-i\omega t}\,\textrm{d}t + \int_{\tau}^{\infty} 0\txte^{-i\omega t}\,\textrm{d}t
    \\
    &= \int_{-\tau}^{\tau} A\left[\frac{\txte^{i(\omega_0 t + \phi)} + \txte^{-i(\omega_0 t + \phi)}}{2}\right]\txte^{-i\omega t}\,\textrm{d}t
    \\
    &= \frac{A}{2}\left[\int_{-\tau}^{\tau}\txte^{i\phi}\txte^{i(\omega_0 - \omega)t}\,\textrm{d}t +  \int_{-\tau}^{\tau} \txte^{-i\phi}\txte^{-i(\omega_0 + \omega)t}\,\textrm{d}t \right]
    \\
    &= \frac{A\txte^{i\phi}}{2} \left[\frac{\txte^{i(\omega_0-\omega)t}}{i(\omega_0 - \omega)} +\txte^{-2i\phi}\, \frac{\txte^{-i(\omega_0+\omega)t}}{-i(\omega_0 +\omega)} \right\vert_{-\tau}^\tau
    \\
    &= A\tau\txte^{i\phi} \left[\frac{\sin(\omega_0 - \omega)\tau}{(\omega_0 - \omega)\tau} + \txte^{-2i\phi}\,\frac{\sin(\omega_0 + \omega)\tau}{(\omega_0 + \omega)\tau}\right].
\end{align*}
Again, since the sine cardinal approaches unity as its argument goes to zero, then this Transform is maximized at when $\omega = \omega_0$ or $\omega = -\omega_0$, for arbitrary $\tau$.  This means that there are two frequencies that dominate the linear combination: $+\omega_0$ and $-\omega_0$.  This is exactly what we expect for an even signal wiggling around at frequency $\omega_0$.
\end{example}
\vspace{0.15in}

It is important to note that as $\tau\rightarrow\infty$,  both sine cardinals approach Dirac-$\delta$ function like behavior, since the zeros of the sine part get closer together, with the exception of when the argument is zero (see the example above for elaboration). Furthermore, as $\tau$ gets larger, the whole Transform increases in magnitude at $\omega = \pm\omega_0$.  Thus, as $\tau$ gets larger, we need fewer and fewer frequency components to describe $f(t)$. And that is exactly what we expect for a sinusoidal function known for an infinite domain! What we can interpret from this is that all of the frequencies other than $\pm\omega_0$ work to completely cancel our signal to exactly $f(t) = 0$ along $ -\infty < t < -\tau$ and $\tau < t< \infty$! 

\subsubsection{Derivatives in Different Representations}
There are many other properties of Fourier Transforms that are important and insightful.  However, as I'm sure you can tell by now, it is pretty easy to keep writing and writing about Fourier Analysis.  If you would like to know more about some of the properties, please see \cite{wang_moore_2009, thefouriertransform}. The last property I want to talk about are the derivative properties of Fourier Transforms.

\vspace{0.15in}
\begin{example}[Derivatives in Fourier Space]{ex: derivs in Fourier space}
Let's start with derivatives.  Suppose I gave you $f(x)$ and its Fourier Transform $F(\kappa)$ given by Eq. \ref{Fourier Transform}.  What is the Fourier Transform of $\textrm{d}f/\textrm{d}x$, namely $G(\kappa)$? We calculate it directly, but using the Inverse Transform instead (Eq. \ref{Inverse Fourier Transform}).
\begin{align*}
    \frac{\textrm{d}}{\textrm{d}x}f(x) &= \frac{\textrm{d}}{\textrm{d}x}\left[  \frac{1}{2\pi}\int_{-\infty}^{\infty} F(\kappa)\txte^{i\kappa x}\,\textrm{d}\kappa,\right]\;\; -\infty < x < \infty,
    \\
    &= \frac{1}{2\pi}\int_{-\infty}^{\infty} \frac{\partial}{\partial x}\left[ F(\kappa)\txte^{i\kappa x}\right]\,\textrm{d}\kappa,\;\; -\infty < x < \infty,
    \\
    &= \frac{1}{2\pi}\int_{-\infty}^{\infty} i\kappa F(\kappa)\txte^{i\kappa x}\,\textrm{d}\kappa,\;\; -\infty < x < \infty,
    \\
    &=\frac{1}{2\pi}\int_{-\infty}^{\infty} G(\kappa)\txte^{i\kappa x}\,\textrm{d}\kappa,\;\; -\infty < x < \infty.
\end{align*}
Thus $G(\kappa) = i\kappa F(\kappa)$. The reason why this works is because all of the $x$-dependence is in the $\txte^{i\kappa x}$ part since $F(\kappa)$ only depends on $\kappa$. Furthermore, if we subract the middle two lines from each other, we have 
\begin{align*}
    0 &= \frac{1}{2\pi}\int_{-\infty}^{\infty} \frac{\partial}{\partial x}\left[ F(\kappa)\txte^{i\kappa x}\right]\,\textrm{d}\kappa - \frac{1}{2\pi}\int_{-\infty}^{\infty} i\kappa F(\kappa)\txte^{i\kappa x}\,\textrm{d}\kappa,\;\; -\infty < x < \infty,
    \\
    &= \frac{1}{2\pi}\int_{-\infty}^{\infty} \left[ \frac{\partial}{\partial x} - i\kappa \right]F(\kappa)\txte^{i\kappa x}\,\textrm{d}\kappa,\;\; -\infty < x < \infty.
\end{align*}
Since this integral must always be zero as long as $f(x)$ is differentiable, then we conclude that the difference in the integral must vanish since nothing about $F(\kappa)$ has been specified (see the first example on uniqueness for details). In other words, with respect to Fourier Transforms, $\partial/\partial x$ is interchangeable with $i\kappa$.  I leave it to you to show that $\partial^n/\partial x^n$ is interchangeable with $i^n\kappa^n$, where $n$ is a nonnegative integer, under Fourier Transformations. Furthermore, using the same argument, show that the real-space representation of $\textrm{d}^nF/\textrm{d}\kappa^n$ is $(-ix)^n$ which implies that $\partial/\partial \kappa$ is interchangeable with $-ix$ under Fourier Transformations.  
\end{example}
\vspace{0.15in}

\section{Concluding Remarks}
As I have said in this document a few times, this chapter by no means can possibly cover all there is to Fourier Analysis.  For example, we have only talked about Fourier Analysis of one independent variable (to see Fourier Analysis in multiple dimensions check out \cite{MultidimensionalTransform}).  It is my hope, however, that the material we have covered will serve as a sufficient introductory resource for you in Fourier Analysis that you may have not otherwise covered in an undergraduate career in physics. I have included \tblref{tab: Fouier} with a bunch of useful formulas that you can refer to later on, as well as when I talk about them in the text.  Remember, in physics our job is to describe the laws of nature as we observe them.  Although Fourier Analysis may look a little mathematically messy, it is ultimately a tool to help us look at different physical phenomena from a new angle, and we can do so just by adding up contributions from distinct wiggling components.

\begin{table}
    \centering
    \caption{A bunch of useful and general formulas for Fourier analysis and their text references.}
    \label{tab: Fouier}
    \scalebox{0.8}{
    \begin{tabular}{ccc}
        \hline\hline
        \textbf{Equation Description} & \textbf{Equation Formula} & \textbf{Text Reference}\\ \hline & & \\
        Linear Combination of Frequencies & $ f(x) = \sum_{\kappa}  a_\kappa \cos(\kappa x) + b_\kappa \sin(\kappa x)$ & \equaref{FourierLinearCombination}
        \\ && \\
        Cosine Orthogonality & $\dfrac{2}{L}\int_0^L \cos\left(\dfrac{j\pi x}{L}\right)\cos\left(\dfrac{k\pi x}{L}\right)\,\textrm{d}x = \begin{cases}
    1, & \textrm{if } j = k \neq 0
    \\
    2, & \textrm{if } j = k = 0
    \\
    0, & \textrm{if } j \neq k
    \end{cases}$ & \equaref{Cosine Orthogonality}
        \\ && \\
        Sine Orthogonality & $\dfrac{2}{L}\int_0^L \sin\left(\dfrac{j\pi x}{L}\right)\sin\left(\dfrac{k\pi x}{L}\right)\,\textrm{d}x = \begin{cases}
    1, & \textrm{if } j = k \neq 0
    \\
    0, & \textrm{if } j \neq k
    \end{cases}$ & \equaref{Sine Orthogonality}
        \\ && \\
        Sine-Cosine Orthogonality & $\dfrac{2}{L}\int_0^L \sin\left(\dfrac{j\pi x}{L}\right)\cos\left(\dfrac{k\pi x}{L}\right)\,\textrm{d}x = 0$ & \equaref{Sine Cosine Orthogonality}
        \\ && \\
        Fourier Constant Coefficient & $a_0 = \dfrac{1}{L} \int_0^L f(x)\,\textrm{d}x,\;\; 0\leq x\leq L$ & \equaref{Value for a_0}
        \\ && \\
        Fourier Cosine Coefficients & $ a_{j\geq 1} = \dfrac{2}{L} \int_0^L f(x)\cos\left(\dfrac{j\pi x}{L}\right)\,\textrm{d}x, \;\; 0\leq x\leq L$ & \equaref{Value for a_j}
        \\ && \\
        Fourier Sine Coefficients & $ b_{j} = \dfrac{2}{L} \int_0^L f(x)\sin\left(\dfrac{j\pi x}{L}\right)\,\textrm{d}x, \;\; 0\leq x\leq L$ & \equaref{Value for b_k}
        \\ && \\
        Discrete Complex Orthogonality & $\int_{-\lambda/2}^{\lambda/2} \exp \left( \frac{2\pi i j x}{\lambda} \right) \exp \left( -\frac{2\pi i k x}{\lambda} \right)\,\textrm{d}x = \begin{cases}
    0,\;\;\textrm{if } j\neq k
    \\
    \lambda,\;\;\textrm{if } j = k
    \end{cases}$ & \equaref{Discrete Complex Orthogonality}
        \\ && \\ 
        Complex Fourier Coefficients & $c_j = \dfrac{1}{\lambda} \int_{-\lambda/2}^{\lambda/2} f(x)\exp\left(-\dfrac{2\pi i j x}{\lambda}\right)\,\textrm{d}x$ & \equaref{eq: complex fourier series coefficients}
        \\ && \\
        The Fourier Transform & $F(\kappa) = \int_{-\infty}^{\infty} f(x)\txte^{-i\kappa x}\,\textrm{d}x,\;\; -\infty < \kappa < \infty$ & \equaref{Fourier Transform}
        \\ && \\
        The Inverse Fourier Transform & $f(x) = \dfrac{1}{2\pi}\int_{-\infty}^{\infty} F(\kappa)\txte^{i\kappa x}\,\textrm{d}\kappa,\;\; -\infty < x < \infty$ & \equaref{Inverse Fourier Transform}
        \\ && \\
        Normalized Fourier Transform & $F(\kappa) = \dfrac{1}{\sqrt{2\pi}}\int_{-\infty}^{\infty} f(x)\txte^{-i\kappa x}\,\textrm{d}x,\;\; -\infty < \kappa < \infty$ & \equaref{Normalized Fourier Transform}
        \\ && \\
        Normalized Inverse Fourier Transform & $f(x) = \dfrac{1}{\sqrt{2\pi}}\int_{-\infty}^{\infty} F(\kappa)\txte^{i\kappa x}\,\textrm{d}\kappa,\;\; -\infty < x < \infty$ & \equaref{Normalized Inverse Fourier Transform}
        \\ && \\ \hline\hline &&
    \end{tabular}
    }
\end{table}

\newpage

\begin{thebibliography}{1}

\bibitem{Schaums}
Murray~R. Spiegel, Seymour Lipschutz, and John Liu.
\newblock {\em Mathematical handbook of formulas and tables}.
\newblock Schaum's Outlines. McGraw-Hill Education, 5 edition, 2018.

\bibitem{bbc_1986}
Jeremy Gray.
\newblock The birth of the calculus (bbc), 1986.
\newblock \url{https://www.youtube.com/watch?v=ObPg3ki9GOI}.

\bibitem{deriv_notations}
Wikipedia.
\newblock Notation for differentiation, Apr 2018.
\newblock Here is a slew of derivative and integral notations.
  \url{https://en.wikipedia.org/wiki/Notation_for_differentiation}.

\bibitem{pauls_derivative_table}
Paul Dawkins.
\newblock Common derivatives and integrals, 2005.
\newblock
  \url{http://tutorial.math.lamar.edu/pdf/Common_Derivatives_Integrals_Reduced.pdf}.

\bibitem{thompson_systems_technology}
Peter~M Thompson and Systems Technology.
\newblock Snap, crackle, and pop, April 2018.
\newblock I pretty much only cited this as an official document recognizing
  Snap, Crackle, and Pop as the names of technical physical quantities. If
  you're interested in reading about them in the context of aerospace
  engineering, go to
  \url{https://info.aiaa.org/Regions/Western/Orange_County/Newsletters/Presentations\%20Posted\%20by\%20Enrique\%20P.\%20Castro/AIAAOC_SnapCracklePop_docx.pdf}.

\end{thebibliography}


\begin{thebibliography}{1}

\bibitem{ComplexNumberHistory}
Orlando Merino.
\newblock {A Short History of Complex Numbers}, Jan 2006.
\newblock
  \url{http://www.math.uri.edu/~merino/spring06/mth562/ShortHistoryComplexNumbers2006.pdf}.

\bibitem{Gauss_Quote}
John~J O'Connor and Edmund~F Robertson.
\newblock Quotations by {Carl Friedrich Gauss}, Dec 2013.
\newblock \url{http://www-history.mcs.st-andrews.ac.uk/Quotations/Gauss.html}.

\bibitem{arnold_complexanalysis}
Douglas~N Arnold.
\newblock {\em Complex Analysis}.
\newblock Originally Pennsylvania State University. Maintained by University of
  Minnesota., 2017.
\newblock These lectures notes are at a pretty high level, but they are freely
  available and show what pure mathematicians do, so I thought I'd include them
  anyway. You can access them here
  \url{http://www-users.math.umn.edu/~arnold/502.s97/complex.pdf}.

\bibitem{beck_marchesi_pixton_sabalka_2017}
Matthias Beck, Gerald Marchesi, Dennis Pixton, and Lucas Sabalka.
\newblock {\em A First Course in Complex Analysis}.
\newblock San Francisco State University, version 1.53 edition, 2017.
\newblock Accessible and downloadable for free at
  \url{http://math.sfsu.edu/beck/papers/complex.pdf}.

\bibitem{complex_analysis_wikipedia_2018}
Wikipedia.
\newblock Complex analysis, Mar 2018.
\newblock This Wikipedia page provides an overview to a lot of what's available
  in Complex Analysis and has some pretty pictures. Use the references at the
  bottom for more information.
  \url{https://en.wikipedia.org/wiki/Complex_analysis}.

\bibitem{spin_interferometry}
Apoorva~G Wagh and Veer~Chand Rakhecha.
\newblock Interfering with the neutron spin, Jul 2004.
\newblock This is a link to an example of a spin interferometry experiment with
  spin-1/2 particles.
  \url{http://www.ias.ac.in/article/fulltext/pram/063/01/0051-0056}.

\end{thebibliography}


\begin{thebibliography}{1}

\bibitem{wikipedia_Product2Sum}
Wikipedia.
\newblock List of trigonometric identities, Dec 2017.
\newblock
  \url{https://en.wikipedia.org/wiki/List_of_trigonometric_identities#Product-to-sum_and_sum-to-product_identities}.

\bibitem{MultidimensionalTransform}
Wikipedia.
\newblock Multidimensional transform, Nov 2017.
\newblock
  \url{https://en.wikipedia.org/wiki/Multidimensional_transform#Multidimensional_Fourier_transform}.

\bibitem{POMN_ConvergenceFourierSeries}
Paul Dawkins.
\newblock Convergence of fourier series, Jan 2018.
\newblock
  \url{http://tutorial.math.lamar.edu/Classes/DE/ConvergenceFourierSeries.aspx}.

\bibitem{stackSincFunction}
Ahmed~S. Attaalla.
\newblock Integration of sinc function, Jul 2016.
\newblock
  \url{https://math.stackexchange.com/questions/891812/integration-of-sinc-function}.

\bibitem{thefouriertransform}
Fourier transforms.
\newblock \url{http://www.thefouriertransform.com/}.

\bibitem{wang_moore_2009}
Ruye Wang and Ross Moore, Jul 2009.
\newblock \url{http://fourier.eng.hmc.edu/e101/lectures/handout3/node2.html}.

\bibitem{wikipedia_GibbsFunctions}
Wikipedia.
\newblock Gibbs phenomenon, Dec 2017.
\newblock \url{https://en.wikipedia.org/wiki/Gibbs_phenomenon}.

\end{thebibliography}


\begin{thebibliography}{10}

\bibitem{axial_vector}
Wikipedia.
\newblock Pseudovector, Apr 2018.
\newblock The mathematical details of this page are advanced for youngling
  physics majors, so don't worry if you don't understand it initially. Just
  know that there are different species of vectors out there that will be of
  importance for you at a more advanced stage | particularly in electrodynamics
  courses. Anyway, to read about axial vectors, go to
  \url{https://en.wikipedia.org/wiki/Pseudovector}.

\bibitem{charap_2011}
John~M. Charap.
\newblock {\em Covariant Electrodynamics: A Concise Guide}.
\newblock J. Hopkins University Press, 2011.

\bibitem{einstein_menendez}
Albert Einstein and Jos\'e Men\'endez.
\newblock {\em Relativity: The Special and General Theory}.
\newblock Elegant Ebooks. Orig. Henry Holt and Company (1920), digital reprint
  edition.
\newblock This book is in the public domain in the United States. I strongly
  recommend this book to anyone interested in physics. Einstein wrote it in a
  conceptual style rather than a mathematical one in the hopes that us mental
  mortals could see relativity as he did. It's a good book. For online (freely
  downloadable) access, go to
  \url{https://ibiblio.org/ebooks/Einstein/Einstein_Relativity.pdf}.

\bibitem{gerlach1922experimental}
W~Gerlach and O~Stern.
\newblock The experimental evidence of direction quantisation in the magnetic
  field.
\newblock {\em Zeitschrift f{\"u}r Physik}, 9:349, 1922.

\bibitem{griffiths_2014}
David~Jeffrey Griffiths.
\newblock {\em {Introduction to Elementary Particles}}.
\newblock Wiley-VCH Verlag, illustrated, reprint edition, 2014.

\bibitem{hyperphys_law_of_cosines}
R~Nave, Department of~Physics, and Astronomy.
\newblock Law of cosines, 2016.
\newblock Produced by Georgia State University.
  \url{http://hyperphysics.phy-astr.gsu.edu/hbase/lcos.html}.

\bibitem{hyperphysics_maxwells_equations}
R~Nave, Department of~Physics, and Astronomy.
\newblock Maxwell's equations, 2016.
\newblock Produced by Georgia State University.
  \url{http://hyperphysics.phy-astr.gsu.edu/hbase/electric/maxeq.html}.

\bibitem{nasa_science_dark_matter}
NASA Science.
\newblock Dark energy, dark matter.
\newblock For a brief astrophysical summary of Dark Matter and Dark Energy,
  visit
  \url{https://science.nasa.gov/astrophysics/focus-areas/what-is-dark-energy}.

\bibitem{phet_dubson_mckagan_wieman_2011}
Michael Dubson, Sam McKagan, and Carl Wieman.
\newblock {S}tern-{G}erlach experiment, Jun 2011.
\newblock To access this simulation online, go to
  \url{https://phet.colorado.edu/sims/stern-gerlach/stern-gerlach_en.html}.

\bibitem{rudin_1976}
Walter Rudin.
\newblock {\em Principles of Mathematical Analysis}.
\newblock The McGraw-Hill Companies, Inc., 3rd edition, 1976.

\bibitem{sakurai_napolitano_2011}
J.~J. Sakurai and Jim Napolitano.
\newblock {\em {Modern Quantum Mechanics}}.
\newblock Cambridge University Press, 2nd edition, 2011.
\newblock This is a harder starting point for quantum mechanics, but it is not
  impossible. It is a graduate text though, so keep that in mind.

\bibitem{stern_gerlach_exp_revisited}
Horst Schmidt-Böcking, Lothar Schmidt, Hans~Jürgen Lüdde, Wolfgang Trageser,
  Alan Templeton, and Tilman Sauer.
\newblock The {S}tern-{G}erlach experiment revisited.
\newblock {\em The European Physical Journal H}, 41(4-5):327–364, 2016.
\newblock For arXiv access, go to \url{https://arxiv.org/pdf/1609.09311.pdf}.

\bibitem{townsend_2012}
John~S. Townsend.
\newblock {\em {A Modern Approach to Quantum Mechanics}}.
\newblock University Science Books, 2nd edition, 2012.
\newblock This is an easier starting point for quantum mechanics than Sakurai.
  The first chapter reads like a novel and I highly recommend it.

\bibitem{wikipedia_functions}
Wikipedia.
\newblock Function (mathematics), Jan 2018.
\newblock \url{https://en.wikipedia.org/wiki/Function_%28mathematics%29}.

\bibitem{wikipedia_law_of_cosines}
Wikipedia.
\newblock Law of cosines, Jan 2018.
\newblock \url{https://en.wikipedia.org/wiki/Law_of_cosines}.

\bibitem{wolfram_function}
Christopher Stover and Eric~W. Weisstein.
\newblock Function, Jan 2018.
\newblock From MathWorld -- A Wolfram Web Resource.
  \url{http://mathworld.wolfram.com/Function.html}.

\end{thebibliography}


\begin{thebibliography}{10}

\bibitem{charap_2011}
John~M. Charap.
\newblock {\em Covariant Electrodynamics: A Concise Guide}.
\newblock J. Hopkins University Press, 2011.

\bibitem{hyperphysics_maxwells_equations}
R~Nave, Department of~Physics, and Astronomy.
\newblock Maxwell's equations, 2016.
\newblock Produced by Georgia State University.
  \url{http://hyperphysics.phy-astr.gsu.edu/hbase/electric/maxeq.html}.

\bibitem{wolfram_function}
Christopher Stover and Eric~W. Weisstein.
\newblock Function, Jan 2018.
\newblock From MathWorld -- A Wolfram Web Resource.
  \url{http://mathworld.wolfram.com/Function.html}.

\bibitem{rudin_1976}
Walter Rudin.
\newblock {\em Principles of Mathematical Analysis}.
\newblock The McGraw-Hill Companies, Inc., 3rd edition, 1976.

\bibitem{wikipedia_functions}
Wikipedia.
\newblock Function (mathematics), Jan 2018.
\newblock \url{https://en.wikipedia.org/wiki/Function_%28mathematics%29}.

\bibitem{einstein_menendez}
Albert Einstein and Jos\'e Men\'endez.
\newblock {\em Relativity: The Special and General Theory}.
\newblock Elegant Ebooks. Orig. Henry Holt and Company, digital reprint
  edition, 1920.
\newblock This book is in the public domain in the United States. I strongly
  recommend this book to anyone interested in physics. Einstein wrote it in a
  conceptual style rather than a mathematical one in the hopes that us mental
  mortals could see relativity as he did. It's a good book. For online (freely
  downloadable) access, go to
  \url{https://ibiblio.org/ebooks/Einstein/Einstein_Relativity.pdf}.

\bibitem{hyperphys_law_of_cosines}
R~Nave, Department of~Physics, and Astronomy.
\newblock Law of cosines, 2016.
\newblock Produced by Georgia State University.
  \url{http://hyperphysics.phy-astr.gsu.edu/hbase/lcos.html}.

\bibitem{wikipedia_law_of_cosines}
Wikipedia.
\newblock Law of cosines, Jan 2018.
\newblock \url{https://en.wikipedia.org/wiki/Law_of_cosines}.

\bibitem{gerlach1922experimental}
W~Gerlach and O~Stern.
\newblock The experimental evidence of direction quantisation in the magnetic
  field.
\newblock {\em Zeitschrift f{\"u}r Physik}, 9:349, 1922.

\bibitem{stern_gerlach_exp_revisited}
Horst Schmidt-Böcking, Lothar Schmidt, Hans~Jürgen Lüdde, Wolfgang Trageser,
  Alan Templeton, and Tilman Sauer.
\newblock The {S}tern-{G}erlach experiment revisited.
\newblock {\em The European Physical Journal H}, 41(4-5):327–364, 2016.
\newblock For arXiv access, go to \url{https://arxiv.org/pdf/1609.09311.pdf}.

\bibitem{phet_dubson_mckagan_wieman_2011}
Michael Dubson, Sam McKagan, and Carl Wieman.
\newblock {S}tern-{G}erlach experiment, Jun 2011.
\newblock To access this simulation online, go to
  \url{https://phet.colorado.edu/sims/stern-gerlach/stern-gerlach_en.html}.

\bibitem{townsend_2012}
John~S. Townsend.
\newblock {\em {A Modern Approach to Quantum Mechanics}}.
\newblock University Science Books, 2nd edition, 2012.
\newblock This is an easier starting point for quantum mechanics than Sakurai.
  The first chapter reads like a novel and I highly recommend it.

\bibitem{sakurai_napolitano_2011}
J.~J. Sakurai and Jim Napolitano.
\newblock {\em {Modern Quantum Mechanics}}.
\newblock Cambridge University Press, 2nd edition, 2011.
\newblock This is a harder starting point for quantum mechanics, but it is not
  impossible. It is a graduate text though, so keep that in mind.

\bibitem{axial_vector}
Wikipedia.
\newblock Pseudovector, Apr 2018.
\newblock The mathematical details of this page are advanced for youngling
  physics majors, so don't worry if you don't understand it initially. Just
  know that there are different species of vectors out there that will be of
  importance for you at a more advanced stage | particularly in electrodynamics
  courses. Anyway, to read about axial vectors, go to
  \url{https://en.wikipedia.org/wiki/Pseudovector}.

\bibitem{griffiths_2014}
David~Jeffrey Griffiths.
\newblock {\em {Introduction to Elementary Particles}}.
\newblock Wiley-VCH Verlag, illustrated, reprint edition, 2014.

\bibitem{nasa_science_dark_matter}
NASA Science.
\newblock Dark energy, dark matter.
\newblock For a brief astrophysical summary of Dark Matter and Dark Energy,
  visit
  \url{https://science.nasa.gov/astrophysics/focus-areas/what-is-dark-energy}.

\bibitem{ComplexNumberHistory}
Orlando Merino.
\newblock {A Short History of Complex Numbers}, Jan 2006.
\newblock
  \url{http://www.math.uri.edu/~merino/spring06/mth562/ShortHistoryComplexNumbers2006.pdf}.

\bibitem{Gauss_Quote}
John~J O'Connor and Edmund~F Robertson.
\newblock Quotations by {Carl Friedrich Gauss}, Dec 2013.
\newblock \url{http://www-history.mcs.st-andrews.ac.uk/Quotations/Gauss.html}.

\bibitem{spin_interferometry}
Apoorva~G Wagh and Veer~Chand Rakhecha.
\newblock Interfering with the neutron spin, Jul 2004.
\newblock This is a link to an example of a spin interferometry experiment with
  spin-1/2 particles.
  \url{http://www.ias.ac.in/article/fulltext/pram/063/01/0051-0056}.

\bibitem{complex_analysis_wikipedia_2018}
Wikipedia.
\newblock Complex analysis, Mar 2018.
\newblock This Wikipedia page provides an overview to a lot of what's available
  in Complex Analysis and has some pretty pictures. Use the references at the
  bottom for more information.
  \url{https://en.wikipedia.org/wiki/Complex_analysis}.

\bibitem{beck_marchesi_pixton_sabalka_2017}
Matthias Beck, Gerald Marchesi, Dennis Pixton, and Lucas Sabalka.
\newblock {\em A First Course in Complex Analysis}.
\newblock San Francisco State University, version 1.53 edition, 2017.
\newblock Accessible and downloadable for free at
  \url{http://math.sfsu.edu/beck/papers/complex.pdf}.

\bibitem{arnold_complexanalysis}
Douglas~N Arnold.
\newblock {\em Complex Analysis}.
\newblock Originally Pennsylvania State University. Maintained by University of
  Minnesota., 2017.
\newblock These lectures notes are at a pretty high level, but they are freely
  available and show what pure mathematicians do, so I thought I'd include them
  anyway. You can access them here
  \url{http://www-users.math.umn.edu/~arnold/502.s97/complex.pdf}.

\bibitem{thompson_systems_technology}
Peter~M Thompson and Systems Technology.
\newblock Snap, crackle, and pop, April 2018.
\newblock I pretty much only cited this as an official document recognizing
  Snap, Crackle, and Pop as the names of technical physical quantities. If
  you're interested in reading about them in the context of aerospace
  engineering, go to
  \url{https://info.aiaa.org/Regions/Western/Orange_County/Newsletters/Presentations\%20Posted\%20by\%20Enrique\%20P.\%20Castro/AIAAOC_SnapCracklePop_docx.pdf}.

\bibitem{deriv_notations}
Wikipedia.
\newblock Notation for differentiation, Apr 2018.
\newblock Here is a slew of derivative and integral notations.
  \url{https://en.wikipedia.org/wiki/Notation_for_differentiation}.

\bibitem{Schaums}
Murray~R. Spiegel, Seymour Lipschutz, and John Liu.
\newblock {\em Mathematical handbook of formulas and tables}.
\newblock Schaum's Outlines. McGraw-Hill Education, 5 edition, 2018.

\bibitem{pauls_derivative_table}
Paul Dawkins.
\newblock Common derivatives and integrals, 2005.
\newblock
  \url{http://tutorial.math.lamar.edu/pdf/Common_Derivatives_Integrals_Reduced.pdf}.

\bibitem{bbc_1986}
Jeremy Gray.
\newblock {The Birth of the Calculus (BBC)}, 1986.
\newblock \url{https://www.youtube.com/watch?v=ObPg3ki9GOI}.

\bibitem{wikipedia_Product2Sum}
Wikipedia.
\newblock List of trigonometric identities, Dec 2017.
\newblock
  \url{https://en.wikipedia.org/wiki/List_of_trigonometric_identities#Product-to-sum_and_sum-to-product_identities}.

\bibitem{POMN_ConvergenceFourierSeries}
Paul Dawkins.
\newblock Convergence of fourier series, Jan 2018.
\newblock
  \url{http://tutorial.math.lamar.edu/Classes/DE/ConvergenceFourierSeries.aspx}.

\bibitem{wikipedia_GibbsFunctions}
Wikipedia.
\newblock Gibbs phenomenon, Dec 2017.
\newblock \url{https://en.wikipedia.org/wiki/Gibbs_phenomenon}.

\bibitem{stackSincFunction}
Ahmed~S. Attaalla.
\newblock Integration of sinc function, Jul 2016.
\newblock
  \url{https://math.stackexchange.com/questions/891812/integration-of-sinc-function}.

\bibitem{wang_moore_2009}
Ruye Wang and Ross Moore, Jul 2009.
\newblock \url{http://fourier.eng.hmc.edu/e101/lectures/handout3/node2.html}.

\bibitem{thefouriertransform}
Fourier transforms.
\newblock \url{http://www.thefouriertransform.com/}.

\bibitem{MultidimensionalTransform}
Wikipedia.
\newblock Multidimensional transform, Nov 2017.
\newblock
  \url{https://en.wikipedia.org/wiki/Multidimensional_transform#Multidimensional_Fourier_transform}.

\end{thebibliography}

\newpage\phantomsection
\renewcommand{\bibname}{References}
\addcontentsline{toc}{chapter}{\bibname}
\bibliographystyle{unsrt}

\end{document}